\pdfoutput=1
\documentclass[a4paper,12pt]{article}
\usepackage{amsfonts}
\usepackage{mathrsfs}
\usepackage{amsmath}
\usepackage{amssymb}
\usepackage{framed}
\usepackage[medium]{titlesec}
\usepackage{bm}
\usepackage{cite}
\usepackage[normalem]{ulem}
\usepackage{extarrows}
\usepackage{slashed}
\usepackage{isodateo}
\usepackage{graphicx}
\usepackage{xcolor}
\usepackage[bookmarksnumbered=true,bookmarksopen=true]{hyperref}
 \hypersetup{colorlinks,%
             linkcolor=[rgb]{0,0.3,0.6}, %
             citecolor=[rgb]{0,0.3,0.6}, %
             urlcolor=[rgb]{0,0.3,0.6}}
\usepackage[hmargin=.7in,vmargin=1.1in]{geometry}
\usepackage{indentfirst}
\usepackage{booktabs}

\newcommand{\FR}[2]{\displaystyle\frac{\,{#1}\,}{#2}}
\newcommand{\fr}[2]{\mbox{$\frac{\,{#1}\,}{#2}$}}
\newcommand{\n}{\nonumber}

\graphicspath{{fig/}}

\def\bge{\begin{equation}}
\def\ede{\end{equation}}
\def\bga{\begin{aligned}}
\def\eda{\end{aligned}}
\def\bgb{\begin{bmatrix}}
\def\edb{\end{bmatrix}}
\def\bgp{\begin{pmatrix}}
\def\edp{\end{pmatrix}}
\def\bgm{\begin{matrix}}
\def\edm{\end{matrix}}
\def\bgs{\begin{subequations}}
\def\eds{\end{subequations}}

\def\di{{\mathrm{d}}}

\def\mb{\mathbf}

\def\to{\rightarrow}

\def\ii{\mathrm{i}}

\def\aa{\mathsf{a}}
\def\bb{\mathsf{b}}

\usepackage{mdframed}

\newmdenv[skipabove=0pt,%
          skipbelow=5pt,%
          leftmargin=0pt,%
          rightmargin=0pt,%
          innertopmargin=-5pt,%
          innerbottommargin=7pt,%
          innerleftmargin=2pt,%
          innerrightmargin=2pt,%
          splittopskip=0pt,%
          splitbottomskip=0pt,%
          linewidth=0pt,%
          nobreak=true]%
          {keyeqn2}

\newmdenv[backgroundcolor=gray!15,%
          skipabove=0pt,%
          skipbelow=5pt,%
          leftmargin=0pt,%
          rightmargin=0pt,%
          innertopmargin=-5pt,%
          innerbottommargin=7pt,%
          innerleftmargin=2pt,%
          innerrightmargin=2pt,%
          splittopskip=0pt,%
          splitbottomskip=0pt,%
          linewidth=0pt,%
          nobreak=true]%
          {keyeqn}

\usepackage{titlesec}          
\titleformat{\section}
{\normalfont\fontsize{15}{20}\bfseries}{\thesection}{1em}{}

\newcommand{\wt}[1]{\mkern 2mu \widetilde{\mkern -2mu #1 \mkern -2mu}\mkern 2mu}
\newcommand{\wh}[1]{\mkern 2mu \widehat{\mkern-2mu#1\mkern-2mu}\mkern 2mu}

\newcommand{\fnemail}[1]{\footnote{Email: \href{mailto:#1}{\nolinkurl{#1}}}}

\allowdisplaybreaks

\newcommand{\ud}{\mathrm{d}}

\newcommand{\pdd}[2]{\frac{\partial{#1}}{\partial{#2}}}

\begin{document}
\title{\Large\textbf{Dimensional Regularization of Bubble Diagrams in de Sitter Spacetime}\\[2mm]}

\author{Hongyu Zhang$^{a\,}$\fnemail{zhy21@mails.tsinghua.edu.cn}\\[5mm]
$^a\,$\normalsize{\emph{Department of Physics, Tsinghua University, Beijing 100084, China} }}

\date{}
\maketitle

\vspace{20mm}
\begin{abstract}
    Correlators of large-scale fluctuations produced during cosmic inflation are major observables of inflationary cosmology. In cosmological collider physics, many interesting correlators are generated through loop processes. However, ultraviolet divergences often appear when computing the loop correlators, and a regularization is required. In this work, K\"all\'en-Lehmann representation and dimensional regularization are used to analytically compute various correlators with a bubble loop of massive bulk propagators in de Sitter spacetime. Examples include 4-point and 2-point correlators with 1-loop bubbe exchanges of derivatively coupled massive scalars or massive spin-1 bosons.
\end{abstract}
\newpage
\tableofcontents
\newpage

\section{Introduction}
It is widely believed that the universe has experienced a nearly exponential expansion in an extremely early epoch, known as inflation. The Hubble parameter $H$ is nearly a constant in the inflationary spacetime. The spacetime of inflation is close to the Poincar\'e patch of $3+1$ dimensional de Sitter spacetime $\text{dS}_4$, which is one of the maximally symmetric spacetimes. The Hubble parameter is the inverse of the curvature radius of dS.

The equal-time correlation functions are important quantities during inflation. These functions can be computed theoretically. The study of processes of quantum field theory in the inflationary spacetime has gained considerable interest. On the other side, quantum processes during cosmic inflation generate large-scale inhomogeneities of the universe. The inhomogeneities can be studied from the measurements including the temperature and the polarizations of the cosmological microwave background \cite{Planck:2018jri,Planck:2019kim}, the large-scale structure survey \cite{Ferraro:2022cmj}, and the more futuristic 21cm tomography \cite{Munoz:2015eqa,Liu:2022iyy}. From the measurements, the correlation functions can be extracted. Therefore, the correlation functions serve as fundamental links between quantum field theory in inflation and observational data. In this work, the correlation functions are called inflation correlators. There have been many works in the study of the inflation correlators \cite{Maldacena:2011nz,Assassi:2012zq,Arkani-Hamed:2017fdk,Baumann:2017jvh,Arkani-Hamed:2018bjr,Sleight:2021iix,Hillman:2019wgh,Pajer:2020wxk,Bonifacio:2021azc,Cabass:2021fnw,Premkumar:2021mlz,Heckelbacher:2022hbq,Cabass:2022jda,Xianyu:2022jwk,Bonifacio:2022vwa,Lee:2022fgr,Lee:2023jby,Loparco:2023rug,De:2023xue,Xianyu:2023ytd,Green:2023ids,Arkani-Hamed:2023bsv,Arkani-Hamed:2023kig,Benincasa:2024leu,Benincasa:2024lxe,Donath:2024utn,Du:2024hol,Fan:2024iek,Grimm:2024mbw,Melville:2024ove,Stefanyszyn:2024msm,Ema:2024hkj,Cohen:2024anu,He:2024olr,Baumann:2024ttn,Goodhew:2024eup,Green:2024fsz,Hang:2024xas,Baumann:2024mvm,Lee:2024sks,Werth:2024mjg,Chen:2024glu,Gasparotto:2024bku,De:2024zic,Qin:2024gtr,Huenupi:2024ztu,Green:2024cmx,Launay:2024trh,Cespedes:2025dnq,Westerdijk:2025ywh}. There are processes from the analytical structures of the correlators \cite{Goodhew:2020hob,Goodhew:2021oqg,Melville:2021lst,Meltzer:2021zin,DiPietro:2021sjt,Tong:2021wai,Salcedo:2022aal,AguiSalcedo:2023nds,Jazayeri:2021fvk}, parity violations \cite{Liu:2019fag,Cabass:2022rhr,Cabass:2022oap,Stefanyszyn:2023qov,Niu:2022fki,Jazayeri:2023kji,Thavanesan:2025kyc}, and numerical methods \cite{Wang:2021qez,Werth:2023pfl,Pinol:2023oux,Werth:2024aui}.

The correlators provide a new way to study the unknown heavy particles and their interactions at the inflation scale \cite{Chen:2009we,Chen:2009zp,Baumann:2011nk,Chen:2012ge,Pi:2012gf,Noumi:2012vr,Gong:2013sma,Arkani-Hamed:2015bza}. Correlators mediated by massive particles leave logarithmic oscillatory shapes in momentum ratios. The logarithmic oscillatory shapes are named as ``cosmological collider signals" \cite{Arkani-Hamed:2015bza}. Various studies focus on CC signals in recent years \cite{Chen:2015lza,Chen:2016nrs,Chen:2016uwp,Chen:2016hrz,An:2017hlx,An:2017rwo,MoradinezhadDizgah:2017szk,Iyer:2017qzw,Kumar:2017ecc,Chen:2017ryl,Tong:2018tqf,Chen:2018sce,Chen:2018xck,Chen:2018cgg,Chua:2018dqh,Wu:2018lmx,Saito:2018omt,Li:2019ves,Lu:2019tjj,Hook:2019zxa,Hook:2019vcn,Kumar:2018jxz,Kumar:2019ebj,Wang:2019gok,Li:2020xwr,Wang:2020ioa,Fan:2020xgh,Aoki:2020zbj,Maru:2021ezc,Lu:2021gso,Lu:2021wxu,Pinol:2021aun,Cui:2021iie,Reece:2022soh,Chen:2022vzh,Maru:2022bhr,Niu:2022quw,Aoki:2023tjm,Aoki:2023wdc,Aoki:2024uyi,Tong:2023krn,Jazayeri:2023xcj,Yin:2023jlv,Ema:2023dxm,Chakraborty:2023qbp,Chakraborty:2023eoq,Craig:2024qgy,McCulloch:2024hiz,Cabass:2024wob,Wu:2024wti,Aoki:2024jha,Bodas:2024hih}. These studies suggest that many properties of heavy particles can be analyzed from the signals, including mass \cite{Lee:2016vti,Meerburg:2016zdz}, spin \cite{Alexander:2019vtb,Wang:2020uic,Kogai:2020vzz,Tong:2022cdz}, chemical potential \cite{Wang:2019gbi,Bodas:2020yho,Sou:2021juh,Chen:2023txq}, etc.

There are also extensive studies on analytical computations of part and full results of the cosmological correlators. Some methods are used to derive the expressions, such as Mellin space representation \cite{Sleight:2019hfp,Sleight:2019mgd,Sleight:2020obc,Sleight:2021plv}, partial Mellin-Barnes representation \cite{Qin:2022lva,Qin:2022fbv,Qin:2023ejc,Qin:2023bjk,Qin:2023nhv}, bootstrap equations \cite{Arkani-Hamed:2018kmz,Baumann:2019oyu,Baumann:2020dch,Pajer:2020wnj,Hillman:2021bnk,Baumann:2021fxj,Hogervorst:2021uvp,Pimentel:2022fsc,Jazayeri:2022kjy,Wang:2022eop,Baumann:2022jpr,Gomez:2021qfd,Gomez:2021ujt,Chen:2023iix,DuasoPueyo:2023kyh,Liu:2024str}, and dispersion relations \cite{Liu:2024xyi}. These techniques make it possible to get the expressions of many tree-level correlators.

The study of correlators with one loop is useful. In many particle models, the leading CC signal appears at the 1-loop level rather than the tree level. In these models, massive particles are created and annihilated in pairs. The tree-level process of these models is simply absent. These include models with most particles of the Standard Model in the symmetric phase \cite{Chen:2016uwp,Chen:2016hrz,Chen:2018xck,Lu:2019tjj,Hook:2019zxa,Hook:2019vcn}, new physical particles such as heavy neutrinos \cite{Cui:2021iie} and Kaluza-Klein \cite{Kumar:2018jxz} gravitons. There are also some models \cite{Wang:2019gbi,Wang:2020ioa,Tong:2022cdz} in which the signals of the 1-loop level are more enhanced than those at the tree level (correlators with more loops remain subdominant so that perturbation theory still works).

The calculation of 1-loop correlators is challenging, due to the time ordering in the computation, the complicated mode functions and the loop momentum integral. The full results of 1-loop correlators are fewer than those of correlators at tree level. 
\cite{Marolf:2010zp} provides the 1-loop bubble correction to the 2-point function from scalar fields of arbitrary mass in Euclidean dS. The 4-point function with 1-loop exchange of conformal scalar in the position space is calculated in \cite{Heckelbacher:2022hbq}. Correlators of 4-point and 3-point with one bubble loop of scalar are calculated in \cite{Xianyu:2022jwk} by using spectral decomposition. The result of the correlators with one loop can also be derived by using dispersion \cite{Liu:2024xyi}.

The UV divergence is expected to appear in the 1-loop correlators. Therefore, regularization scheme should be applied to the correlators. There are several works on regularization in de Sitter spacetime, such as \cite{Weinberg:2005vy,Senatore:2009cf,Goswami:2013tya,Chen:2016nrs,Tan:2019czo,Lee:2023jby,Braglia:2025qrb,Kristiano:2025ajj}. These works mainly focus on correlators with internal lines of massless scalars. \cite{Braglia:2025cee} deals with loops from a conformal ($m^2=2H^2$) scalar field and provides a full calculation including renormalization. \cite{Huenupi:2024ksc} studies dimensional regularization for daisy loops. It is necessary to find out a way to regularize correlators with internal lines of arbitrary mass.

There are several regularization schemes in flat spacetime, such as cutoff regularization, dimensional regularization and Pauli-Villars regularization. In this work, the correlators with a bubble loop are computed by using K\"all\'en-Lehmann representation. The integral in the representation is hard to calculate if there is a cutoff in the integration. If one tries to finish the Pauli-Villars regularization, it is difficult to take the asymptotic expansions of the cutoff $\Lambda$ and subtract the divergences for $\Lambda\rightarrow\infty$ from the correlators. Therefore, dimensional regularization is used in de Sitter spacetime in this work.

Dimensional regularization, introduced by 't Hooft and Veltman in 1972 \cite{tHooft:1972tcz}, is a widely used method for regulating divergent integrals. The advantage of dimensional regularization is that gauge invariance is maintained. In this regularization, the dimension of the spacetime is continued from 4 to an arbitrary number $D$. After the expressions of correlation functions and $S$-matrix elements are calculated in $D$ dimensions, the limit $D\rightarrow4$ is taken. Then the divergences that occur when $D\rightarrow4$ are canceled by MS subtraction or $\overline{\text{MS}}$ subtraction.

The difficulty is how to compute the divergence of $D\rightarrow4$. The expressions of the 1-loop correlators are usually composed of divergent series. Each term of the series converges when $D\rightarrow4$. Therefore, to take the divergence of the series, the properties of the series need to be clear, rather than each term.

The dimensional regularization of 1-loop 4-point and 3-point correlators with directly coupled massive scalars is finished in \cite{Xianyu:2022jwk}. In this work, the dimensional regularization of 1-loop 4-point correlators with directly coupled massive scalars is done in another way. The expressions of the divergent series are regularized with the help of Riemann zeta functions. The asymptotic expansions of each term of the series regarding the summation index are used to subtract the divergences of the series. The sums of asymptotic expansions can be summed into sums of several Riemann zeta functions. After taking the series expansions of the Riemann zeta function with respect to $(D-4)$, the divergences of the correlators can be found. The method is also applied to get the analytical expressions of 1-loop 2-point correlators, 1-loop 4-point correlators with derivatively coupled massive scalars, and 1-loop correlators with a massive vector loop.

\paragraph{Outline of this work} The rest of the paper is organized as follows. In Sec. \ref{Sec2}, the loop seed integral defined in \cite{Xianyu:2022jwk} and the spectral decomposition of the seed integral are reviewed. The expression of the loop seed integral is computed with some improvement to obtain a simpler result. Then the divergence that appears when the spatial dimension $d\rightarrow3$ is studied to make it possible to take dimensional regularization. The MS subtraction is applied to the seed integral after dimensional regularization.

In Sec. \ref{Sec3}, dimensional regularization is used to take the expressions of more difficult cases, such as 1-loop 2-point correlator, 1-loop correlator with derivative coupling, and 1-loop correlator with massive vector loop. The expressions of these models in flat spacetime are also calculated, and the coefficients of the divergences are compared with those in de Sitter spacetime. For loops composed of time-ordered propagators, note that the K\"all\'en-Lehmann representation cannot be applied directly to either the derivative coupling loop or the massive vector loop. The conclusions and outlooks are given in Sec. \ref{Sec4}.

There are several appendices in this work. App. \ref{AppA} collects some useful functions that are used in this work. The pole structure of the tree seed integral is studied in App. \ref{AppB}. This study can simplify the calculation of the spectral integral in this work. App. \ref{AppC} provides the details of finding the expression of the 2-point correlator at tree level. The expression of the 2-point correlator with one loop is obtained by integrating the spectral integral of the correlator at tree level. The flat spacetime limit of two models introduced in Sec. \ref{Sec3} is computed in App. \ref{AppD}. The coefficient of the divergences for $D\rightarrow4(d\rightarrow3)$ are taken to compare with those obtained in de Sitter spacetime. It is expected that the coefficients of the divergences should agree with those in flat spacetime if $H\rightarrow0$. In App. \ref{AppE}, the spectral functions of K\"all\'en-Lehmann representation of the two models in this work are calculated. Embedding formalism is introduced in this appendix to finish the calculation. App. \ref{AppF} collects some expressions which are omitted in the main text. \cite{Marolf:2010zp} and \cite{Loparco:2023rug} give two different kinds of spectral functions to calculate the correlators with one bubble loop. App. \ref{AppG} studies the relation between the two kinds of spectral functions. App. \ref{AppH} proves the convergence of the series of 1-loop 2-point correlators.

\paragraph{Notations and conventions} In this work, the metric is $\ud s^2=a^2(\tau)(-\ud\tau^2+\ud\mathbf{y}^2)$ where $\tau\in(-\infty,0)$ is the conformal time, $\mathbf{y}\in\mathbb{R}^d$ is the conformal coordinate. In the Poincar\'e patch of the dS spacetime, the scale factor $a(\tau)=-1/(H\tau)$. In this work, the Hubble parameter $H$ is set to 1.

The external spatial momenta of a 4-point correlator are labeled as $\mathbf{k}_i\ (i=1,2,3)$. Their magnitudes are $k_i\equiv|\mathbf{k}_i|$. The $s$-channel momentum is defined by $\mathbf{k}_s\equiv\mathbf{k}_1+\mathbf{k}_2$. The shorthand notations for sums of several indexed quantities are used in this work. For example,
\begin{align}
    &k_{ij}\equiv k_i+k_j,&& n_{12}\equiv n_1+n_2,&& p_{12}\equiv p_1+p_2.
\end{align}
The shorthand $\bar{p}_{12}\equiv p_1-p_2$ is also used. The momentum ratios are $r_1\equiv k_s/k_{12}, r_2\equiv k_s/k_{34}$.
\section{Dimensional Regularization of the Basic Bubble Diagram}\label{Sec2}
\begin{figure}[!htp]
    \centering
    \includegraphics{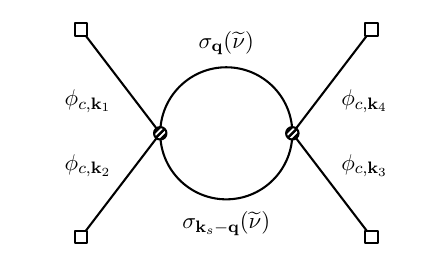}
    \caption{Feynman diagram of 1-loop 4-point correlator with a scalar loop.}
    \label{Fig1}
\end{figure}

Consider the 1-loop correlator shown in Fig. \ref{Fig1}.
The diagrammatic notations are introduced in \cite{Chen:2017ryl}. The external legs are bulk-to-boundary propagators of a conformal scalar field $\phi_c$, i.e., a scalar field with mass $m_c^2=(d^2-1)/4$. The two internal lines represent the bulk propagators of a real scalar field $\sigma$ of mass $m$. It is convenient to introduce the mass parameter $\wt{\nu}=\sqrt{m^2-d^2/4}$. Suppose the interaction is
\begin{equation}
    \Delta\mathscr{L}=-\frac14a^{d+1}\phi_c^2\sigma^2.
\end{equation}
The model with the above interaction is called $\sigma\sigma$ model in this work. The bulk-to-boundary propagator $C_{\mathsf{a}}(k;\tau)$ of the conformal scalar field reads:
\begin{equation}
    C_{\mathsf{a}}(k;\tau)=\frac{(\tau\tau_f)^{(d-1)/2}}{2k}e^{\ii\mathsf{a}k\tau}.
\end{equation}
Here $\tau_f$ is a final-time cutoff, and $\aa=\pm$ is the Schwinger-Keldysh (SK) index of the bulk point. The expression of the bulk-propagator $D_{\wt\nu,\aa\bb}(k;\tau_1,\tau_2)$ is
\begin{align}
    D_{\wt\nu,\pm\pm}(k;\tau_1,\tau_2)
    =&~D_{\wt\nu,\gtrless}(k;\tau_1,\tau_2)\theta(\tau_1-\tau_2)+D_{\wt\nu,\lessgtr}(k;\tau_1,\tau_2)\theta(\tau_2-\tau_1),\\
    D_{\wt\nu,\pm\mp}(k;\tau_1,\tau_2)
    =&~D_{\wt\nu,\lessgtr}(k;\tau_1,\tau_2).
\end{align}
Here two ``homogeneous'' propagators are introduced,
\begin{align}
    D_{\wt\nu,>}(k;\tau_1,\tau_2)=\FR{\pi}{4}e^{-\pi(\wt\nu+\wt{\nu}^*)/2}(\tau_1\tau_2)^{d/2}\text{H}_{\ii\wt\nu}^{(1)}(-k\tau_1)\text{H}_{-\ii\wt\nu^*}^{(2)}(-k\tau_2),
\end{align}
and, $D_{\wt\nu,<}(k;\tau_1,\tau_2)=D_{\wt\nu,>}^*(k;\tau_1,\tau_2)$.

With the expressions of bulk-to-boundary and bulk propagators given above, the expression of $s$-channel of the 1-loop 4-point function is built with the SK formalism \cite{Weinberg:2005vy,Chen:2017ryl}. It can be written as:
\begin{equation}
    \langle\phi_{c,\mathbf{k}_1}\phi_{c,\mathbf{k}_2}\phi_{c,\mathbf{k}_3}\phi_{c,\mathbf{k}_4}\rangle_s=(2\pi)^d\delta^{(d)}(\mathbf{k}_1+\mathbf{k}_2+\mathbf{k}_3+\mathbf{k}_4)\mathcal{L}_{\phi_c,\sigma\sigma}(\mathbf{k}_1,\mathbf{k}_2,\mathbf{k}_3,\mathbf{k}_4),
\end{equation}
where the loop amplitude $\mathcal{L}_{\phi_c,\sigma\sigma}$ is
\begin{align}\label{8}
    \mathcal{L}_{\phi_c,\sigma\sigma}=&-\frac12\sum_{\mathsf{a},\mathsf{b}=\pm}\mathsf{ab}\int_{-\infty}^0\frac{\ud\tau_1}{(-\tau_1)^{d+1}}\frac{\ud\tau_2}{(-\tau_2)^{d+1}}\n \\
    &\times C_{\mathsf{a}}(k_1,\tau_1)C_{\mathsf{a}}(k_2,\tau_1)C_{\mathsf{b}}(k_3,\tau_2)C_{\mathsf{b}}(k_4,\tau_2)\mathcal{Q}_{\sigma\sigma,\mathsf{ab}}\big(k_s;\tau_1,\tau_2\big).
\end{align}
Here the loop momentum integral $\mathcal{Q}_{\sigma\sigma,\wt{\nu},\mathsf{ab}}$ is defined as:
\begin{align}
    \mathcal{Q}_{\sigma\sigma,\mathsf{ab}}\big(k_s;\tau_1,\tau_2\big)\equiv&\int\frac{\ud^d\mathbf{q}}{(2\pi)^d}D_{\wt{\nu},\mathsf{ab}}\big(q;\tau_1,\tau_2\big)D_{\wt{\nu},\mathsf{ab}}\big(|\mathbf{k}_s-\mathbf{q}|;\tau_1,\tau_2\big).
\end{align}

It is found that many massive 1-loop correlators have similar structures. They can be generated from the loop seed integral \cite{Xianyu:2022jwk}:
\begin{align}
    \mathcal{J}^{p_1p_2}_{\sigma\sigma}(r_1,r_2)\equiv -\FR{1}{2}\!\sum_{\aa,\bb=\pm}\!\!\aa\bb\, k_s^{d+2+p_{12}}\!\int_{-\infty}^0\!\di\tau_1\di\tau_2(-\tau_1)^{p_1}(-\tau_2)^{p_2}e^{\ii\aa k_{12}\tau_1+\ii \bb k_{34}\tau_2}\mathcal{Q}_{\sigma\sigma,\aa\bb}\big(k_s;\tau_1,\tau_2\big).
\end{align}
The factor $k_s^{d+2+p_{12}}$ is inserted to make the integral dimensionless. It can be seen that the loop seed integral depends on various external momenta only through two ratios $r_1\equiv k_s/k_{12}$ and $r_2\equiv k_s/k_{34}$. The seed integral can be generated by the following interaction:
\begin{equation}
    \Delta\mathscr{L}=-\frac14a^{d-1-p_1}\phi_c^2\sigma^2-\frac14a^{d-1-p_2}\phi_c^2\sigma^2.
\end{equation}

The seed integral diverges at $\tau_{1,2}\rightarrow0$ if $p_{1,2}$ are negative numbers with large absolute values. In this case, the seed integral is inferred (IR) divergent. However, it is difficult to determine the values of $p_{1,2}$ below which the seed integral diverges. In this work, $p_{1,2}$ are assumed to ensure there is no IR divergence in the seed integral.

The amplitude $\mathcal{L}_{\phi_c,\sigma\sigma}$ defined as (\ref{8}) can be expressed as
\begin{align}
    \mathcal{L}_{\phi_c,\wt\nu}=\FR{(-\tau_f)^{2(d-1)}}{16k_1k_2k_3k_4k_s^{d-2}}\mathcal{J}_{\sigma\sigma}^{-2,-2}(r_1,r_2).
\end{align}

The seed integral can be used to express other kinds of correlators. For example, consider the following interaction:
\begin{equation}
    \Delta\mathscr{L} = -\frac14a^{d-1}\varphi'^2\sigma^2.
\end{equation}
Here, $\varphi$ represents the nearly massless inflaton field, and the prime denotes conformal time derivative: $\varphi'\equiv \mathrm{d}\varphi/\mathrm{d}\tau$. The $s$-channel of the 1-loop 4-point amplitude $\mathcal{L}_{\varphi,\sigma\sigma}$ of the inflaton is
\begin{align}
    \mathcal{L}_{\varphi,\sigma\sigma}=&-\frac12\sum_{\mathsf{a,b}=\pm}\mathsf{ab}\int_{-\infty}^{\tau_f}\frac{\ud\tau_1}{(-\tau_1)^{d-1}}\frac{\ud\tau_2}{(-\tau_2)^{d-1}}\n \\
    &\times\partial_{\tau_1}G_{\mathsf{a}}(k_1,\tau_1)\partial_{\tau_1}G_{\mathsf{a}}(k_2,\tau_1)\partial_{\tau_2}G_{\mathsf{b}}(k_3,\tau_2)\partial_{\tau_2}G_{\mathsf{b}}(k_4,\tau_2)\mathcal{Q}_{\sigma\sigma,\mathsf{ab}}\big(k_s;\tau_1,\tau_2\big).
\end{align}
The bulk-to-boundary propagator $G_{\mathsf{a}}(k,\tau)$ of a massless scalar field in general $d$ spatial dimensions is complicated. Specialized to $d=3$, the bulk-to-boundary propagator reads:
\begin{equation}
    G_{\mathsf{a}}(k,\tau)=\frac1{2k^3}(1-\ii\mathsf{a}k\tau)e^{\ii\mathsf{a}k\tau}.
\end{equation}
In $d=3$, the inflaton correlators can be expressed in terms of the loop seed integral:
\begin{equation}
    \mathcal{L}_{\varphi,\sigma\sigma}=\frac1{16k_1k_2k_3k_4k_s^5}\Big[\mathcal{J}_{\wt\nu}^{00}(r_1,r_2)\Big]_{\text{MS}} .~~~~~(d=3)
\end{equation}
Here the notation $[\cdots]_{\text{MS}}$ means that the divergence of the loop seed integral at $d=3$ is subtracted by the $\text{MS}$ scheme. See \cite{Xianyu:2022jwk} for more details.

In the rest of this section, the expression of the loop seed integral is listed and the divergence of $d\rightarrow3$ is studied.
\subsection{Expression of the Seed Integral}\label{Sec2.1}
The loop in the seed integral can be written in terms of the K\"all\'en-Lehmann integration of the propagator of a free scalar field \cite{Loparco:2023rug}. Therefore, the correlator with a bubble can be derived from the integration of the correlator of tree level introduced in \cite{Qin:2022fbv}:
\begin{equation}\label{13}
    \mathcal{J}_{\sigma\sigma}^{p_1p_2}(r_1,r_2)=\frac12\int_{-\infty}^\infty\ud\wt{\nu}'\rho_{\sigma\sigma}^{\mathcal{P},0}(\wt{\nu}')\mathcal{I}_{\wt{\nu}'}^{p_1p_2}(r_1,r_2).
\end{equation}

The spectral function $\rho_{\sigma\sigma}^{\mathcal{P},0}(\wt{\nu}')$ of K\"all\'en-Lehmann representation is \cite{Loparco:2023rug}
\begin{equation}\label{20}
    \rho_{\sigma\sigma}^{\mathcal{P},0}(\wt{\nu}')=\frac{\Gamma\big(\frac{d+1}2\big)\wt{\nu}'\sinh(\pi\wt{\nu}')}{2^{6-d}\pi^{\frac{d+7}2}\Gamma(d)\Gamma\big(\frac d2+\ii\wt{\nu}'\big)\Gamma\big(\frac d2-\ii\wt{\nu}'\big)}\prod_{\pm,\pm,\pm}\Gamma\bigg(\frac{\frac d2\pm\ii\wt{\nu}\pm\ii\wt{\nu}\pm\ii\wt{\nu}'}2\bigg).
\end{equation}
The superscript $\mathcal{P}$ stands for the spectral function corresponding to principal series, and the $0$ denotes that the bulk propagator in the spectral integration is that of the scalar field. The expression above is quite different from that labeled as $\rho_{\wt{\nu}}^\text{dS}(\wt{\nu}')$ in \cite{Xianyu:2022jwk}. It can be shown that the relation between these two kinds of spectral functions is (App. \ref{AppG})
\begin{equation}
    \rho_{\sigma\sigma}^{\mathcal{P},0}(\wt{\nu}')=\frac{\wt{\nu}'}{2\pi\ii}\big(\rho_{\wt{\nu}}^\text{dS}(\wt{\nu}')-\rho_{\wt{\nu}}^\text{dS}(-\wt{\nu}')\big).
\end{equation}

It is convenient to break the tree seed integral into three pieces according to their analytic properties at $r_{1,2}=0$:
\begin{align}
    \mathcal{I}^{p_1p_2}_{\wt\nu}(r_1,r_2)=\mathcal{I}_{\text{NL},\wt\nu}^{p_1p_2}(r_1,r_2)+\mathcal{I}_{\text{L},\wt\nu}^{p_1p_2}(r_1,r_2)+\mathcal{I}_{\text{BG},\wt\nu}^{p_1p_2}(r_1,r_2).
\end{align}
The three terms are the nonlocal-signal piece (NL), the local signal piece (L), and the background piece (BG). The expressions for the three pieces when $r_1<r_2$ are given below \cite{Qin:2022fbv,Xianyu:2022jwk}:
\begin{align}
    \label{15}
    \mathcal{I}_{\text{NL},\wt\nu}^{p_1p_2}(r_1,r_2)=&\mathcal{C}_{\ii\wt\nu,d}^{p_1p_2}\mathbf{F}_{\ii\wt\nu,d}^{p_1}(r_1)\mathbf{F}_{\ii\wt\nu,d}^{p_2}(r_2)(r_1r_2)^{+\ii \wt{\nu}}+(\wt{\nu}\rightarrow(-\wt{\nu})), \\
    \label{16}
    \mathcal{I}_{\text{L},\wt\nu}^{p_1p_2}(r_1,r_2)=&-\mathcal{C}_{\ii\wt\nu,d}^{p_1p_2}\mathbf{F}_{\ii\wt\nu,d}^{p_1}(r_1)\mathbf{F}_{-\ii\wt\nu,d}^{p_2}(r_2)\Big(\frac{r_1}{r_2}\Big)^{+\ii \wt{\nu}}+(\wt{\nu}\rightarrow(-\wt{\nu})), \\
    \label{17}
    \mathcal{I}_{\text{BG},\wt\nu}^{p_1p_2}(r_1,r_2)=&\sum_{\ell,j=0}^\infty\frac{(-1)^{\ell+1}\sin[\frac{\pi}{2}(p_{12}+d)](\ell+1)_{2j+d+p_{12}+1}}{2^{2j+1}\big(\tfrac{\ell-\ii \wt{\nu}+p_2+1}2+\tfrac d4\big)_{j+1}\big(\tfrac{\ell+\ii \wt{\nu}+p_2+1}2+\tfrac d4\big)_{j+1}}r_1^{2j+d+p_{12}+2}\Big(\frac{r_1}{r_2}\Big)^{\ell}\n \\
    =&-\sum_{\ell,j=0}^\infty\frac{(-1)^{\ell+1}\mathcal{C}_{\ii\wt{\nu},d}^{p_1p_2}\sin(\pi\ii\wt{\nu})(\ell+1)_{2j+d+p_{12}+1}}{2^{2j-1}\big(\tfrac{\ell-\ii \wt{\nu}+p_2+1}2+\tfrac d4\big)_{j+1}\big(\tfrac{\ell+\ii \wt{\nu}+p_2+1}2+\tfrac d4\big)_{j+1}}r_1^{2j+d+p_{12}+2}\Big(\frac{r_1}{r_2}\Big)^{\ell}+(\wt{\nu}\rightarrow(-\wt{\nu})),
\end{align}
where the coefficient $\mathcal{C}_{\ii\wt\nu,d}^{p_1p_2}$ and the function $\mb{F}_{\ii\wt\nu,d}^{p}(r)$ are defined by:
\begin{align}
    \mathcal{C}_{\ii\wt\nu,d}^{p_1p_2}\equiv&~\FR{1}{8}\csc^2(\pi\ii\wt\nu)\Big\{\cos \FR{\pi \bar p_{12}}2 +\cos\Big[\pi\Big(\ii \wt{\nu}+\FR{p_{12}+d}2\Big)\Big]\Big\},\\
    \mb{F}_{\ii\wt\nu,d}^{p}(r)\equiv &~(2r)^{p+d/2+1}\times{}_2\mathcal{F}_1\left[ \begin{matrix}
        \tfrac d4+\tfrac12+\tfrac{p}2+\tfrac{\ii \wt{\nu}}2, \tfrac d4+1+\tfrac{p}2+\tfrac{\ii \wt{\nu}}2 \\
        1+\ii \wt{\nu}
    \end{matrix}\middle|r^2\right].
\end{align}
The shorthands $p_{12}\equiv p_1+p_2$ and $\bar{p}_{12}\equiv p_1-p_2$ are introduced. The function ${}_2\mathcal{F}_1$ is the dressed hypergeometric function defined in (\ref{162}).

For the case $0<r_1<r_2<1$, it can be derived that $0<r_1r_2<1$ and $0<r_1/r_2<1$. Therefore, the integral contour in (\ref{13}) should be closed with a large semi-circle in the lower-half $\wt{\nu}'$-plane for the term proportional to $(r_1r_2)^{+\ii\wt{\nu}'}$ in $\mathcal{I}_{\text{NL},\wt\nu}^{p_1p_2}$ and the term proportional to $(r_1/r_2)^{+\ii\wt{\nu}'}$ in $\mathcal{I}_{\text{L},\wt\nu}^{p_1p_2}$. For the $(\wt{\nu}\rightarrow(-\wt{\nu}))$ terms of the terms mentioned before, the integral contour should be closed with a large semi-circle in the upper-half $\wt{\nu}'$-plane. For the terms of the background $\mathcal{I}_{\text{BG},\wt\nu}^{p_1p_2}$, the integral contour can be closed with a semi-circle in an arbitrary half $\wt{\nu}'$-plane. However, it is better to close the contour in the lower plane for the term shown in (\ref{17}) directly while in the upper plane for the $(\wt{\nu}\rightarrow(-\wt{\nu}))$ term. The reason is that there are no poles of the integrand from the sum of $(r_1r_2)^{+\ii\wt{\nu}'}$ term in $\mathcal{I}_{\text{NL},\wt\nu}^{p_1p_2}$, $(r_1/r_2)^{+\ii\wt{\nu}'}$ term in $\mathcal{I}_{\text{L},\wt\nu}^{p_1p_2}$ and the term shown in (\ref{17}) directly in the lower $\wt{\nu}'$-plane. This is right for the other three terms. The poles of the integrand of (\ref{13}) can be classified into three sets\footnote{There are some other poles of the signals and the background separately. However, the poles are canceled if take the sum of the signals and the background. See App. \ref{AppB} for more details.}:
\begin{align}
    &\text{Poles}   
    &&\text{terms shown directly} 
    &&\text{hidden terms} \n \\
    &\text{Set A:} &&\wt{\nu}'=-(\ii d/2+2\wt{\nu}+2\ii n)  &&\wt{\nu}'=\ii d/2+2\wt{\nu}+2\ii n, \\
    &\text{Set B:} &&\wt{\nu}'=-(\ii d/2+2\ii n)  &&\wt{\nu}'=\ii d/2+2\ii n,\\
    &\text{Set C:} &&\wt{\nu}'=-(\ii d/2-2\wt{\nu}+2\ii n)  &&\wt{\nu}'=\ii d/2-2\wt{\nu}+2\ii n.
\end{align}
Here ``hidden terms" refer to terms given in Eqs. (\ref{15}), (\ref{16}), (\ref{17}) as $(\wt{\nu}\rightarrow(-\wt{\nu}))$ terms. In these expressions $n$ goes over all nonnegative integers.

The full result of the loop seed integral can be divided into several pieces in terms of their analytic properties in the squeezed limit $r_{1,2}\rightarrow0$:
\begin{align}
    \mathcal{J}_{\sigma\sigma}^{p_1p_2}(r_1,r_2)=\mathcal{J}_{\sigma\sigma,\text{NS}}^{p_1p_2}(r_1,r_2)+\mathcal{J}_{\sigma\sigma,\text{LS}}^{p_1p_2}(r_1,r_2)+\mathcal{J}_{\sigma\sigma,\text{BG}}^{p_1p_2}(r_1,r_2).
\end{align}
The results of nonlocal and local signals are:
\begin{align}\label{26}
    \mathcal{J}_{\sigma\sigma,\text{NS}}^{p_1p_2}=&-\frac{(r_1r_2)^{d/2+2\ii\wt{\nu}}\sin[\pi(\fr{d}{2}+2\ii\wt{\nu})]}{8\pi^{d/2}\Gamma\big(\tfrac d2\big)\sin^2(\pi\ii\wt{\nu})}\sum_{n=0}^\infty\FR{(1+n)_{\frac d2-1} \big[(1+\ii\wt\nu+n)_{\frac d2-1}\big]^2(1+2\ii\wt\nu+n)_{\frac d2-1}}{(1+2\ii\wt\nu+2n)_{d-1}}\n\\
    &\times(\fr{d}{2}+2\ii\wt{\nu}+2n)\mathcal{C}_{2\ii\wt{\nu}+d/2+2n,d}^{p_1p_2}\mb{F}_{2\ii\wt{\nu}+d/2+2n,d}^{p_1}(r_1)\mb{F}_{2\ii\wt{\nu}+d/2+2n,d}^{p_2}(r_2)(r_1r_2)^{2n}\n \\
    &+(\wt{\nu}\rightarrow(-\wt{\nu})),
\end{align}
\begin{align}\label{27}
    \mathcal{J}_{\sigma\sigma,\text{LS}}^{p_1p_2}=&\frac{(r_1/r_2)^{d/2+2\ii\wt{\nu}}\sin[\pi(\fr{d}{2}+2\ii \wt{\nu})]}{8\pi^{d/2}\Gamma\big(\tfrac d2\big)\sin^2(\pi\ii\wt{\nu})}\sum_{n=0}^\infty\FR{(1+n)_{\frac d2-1} \big[(1+\ii\wt\nu+n)_{\frac d2-1}\big]^2(1+2\ii\wt\nu+n)_{\frac d2-1}}{(1+2\ii\wt\nu+2n)_{d-1}}\n\\
    &\times(\fr{d}{2}+2\ii\wt{\nu}+2n)\mathcal{C}_{2\ii\wt{\nu}+d/2+2n,d}^{p_1p_2}\mb{F}_{2\ii\wt{\nu}+d/2+2n,d}^{p_1}(r_1)\mb{F}_{-2\ii\wt{\nu}-d/2-2n,d}^{p_2}(r_2)\bigg(\frac{r_1}{r_2}\bigg)^{2n}\n \\
    &+(\wt{\nu}\rightarrow(-\wt{\nu})).
\end{align}
They are from the integration of signals of the tree seed integral, while the poles are Set A and Set C.

The expression of the background consists of multiple parts:
\begin{equation}
    \mathcal{J}_{\sigma\sigma,\text{BG}}^{p_1p_2}=\mathcal{J}_{\sigma\sigma,\text{(B)}}^{p_1p_2}+\mathcal{J}_{\sigma\sigma,(\text{3A})}^{p_1p_2}+\mathcal{J}_{\sigma\sigma,(\text{3B})}^{p_1p_2}+\mathcal{J}_{\sigma\sigma,(\text{3C})}^{p_1p_2}.
\end{equation}
$\mathcal{J}_{\sigma\sigma,\text{(B)}}^{p_1p_2}$ comes from the Set B poles of the integrand of (\ref{13}), while the signals of the integrand are considered. The expression is:
\begin{align}\label{29}
    \mathcal{J}_{\sigma\sigma,\text{(B)}}^{p_1p_2}=&~\frac{\sin\big(\frac{\pi d}2\big)}{4\pi^{d/2}\Gamma\big(\tfrac d2\big)\sin^2(\pi\ii\wt{\nu})}\sum_{n=0}^\infty\FR{\big[(1+n)_{\frac d2-1}\big]^2 (1+\ii\wt\nu+n)_{\frac d2-1} (1-\ii\wt\nu+n)_{\frac d2-1}}{(1+2n)_{d-1}}(\fr{d}{2}+2n)\mathcal{C}_{d/2+2n,d}^{p_1p_2}\n\\
    &\times\mb{F}_{d/2+2n,d}^{p_1}(r_1)r_1^{2n+d/2}\Big[\mb{F}_{d/2+2n,d}^{p_2}(r_2)r_2^{2n+d/2}+\mb{F}_{-d/2-2n,d}^{p_2}(r_2)r_2^{-2n-d/2}\Big] . 
\end{align} 
The pieces $\mathcal{J}_{\sigma\sigma,(\text{3A})}^{p_1p_2}$, $\mathcal{J}_{\sigma\sigma,(\text{3B})}^{p_1p_2}$, $\mathcal{J}_{\sigma\sigma,(\text{3C})}^{p_1p_2}$ are from the integration of the tree background $\mathcal{I}_{\text{BG},\wt\nu}^{p_1p_2}$. Their expressions are
\begin{align}
    \mathcal{J}_{\sigma\sigma,(\text{3A})}^{p_1p_2}=&-\sum_{n=0}^\infty\sum_{\ell,j=0}^\infty\frac{(-1)^\ell\big(\frac d2-2\ii\wt{\nu}+2n\big)\mathcal{C}_{\frac d2-2\ii\wt{\nu}+2n,d}^{p_1p_2}\sin^2\big[\pi\big(\frac d2-2\ii\wt{\nu}\big)\big](\ell+1)_{2j+d+p_{12}+1}}{2^{2j+d+1}\pi^{\frac d2}\sin^2(\pi\ii\wt{\nu})\Gamma\big(\frac d2\big)n!\big(\frac{\ell+2\ii\wt{\nu}+p_2+1}2-n\big)_{j+1}\big(\frac{\ell+d-2\ii\wt{\nu}+p_2+1}2+n\big)_{j+1}}\n \\
    &\times\Gamma\Bigg[\begin{matrix}
        \frac d2+n, \frac d2-\ii\wt{\nu}+n, \frac d2-2\ii\wt{\nu}+n, -\ii\wt{\nu}+n+\frac12 \\
        \frac{d+1}2-\ii\wt{\nu}+n, -2\ii\wt{\nu}+n+1, -\ii\wt{\nu}+n+1
    \end{matrix}\Bigg]\n \\
    &\times r_1^{2j+d+p_{12}+2}\bigg(\frac{r_1}{r_2}\bigg)^\ell,
\end{align}
\begin{align}
    \mathcal{J}_{\sigma\sigma,(\text{3B})}^{p_1p_2}=&\sum_{n=0}^\infty\sum_{\ell,j=0}^\infty\frac{(-1)^\ell\big(\frac d2+2n\big)\mathcal{C}_{\frac d2+2n,d}^{p_1p_2}\sin^2\big(\frac{\pi d}2\big)(\ell+1)_{2j+d+p_{12}+1}}{2^{2j+d}\pi^{\frac d2}\sin^2(\pi\ii\wt{\nu})\Gamma\big(\frac d2\big)n!\big(\frac{\ell+p_2+1}2-n\big)_{j+1}\big(\frac{\ell+d+p_2+1}2+n\big)_{j+1}}\n \\
    &\times\Gamma\Bigg[\begin{matrix}
        \frac d2+\ii\wt{\nu}+n, \frac d2+n, \frac d2-\ii\wt{\nu}+n, n+\frac12 \\
        \frac{d+1}2+n, 1-\ii\wt{\nu}+n, 1+\ii\wt{\nu}+n
    \end{matrix}\Bigg]\n \\
    &\times r_1^{2j+d+p_{12}+2}\bigg(\frac{r_1}{r_2}\bigg)^\ell,
\end{align}
\begin{align}
    \mathcal{J}_{\sigma\sigma,(\text{3C})}^{p_1p_2}=&-\sum_{n=0}^\infty\sum_{\ell,j=0}^\infty\frac{(-1)^\ell\big(\frac d2+2\ii\wt{\nu}+2n\big)\mathcal{C}_{\frac d2+2\ii\wt{\nu}+2n,d}^{p_1p_2}\sin^2\big[\pi\big(\frac d2+2\ii\wt{\nu}\big)\big](\ell+1)_{2j+d+p_{12}+1}}{2^{2j+d+1}\pi^{\frac d2}\sin^2(\pi\ii\wt{\nu})\Gamma\big(\frac d2\big)n!\big(\frac{\ell-2\ii\wt{\nu}+p_2+1}2-n\big)_{j+1}\big(\frac{\ell+d+2\ii\wt{\nu}+p_2+1}2+n\big)_{j+1}}\n \\
    &\times\Gamma\Bigg[\begin{matrix}
        \frac d2+2\ii\wt{\nu}+n, \frac d2+\ii\wt{\nu}+n, \frac d2+n, \ii\wt{\nu}+n+\frac12 \\
        \frac{d+1}2+\ii\wt{\nu}+n, 1+\ii\wt{\nu}+n, 1+2\ii\wt{\nu}+n
    \end{matrix}\Bigg]\n \\
    &\times r_1^{2j+d+p_{12}+2}\bigg(\frac{r_1}{r_2}\bigg)^\ell.
\end{align}
$\mathcal{J}_{\sigma\sigma,(\text{3C})}^{p_1p_2}$ can be derived from $\mathcal{J}_{\sigma\sigma,(\text{3A})}^{p_1p_2}$ by changing the variable $\wt{\nu}$ to $(-\wt{\nu}$).

\subsection{Divergence of the Seed Integral}
The signals $\mathcal{J}_{\sigma\sigma,\text{NS}}^{p_1p_2}$ and $\mathcal{J}_{\sigma\sigma,\text{LS}}^{p_1p_2}$ and one piece of background $\mathcal{J}_{\sigma\sigma,\text{(B)}}^{p_1p_2}$ converge for arbitrary dimension $d$, while $\mathcal{J}_{\sigma\sigma,(\text{3A})}^{p_1p_2}$, $\mathcal{J}_{\sigma\sigma,(\text{3B})}^{p_1p_2}$, $\mathcal{J}_{\sigma\sigma,(\text{3C})}^{p_1p_2}$ do not. For example, consider the following summation in $\mathcal{J}_{\sigma\sigma,(\text{3A})}^{p_1p_2}$ when $j=0$. As $(n+1)\rightarrow\infty$, the summation tends to
\begin{align}\label{33}
    &\sum_{\ell=0}^\infty\frac{(-1)^\ell\big(\frac d2-2\ii\wt{\nu}+2n\big)\mathcal{C}_{\frac d2-2\ii\wt{\nu}+2n,d}^{p_1p_2}\sin^2\big[\pi\big(\frac d2-2\ii\wt{\nu}\big)\big](\ell+1)_{d+p_{12}+1}}{2^{d+1}\pi^{\frac d2}\sin^2(\pi\ii\wt{\nu})\Gamma\big(\frac d2\big)n!\big(\frac{\ell+2\ii\wt{\nu}+p_2+1}2-n\big)\big(\frac{\ell+d-2\ii\wt{\nu}+p_2+1}2+n\big)}\n \\
    &\times\Gamma\Bigg[\begin{matrix}
        \frac d2+n, \frac d2-\ii\wt{\nu}+n, \frac d2-2\ii\wt{\nu}+n, -\ii\wt{\nu}+n+\frac12 \\
        \frac{d+1}2-\ii\wt{\nu}+n, -2\ii\wt{\nu}+n+1, -\ii\wt{\nu}+n+1
    \end{matrix}\Bigg]r_1^{d+p_{12}+2}\bigg(\frac{r_1}{r_2}\bigg)^\ell\n \\
    \sim&-\sum_{\ell=0}^\infty\frac{(-1)^\ell\mathcal{C}_{\frac d2-2\ii\wt{\nu},d}^{p_1p_2}\sin^2\big[\pi\big(\frac d2-2\ii\wt{\nu}\big)\big](\ell+1)_{d+p_{12}+1}}{2^{d}\pi^{\frac d2}\sin^2(\pi\ii\wt{\nu})\Gamma\big(\frac d2\big)}r_1^{d+p_{12}+2}\bigg(\frac{r_1}{r_2}\bigg)^\ell(n+1)^{d-4}.
\end{align}
It can be seen that the sum over $n$ is divergent if $d\rightarrow3$. If $j>0$, the leading term as $(n+1)\rightarrow\infty$ in the asymptotic expansion of the above summation is proportional to $(n+1)^{d-4-2j}$. Therefore, divergence arises solely from the $j=0$ contribution when $d\rightarrow3$. For $\text{Re}\;d<3$, the sum over $n$ can be written as
\begin{equation}
    \sum_{n=0}^\infty(n+1)^{d-4}=\zeta(-d+4).
\end{equation}
Therefore, $\mathcal{J}_{\sigma\sigma,(\text{3A})}^{p_1p_2}$ can be regularized as
\begin{align}\label{35}
    \mathcal{J}_{\sigma\sigma,(\text{3A})}^{p_1p_2}=&-\sum_{n=0}^\infty\sum_{\ell,j=0}^\infty\frac{(-1)^\ell\big(\frac d2-2\ii\wt{\nu}+2n\big)\mathcal{C}_{\frac d2-2\ii\wt{\nu}+2n,d}^{p_1p_2}\sin^2\big[\pi\big(\frac d2-2\ii\wt{\nu}\big)\big](\ell+1)_{2j+d+p_{12}+1}}{2^{2j+d+1}\pi^{\frac d2}\sin^2(\pi\ii\wt{\nu})\Gamma\big(\frac d2\big)n!\big(\frac{\ell+2\ii\wt{\nu}+p_2+1}2-n\big)_{j+1}\big(\frac{\ell+d-2\ii\wt{\nu}+p_2+1}2+n\big)_{j+1}}\n \\
    &\times\Gamma\Bigg[\begin{matrix}
        \frac d2+n, \frac d2-\ii\wt{\nu}+n, \frac d2-2\ii\wt{\nu}+n, -\ii\wt{\nu}+n+\frac12 \\
        \frac{d+1}2-\ii\wt{\nu}+n, -2\ii\wt{\nu}+n+1, -\ii\wt{\nu}+n+1
    \end{matrix}\Bigg]\n \\
    &\times r_1^{2j+d+p_{12}+2}\bigg(\frac{r_1}{r_2}\bigg)^\ell\n \\
    =&-\sum_{n,\ell=0}^\infty\sum_{j=1}^\infty\frac{(-1)^\ell\big(\frac d2-2\ii\wt{\nu}+2n\big)\mathcal{C}_{\frac d2-2\ii\wt{\nu}+2n,d}^{p_1p_2}\sin^2\big[\pi\big(\frac d2-2\ii\wt{\nu}\big)\big](\ell+1)_{2j+d+p_{12}+1}}{2^{2j+d+1}\pi^{\frac d2}\sin^2(\pi\ii\wt{\nu})\Gamma\big(\frac d2\big)n!\big(\frac{\ell+2\ii\wt{\nu}+p_2+1}2-n\big)_{j+1}\big(\frac{\ell+d-2\ii\wt{\nu}+p_2+1}2+n\big)_{j+1}}\n \\
    &\times\Gamma\Bigg[\begin{matrix}
        \frac d2+n, \frac d2-\ii\wt{\nu}+n, \frac d2-2\ii\wt{\nu}+n, -\ii\wt{\nu}+n+\frac12 \\
        \frac{d+1}2-\ii\wt{\nu}+n, -2\ii\wt{\nu}+n+1, -\ii\wt{\nu}+n+1
    \end{matrix}\Bigg]\n \\
    &\times r_1^{2j+d+p_{12}+2}\bigg(\frac{r_1}{r_2}\bigg)^\ell\n \\
    &-\sum_{\ell=0}^\infty\frac{(-1)^\ell\mathcal{C}_{\frac d2-2\ii\wt{\nu},d}^{p_1p_2}\sin^2\big[\pi\big(\frac d2-2\ii\wt{\nu}\big)\big](\ell+1)_{d+p_{12}+1}}{2^{d+1}\pi^{\frac d2}\sin^2(\pi\ii\wt{\nu})\Gamma\big(\frac d2\big)}r_1^{d+p_{12}+2}\bigg(\frac{r_1}{r_2}\bigg)^\ell\n \\
    &\times\Bigg[\sum_{n=0}^\infty\Bigg(\frac{\frac d2-2\ii\wt{\nu}+2n}{n!\big(\frac{\ell+2\ii\wt{\nu}+p_2+1}2-n\big)\big(\frac{\ell+d-2\ii\wt{\nu}+p_2+1}2+n\big)}\n \\
    &\times\Gamma\Bigg[\begin{matrix}
        \frac d2+n, \frac d2-\ii\wt{\nu}+n, \frac d2-2\ii\wt{\nu}+n, -\ii\wt{\nu}+n+\frac12 \\
        \frac{d+1}2-\ii\wt{\nu}+n, -2\ii\wt{\nu}+n+1, -\ii\wt{\nu}+n+1
    \end{matrix}\Bigg]+2(n+1)^{d-4}\Bigg)-2\zeta(-d+4)\Bigg].
\end{align}
The sums over $n$ in eq. (\ref{35}) are finite and it can be continued to arbitrary dimension $d$. A divergence appears when $d\rightarrow3$ in the Riemann zeta function. The Riemann zeta function can be expanded as a series of $(3-d)$:
\begin{equation}
    \zeta(-d+4)=\frac1{3-d}+\gamma_E+\mathcal{O}((3-d)),
\end{equation}
with $\gamma_E$ being Euler's constant. Then the divergence of $\mathcal{J}_{\sigma\sigma,(\text{3A})}^{p_1p_2}$ is:
\begin{align}
    \text{Div}\mathcal{J}_{\sigma\sigma,(\text{3A})}^{p_1p_2}=&\sum_{\ell=0}^\infty\frac{(-1)^\ell\mathcal{C}_{\frac d2-2\ii\wt{\nu},3}^{p_1p_2}\sin^2\big[\pi\big(\frac d2-2\ii\wt{\nu}\big)\big](\ell+1)_{d+p_{12}+1}}{8\pi^2\sin^2(\pi\ii\wt{\nu})}r_1^{d+p_{12}+2}\bigg(\frac{r_1}{r_2}\bigg)^\ell\frac2{3-d}\n \\
    =&\frac{\mathcal{C}_{\frac d2-2\ii\wt{\nu},3}^{p_1p_2}\sin^2\big[\pi\big(\frac d2-2\ii\wt{\nu}\big)\big]\Gamma(d+p_{12}+2)}{8\pi^2\sin^2(\pi\ii\wt{\nu})}\bigg(\frac{r_1r_2}{r_1+r_2}\bigg)^{d+p_{12}+2}\frac2{3-d}.
\end{align}
It should be emphasized that the mass parameter $\wt{\nu}$ in the above formula is defined in the fixed dimension $d=3$. This means that $\wt{\nu}$ in the expressions for arbitrary $d$ should be expanded as a series of $(3-d)$ while making sure the mass $m$ is fixed for arbitrary $d$. The same procedure can be taken to $\mathcal{J}_{\sigma\sigma,(\text{3B})}^{p_1p_2}$ and $\mathcal{J}_{\sigma\sigma,(\text{3C})}^{p_1p_2}$. Consider the contribution of $\mathcal{J}_{\sigma\sigma,(\text{3A})}^{p_1p_2}$, $\mathcal{J}_{\sigma\sigma,(\text{3B})}^{p_1p_2}$ and $\mathcal{J}_{\sigma\sigma,(\text{3C})}^{p_1p_2}$, the divergence of background is
\begin{align}\label{38}
    \text{Div}\mathcal{J}_{\sigma\sigma,\text{BG}}^{p_1p_2}=-\frac1{(4\pi)^2}\frac2{(3-d)}\sin\big[\tfrac\pi2(d+p_{12})\big]\Gamma(d+p_{12}+2)\bigg(\frac{r_1r_2}{r_1+r_2}\bigg)^{d+p_{12}+2}.
\end{align}
This is nothing but the 4-point correlator generated by a contact interaction. The relative coefficients of $\mathcal{J}_{\sigma\sigma,(\text{3A})}^{p_1p_2}$, $\mathcal{J}_{\sigma\sigma,(\text{3B})}^{p_1p_2}$ and $\mathcal{J}_{\sigma\sigma,(\text{3C})}^{p_1p_2}$ give the right coefficient of the divergence of the background, which is essential for renormalizability. Consider the following Lagrangian of interaction:
\begin{equation}
    \Delta\mathscr{L}=-\frac1{24}\delta_\lambda a^{d-3-p_{12}}\phi_c^4.
\end{equation}
The seed integral generated by the interaction is:
\begin{align}\label{40}
    \delta\mathcal{J}_{\sigma\sigma}^{p_1p_2}(r_1,r_2)=&-\delta_\lambda\sum_{\mathsf{a}=\pm}\mathsf{a}\ii k_s^{d+2+p_{12}}\int_{-\infty}^0\ud\tau\left(-\tau\right)^{d+1+p_{12}}\text{e}^{\ii\mathsf{a}k_{1234}\tau}\n \\
    =&-\delta_\lambda\sum_{\mathsf{a}=\pm}\mathsf{a}\frac{\ii k_s^{d+2+p_{12}}}{k_{1234}^{d+2+p_{12}}}\text{e}^{-\ii\mathsf{a}\pi(d+2+p_{12})/2}\Gamma(d+2+p_{12})\n \\
    =&2\delta_\lambda\sin\big[\tfrac\pi2(d+p_{12})\big]\Gamma(d+p_{12}+2)\bigg(\frac{r_1r_2}{r_1+r_2}\bigg)^{d+p_{12}+2}.
\end{align}
If the coefficient $\delta_\lambda$ is
\begin{equation}\label{41}
    \delta_\lambda=\frac1{(4\pi)^2}\frac1{3-d},
\end{equation}
then the UV divergence (\ref{38}) of the 1-loop correlator can be subtracted by the local term (\ref{40}).

If the MS subtraction is taken to remove the divergence of the background, the expression of the background becomes
\begin{align}
    \wh{\mathcal{J}}_{\sigma\sigma,(\text{3A})}^{p_1p_2}=&-\sum_{n,\ell=0}^\infty\sum_{j=1}^\infty\frac{(-1)^\ell\big(\frac32-2\ii\wt{\nu}+2n\big)\mathcal{C}_{\frac32-2\ii\wt{\nu}+2n,3}^{p_1p_2}\cos^2\big(2\pi\ii\wt{\nu}\big)(\ell+1)_{2j+p_{12}+4}}{2^{2j+3}\pi^2\sin^2(\pi\ii\wt{\nu})n!\big(\frac{\ell+2\ii\wt{\nu}+p_2+1}2-n\big)_{j+1}\big(\frac{\ell-2\ii\wt{\nu}+p_2+4}2+n\big)_{j+1}}\n \\
    &\times\Gamma\Bigg[\begin{matrix}
        \frac32+n, -\ii\wt{\nu}+\frac32+n, -2\ii\wt{\nu}+\frac32+n, -\ii\wt{\nu}+\frac12+n \\
        -\ii\wt{\nu}+2+n, -2\ii\wt{\nu}+1+n, -\ii\wt{\nu}+1+n
    \end{matrix}\Bigg]r_1^{2j+p_{12}+5}\bigg(\frac{r_1}{r_2}\bigg)^\ell\n \\
    &-\sum_{\ell=0}^\infty\frac{(-1)^\ell\mathcal{C}_{\frac32-2\ii\wt{\nu},3}^{p_1p_2}\cos^2\big(2\pi\ii\wt{\nu}\big)(\ell+1)_{p_{12}+4}}{8\pi^2\sin^2(\pi\ii\wt{\nu})}r_1^{p_{12}+5}\bigg(\frac{r_1}{r_2}\bigg)^\ell\n \\
    &\times\Bigg[\sum_{n=0}^\infty\Bigg(\frac{\frac32-2\ii\wt{\nu}+2n}{n!\big(\frac{\ell+2\ii\wt{\nu}+p_2+1}2-n\big)\big(\frac{\ell-2\ii\wt{\nu}+p_2+4}2+n\big)}\n \\
    &\times\Gamma\Bigg[\begin{matrix}
        \frac32+n, -\ii\wt{\nu}+\frac32+n, -2\ii\wt{\nu}+\frac32+n, -\ii\wt{\nu}+\frac12+n \\
        -\ii\wt{\nu}+2+n, -2\ii\wt{\nu}+1+n, -\ii\wt{\nu}+1+n
    \end{matrix}\Bigg]+\frac2{n+1}\Bigg)\n \\
    &-2\bigg(\gamma_E+\log\mu_R+\tfrac12\log4\pi+\psi\big(\tfrac32\big)-\frac{\pi\sin\big[\pi\big(-2\ii\wt{\nu}+\tfrac{p_{12}}2\big)\big]}{16\cos^2(2\pi\ii\wt{\nu})\mathcal{C}_{\frac32-2\ii\wt{\nu},3}^{p_1p_2}}\bigg)\Bigg],
\end{align}
\begin{align}
    \wh{\mathcal{J}}_{\sigma\sigma,(\text{3B})}^{p_1p_2}=&\sum_{n,\ell=0}^\infty\sum_{j=1}^\infty\frac{(-1)^\ell\big(\frac32+2n\big)\mathcal{C}_{\frac32+2n,3}^{p_1p_2}(\ell+1)_{2j+p_{12}+4}}{2^{2j+2}\pi^2\sin^2(\pi\ii\wt{\nu})n!\big(\frac{\ell+p_2+1}2-n\big)_{j+1}\big(\frac{\ell+p_2+4}2+n\big)_{j+1}}\n \\
    &\times\Gamma\Bigg[\begin{matrix}
        \ii\wt{\nu}+\frac32+n, \frac32+n, -\ii\wt{\nu}+\frac32+n, \frac12+n \\
        2+n, -\ii\wt{\nu}+1+n, \ii\wt{\nu}+1+n
    \end{matrix}\Bigg]r_1^{2j+p_{12}+5}\bigg(\frac{r_1}{r_2}\bigg)^\ell\n \\
    &+\sum_{\ell=0}^\infty\frac{(-1)^\ell\mathcal{C}_{\frac32,3}^{p_1p_2}(\ell+1)_{p_{12}+4}}{4\pi^2\sin^2(\pi\ii\wt{\nu})}r_1^{p_{12}+5}\bigg(\frac{r_1}{r_2}\bigg)^\ell\Bigg[\sum_{n=0}^\infty\Bigg(\frac{\frac32+2n}{n!\big(\frac{\ell+p_2+1}2-n\big)\big(\frac{\ell+p_2+4}2+n\big)}\n \\
    &\times\Gamma\Bigg[\begin{matrix}
        \ii\wt{\nu}+\frac32+n, \frac32+n, -\ii\wt{\nu}+\frac32+n, \frac12+n \\
        2+n, -\ii\wt{\nu}+1+n, \ii\wt{\nu}+1+n
    \end{matrix}\Bigg]+\frac2{n+1}
    \Bigg)\n \\
    &-2\bigg(\gamma_E+\log\mu_R+\tfrac12\log4\pi+\psi\big(\tfrac32\big)-\frac{\pi\sin\big(\frac{\pi p_{12}}2\big)}{16\mathcal{C}_{\frac32,3}^{p_1p_2}}\bigg)\Bigg].
\end{align}
\begin{align}
    \wh{\mathcal{J}}_{\sigma\sigma,(\text{3C})}^{p_1p_2}=&-\sum_{n,\ell=0}^\infty\sum_{j=1}^\infty\frac{(-1)^\ell\big(\frac32+2\ii\wt{\nu}+2n\big)\mathcal{C}_{\frac32+2\ii\wt{\nu}+2n,3}^{p_1p_2}\cos^2\big(2\pi\ii\wt{\nu}\big)(\ell+1)_{2j+p_{12}+4}}{2^{2j+3}\pi^2\sin^2(\pi\ii\wt{\nu})n!\big(\frac{\ell-2\ii\wt{\nu}+p_2+1}2-n\big)_{j+1}\big(\frac{\ell+2\ii\wt{\nu}+p_2+4}2+n\big)_{j+1}}\n \\
    &\times\Gamma\Bigg[\begin{matrix}
        2\ii\wt{\nu}+\frac32+n, \ii\wt{\nu}+\frac32+n, \frac32+n, \ii\wt{\nu}+\frac12+n \\
        \ii\wt{\nu}+2+n, \ii\wt{\nu}+1+n, 2\ii\wt{\nu}+1+n
    \end{matrix}\Bigg]r_1^{2j+p_{12}+5}\bigg(\frac{r_1}{r_2}\bigg)^\ell\n \\
    &-\sum_{\ell=0}^\infty\frac{(-1)^\ell\mathcal{C}_{\frac32+2\ii\wt{\nu},3}^{p_1p_2}\cos^2\big(2\pi\ii\wt{\nu}\big)(\ell+1)_{p_{12}+4}}{8\pi^2\sin^2(\pi\ii\wt{\nu})}r_1^{p_{12}+5}\bigg(\frac{r_1}{r_2}\bigg)^\ell\n \\
    &\times\Bigg[\sum_{n=0}^\infty\Bigg(\frac{\frac32+2\ii\wt{\nu}+2n}{n!\big(\frac{\ell-2\ii\wt{\nu}+p_2+1}2-n\big)\big(\frac{\ell+2\ii\wt{\nu}+p_2+4}2+n\big)}\n \\
    &\times\Gamma\Bigg[\begin{matrix}
        2\ii\wt{\nu}+\frac32+n, \ii\wt{\nu}+\frac32+n, \frac32+n, \ii\wt{\nu}+\frac12+n \\
        \ii\wt{\nu}+2+n, \ii\wt{\nu}+1+n, 2\ii\wt{\nu}+1+n
    \end{matrix}\Bigg]+\frac2{n+1}\Bigg)\n \\
    &-2\bigg(\gamma_E+\log\mu_R+\tfrac12\log4\pi+\psi\big(\tfrac32\big)-\frac{\pi\sin\big[\pi\big(2\ii\wt{\nu}+\tfrac{p_{12}}2\big)\big]}{16\cos^2(2\pi\ii\wt{\nu})\mathcal{C}_{\frac32+2\ii\wt{\nu},3}^{p_1p_2}}\bigg)\Bigg].
\end{align}
Here $\mu_R$ is introduced as the renormalization scale. With the expressions of signals (\ref{26}), (\ref{27}) and $\mathcal{J}_{\sigma\sigma,\text{(B)}}^{p_1p_2}$ (\ref{29}), the full result of the loop seed integral is computed.

In \cite{Xianyu:2022jwk}, the dimensional regularization of 1-loop 4-point and 3-point correlators of $\sigma\sigma$ model has been performed in a different way. The divergence of the background in \cite{Xianyu:2022jwk} is from the expression of the spectral function $\rho_{\wt{\nu}}^\text{dS}(\wt{\nu}')$. When $d\rightarrow3$, $\rho_{\wt{\nu}}^\text{dS}(\wt{\nu}')$ is divergent, which results in the divergence of the background. However, in this work, the summand of the expression of the background converges when $d\rightarrow3$, while the divergence appears if the summation is taken. This is the reason the dimensional regularization is taken in different ways between \cite{Xianyu:2022jwk} and this work.

\subsection{Summary of the Procedure of Dimensional Regularization}
In the previous subsection, the dimensional regularization is completed, and the divergence of $d\rightarrow3$ is subtracted. The main procedure is listed below:
\begin{itemize}
    \item Take the asymptotic expansions of the summation which is divergent when $d\rightarrow3$ regarding the summation index.
    \item Subtract the asymptotic expansions from the expressions of the correlators and add several Riemann zeta functions related to the asymptotic expansions to make sure the expressions are analytically continued to arbitrary $d$.
    \item Take the series expansions of the Riemann zeta function with respect to $(3-d)$ and subtract the $(3-d)^{-1}$ terms.
\end{itemize}
\section{Applications}\label{Sec3}
In the previous section, the divergent summation (\ref{33}) is regularized with the Riemann Zeta function. Several divergent summations in de Sitter spacetime can be regularized in this way. Here are some examples.
\begin{figure}[!htp]
    \centering
    \includegraphics{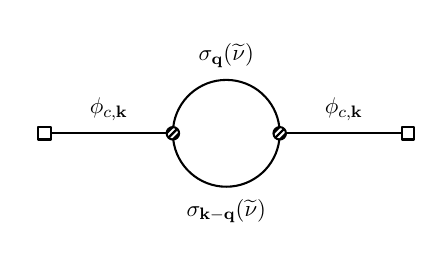}
    \caption{Feynman diagram of 1-loop 2-point correlator with a scalar loop.}
    \label{Fig2}
\end{figure}
\subsection{2-point Correlator}
The first example is the 1-loop 2-point correlator whose Feynman diagram is Fig. \ref{Fig2}. To get the expression of the 2-point correlator, the double folded limit $(r_{1,2}\rightarrow1)$ is taken to the 4-point correlator \cite{Qin:2023ejc}. It is better to use the following expression of the background of the 4-point correlator to take the double folded limit \cite{Qin:2023ejc}:
\begin{align}
    \mathcal{I}_{\text{BG},\wt\nu}^{p_1p_2}(r_1(u_1),r_2(u_2))=&-\sum_{\ell,j=0}^\infty\frac{\sin\big[\frac\pi2(p_{12}+d)\big]\Gamma(p_{12}+d+2+\ell+j)\big(\ell+p_2+\frac{d+3}2\big)_j}{2^{p_{12}+d+1}\ell!\big(\ell+\ii\wt{\nu}+p_2+1+\frac d2\big)_{j+1}\big(\ell-\ii\wt{\nu}+p_2+1+\frac d2\big)_{j+1}}\n \\
    &\times u_1^{j+\ell+p_{12}+d+2}\bigg(1-\frac1{u_2}\bigg)^\ell\n \\
    =&\sum_{\ell,j=0}^\infty\frac{\mathcal{C}_{\ii\wt{\nu},d}^{p_1p_2}\sin(\pi\ii\wt{\nu})\Gamma(p_{12}+d+2+\ell+j)\big(\ell+p_2+\frac{d+3}2\big)_j}{2^{p_{12}+d-1}\ell!\big(\ell+\ii\wt{\nu}+p_2+1+\frac d2\big)_{j+1}\big(\ell-\ii\wt{\nu}+p_2+1+\frac d2\big)_{j+1}}\n \\
    &\times u_1^{j+\ell+p_{12}+d+2}\bigg(1-\frac1{u_2}\bigg)^\ell+(\wt{\nu}\rightarrow(-\wt{\nu})),
\end{align}
where $u_{1,2}\equiv2r_{1,2}/(1+r_{1,2})$. Then $u_2\rightarrow1$ is taken to get the seed integral which can be used to get the expressions of 3-point correlators \cite{Qin:2023ejc}:
\begin{align}
    \mathcal{I}_{\text{BG},\wt\nu}^{p_1p_2}(r(u),1)=&2^{-p_{12}-d+1}\mathcal{C}_{\ii\wt{\nu},d}^{p_1p_2}\sin(\pi\ii\wt{\nu})\Gamma\big[p_{12}+d+2, \ii\wt{\nu}+p_2+1+\tfrac d2, -\ii\wt{\nu}+p_2+1+\tfrac d2\big]\n \\
    &\times u^{p_{12}+d+2}{}_3\wt{\mathrm{F}}_2\left[\begin{matrix}
        p_{12}+d+2, p_2+\tfrac{d+3}2, 1 \\
        \ii\wt{\nu}+p_2+2+\tfrac d2, -\ii\wt{\nu}+p_2+2+\tfrac d2
    \end{matrix}\middle|u\right]+(\wt{\nu}\rightarrow(-\wt{\nu})).
\end{align}
The expression of nonlocal and local signals diverge separately. However, the sum of the two signals converge when $r_2\rightarrow1$. To prove this, it is better to change the variable of signals (\ref{15}), (\ref{16}) from $r_i$ to $u_i$, with the help of (\ref{164}):
\begin{align}
    r^{\ii\wt{\nu}}\mathbf{F}_{\ii\wt\nu,d}^{p}(r)
  \equiv &r^{\ii\wt{\nu}}(2r)^{p+d/2+1}\times{}_2\mathcal{F}_1\left[ \begin{matrix}
        \tfrac d4+\tfrac12+\tfrac{p}2+\tfrac{\ii \wt{\nu}}2, \tfrac d4+1+\tfrac{p}2+\tfrac{\ii \wt{\nu}}2 \\
        1+\ii \wt{\nu}
    \end{matrix}\middle|r^2\right]\n \\
    =&2^{1+\ii\wt{\nu}}\bigg(\frac{u}2\bigg)^{p+d/2+1+\ii\wt{\nu}}{}_2\mathcal{F}_1\left[\begin{matrix}
        p+\frac d2+1+\ii\wt{\nu}, \frac12+\ii\wt{\nu} \\
        1+2\ii\wt{\nu}
    \end{matrix}\middle|u\right].
\end{align}
Then by using (\ref{163}), one may get the following formula:
\begin{equation}
    \lim_{r\rightarrow1}\big(r^{\ii\wt{\nu}}\mathbf{F}_{\ii\wt\nu,d}^{p}(r)-r^{-\ii\wt{\nu}}\mathbf{F}_{-\ii\wt\nu,d}^{p}(r)\big)=-2^{1-p-d/2}\sin(\pi\ii\wt{\nu})\Gamma\Bigg[\begin{matrix}
        p+\frac d2+1+\ii\wt{\nu}, p+\frac d2+1-\ii\wt{\nu} \\
        p+\frac d2+\frac32
    \end{matrix}\Bigg].
\end{equation}
Therefore, the sum of nonlocal and local signals in the limit $u_2\rightarrow1$ is
\begin{equation}
    \mathcal{I}_{\text{S},\wt\nu}^{p_1p_2}(r,1)=\Bigg(-2^{1-p_2-d/2}\sin(\pi\ii\wt{\nu})\Gamma\Bigg[\begin{matrix}
        p_2+\frac d2+1+\ii\wt{\nu}, p_2+\frac d2+1-\ii\wt{\nu} \\
        p_2+\frac d2+\frac32
    \end{matrix}\Bigg]\mathcal{C}_{\ii\wt\nu,d}^{p_1p_2}r^{\ii\wt{\nu}}\mb{F}_{\ii\wt\nu,d}^{p_1}(r)\Bigg)+(\wt{\nu}\rightarrow(-\wt{\nu})).
\end{equation}
The double folded limit of the seed integral is quite difficult. It can be proved that the expression of the 2-point seed integral is (See App. \ref{AppC})
\begin{align}\label{50}
    \mathcal{I}_{\wt\nu}^{p_1p_2}(1,1)=&\Bigg(-2^{-p_{12}-d+1}\mathcal{C}_{\ii\wt\nu,d}^{p_1p_2}\sin(\pi\ii\wt{\nu})\Gamma\big[p_{12}+d+2, \ii\wt{\nu}+p_1+1+\tfrac d2, \ii\wt{\nu}+p_2+1+\tfrac d2\big]\n \\
    &\times{}_3\mathrm{\wt{F}}_2\left[\begin{matrix}
        p_{12}+d+2, \frac12+\ii\wt{\nu}, 1 \\
        \ii\wt{\nu}+p_1+2+\tfrac d2, \ii\wt{\nu}+p_2+2+\tfrac d2
    \end{matrix}\middle|1\right]\Bigg)+(\wt{\nu}\rightarrow(-\wt{\nu})).
\end{align}

While taking the K\"all\'en-Lehmann integral, the integral contour can be closed with a semi-circle in the proper half $\wt{\nu}'$-plane to avoid poles from (\ref{50}). Therefore, only poles from the spectral function (\ref{20}) are considered. The result of the double folded limit of the loop seed integral $\mathcal{J}_{\sigma\sigma}^{p_1p_2}(1,1)$ is broken into three pieces:
\begin{equation}
    \mathcal{J}_{\sigma\sigma}^{p_1p_2}(1,1)=\mathcal{J}_{\sigma\sigma,(\text{A})}^{p_1p_2}(1,1)+\mathcal{J}_{\sigma\sigma,(\text{B})}^{p_1p_2}(1,1)+\mathcal{J}_{\sigma\sigma,(\text{C})}^{p_1p_2}(1,1).
\end{equation}
The expressions of the three pieces are:
\begin{align}
    \mathcal{J}_{\sigma\sigma,(\text{A})}^{p_1p_2}(1,1)=&\sum_{n,j=0}^\infty\frac{\big(\frac d2-2\ii\wt{\nu}+2n\big)\sin^2\big[\pi\big(\frac d2-2\ii\wt{\nu}\big)\big]\Gamma(p_{12}+d+2+j)\big(\frac12+\frac d2-2\ii\wt{\nu}+2n\big)_j}{n!(d-2\ii\wt{\nu}+p_1+1+2n)_{j+1}(d-2\ii\wt{\nu}+p_2+1+2n)_{j+1}}\n \\
    &\times\frac{\mathcal{C}_{\frac d2-2\ii\wt{\nu}+2n,d}^{p_1p_2}}{2^{p_{12}+2d+1}\pi^{\frac d2}\sin^2(\pi\ii\wt{\nu})\Gamma\big(\frac d2\big)}\Gamma\Bigg[\begin{matrix}
        \frac d2+n, \frac d2-\ii\wt{\nu}+n, \frac d2-2\ii\wt{\nu}+n, -\ii\wt{\nu}+n+\frac12 \\
        \frac{d+1}2-\ii\wt{\nu}+n, -2\ii\wt{\nu}+n+1, -\ii\wt{\nu}+n+1
    \end{matrix}\Bigg],
\end{align}
\begin{align}
    \mathcal{J}_{\sigma\sigma,(\text{B})}^{p_1p_2}(1,1)=&-\sum_{n,j=0}^\infty\frac{\big(\frac d2+2n\big)\sin^2\big(\frac{\pi d}2\big)\Gamma(p_{12}+d+2+j)\big(\frac12+\frac d2+2n\big)_j}{n!(d+p_1+1+2n)_{j+1}(d+p_2+1+2n)_{j+1}}\n \\
    &\times\frac{\mathcal{C}_{\frac d2+2n,d}^{p_1p_2}}{2^{p_{12}+2d}\pi^{\frac d2}\sin^2(\pi\ii\wt{\nu})\Gamma\big(\frac d2\big)}\Gamma\Bigg[\begin{matrix}
        \frac d2+\ii\wt{\nu}+n, \frac d2+n, \frac d2-\ii\wt{\nu}+n, n+\frac12 \\
        \frac{d+1}2+n, 1-\ii\wt{\nu}+n, 1+\ii\wt{\nu}+n
    \end{matrix}\Bigg],
\end{align}
\begin{align}
    \mathcal{J}_{\sigma\sigma,(\text{C})}^{p_1p_2}(1,1)=&\sum_{n,j=0}^\infty\frac{\big(\frac d2+2\ii\wt{\nu}+2n\big)\sin^2\big[\pi\big(\frac d2+2\ii\wt{\nu}\big)\big]\Gamma(p_{12}+d+2+j)\big(\frac12+\frac d2+2\ii\wt{\nu}+2n\big)_j}{n!(d+2\ii\wt{\nu}+p_1+1+2n)_{j+1}(d+2\ii\wt{\nu}+p_2+1+2n)_{j+1}}\n \\
    &\times\frac{\mathcal{C}_{\frac d2+2\ii\wt{\nu}+2n,d}^{p_1p_2}}{2^{p_{12}+2d+1}\pi^{\frac d2}\sin^2(\pi\ii\wt{\nu})}\Gamma\Bigg[\begin{matrix}
        \frac d2+2\ii\wt{\nu}+n, \frac d2+\ii\wt{\nu}+n, \frac d2+n, \ii\wt{\nu}+n+\frac12 \\
        \frac{d+1}2+\ii\wt{\nu}+n, 1+\ii\wt{\nu}+n, 1+2\ii\wt{\nu}+n
    \end{matrix}\Bigg].
\end{align}
It can be seen that the summation over $n$ is divergent when $d\rightarrow3$:
\begin{align}
    &\sum_{n=0}^\infty\frac{\big(\frac d2-2\ii\wt{\nu}+2n\big)}{(d-2\ii\wt{\nu}+p_1+1+2n)(d-2\ii\wt{\nu}+p_2+1+2n)}\Gamma\Bigg[\begin{matrix}
        \frac d2+n, \frac d2-\ii\wt{\nu}+n, \frac d2-2\ii\wt{\nu}+n, -\ii\wt{\nu}+n+\frac12 \\
        \frac{d+1}2-\ii\wt{\nu}+n, -2\ii\wt{\nu}+n+1, -\ii\wt{\nu}+n+1
    \end{matrix}\Bigg]\n \\
    \sim&\sum_{n=0}^\infty\frac12(n+1)^{d-4}=\frac12\zeta(-d+4)\sim\frac1{2(3-d)}.
\end{align}
$\mathcal{J}_{\sigma\sigma,(\text{A})}^{p_1p_2}(1,1)$ can be regularized as
\begin{align}
    \mathcal{J}_{\sigma\sigma,(\text{A})}^{p_1p_2}(1,1)=&\sum_{n=0}^\infty\sum_{j=1}^\infty\frac{\big(\frac d2-2\ii\wt{\nu}+2n\big)\sin^2\big[\pi\big(\frac d2-2\ii\wt{\nu}\big)\big]\Gamma(p_{12}+d+2+j)\big(\frac12+\frac d2-2\ii\wt{\nu}+2n\big)_j}{n!(d-2\ii\wt{\nu}+p_1+1+2n)_{j+1}(d-2\ii\wt{\nu}+p_2+1+2n)_{j+1}}\n \\
    &\times\frac{\mathcal{C}_{\frac d2-2\ii\wt{\nu}+2n,d}^{p_1p_2}}{2^{p_{12}+2d+1}\pi^{\frac d2}\sin^2(\pi\ii\wt{\nu})\Gamma\big(\frac d2\big)}\Gamma\Bigg[\begin{matrix}
        \frac d2+n, \frac d2-\ii\wt{\nu}+n, \frac d2-2\ii\wt{\nu}+n, -\ii\wt{\nu}+n+\frac12 \\
        \frac{d+1}2-\ii\wt{\nu}+n, -2\ii\wt{\nu}+n+1, -\ii\wt{\nu}+n+1
    \end{matrix}\Bigg]\n \\
    &+\frac{\mathcal{C}_{\frac d2-2\ii\wt{\nu},d}^{p_1p_2}\sin^2\big[\pi\big(\frac d2-2\ii\wt{\nu}\big)\big]\Gamma(p_{12}+d+2)}{2^{p_{12}+2d+1}\pi^{\frac d2}\sin^2(\pi\ii\wt{\nu})\Gamma\big(\frac d2\big)}\n \\
    &\times\Bigg[\sum_{n=0}^\infty\Bigg(\frac{\big(\frac d2-2\ii\wt{\nu}+2n\big)}{(d-2\ii\wt{\nu}+p_1+1+2n)(d-2\ii\wt{\nu}+p_2+1+2n)}\n \\
    &\times\Gamma\Bigg[\begin{matrix}
        \frac d2+n, \frac d2-\ii\wt{\nu}+n, \frac d2-2\ii\wt{\nu}+n, -\ii\wt{\nu}+n+\frac12 \\
        1+n, \frac{d+1}2-\ii\wt{\nu}+n, -2\ii\wt{\nu}+n+1, -\ii\wt{\nu}+n+1
    \end{matrix}\Bigg]-\tfrac12(n+1)^{d-4}\Bigg)+\tfrac12\zeta(-d+4)\Bigg].
\end{align}
The divergence of $\mathcal{J}_{\sigma\sigma,(\text{A})}^{p_1p_2}(1,1)$ is
\begin{equation}
    \text{Div}\mathcal{J}_{\sigma\sigma,(\text{A})}^{p_1p_2}(1,1)=\frac{\mathcal{C}_{\frac d2-2\ii\wt{\nu},3}^{p_1p_2}\sin^2\big[\pi\big(\frac d2-2\ii\wt{\nu}\big)\big]\Gamma(d+p_{12}+2)}{2^{p_{12}+d+4}\pi^2\sin^2(\pi\ii\wt{\nu})}\frac1{3-d}.
\end{equation}
The divergences of $\mathcal{J}_{\sigma\sigma,(\text{B})}^{p_1p_2}(1,1)$ and $\mathcal{J}_{\sigma\sigma,(\text{C})}^{p_1p_2}(1,1)$ can be calculated in the same way. Taking the sum of the divergences, the divergence of the seed integral is found as
\begin{equation}\label{58}
    \text{Div}\mathcal{J}_{\sigma\sigma}^{p_1p_2}(1,1)=-\frac{\sin\big[\tfrac\pi2(d+p_{12})\big]\Gamma(d+p_{12}+2)}{2^{p_{12}+d+5}\pi^2}\frac1{3-d}.
\end{equation}
After taking the double folded limit, (\ref{40}) becomes
\begin{equation}\label{59}
    \delta\mathcal{J}_{\sigma\sigma}^{p_1p_2}(1,1)=2^{-d-p_{12}-1}\delta_\lambda\sin\big[\tfrac\pi2(d+p_{12})\big]\Gamma(d+p_{12}+2).
\end{equation}
It is expected that the divergence of $\mathcal{J}_{\sigma\sigma}^{p_1p_2}(1,1)$ can be subtracted by (\ref{59}). After comparing (\ref{58}) and (\ref{59}), it can be found that the coefficient $\delta_\lambda$ is
\begin{equation}
    \delta_\lambda=\frac1{(4\pi)^2}\frac1{3-d},
\end{equation}
which is the same as (\ref{41}).

The full result of the loop seed integral is
\begin{align}\label{61}
    \wh{\mathcal{J}}_{\sigma\sigma,(\text{A})}^{p_1p_2}(1,1)=&\sum_{n=0}^\infty\sum_{j=1}^\infty\frac{\big(\frac32-2\ii\wt{\nu}+2n\big)\cos^2(2\pi\ii\wt{\nu})\Gamma(p_{12}+5+j)\big(2-2\ii\wt{\nu}+2n\big)_j}{n!(-2\ii\wt{\nu}+p_1+4+2n)_{j+1}(-2\ii\wt{\nu}+p_2+4+2n)_{j+1}}\n \\
    &\times\frac{\mathcal{C}_{\frac32-2\ii\wt{\nu}+2n,3}^{p_1p_2}}{2^{p_{12}+6}\pi^2\sin^2(\pi\ii\wt{\nu})}\Gamma\Bigg[\begin{matrix}
        \frac32+n, -\ii\wt{\nu}+\frac32+n, -2\ii\wt{\nu}+\frac32+n, -\ii\wt{\nu}+\frac12+n \\
        -\ii\wt{\nu}+2+n, -2\ii\wt{\nu}+1+n, -\ii\wt{\nu}+1+n
    \end{matrix}\Bigg]\n \\
    &+\frac{\mathcal{C}_{\frac32-2\ii\wt{\nu},3}^{p_1p_2}\cos^2(2\pi\ii\wt{\nu})\Gamma(p_{12}+5)}{2^{p_{12}+6}\pi^2\sin^2(\pi\ii\wt{\nu})}\Bigg[\sum_{n=0}^\infty\Bigg(\frac{\big(\frac32-2\ii\wt{\nu}+2n\big)}{(-2\ii\wt{\nu}+p_1+4+2n)(-2\ii\wt{\nu}+p_2+4+2n)}\n \\
    &\times\Gamma\Bigg[\begin{matrix}
        \frac32+n, -\ii\wt{\nu}+\frac32+n, -2\ii\wt{\nu}+\frac32+n, -\ii\wt{\nu}+\frac12+n \\
        1+n, -\ii\wt{\nu}+2+n, -2\ii\wt{\nu}+1+n, -\ii\wt{\nu}+1+n
    \end{matrix}\Bigg]-\frac1{2(n+1)}\Bigg)\n \\
    &+\frac12\bigg(\gamma_E+\log\mu_R+\tfrac12\log4\pi+\psi\big(\tfrac32\big)-\frac{\pi\sin\big[\pi\big(-2\ii\wt{\nu}+\tfrac{p_{12}}2\big)\big]}{16\cos^2(2\pi\ii\wt{\nu})\mathcal{C}_{\frac32-2\ii\wt{\nu},3}^{p_1p_2}}\bigg)\Bigg],
\end{align}
\begin{align}\label{62}
    \wh{\mathcal{J}}_{\sigma\sigma,(\text{B})}^{p_1p_2}(1,1)=&-\sum_{n=0}^\infty\sum_{j=1}^\infty\frac{\big(\frac32+2n\big)\mathcal{C}_{\frac32+2n,3}^{p_1p_2}\Gamma(p_{12}+5+j)\big(2+2n\big)_j}{2^{p_{12}+5}\pi^2\sin^2(\pi\ii\wt{\nu})n!(p_1+4+2n)_{j+1}(p_2+4+2n)_{j+1}}\n \\
    &\times\Gamma\Bigg[\begin{matrix}
        \ii\wt{\nu}+\frac32+n, \frac32+n, -\ii\wt{\nu}+\frac32+n, \frac12+n \\
        2+n, -\ii\wt{\nu}+1+n, \ii\wt{\nu}+1+n
    \end{matrix}\Bigg]\n \\
    &-\frac{\mathcal{C}_{\frac32,3}^{p_1p_2}\Gamma(p_{12}+5)}{2^{p_{12}+5}\pi^2\sin^2(\pi\ii\wt{\nu})}\Bigg[\sum_{n=0}^\infty\Bigg(\frac{\big(\frac32+2n\big)}{(p_1+4+2n)(p_2+4+2n)}\n \\
    &\times\Gamma\Bigg[\begin{matrix}
        \ii\wt{\nu}+\frac32+n, \frac32+n, -\ii\wt{\nu}+\frac32+n, \frac12+n \\
        1+n, 2+n, -\ii\wt{\nu}+1+n, \ii\wt{\nu}+1+n
    \end{matrix}\Bigg]-\frac1{2(n+1)}\Bigg)\n \\
    &+\frac12\bigg(\gamma_E+\log\mu_R+\tfrac12\log4\pi+\psi\big(\tfrac32\big)-\frac{\pi\sin\big(\frac{\pi p_{12}}2\big)}{16\mathcal{C}_{\frac32,3}^{p_1p_2}}\bigg)\Bigg],
\end{align}
\begin{align}\label{63}
    \wh{\mathcal{J}}_{\sigma\sigma,(\text{C})}^{p_1p_2}(1,1)=&\sum_{n=0}^\infty\sum_{j=1}^\infty\frac{\big(\frac32+2\ii\wt{\nu}+2n\big)\cos^2(2\pi\ii\wt{\nu})\Gamma(p_{12}+5+j)\big(2+2\ii\wt{\nu}+2n\big)_j}{n!(2\ii\wt{\nu}+p_1+4+2n)_{j+1}(2\ii\wt{\nu}+p_2+4+2n)_{j+1}}\n \\
    &\times\frac{\mathcal{C}_{\frac32+2\ii\wt{\nu}+2n,3}^{p_1p_2}}{2^{p_{12}+6}\pi^2\sin^2(\pi\ii\wt{\nu})}\Gamma\Bigg[\begin{matrix}
        2\ii\wt{\nu}+\frac32+n, \ii\wt{\nu}+\frac32+n, \frac32+n, \ii\wt{\nu}+\frac12+n \\
        \ii\wt{\nu}+2+n, \ii\wt{\nu}+1+n, 2\ii\wt{\nu}+1+n
    \end{matrix}\Bigg]\n \\
    &+\frac{\mathcal{C}_{\frac32+2\ii\wt{\nu},3}^{p_1p_2}\cos^2(2\pi\ii\wt{\nu})\Gamma(p_{12}+5)}{2^{p_{12}+6}\pi^2\sin^2(\pi\ii\wt{\nu})}\Bigg[\sum_{n=0}^\infty\Bigg(\frac{\big(\frac32+2\ii\wt{\nu}+2n\big)}{(2\ii\wt{\nu}+p_1+4+2n)(2\ii\wt{\nu}+p_2+4+2n)}\n \\
    &\times\Gamma\Bigg[\begin{matrix}
        2\ii\wt{\nu}+\frac32+n, \ii\wt{\nu}+\frac32+n, \frac32+n, \ii\wt{\nu}+\frac12+n \\
        1+n, \ii\wt{\nu}+2+n, \ii\wt{\nu}+1+n, 2\ii\wt{\nu}+1+n
    \end{matrix}\Bigg]-\frac1{2(n+1)}\Bigg)\n \\
    &+\frac12\bigg(\gamma_E+\log\mu_R+\tfrac12\log4\pi+\psi\big(\tfrac32\big)-\frac{\pi\sin\big[\pi\big(2\ii\wt{\nu}+\tfrac{p_{12}}2\big)\big]}{16\cos^2(2\pi\ii\wt{\nu})\mathcal{C}_{\frac32+2\ii\wt{\nu},3}^{p_1p_2}}\bigg)\Bigg].
\end{align}

It can be proved that the series in (\ref{61}), (\ref{62}), (\ref{63}) are convergent if $\wt{\nu}$ is real or $\wt{\nu}$ is complex and $0<\ii\wt{\nu}\leqslant1$. For $1<\ii\wt{\nu}<\frac32$, the expressions can be written in another way. See App. \ref{AppH} for more details.

\subsection{1-loop 4-point Functions in $\nabla\sigma\nabla\sigma$ Model}\label{Sec3.2}
Consider a model which is slightly difficult. The interaction is
\begin{equation}\label{64}
    \Delta\mathscr{L}=-\frac14a^{d-1-p_1}\phi_c^2g^{\mu\nu}(\nabla_\mu\sigma)(\nabla_\nu\sigma)-\frac14a^{d-1-p_2}\phi_c^2g^{\mu\nu}(\nabla_\mu\sigma)(\nabla_\nu\sigma).
\end{equation}
In this work, it is called $\nabla\sigma\nabla\sigma$ model. Before calculating the correlator, note that there is a quadratic term of $\nabla_\mu\sigma$ in the interaction. Therefore, there is a determinant $\big(\text{Det}[1+\frac12a^{-2-p_1}\phi_{c,+}^2+\frac12a^{-2-p_2}\phi_{c,+}^2][1+\frac12a^{-2-p_1}\phi_{c,-}^2+\frac12a^{-2-p_2}\phi_{c,-}^2]\big)^{-1/2}$ in the path integral of SK formalism. Using the relation $\text{Det}\,\mathscr{A}=\exp\text{Tr}\,\log\mathscr{A}$, the determinant can be recognized as a contribution to the effective Lagrangian \cite{Weinberg:1995mt}:
\begin{align}
    &\big(\text{Det}[1+\tfrac12a^{-2-p_1}\phi_{c,+}^2+\tfrac12a^{-2-p_2}\phi_{c,+}^2][1+\tfrac12a^{-2-p_1}\phi_{c,-}^2+\tfrac12a^{-2-p_2}\phi_{c,-}^2]\big)^{-1/2}\n \\
    &\propto\exp\bigg[-\Omega^{-1}\int\ud^{d+1}x\log\Big[1+\tfrac12a^{-2-p_1}\phi_{c,+}^2+\tfrac12a^{-2-p_2}\phi_{c,+}^2\Big]\n \\
    &-\Omega^{-1}\int\ud^{d+1}x\log\Big[1+\tfrac12a^{-2-p_1}\phi_{c,-}^2+\tfrac12a^{-2-p_2}\phi_{c,-}^2\Big]\bigg],
\end{align}
where $\Omega^{-1}$ can be written as $\delta^{d+1}(y-y)$, which is divergent. However, the delta function is recognized as 0 in dimensional regularization. There is no need to consider the contribution of this term.

One may attempt to apply K\"all\'en-Lehmann representation to the bubble of correlators of $\nabla\sigma\nabla\sigma$ model. However, it leads to an incorrect result. To find out this, correlators of $\nabla\sigma\nabla\sigma$ model will be calculated by using K\"all\'en-Lehmann representation in this subsection.

For the $\nabla\sigma\nabla\sigma$ model, $s$-channel of 4-point correlator of conformal scalar $\phi_c$ is
\begin{equation}
    \langle\phi_{c,\mathbf{k}_1}\phi_{c,\mathbf{k}_2}\phi_{c,\mathbf{k}_3}\phi_{c,\mathbf{k}_4}\rangle_s=(2\pi)^d\delta^{(d)}(\mathbf{k}_1+\mathbf{k}_2+\mathbf{k}_3+\mathbf{k}_4)\mathcal{L}_{\phi_c,\nabla\sigma\nabla\sigma}(\mathbf{k}_1,\mathbf{k}_2,\mathbf{k}_3,\mathbf{k}_4),
\end{equation}
where the loop amplitude $\mathcal{L}_{\phi_c,\nabla\sigma\nabla\sigma}$ is
\begin{align}
    \mathcal{L}_{\phi_c,\nabla\sigma\nabla\sigma}=&-\frac12\sum_{\mathsf{a},\mathsf{b}=\pm}\mathsf{ab}\int_{-\infty}^{\tau_f}\frac{\ud\tau_1}{(-\tau_1)^{d+1}}\frac{\ud\tau_2}{(-\tau_2)^{d+1}}\n \\
    &\times C_{\mathsf{a}}(k_1,\tau_1)C_{\mathsf{a}}(k_2,\tau_1)C_{\mathsf{b}}(k_3,\tau_2)C_{\mathsf{b}}(k_4,\tau_2)\mathcal{Q}_{\nabla\sigma\nabla\sigma,\wt{\nu},\mathsf{ab}}\big(k_s;\tau_1,\tau_2\big).
\end{align}
The loop momentum integral $\mathcal{Q}_{\nabla\sigma\nabla\sigma,\wt{\nu},\mathsf{ab}}$ is
\begin{align}
    \mathcal{Q}_{\nabla\sigma\nabla\sigma,\mathsf{ab}}\big(k_s;\tau_1,\tau_2\big)\equiv&\int\frac{\ud^d\mathbf{q}}{(2\pi)^d}\big[q_i q_{i'}D_{\wt{\nu},\mathsf{ab}}\big(q;\tau_1,\tau_2\big)(k_s-q)_i(k_s-q)_{i'}D_{\wt{\nu},\mathsf{ab}}\big(|\mathbf{k}_s-\mathbf{q}|;\tau_1,\tau_2\big)\n \\
    &+\partial_{\tau_1} \partial_{\tau_2}D_{\wt{\nu},\mathsf{ab}}\big(q;\tau_1,\tau_2\big)\partial_{\tau_1}\partial_{\tau_2}D_{\wt{\nu},\mathsf{ab}}\big(|\mathbf{k}_s-\mathbf{q}|;\tau_1,\tau_2\big)\big](-\tau_1)^2(-\tau_2)^2.
\end{align}
The amplitude $\mathcal{L}_{\phi_c,\nabla\sigma\nabla\sigma}$ can be generated from the following seed integral which comes from the interaction (\ref{64}):
\begin{equation}
    \mathcal{J}_{\nabla\sigma\nabla\sigma}^{p_1p_2}(r_1,r_2)\equiv-\frac12\sum_{\mathsf{a},\mathsf{b}=\pm}\mathsf{ab}k_s^{d+2+p_{12}}\int_{-\infty}^0\ud\tau_1\ud\tau_2(-\tau_1)^{p_1}(-\tau_2)^{p_2}e^{\ii\mathsf{a}k_{12}\tau_1+\ii\mathsf{a}k_{34}\tau_2}\mathcal{Q}_{\nabla\sigma\nabla\sigma,\mathsf{ab}}\big(k_s;\tau_1,\tau_2\big),
\end{equation}
in this way
\begin{equation}
    \mathcal{L}_{\phi_c,\nabla\sigma\nabla\sigma}=\frac{(-\tau_f)^{2(d-1)}}{16k_1k_2k_3k_4k_s^{d-2}}\mathcal{J}_{\nabla\sigma\nabla\sigma}^{-2,-2}(r_1,r_2).
\end{equation}

The spectral function of K\"all\'en-Lehmann representation in $\nabla\sigma\nabla\sigma$ model is
\begin{equation}
    \rho_{\nabla\sigma\nabla\sigma}^{\mathcal{P},0}(\wt{\nu}')=\frac{\Gamma\big(\frac{d+1}2\big)\wt{\nu}'\sinh(\pi\wt{\nu}')}{2^{8-d}\pi^{\frac{d+7}2}\Gamma(d)\Gamma\big(\frac d2+\ii\wt{\nu}'\big)\Gamma\big(\frac d2-\ii\wt{\nu}'\big)}\big(\tfrac{d^2}4+2\wt{\nu}^2-\wt{\nu}'^2\big)^2\prod_{\pm,\pm,\pm}\Gamma\bigg(\frac{\frac d2\pm\ii\wt{\nu}\pm\ii\wt{\nu}\pm\ii\wt{\nu}'}2\bigg).
\end{equation}
The detail of the calculation is in App. \ref{AppE}. Someone may expect that
\begin{equation}
    \mathcal{Q}_{\nabla\sigma\nabla\sigma,\mathsf{ab}}\big(k_s;\tau_1,\tau_2\big)=\int_{-\infty}^\infty\ud\wt{\nu}'\rho_{\nabla\sigma\nabla\sigma}^{\mathcal{P},0}(\wt{\nu}')D_{\wt{\nu}',\mathsf{ab}}(k_s;\tau_1,\tau_2),
\end{equation}
and then
\begin{equation}\label{73}
    \mathcal{J}_{\nabla\sigma\nabla\sigma}^{p_1p_2}(r_1,r_2)=\frac12\int_{-\infty}^\infty\ud\wt{\nu}'\rho_{\nabla\sigma\nabla\sigma}^{\mathcal{P},0}(\wt{\nu}')\mathcal{I}_{\wt{\nu}'}^{p_1p_2}(r_1,r_2).
\end{equation}

With the expression of the spectral function, it seems that the seed integral of $\nabla\sigma\nabla\sigma$ model can be computed, in the way which is similar to the method in $\sigma\sigma$ model. The expression is
\begin{equation}
    \mathcal{J}_{\nabla\sigma\nabla\sigma}^{p_1p_2}(r_1,r_2)=\mathcal{J}_{\nabla\sigma\nabla\sigma,\text{NS}}^{p_1p_2}(r_1,r_2)+\mathcal{J}_{\nabla\sigma\nabla\sigma,\text{LS}}^{p_1p_2}(r_1,r_2)+\mathcal{J}_{\nabla\sigma\nabla\sigma,\text{BG}}^{p_1p_2}(r_1,r_2).
\end{equation}
The signals are
\begin{align}
    \mathcal{J}_{\nabla\sigma\nabla\sigma,\text{NS}}^{p_1p_2}=&-\frac{(r_1r_2)^{d/2+2\ii\wt{\nu}}\sin[\pi(\fr{d}{2}+2\ii\wt{\nu})]}{32\pi^{d/2}\Gamma\big(\tfrac d2\big)\sin^2(\pi\ii\wt{\nu})}\sum_{n=0}^\infty\FR{(1+n)_{\frac d2-1} \big[(1+\ii\wt\nu+n)_{\frac d2-1}\big]^2(1+2\ii\wt\nu+n)_{\frac d2-1}}{(1+2\ii\wt\nu+2n)_{d-1}}\n\\
    &\times(\fr{d}{2}+2\ii\wt{\nu}+2n)\mathcal{C}_{2\ii\wt{\nu}+d/2+2n,d}^{p_1p_2}\mb{F}_{2\ii\wt{\nu}+d/2+2n,d}^{p_1}(r_1)\mb{F}_{2\ii\wt{\nu}+d/2+2n,d}^{p_2}(r_2)(r_1r_2)^{2n}\n \\
    &\times\big[\tfrac{d^2}4+2\wt{\nu}^2+\big(\tfrac d2+2\ii\wt{\nu}+2n\big)^2\big]^2\n \\
    &+(\wt{\nu}\rightarrow(-\wt{\nu})),
\end{align}
and
\begin{align}
    \mathcal{J}_{\nabla\sigma\nabla\sigma,\text{LS}}^{p_1p_2}=&\frac{(r_1/r_2)^{d/2+2\ii\wt{\nu}}\sin[\pi(\fr{d}{2}+2\ii \wt{\nu})]}{32\pi^{d/2}\Gamma\big(\tfrac d2\big)\sin^2(\pi\ii\wt{\nu})}\sum_{n=0}^\infty\FR{(1+n)_{\frac d2-1} \big[(1+\ii\wt\nu+n)_{\frac d2-1}\big]^2(1+2\ii\wt\nu+n)_{\frac d2-1}}{(1+2\ii\wt\nu+2n)_{d-1}}\n\\
    &\times(\fr{d}{2}+2\ii\wt{\nu}+2n)\mathcal{C}_{2\ii\wt{\nu}+d/2+2n,d}^{p_1p_2}\mb{F}_{2\ii\wt{\nu}+d/2+2n,d}^{p_1}(r_1)\mb{F}_{-2\ii\wt{\nu}-d/2-2n,d}^{p_2}(r_2)\bigg(\frac{r_1}{r_2}\bigg)^{2n}\n \\
    &\times\big[\tfrac{d^2}4+2\wt{\nu}^2+\big(\tfrac d2+2\ii\wt{\nu}+2n\big)^2\big]^2\n \\
    &+(\wt{\nu}\rightarrow(-\wt{\nu})).
\end{align}
The background can be divided into several terms:
\begin{equation}
    \mathcal{J}_{\nabla\sigma\nabla\sigma,\text{BG}}^{p_1p_2}=\mathcal{J}_{\nabla\sigma\nabla\sigma,(\text{B})}^{p_1p_2}+\mathcal{J}_{\nabla\sigma\nabla\sigma,(\text{3A})}^{p_1p_2}+\mathcal{J}_{\nabla\sigma\nabla\sigma,(\text{3B})}^{p_1p_2}+\mathcal{J}_{\nabla\sigma\nabla\sigma,(\text{3C})}^{p_1p_2}.
\end{equation}
The label (B) means the term is from the poles of Set B listed in Sec. \ref{Sec2.1} of the nonlocal and local tree integrals, while the labels (3A), (3B) and (3C) mean these terms come from the poles of Set A, B and C of the background tree integral. It can be proved that $\mathcal{J}_{\nabla\sigma\nabla\sigma,(\text{B})}^{p_1p_2}$ converges when $d\rightarrow3$, while the other terms diverge. The expressions of these terms are:
\begin{align}
    \mathcal{J}_{\nabla\sigma\nabla\sigma,(\text{B})}^{p_1p_2}=&\frac{\sin\big(\frac{\pi d}2\big)}{16\pi^{d/2}\Gamma\big(\tfrac d2\big)\sin^2(\pi\ii\wt{\nu})}\sum_{n=0}^\infty\FR{\big[(1+n)_{\frac d2-1}\big]^2 (1+\ii\wt\nu+n)_{\frac d2-1}(1-\ii\wt\nu+n)_{\frac d2-1}}{(1+2n)_{d-1}}(\fr{d}{2}+2n)\mathcal{C}_{d/2+2n,d}^{p_1p_2}\n\\
    &\times\mb{F}_{d/2+2n,d}^{p_1}(r_1)r_1^{2n+d/2}\big(\mb{F}_{d/2+2n,d}^{p_2}(r_2)r_2^{2n+d/2}-\mb{F}_{-d/2-2n,d}^{p_2}(r_2)r_2^{-2n-d/2}\big)\n \\
    &\times\bigg(\frac{d^2}4+2\wt{\nu}^2+\bigg(\frac d2+2n\bigg)^2\bigg)^2,
\end{align}
\begin{align}
    \mathcal{J}_{\nabla\sigma\nabla\sigma,(\text{3A})}^{p_1p_2}=&-\sum_{n=0}^\infty\sum_{\ell,j=0}^\infty\frac{(-1)^\ell\big(\frac d2-2\ii\wt{\nu}+2n\big)\mathcal{C}_{\frac d2-2\ii\wt{\nu}+2n,d}^{p_1p_2}\sin^2\big[\pi\big(\frac d2-2\ii\wt{\nu}\big)\big](\ell+1)_{2j+d+p_{12}+1}}{2^{2j+d+3}\pi^{\frac d2}\sin^2(\pi\ii\wt{\nu})\Gamma\big(\frac d2\big)n!\big(\frac{\ell+2\ii\wt{\nu}+p_2+1}2-n\big)_{j+1}\big(\frac{\ell+d-2\ii\wt{\nu}+p_2+1}2+n\big)_{j+1}}\n \\
    &\times\Gamma\Bigg[\begin{matrix}
        \frac d2+n, \frac d2-\ii\wt{\nu}+n, \frac d2-2\ii\wt{\nu}+n, -\ii\wt{\nu}+\frac12+n \\
        \frac{d+1}2-\ii\wt{\nu}+n, -2\ii\wt{\nu}+1+n, -\ii\wt{\nu}+1+n
    \end{matrix}\Bigg]\n \\
    &\times r_1^{2j+d+p_{12}+2}\bigg(\frac{r_1}{r_2}\bigg)^\ell\big[\tfrac{d^2}4+2\wt{\nu}^2+\big(\tfrac d2-2\ii\wt{\nu}+2n\big)^2\big]^2,
\end{align}
\begin{align}
    \mathcal{J}_{\nabla\sigma\nabla\sigma,(\text{3B})}^{p_1p_2}=&\sum_{n=0}^\infty\sum_{\ell,j=0}^\infty\frac{(-1)^\ell\big(\frac d2+2n\big)\mathcal{C}_{\frac d2+2n,d}^{p_1p_2}\sin^2\big(\frac{\pi d}2\big)(\ell+1)_{2j+d+p_{12}+1}}{2^{2j+d+2}\pi^{\frac d2}\sin^2(\pi\ii\wt{\nu})\Gamma\big(\frac d2\big)n!\big(\frac{\ell+p_2+1}2-n\big)_{j+1}\big(\frac{\ell+d+p_2+1}2+n\big)_{j+1}}\n \\
    &\times\Gamma\Bigg[\begin{matrix}
        \frac d2+\ii\wt{\nu}+n, \frac d2+n, \frac d2-\ii\wt{\nu}+n, \frac12+n \\
        \frac{d+1}2+n, -\ii\wt{\nu}+1+n, \ii\wt{\nu}+1+n
    \end{matrix}\Bigg]\n \\
    &\times r_1^{2j+d+p_{12}+2}\bigg(\frac{r_1}{r_2}\bigg)^\ell\big[\tfrac{d^2}4+2\wt{\nu}^2+\big(\tfrac d2+2n\big)^2\big]^2,
\end{align}
\begin{align}
    \mathcal{J}_{\nabla\sigma\nabla\sigma,(\text{3C})}^{p_1p_2}=&-\sum_{n=0}^\infty\sum_{\ell,j=0}^\infty\frac{(-1)^\ell\big(\frac d2+2\ii\wt{\nu}+2n\big)\mathcal{C}_{\frac d2+2\ii\wt{\nu}+2n,d}^{p_1p_2}\sin^2\big[\pi\big(\frac d2+2\ii\wt{\nu}\big)\big](\ell+1)_{2j+d+p_{12}+1}}{2^{2j+d+3}\pi^{\frac d2}\sin^2(\pi\ii\wt{\nu})\Gamma\big(\frac d2\big)n!\big(\frac{\ell-2\ii\wt{\nu}+p_2+1}2-n\big)_{j+1}\big(\frac{\ell+d+2\ii\wt{\nu}+p_2+1}2+n\big)_{j+1}}\n \\
    &\times\Gamma\Bigg[\begin{matrix}
        \frac d2+2\ii\wt{\nu}+n, \frac d2+\ii\wt{\nu}+n, \frac d2+n, \ii\wt{\nu}+\frac12+n \\
        \frac{d+1}2+\ii\wt{\nu}+n, \ii\wt{\nu}+1+n, 2\ii\wt{\nu}+1+n
    \end{matrix}\Bigg]\n \\
    &\times r_1^{2j+d+p_{12}+2}\bigg(\frac{r_1}{r_2}\bigg)^\ell\big[\tfrac{d^2}4+2\wt{\nu}^2+\big(\tfrac d2+2\ii\wt{\nu}+2n\big)^2\big]^2.
\end{align}

These series of $n$ in $\mathcal{J}_{\nabla\sigma\nabla\sigma,(\text{3A})}^{p_1p_2}$, $\mathcal{J}_{\nabla\sigma\nabla\sigma,(\text{3B})}^{p_1p_2}$ and $\mathcal{J}_{\nabla\sigma\nabla\sigma,(\text{3C})}^{p_1p_2}$ diverge when $d\rightarrow3$ if $j=0,1,2$. To analyze the divergence of the series, the asymptotic expansions for large $(n+1)$ are computed. The asymptotic expansions are:
\begin{align}\label{82}
    &\sum_{\ell=0}^\infty\frac{(-1)^\ell\big(\frac d2-2\ii\wt{\nu}+2n\big)\mathcal{C}_{\frac d2-2\ii\wt{\nu},d}^{p_1p_2}\sin^2\big[\pi\big(\frac d2-2\ii\wt{\nu}\big)\big](\ell+1)_{d+p_{12}+1}}{2^{d+3}\pi^{\frac d2}\sin^2(\pi\ii\wt{\nu})\Gamma\big(\frac d2\big)n!\big(\frac{\ell+2\ii\wt{\nu}+p_2+1}2-n\big)\big(\frac{\ell+d-2\ii\wt{\nu}+p_2+1}2+n\big)}\n \\
    &\times\Gamma\Bigg[\begin{matrix}
        \frac d2+n, \frac d2-\ii\wt{\nu}+n, \frac d2-2\ii\wt{\nu}+n, -\ii\wt{\nu}+\frac12+n \\
        \frac{d+1}2-\ii\wt{\nu}+n, -2\ii\wt{\nu}+n+1, -\ii\wt{\nu}+1+n
    \end{matrix}\Bigg]\n \\
    &\times r_1^{d+p_{12}+2}\bigg(\frac{r_1}{r_2}\bigg)^\ell\big[\tfrac{d^2}4+2\wt{\nu}^2+\big(\tfrac d2-2\ii\wt{\nu}+2n\big)^2\big]^2\n \\
    \sim&\sum_{\ell=0}^\infty\frac{(-1)^\ell\mathcal{C}_{\frac d2-2\ii\wt{\nu},d}^{p_1p_2}\sin^2\big[\pi\big(\frac d2-2\ii\wt{\nu}\big)\big](\ell+1)_{d+p_{12}+1}}{2^{d-2}\pi^{\frac d2}\sin^2(\pi\ii\wt{\nu})\Gamma\big(\frac d2\big)}r_1^{d+p_{12}+2}\bigg(\frac{r_1}{r_2}\bigg)^\ell\n \\
    &\times\bigg(-(n+1)^{d-3}\sum_{\iota=-1}^3\Upsilon_{\nabla\sigma\nabla\sigma,d,\text{A},\ell,0,\iota}^{p_2}(n+1)^\iota\bigg),
\end{align}
\begin{align}\label{83}
    &\sum_{\ell=0}^\infty\frac{(-1)^\ell\big(\frac d2-2\ii\wt{\nu}+2n\big)\mathcal{C}_{\frac d2-2\ii\wt{\nu},d}^{p_1p_2}\sin^2\big[\pi\big(\frac d2-2\ii\wt{\nu}\big)\big](\ell+1)_{d+p_{12}+3}}{2^{d+5}\pi^{\frac d2}\sin^2(\pi\ii\wt{\nu})\Gamma\big(\frac d2\big)n!\big(\frac{\ell+2\ii\wt{\nu}+p_2+1}2-n\big)_2\big(\frac{\ell+d-2\ii\wt{\nu}+p_2+1}2+n\big)_2}\n \\
    &\times\Gamma\Bigg[\begin{matrix}
        \frac d2+n, \frac d2-\ii\wt{\nu}+n, \frac d2-2\ii\wt{\nu}+n, -\ii\wt{\nu}+\frac12+n \\
        \frac{d+1}2-\ii\wt{\nu}+n, -2\ii\wt{\nu}+n+1, -\ii\wt{\nu}+1+n
    \end{matrix}\Bigg]\n \\
    &\times r_1^{d+p_{12}+4}\bigg(\frac{r_1}{r_2}\bigg)^\ell\big[\tfrac{d^2}4+2\wt{\nu}^2+\big(\tfrac d2-2\ii\wt{\nu}+2n\big)^2\big]^2\n \\
    \sim&\sum_{\ell=0}^\infty\frac{(-1)^\ell\mathcal{C}_{\frac d2-2\ii\wt{\nu},d}^{p_1p_2}\sin^2\big[\pi\big(\frac d2-2\ii\wt{\nu}\big)\big](\ell+1)_{d+p_{12}+3}}{2^d\pi^{\frac d2}\sin^2(\pi\ii\wt{\nu})\Gamma\big(\frac d2\big)}r_1^{d+p_{12}+4}\bigg(\frac{r_1}{r_2}\bigg)^\ell\n \\
    &\times\bigg((n+1)^{d-3}\Big[\Upsilon_{\nabla\sigma\nabla\sigma,d,\text{A},\ell,1,1}^{p_2}(n+1)+\Upsilon_{\nabla\sigma\nabla\sigma,d,\text{A},\ell,1,0}^{p_2}+\Upsilon_{\nabla\sigma\nabla\sigma,d,\text{A},\ell,1,-1}^{p_2}/(n+1)\Big]\bigg),
\end{align}
\begin{align}\label{84}
    &\sum_{\ell=0}^\infty\frac{(-1)^\ell\big(\frac d2-2\ii\wt{\nu}+2n\big)\mathcal{C}_{\frac d2-2\ii\wt{\nu},d}^{p_1p_2}\sin^2\big[\pi\big(\frac d2-2\ii\wt{\nu}\big)\big](\ell+1)_{d+p_{12}+5}}{2^{d+7}\pi^{\frac d2}\sin^2(\pi\ii\wt{\nu})\Gamma\big(\frac d2\big)n!\big(\frac{\ell+2\ii\wt{\nu}+p_2+1}2-n\big)_3\big(\frac{\ell+d-2\ii\wt{\nu}+p_2+1}2+n\big)_3}\n \\
    &\times\Gamma\Bigg[\begin{matrix}
        \frac d2+n, \frac d2-\ii\wt{\nu}+n, \frac d2-2\ii\wt{\nu}+n, -\ii\wt{\nu}+\frac12+n \\
        \frac{d+1}2-\ii\wt{\nu}+n, -2\ii\wt{\nu}+n+1, -\ii\wt{\nu}+1+n
    \end{matrix}\Bigg]\n \\
    &\times r_1^{d+p_{12}+6}\bigg(\frac{r_1}{r_2}\bigg)^\ell\big[\tfrac{d^2}4+2\wt{\nu}^2+\big(\tfrac d2-2\ii\wt{\nu}+2n\big)^2\big]^2\n \\
    \sim&\sum_{\ell=0}^\infty\frac{(-1)^\ell\mathcal{C}_{\frac d2-2\ii\wt{\nu},d}^{p_1p_2}\sin^2\big[\pi\big(\frac d2-2\ii\wt{\nu}\big)\big](\ell+1)_{d+p_{12}+5}}{2^{d+2}\pi^{\frac d2}\sin^2(\pi\ii\wt{\nu})\Gamma\big(\frac d2\big)}r_1^{d+p_{12}+6}\bigg(\frac{r_1}{r_2}\bigg)^\ell\bigg(-(n+1)^{d-4}\bigg),
\end{align}
\begin{align}\label{85}
    &\sum_{\ell=0}^\infty\frac{(-1)^\ell\big(\frac d2+2n\big)\mathcal{C}_{\frac d2,d}^{p_1p_2}\sin^2\big(\frac{\pi d}2\big)(\ell+1)_{d+p_{12}+1}}{2^{d+2}\pi^{\frac d2}\sin^2(\pi\ii\wt{\nu})\Gamma\big(\frac d2\big)n!\big(\frac{\ell+p_2+1}2-n\big)\big(\frac{\ell+d+p_2+1}2+n\big)}\n \\
    &\times\Gamma\Bigg[\begin{matrix}
        \frac d2+\ii\wt{\nu}+n, \frac d2+n, \frac d2-\ii\wt{\nu}+n, \frac12+n \\
        \frac{d+1}2+n, -\ii\wt{\nu}+1+n, \ii\wt{\nu}+1+n
    \end{matrix}\Bigg]\n \\
    &\times r_1^{d+p_{12}+2}\bigg(\frac{r_1}{r_2}\bigg)^\ell\big[\tfrac{d^2}4+2\wt{\nu}^2+\big(\tfrac d2+2n\big)^2\big]^2\n \\
    \sim&\sum_{\ell=0}^\infty\frac{(-1)^\ell\mathcal{C}_{\frac d2,d}^{p_1p_2}\sin^2\big(\frac{\pi d}2\big)(\ell+1)_{d+p_{12}+1}}{2^{d-3}\pi^{\frac d2}\sin^2(\pi\ii\wt{\nu})\Gamma\big(\frac d2\big)}r_1^{d+p_{12}+2}\bigg(\frac{r_1}{r_2}\bigg)^\ell\n \\
    &\times\bigg(-(n+1)^{d-3}\Big[\Upsilon_{\nabla\sigma\nabla\sigma,d,\text{B},\ell,0,3}^{p_2}(n+1)^3+\Upsilon_{\nabla\sigma\nabla\sigma,d,\text{B},\ell,0,2}^{p_2}(n+1)^2+\Upsilon_{\nabla\sigma\nabla\sigma,d,\text{B},\ell,0,1}^{p_2}(n+1)\n \\
    &+\Upsilon_{\nabla\sigma\nabla\sigma,d,\text{B},\ell,0,0}^{p_2}+\Upsilon_{\nabla\sigma\nabla\sigma,d,\text{B},\ell,0,-1}^{p_2}/(n+1)\Big]\bigg),
\end{align}
\begin{align}\label{86}
    &\sum_{\ell=0}^\infty\frac{(-1)^\ell\big(\frac d2+2n\big)\mathcal{C}_{\frac d2,d}^{p_1p_2}\sin^2\big(\frac{\pi d}2\big)(\ell+1)_{d+p_{12}+3}}{2^{d+4}\pi^{\frac d2}\sin^2(\pi\ii\wt{\nu})\Gamma\big(\frac d2\big)n!\big(\frac{\ell+p_2+1}2-n\big)_2\big(\frac{\ell+d+p_2+1}2+n\big)_2}\n \\
    &\times\Gamma\Bigg[\begin{matrix}
        \frac d2+\ii\wt{\nu}+n, \frac d2+n, \frac d2-\ii\wt{\nu}+n, \frac12+n \\
        \frac{d+1}2+n, -\ii\wt{\nu}+1+n, \ii\wt{\nu}+1+n
    \end{matrix}\Bigg]\n \\
    &\times r_1^{d+p_{12}+4}\bigg(\frac{r_1}{r_2}\bigg)^\ell\big[\tfrac{d^2}4+2\wt{\nu}^2+\big(\tfrac d2+2n\big)^2\big]^2\n \\
    \sim&\sum_{\ell=0}^\infty\frac{(-1)^\ell\mathcal{C}_{\frac d2,d}^{p_1p_2}\sin^2\big(\frac{\pi d}2\big)(\ell+1)_{d+p_{12}+3}}{2^{d-1}\pi^{\frac d2}\sin^2(\pi\ii\wt{\nu})\Gamma\big(\frac d2\big)}r_1^{d+p_{12}+4}\bigg(\frac{r_1}{r_2}\bigg)^\ell\n \\
    &\times\bigg((n+1)^{d-3}\Big[\Upsilon_{\nabla\sigma\nabla\sigma,d,\text{B},\ell,1,1}^{p_2}(n+1)+\Upsilon_{\nabla\sigma\nabla\sigma,d,\text{B},\ell,1,0}^{p_2}+\Upsilon_{\nabla\sigma\nabla\sigma,d,\text{B},\ell,1,-1}^{p_2}/(n+1)\Big]\bigg),
\end{align}
\begin{align}\label{87}
    &\sum_{\ell=0}^\infty\frac{(-1)^\ell\big(\frac d2+2n\big)\mathcal{C}_{\frac d2,d}^{p_1p_2}\sin^2\big(\frac{\pi d}2\big)(\ell+1)_{d+p_{12}+5}}{2^{d+6}\pi^{\frac d2}\sin^2(\pi\ii\wt{\nu})\Gamma\big(\frac d2\big)n!\big(\frac{\ell+p_2+1}2-n\big)_3\big(\frac{\ell+d+p_2+1}2+n\big)_3}\n \\
    &\times\Gamma\Bigg[\begin{matrix}
        \frac d2+\ii\wt{\nu}+n, \frac d2+n, \frac d2-\ii\wt{\nu}+n, \frac12+n \\
        \frac{d+1}2+n, -\ii\wt{\nu}+1+n, \ii\wt{\nu}+1+n
    \end{matrix}\Bigg]\n \\
    &\times r_1^{d+p_{12}+6}\bigg(\frac{r_1}{r_2}\bigg)^\ell\big[\tfrac{d^2}4+2\wt{\nu}^2+\big(\tfrac d2+2n\big)^2\big]^2\n \\
    \sim&\sum_{\ell=0}^\infty\frac{(-1)^\ell\mathcal{C}_{\frac d2,d}^{p_1p_2}\sin^2\big(\frac{\pi d}2\big)(\ell+1)_{d+p_{12}+5}}{2^{d+1}\pi^{\frac d2}\sin^2(\pi\ii\wt{\nu})\Gamma\big(\frac d2\big)}r_1^{d+p_{12}+6}\bigg(\frac{r_1}{r_2}\bigg)^\ell\bigg(-(n+1)^{d-3}\bigg).
\end{align}
The explict expressions of $\Upsilon_{\nabla\sigma\nabla\sigma,d,\text{A/B},\ell,i,j}^{p_2}$ can be found in App. \ref{AppF}. The summations in $\mathcal{J}_{\nabla\sigma\nabla\sigma,(\text{3C})}$ can be derived from the summations (\ref{82}), (\ref{83}), (\ref{84}) in $\mathcal{J}_{\nabla\sigma\nabla\sigma,(\text{3A})}$ by taking the transformation $\wt{\nu}\rightarrow(-\wt{\nu})$. After using the definition of Riemann zeta function, $\mathcal{J}_{\nabla\sigma\nabla\sigma,(\text{3A})}$ is regularized as
\begin{align}
    \mathcal{J}_{\nabla\sigma\nabla\sigma,(\text{3A})}^{p_1p_2}=&-\sum_{n=0}^\infty\sum_{\ell,j=0}^\infty\frac{(-1)^\ell\big(\frac d2-2\ii\wt{\nu}+2n\big)\mathcal{C}_{\frac d2-2\ii\wt{\nu}+2n,d}^{p_1p_2}\sin^2\big[\pi\big(\frac d2-2\ii\wt{\nu}\big)\big](\ell+1)_{2j+d+p_{12}+1}}{2^{2j+d+3}\pi^{\frac d2}\sin^2(\pi\ii\wt{\nu})\Gamma\big(\frac d2\big)n!\big(\frac{\ell+2\ii\wt{\nu}+p_2+1}2-n\big)_{j+1}\big(\frac{\ell+d-2\ii\wt{\nu}+p_2+1}2+n\big)_{j+1}}\n \\
    &\times\Gamma\Bigg[\begin{matrix}
        \frac d2+n, \frac d2-\ii\wt{\nu}+n, \frac d2-2\ii\wt{\nu}+n, -\ii\wt{\nu}+n+\frac12 \\
        \frac{d+1}2-\ii\wt{\nu}+n, -2\ii\wt{\nu}+n+1, -\ii\wt{\nu}+n+1
    \end{matrix}\Bigg]\n \\
    &\times r_1^{2j+d+p_{12}+2}\bigg(\frac{r_1}{r_2}\bigg)^\ell\big[\tfrac{d^2}4+2\wt{\nu}^2+\big(\tfrac d2-2\ii\wt{\nu}+2n\big)^2\big]^2\n \\
    =&\sum_{\ell=0}^\infty\Gamma\Bigg[\begin{matrix}
        \frac d2+n, \frac d2-\ii\wt{\nu}+n, \frac d2-2\ii\wt{\nu}+n, -\ii\wt{\nu}+n+\frac12 \\
        \frac{d+1}2-\ii\wt{\nu}+n, -2\ii\wt{\nu}+n+1, -\ii\wt{\nu}+n+1
    \end{matrix}\Bigg]\frac{(-1)^\ell\mathcal{C}_{\frac d2-2\ii\wt{\nu},d}^{p_1p_2}\sin^2\big[\pi\big(\frac d2-2\ii\wt{\nu}\big)\big]}{2^{d-2}\pi^{\frac d2}\sin^2(\pi\ii\wt{\nu})\Gamma\big(\frac d2\big)}\n \\
    &\times\Bigg\{-\Bigg[\sum_{n=0}^\infty\sum_{j=3}^\infty\frac{\big(\frac d4-\ii\wt{\nu}+n\big)(\ell+1)_{2j+d+p_{12}+1}}{2^{2j}n!\big(\frac{\ell+2\ii\wt{\nu}+p_2+1}2-n\big)_{j+1}\big(\frac{\ell+d-2\ii\wt{\nu}+p_2+1}2+n\big)_{j+1}}\n \\
    &\times r_1^{2j+d+p_{12}+2}\bigg(\frac{r_1}{r_2}\bigg)^\ell\big[\tfrac{d^2}{16}+\tfrac{\wt{\nu}^2}2+\big(\tfrac d4-\ii\wt{\nu}+n\big)^2\big]^2\Bigg]\n \\
    &-r_1^{d+p_{12}+2}\bigg(\frac{r_1}{r_2}\bigg)^\ell(\ell+1)_{d+p_{12}+1}\n \\
    &\times\Bigg[\sum_{n=0}^\infty\Bigg(\frac{\big(\frac d4-\ii\wt{\nu}+n\big)}{n!\big(\frac{\ell+2\ii\wt{\nu}+p_2+1}2-n\big)\big(\frac{\ell+d-2\ii\wt{\nu}+p_2+1}2+n\big)}\big[\tfrac{d^2}{16}+\tfrac{\wt{\nu}^2}{2}+\big(\tfrac d4-\ii\wt{\nu}+n\big)^2\big]^2\n \\
    &+(n+1)^{d-3}\sum_{\iota=-1}^3\Upsilon_{\nabla\sigma\nabla\sigma,d,\text{A},\ell,0,\iota}^{p_2}(n+1)^\iota\Bigg)-\sum_{\iota=-1}^3\Upsilon_{\nabla\sigma\nabla\sigma,d,\text{A},\ell,0,\iota}^{p_2}\zeta(-d+3-\iota)\Bigg]\n \\
    &-\frac14r_1^{d+p_{12}+4}\bigg(\frac{r_1}{r_2}\bigg)^\ell(\ell+1)_{d+p_{12}+3}\n \\
    &\times\Bigg[\sum_{n=0}^\infty\Bigg(\frac{\big(\frac d4-\ii\wt{\nu}+n\big)}{n!\big(\frac{\ell+2\ii\wt{\nu}+p_2+1}2-n\big)_2\big(\frac{\ell+d-2\ii\wt{\nu}+p_2+1}2+n\big)_2}\big[\tfrac{d^2}{16}+\tfrac{\wt{\nu}^2}2+\big(\tfrac d4-\ii\wt{\nu}+n\big)^2\big]^2\n \\
    &-(n+1)^{d-3}\sum_{\iota=-1}^1\Upsilon_{\nabla\sigma\nabla\sigma,d,\text{A},\ell,1,\iota}^{p_2}(n+1)^\iota\Bigg)+\sum_{\iota=-1}^1\Upsilon_{\nabla\sigma\nabla\sigma,d,\text{A},\ell,1,\iota}^{p_2}\zeta(-d+3-\iota)\Bigg]\n \\
    &-\frac1{16}r_1^{d+p_{12}+6}\bigg(\frac{r_1}{r_2}\bigg)^\ell(\ell+1)_{d+p_{12}+5}\Bigg[\sum_{n=0}^\infty\Bigg(\frac{\big(\frac d4-\ii\wt{\nu}+n\big)}{n!\big(\frac{\ell+2\ii\wt{\nu}+p_2+1}2-n\big)_3\big(\frac{\ell+d-2\ii\wt{\nu}+p_2+1}2+n\big)_3}\n \\
    &\times\big[\tfrac{d^2}{16}+\tfrac{\wt{\nu}^2}2+\big(\tfrac d4-\ii\wt{\nu}+n\big)^2\big]^2+(n+1)^{d-4}\Bigg)-\zeta(-d+4)\Bigg]\Bigg\}.
\end{align}
It can be seen that the divergence of $\mathcal{J}_{\nabla\sigma\nabla\sigma,(\text{3A})}$ when $d\rightarrow3$ is:
\begin{align}
    \text{Div}\mathcal{J}_{\nabla\sigma\nabla\sigma,(\text{3A})}^{p_1p_2}=&\sum_{\ell=0}^\infty\frac{(-1)^\ell\mathcal{C}_{\frac d2-2\ii\wt{\nu},3}^{p_1p_2}\sin^2\big[\pi\big(\frac d2-2\ii\wt{\nu}\big)\big](\ell+1)_{d+p_{12}+1}}{\pi^2\sin^2(\pi\ii\wt{\nu})}r_1^{d+p_{12}+2}\bigg(\frac{r_1}{r_2}\bigg)^\ell\n \\
    &\times\Bigg[\bigg(8\bigg(\Big(\tfrac d2+p_2+\ell+1\Big)^4-\tfrac{d^2}2\Big(\tfrac d2+p_2+\ell+1\Big)^2+\tfrac{d^4}{16}\bigg)\n \\
    &+(76+48\wt{\nu}^2)\bigg(\Big(\tfrac d2+p_2+\ell+1\Big)^2-\tfrac{d^2}4\bigg)+189+264\wt{\nu}^2+80\wt{\nu}^4\bigg)/(128(3-d))\Bigg]\n \\
    &+\sum_{\ell=0}^\infty\frac{(-1)^\ell\mathcal{C}_{\frac d2-2\ii\wt{\nu},3}^{p_1p_2}\sin^2\big[\pi\big(\frac d2-2\ii\wt{\nu}\big)\big](\ell+1)_{d+p_{12}+3}}{4\pi^2\sin^2(\pi\ii\wt{\nu})}r_1^{d+p_{12}+4}\bigg(\frac{r_1}{r_2}\bigg)^\ell\n \\
    &\times\Bigg[-\bigg(4\Big[\Big(\tfrac d2+p_2+\ell+2\Big)^2+1-\tfrac{d^2}4\Big]+19+12\wt{\nu}^2\bigg)/(8(3-d))\Bigg]\n \\
    &+\sum_{\ell=0}^\infty\frac{(-1)^\ell\mathcal{C}_{\frac d2-2\ii\wt{\nu},3}^{p_1p_2}\sin^2\big[\pi\big(\frac d2-2\ii\wt{\nu}\big)\big](\ell+1)_{d+p_{12}+5}}{16\pi^2\sin^2(\pi\ii\wt{\nu})}r_1^{d+p_{12}+6}\bigg(\frac{r_1}{r_2}\bigg)^\ell\frac1{3-d}.
\end{align}
The same procedure can be applied to $\mathcal{J}_{\nabla\sigma\nabla\sigma,(\text{3B})}$ and $\mathcal{J}_{\nabla\sigma\nabla\sigma,(\text{3C})}$. After calculating the divergences of $\mathcal{J}_{\nabla\sigma\nabla\sigma,(\text{3A})}$, $\mathcal{J}_{\nabla\sigma\nabla\sigma,(\text{3B})}$ and $\mathcal{J}_{\nabla\sigma\nabla\sigma,(\text{3C})}$, the divergence of the loop seed integral is
\begin{align}
    \text{Div}\mathcal{J}_{\nabla\sigma\nabla\sigma}^{p_1p_2}=&-\sum_{\ell=0}^\infty\frac{(-1)^\ell\sin\big[\tfrac\pi2(d+p_{12})\big](\ell+1)_{d+p_{12}+1}}{2\pi^2}r_1^{d+p_{12}+2}\bigg(\frac{r_1}{r_2}\bigg)^\ell\n \\
    &\times\Bigg[\bigg(8\bigg(\Big(\tfrac d2+p_2+\ell+1\Big)^4-\tfrac{d^2}2\Big(\tfrac d2+p_2+\ell+1\Big)^2+\tfrac{d^4}{16}\bigg)\n \\
    &+(76+48\wt{\nu}^2)\bigg(\Big(\tfrac d2+p_2+\ell+1\Big)^2-\tfrac{d^2}4\bigg)+189+264\wt{\nu}^2+80\wt{\nu}^4\bigg)/(128(3-d))\Bigg]\n \\
    &-\sum_{\ell=0}^\infty\frac{(-1)^\ell\sin\big[\tfrac\pi2(d+p_{12})\big](\ell+1)_{d+p_{12}+3}}{8\pi^2}r_1^{d+p_{12}+4}\bigg(\frac{r_1}{r_2}\bigg)^\ell\n \\
    &\times\Bigg[-\bigg(4\Big[\Big(\tfrac d2+p_2+\ell+2\Big)^2+1-\tfrac{d^2}4\Big]+19+12\wt{\nu}^2\bigg)/(8(3-d))\Bigg]\n \\
    &-\sum_{\ell=0}^\infty\frac{(-1)^\ell(\ell+1)_{d+p_{12}+5}}{32\pi^2}r_1^{d+p_{12}+6}\bigg(\frac{r_1}{r_2}\bigg)^\ell\frac1{3-d}.
\end{align}

Consider the following interaction as the counter term of $\nabla\sigma\nabla\sigma$ model:
\begin{align}\label{91}
    \Delta\mathscr{L}=&-\frac1{24}\delta_4a^{d+1}\nabla^2(a^{-2-p_1}\phi_c^2)\nabla^2(a^{-2-p_2}\phi_c^2)-\frac1{24}\delta_2a^{d-1-p_1}\phi_c^2\nabla^2(a^{-2-p_2}\phi_c^2)-\frac1{24}\delta_0a^{d-3-p_{12}}\phi_c^4.
\end{align}
If only the contribution of $s$-channel is considered, the correlator generated by the interaction is:
\begin{align}
    \delta\mathcal{J}_{\nabla\sigma\nabla\sigma}^{p_1p_2}=&-\frac{\ii}3\delta_4\sum_{\mathsf{a}=\pm}k_s^{d+2+p_{12}}\mathsf{a}\int_{-\infty}^0\frac{\ud\tau}{(-\tau)^{d+1}}\big[-\tau^2\partial_\tau^2+(d-1)\tau\partial_\tau-\tau^2k_s^2\big]\big[(-\tau)^{d+1+p_1}e^{\ii\mathsf{a}k_{12}\tau}\big]\n \\
    &\times\big[-\tau^2\partial_\tau^2+(d-1)\tau\partial_\tau-\tau^2k_s^2\big]\big[(-\tau)^{d+1+p_2}e^{\ii\mathsf{a}k_{34}\tau}\big]\n \\
    &-\frac{\ii}6\delta_2\Bigg[\sum_{\mathsf{a}=\pm}k_s^{d+2+p_{12}}\mathsf{a}\int_{-\infty}^0\frac{\ud\tau}{(-\tau)^{d+1}}\big[(-\tau)^{d+1+p_1}e^{\ii\mathsf{a}k_{12}\tau}\big]\n \\
    &\times\big[-\tau^2\partial_\tau^2+(d-1)\tau\partial_\tau-\tau^2k_s^2\big]\big[(-\tau)^{d+1+p_2}e^{\ii\mathsf{a}k_{34}\tau}\big]+(k_{12}\leftrightarrow k_{34})\Bigg]\n \\
    &-\frac{\ii}3\delta_0\sum_{\mathsf{a}=\pm}k_s^{d+2+p_{12}}\mathsf{a}\int_{-\infty}^0\frac{\ud\tau}{(-\tau)^{d+1}}\big[(-\tau)^{d+1+p_1}e^{\ii\mathsf{a}k_{12}\tau}\big]\big[(-\tau)^{d+1+p_2}e^{\ii\mathsf{a}k_{34}\tau}\big]\n \\
    =&\sum_{\ell=0}^\infty2(-1)^\ell\sin\big[\tfrac\pi2(d+p_{12})\big](\ell+1)_{d+p_{12}+1}r_1^{d+p_{12}+2}\bigg(\frac{r_1}{r_2}\bigg)^\ell\n \\
    &\times\Big(\tfrac13\delta_4\big[\big(\tfrac d2+p_2+\ell+1\big)^4-\tfrac{d^2}2\big(\tfrac d2+p_2+\ell+1\big)^2+\tfrac{d^4}{16}\big]\n \\
    &+\tfrac13\delta_2\big[\big(\tfrac d2+p_2+\ell+1\big)^2-\tfrac{d^2}4\big]+\tfrac13\delta_0\Big)\n \\
    &-\sum_{\ell=0}^\infty2(-1)^\ell\sin\big[\tfrac\pi2(d+p_{12})\big](\ell+1)_{d+p_{12}+3}r_1^{d+p_{12}+4}\bigg(\frac{r_1}{r_2}\bigg)^\ell\n \\
    &\times\Big(\tfrac23\delta_4\big[\big(\tfrac d2+p_2+\ell+2\big)^2+1-\tfrac{d^2}4\big]+\tfrac13\delta_2\Big)\n \\
    &+\sum_{\ell=0}^\infty\tfrac23\delta_{4}(-1)^\ell\sin\big[\tfrac\pi2(d+p_{12})\big](\ell+1)_{d+p_{12}+5}r_1^{d+p_{12}+6}\bigg(\frac{r_1}{r_2}\bigg)^\ell.
\end{align}
To subtract the divergence of the seed integral, the coefficients $\delta_4,\delta_2,\delta_0$ should be
\begin{equation}
    \frac13\delta_4=\frac1{64\pi^2}\frac1{3-d},\quad\frac13\delta_2=\frac{19+12\wt{\nu}^2}{128\pi^2}\frac1{3-d},\quad\frac13\delta_0=\frac{189+264\wt{\nu}^2+80\wt{\nu}^4}{512\pi^2}\frac1{3-d}.
\end{equation}
If the flat space limit $H\rightarrow0$ is taken, the coefficients of the three divergent contact terms are:
\begin{equation}
    \frac13\delta_4=\frac1{64\pi^2}\frac1{3-d},\quad\frac13\delta_2=\frac{3m^2}{32\pi^2}\frac1{3-d},\quad\frac13\delta_0=\frac{5m^4}{32\pi^2}\frac1{3-d}.
\end{equation}
Here $m$ is the mass of the scalar field $\sigma$.

The result is different from (\ref{202}). The reason is that the spectral representation
\begin{align}\label{95}
    \int_{-\infty}^\infty\ud\wt{\nu}'\rho_{\nabla\sigma\nabla\sigma}^{\mathcal{P},0}(\wt{\nu}')D_{\wt{\nu}',\pm\pm}(y,y')
\end{align}
is not equivalent to the bubble 
\begin{equation}\label{96}
    g^{\mu\nu}(y)g^{\mu'\nu'}(y')\nabla_\mu\nabla_{\mu'}D_{\wt{\nu},\pm\pm}(y,y')\nabla_\nu\nabla_{\nu'}D_{\wt{\nu},\pm\pm}(y,y'). 
\end{equation}
The derivative acting in the step funtion $\theta$ is crucial for obtaining the correct result. See \cite{Chen:2017ryl} for more details. The integral (\ref{95}) is actually the spectral representation of the following formula:
\begin{align}\label{97}
    &g^{\mu\nu}(y)g^{\mu'\nu'}(y')\nabla_\mu\nabla_{\mu'}D_{\wt{\nu},\mp\pm}(y,y')\nabla_\nu\nabla_{\nu'}D_{\wt{\nu},\mp\pm}(y,y')\theta(\tau-\tau')\n \\
    +&g^{\mu\nu}(y)g^{\mu'\nu'}(y')\nabla_\mu\nabla_{\mu'}D_{\wt{\nu},\pm\mp}(y,y')\nabla_\nu\nabla_{\nu'}D_{\wt{\nu},\pm\mp}(y,y')\theta(\tau'-\tau).
\end{align}
The difference between (\ref{96}) and (\ref{97}) is
\begin{align}
    &\pm2\ii(-\tau)^{d+3}[\nabla_0\nabla_{0'}D_{\wt{\nu},\mp\pm}(y,y')\delta^{d+1}(y-y')\theta(\tau-\tau')+\nabla_0\nabla_{0'}D_{\wt{\nu},\pm\mp}(y,y')\delta^{d+1}(y-y')\theta(\tau'-\tau)]\n \\
    &-(-\tau)^{2d+2}[\delta(y-y')]^2.
\end{align}
Therefore, a correction should be added to the formula (\ref{73}):
\begin{align}\label{99}
    &-\frac12k_s^{d+2+p_{12}}\sum_{\mathsf{a}=\pm}\int_{-\infty}^0\ud\tau(-\tau)^{d+3+p_{12}}e^{\ii k_{1234}\tau}[2\mathsf{a}\ii\nabla_0\nabla_{0'}D_{\wt{\nu},\bar{\mathsf{a}}\mathsf{a}}(y,y'=y)\theta(\tau-\tau')\n \\
    &+2\mathsf{a}\ii\nabla_0\nabla_{0'}D_{\wt{\nu},\mathsf{a}\bar{\mathsf{a}}}(y,y'=y)\theta(\tau'-\tau)-(-\tau)^{d-1}\delta(y-y)].
\end{align}
Delta function $\delta(y,y)$ is recognized as $0$ in dimensional regularization. To regularize $\nabla_0\nabla_{0'}D_{\wt{\nu},\bar{\mathsf{a}}\mathsf{a}}(y,y'=y)$, the expression of $D_{\wt{\nu},\bar{\mathsf{a}}\mathsf{a}}(y,y')$ in $(d+1)$ dimension is written:
\begin{equation}
    D_{\wt{\nu},\bar{\mathsf{a}}\mathsf{a}}(y,y')=\frac1{(4\pi)^{\frac{d+1}2}}{}_2\mathcal{F}_1\left[\begin{matrix}
        \frac d2+\ii\wt{\nu}, \frac d2-\ii\wt{\nu} \\
        \frac{d+1}2
    \end{matrix}\middle|\frac{1+Z(y,y')}2\right],
\end{equation}
Here $Z(y,y')$ is the imbedding distance between $y$ and $y'$:
\begin{equation}\label{101}
    Z(y,y')=\frac{\tau^2+\tau'^2-|\mathbf{y}-\mathbf{y}'|^2}{2\tau\tau'}.
\end{equation}
Then the derivation $\nabla_\mu\nabla_{\nu'}D_{\wt{\nu},\bar{\mathsf{a}}\mathsf{a}}(y,y')$ is
\begin{equation}
    [g_{\mu\nu'}+n_\mu n_{\nu'}(1-Z)]D'_{\wt{\nu},\bar{\mathsf{a}}\mathsf{a}}(y,y')+n_\mu n_{\nu'}(1-Z^2)D''_{\wt{\nu},\bar{\mathsf{a}}\mathsf{a}}(y,y').
\end{equation}
Because $Z(y,y'=y)=1$, the above formula is simplified to
\begin{align}
    \nabla_\mu\nabla_{\nu'}D_{\wt{\nu},\bar{\mathsf{a}}\mathsf{a}}(y,y'=y)=&g_{\mu\nu'}D'_{\wt{\nu},\bar{\mathsf{a}}\mathsf{a}}(y,y'=y)\n \\
    =&\frac1{2(4\pi)^{\frac{d+1}2}}g_{\mu\nu'}{}_2\mathcal{F}_1\left[\begin{matrix}
        \frac d2+\ii\wt{\nu}+1, \frac d2-\ii\wt{\nu}+1 \\
        \frac{d+3}2
    \end{matrix}\middle|1\right]
\end{align}
The $00$ component of $g_{\mu{\nu}'}$ at $y=y'$ is $-1/(-\tau)^2$. For the hypergeometric function, dimensional regularization gives
\begin{align}
    &\frac1{2(4\pi)^{\frac{d+1}2}}{}_2\mathcal{F}_1\left[\begin{matrix}
        \frac d2+\ii\wt{\nu}+1, \frac d2-\ii\wt{\nu}+1 \\
        \frac{d+3}2
    \end{matrix}\middle|1\right]\n \\
    =&\frac1{2(4\pi)^{\frac{d+1}2}}\Gamma\Bigg[\begin{matrix}
        \frac d2+\ii\wt{\nu}+1, \frac d2-\ii\wt{\nu}+1, -\frac{d+1}2 \\
        \ii\wt{\nu}+\frac12, -\ii\wt{\nu}+\frac12
    \end{matrix}\Bigg]\n \\
    \sim&\frac1{32\pi^2}\frac1{16(3-d)}(9+40\wt{\nu}^2+16\wt{\nu}^4)-\frac1{32\pi^2}\frac1{64}(9+40\wt{\nu}^2+16\wt{\nu}^4)\n \\
    &\times\big[-3+2\gamma_E-\log(4\pi)+2\psi\big(\ii\wt{\nu}+\tfrac52\big)+2\psi\big(-\ii\wt{\nu}+\tfrac52\big)\big].
\end{align}
Equation (\ref{168}) is used in the second line. Therefore, the function (\ref{99}) gives
\begin{align}
    &-2\sin\big[\tfrac\pi2(d+p_{12})\big]\Gamma(d+p_{12}+2)\bigg(\frac{r_1r_2}{r_1+r_2}\bigg)^{d+p_{12}+2}\n \\
    &\times\bigg(\frac1{32\pi^2}\frac1{16(3-d)}(9+40\wt{\nu}^2+16\wt{\nu}^4)-\frac1{32\pi^2}\frac1{64}(9+40\wt{\nu}^2+16\wt{\nu}^4)\n \\
    &\times\big[-3+2\gamma_E-2\log(4\pi)+2\psi\big(\ii\wt{\nu}+\tfrac52\big)+2\psi\big(-\ii\wt{\nu}+\tfrac52\big)\big]\bigg).
\end{align}
After adding the contribution of (\ref{99}), the coefficients of divergence in the limit $d\rightarrow3$ should be:
\begin{equation}\label{106}
    \frac13\delta_4=\frac1{64\pi^2}\frac1{3-d},\quad\frac13\delta_2=\frac{19+12\wt{\nu}^2}{128\pi^2}\frac1{3-d},\quad\frac13\delta_0=\frac{69+152\wt{\nu}^2+48\wt{\nu}^4}{256\pi^2}\frac1{3-d}.
\end{equation}
In flat space limit, they are
\begin{equation}
    \frac13\delta_4=\frac1{64\pi^2}\frac1{3-d},\quad\frac13\delta_2=\frac{3m^2}{32\pi^2}\frac1{3-d},\quad\frac13\delta_0=\frac{3m^4}{16\pi^2}\frac1{3-d}.
\end{equation}
This agrees with (\ref{202}). It can be seen that $\delta_2$ and $\delta_4$ rely on the Hubble parameter $H$. This means that the Lagrangian of the $\nabla\sigma\nabla\sigma$ model should contain interactions between $\phi_c$ and the curvature of spacetime, such as $R(\nabla\varphi_c)^2\varphi_c^2$ and $R^2\varphi_c^4$ with $R$ the Ricci scalar.

The expression of background $\wh{\mathcal{J}}_{\nabla\sigma\nabla\sigma,(\text{3A})}^{p_1p_2}$ in $d=3$ is quite complicated. It is shown in App. \ref{AppF}. If the folded limit $r_2\rightarrow1$ or double folded limit $r_{1,2}\rightarrow1$ is taken, the seed integral for 3-point correlators or 2-point correlators will be derived.

\subsection{1-loop 4-point Functions in $AA$ Model}\label{Sec3.3}
Next, consider the interaction between $\phi_c$ and a massive vector field $A$ ($AA$ model):
\begin{equation}
    \Delta\mathscr{L}=-\frac14a^{d-1-p_1}\phi_c^2g^{\mu\nu}A_\mu A_\nu-\frac14a^{d-1-p_2}\phi_c^2g^{\mu\nu}A_\mu A_\nu.
\end{equation}
The mass of the vector field is $m_A$ and its mass parameter is defined as $\wt{\nu}_A\equiv\sqrt{m_A^2-(d/2-1)^2}$. The loop seed integral related to the interaction is
\begin{equation}
    \mathcal{J}_{AA,\wt{\nu}_A}^{p_1p_2}(r_1,r_2)\equiv-\frac12\sum_{\mathsf{a},\mathsf{b}=\pm}\mathsf{ab}k_s^{d+2+p_{12}}\int_{-\infty}^0\ud\tau_1\ud\tau_2(-\tau_1)^{p_1}(-\tau_2)^{p_2}e^{\ii\mathsf{a}k_{12}\tau_1+\ii\mathsf{a}k_{34}\tau_2}\mathcal{Q}_{AA,\wt{\nu}_A,\mathsf{ab}}\big(k_s;\tau_1,\tau_2\big).
\end{equation}

In this model, the $s$-channel of 4-point correlator of $\phi_c$ is
\begin{equation}
    \langle\phi_{c,\mathbf{k}_1}\phi_{c,\mathbf{k}_2}\phi_{c,\mathbf{k}_3}\phi_{c,\mathbf{k}_4}\rangle_s=(2\pi)^d\delta^{(d)}(\mathbf{k}_1+\mathbf{k}_2+\mathbf{k}_3+\mathbf{k}_4)\mathcal{L}_{\phi_c,AA}(\mathbf{k}_1,\mathbf{k}_2,\mathbf{k}_3,\mathbf{k}_4),
\end{equation}
with the loop amplitude $\mathcal{L}_{\phi_c,AA}$
\begin{align}
    \mathcal{L}_{\phi_c,AA}=&-\frac12\sum_{\mathsf{a},\mathsf{b}=\pm}\mathsf{ab}\int_{-\infty}^{\tau_f}\frac{\ud\tau_1}{(-\tau_1)^{d+1}}\frac{\ud\tau_2}{(-\tau_2)^{d+1}}\n \\
    &\times C_{\mathsf{a}}(k_1,\tau_1)C_{\mathsf{a}}(k_2,\tau_1)C_{\mathsf{b}}(k_3,\tau_2)C_{\mathsf{b}}(k_4,\tau_2)\mathcal{Q}_{AA,\mathsf{ab}}\big(k_s;\tau_1,\tau_2\big).
\end{align}
Here the loop momentum integral $\mathcal{Q}_{AA,\mathsf{ab}}$ is
\begin{equation}
    \mathcal{Q}_{AA,\mathsf{ab}}\big(k_s;\tau_1,\tau_2\big)=\int\frac{\ud^d\mathbf{q}}{(2\pi)^d}g^{\mu\nu}g^{\mu'\nu'}D_{\wt{\nu}_A,\mu\mu',\mathsf{ab}}\big(\mathbf{q};\tau_1,\tau_2\big)D_{\wt{\nu}_A,\nu\nu',\mathsf{ab}}\big(\mathbf{k}_s-\mathbf{q};\tau_1,\tau_2\big).
\end{equation}
The amplitude can be represented as
\begin{equation}
    \mathcal{L}_{\phi_c,AA}=\frac{(-\tau_f)^{2(d-1)}}{16k_1k_2k_3k_4k_s^{d-2}}\mathcal{J}_{AA}^{-2,-2}(r_1,r_2).
\end{equation}
The expressions of bulk-propagator of massive vector field are
\begin{align}
    D_{\wt{\nu}_A,\mu\nu',-+}(\mathbf{k};\tau_1,\tau_2)=&\sum_{n=0}^dA_{(n),\mu}(\mathbf{k},\tau_1)A_{(n),\nu'}^*(\mathbf{k},\tau_2), \\
    D_{\wt{\nu}_A,\mu\nu',+-}(\mathbf{k};\tau_1,\tau_2)=&D_{\wt{\nu}_A,\mu\nu',-+}^*(k;\tau_1,\tau_2),
\end{align}
where mode function $A_{(n),\mu}(\mathbf{k},\tau)$ is defined in \cite{10.1063/1.4879496}. Note that $D_{\wt{\nu}_A,\mu\nu',\pm\pm}(k;\tau_1,\tau_2)$ is not equivalent to $D_{\wt{\nu}_A,\mu\nu',\mp\pm}(k;\tau_1,\tau_2)\theta(\tau_1-\tau_2)+D_{\wt{\nu}_A,\mu\nu',\pm\mp}(k;\tau_1,\tau_2)\theta(\tau_2-\tau_1)$. Instead, the propagator $D_{\wt{\nu}_A,\mu\nu',++}(y,y')$ in the position space is
\begin{align}\label{119}
    D_{\wt{\nu}_A,\mu\nu',++}(y,y')=&D_{\wt{\nu}_A,\mu\nu',-+}(y,y')\theta(\tau-\tau')+D_{\wt{\nu}_A,\mu\nu',+-}(y,y')\theta(\tau'-\tau)\n \\
    &+\frac{\ii}{m_A^2}(-\tau)^{d-1}\delta^0_\mu\delta^0_{\nu'}\delta(y-y').
\end{align}
The expression of $D_{\wt{\nu}_A,\mu\nu',--}$ is the complex conjugate of $D_{\wt{\nu}_A,\mu\nu',++}$. In the calculation of this subsection, the local term $\ii(-\tau)^{d-1}\delta^0_\mu\delta^0_{\nu'}\delta(y-y')/m_A^2$ is ignored at first. After computing the spectral integral, the contribution of local term will be added.

The spectral function of K\"all\'en-Lehmann representation of $AA$ model is (See App. \ref{AppE} for the details of the calculation)
\begin{align}
    \rho^{\mathcal{P},0}_{AA}(\wt{\nu}')=&\frac{\wt{\nu}'\sinh(\pi\wt{\nu}')\Gamma\big(\frac{d+1}2\big)}{2^{12-d}\pi^{\frac{d+7}2}\Gamma(d)\big(\frac d2+\ii\wt{\nu}_A-1\big)^2\big(\frac d2-\ii\wt{\nu}_A-1\big)^2\Gamma\big(\frac d2+\ii\wt{\nu}'\big)\Gamma\big(\frac d2-\ii\wt{\nu}'\big)}\n \\
    &\times\prod_{\pm,\pm,\pm}\Gamma\bigg(\frac{\frac d2\pm\ii\wt{\nu}_A\pm\ii\wt{\nu}_A\pm\ii\wt{\nu}'}2\bigg)(8d^3-11d^4+4d^5-32d\wt{\nu}'^2+24d^2\wt{\nu}'^2\n \\
    &+16\wt{\nu}'^4+64d\wt{\nu}_A^2-80d^2\wt{\nu}_A^2+32d^3\wt{\nu}_A^2-64\wt{\nu}'^2\wt{\nu}_A^2+64d\wt{\nu}_A^4)\n \\
    \equiv&\frac{\wt{\nu}'\sinh(\pi\wt{\nu}')\Gamma\big(\frac{d+1}2\big)}{2^{12-d}\pi^{\frac{d+7}2}\Gamma(d)\big(\frac d2+\ii\wt{\nu}_A-1\big)^2\big(\frac d2-\ii\wt{\nu}_A-1\big)^2\Gamma\big(\frac d2+\ii\wt{\nu}'\big)\Gamma\big(\frac d2-\ii\wt{\nu}'\big)}\n \\
    &\times\prod_{\pm,\pm,\pm}\Gamma\bigg(\frac{\frac d2\pm\ii\wt{\nu}_A\pm\ii\wt{\nu}_A\pm\ii\wt{\nu}'}2\bigg)f_{AA}(\wt{\nu}'),
\end{align}
where
\begin{equation}
    f_{AA}(\wt{\nu}')\equiv f_{AA,4}\wt{\nu}'^4+f_{AA,2}\wt{\nu}'^2+f_{AA,0},
\end{equation}
with
\begin{align}
    f_{AA,4}=&16, \\
    f_{AA,2}=&-32d+24d^2-64\wt{\nu}_A^2, \\
    f_{AA,0}=&8d^3-11d^4+4d^5+64d\wt{\nu}_A^2-80d^2\wt{\nu}_A^2+32d^3\wt{\nu}_A^2+64d\wt{\nu}_A^4.
\end{align}
With the expression of the spectral function, the seed integral becomes
\begin{align}\label{125}
    \mathcal{J}_{AA}^{p_1p_2}(r_1,r_2)\equiv&\frac12\int_{-\infty}^\infty\ud\wt{\nu}'\rho_{AA}^{\mathcal{P},0}(\wt{\nu}')\mathcal{I}_{\wt{\nu}'}^{p_1p_2}(r_1,r_2)+\text{local term}.
\end{align}

The result of the integration can be divided into three parts:
\begin{equation}
    \mathcal{J}_{AA}^{p_1p_2}(r_1,r_2)=\mathcal{J}_{AA,\text{NS}}^{p_1p_2}(r_1,r_2)+\mathcal{J}_{AA,\text{LS}}^{p_1p_2}(r_1,r_2)+\mathcal{J}_{AA,\text{BG}}^{p_1p_2}(r_1,r_2).
\end{equation}
The nonlocal signal and local signal are
\begin{align}
    \mathcal{J}_{AA,\text{NS}}^{p_1p_2}=&-\frac{(r_1r_2)^{d/2+2\ii\wt{\nu}_A}\sin[\pi(\fr{d}{2}+2\ii \wt{\nu}_A)]}{512\pi^{d/2}\Gamma\big(\tfrac d2\big)\sin^2(\pi\ii\wt{\nu}_A)}\sum_{n=0}^\infty\FR{(1+n)_{\frac d2-1} \big[(1+\ii\wt\nu_A+n)_{\frac d2-1}\big]^2(1+2\ii\wt\nu_A+n)_{\frac d2-1}}{(1+2\ii\wt\nu_A+2n)_{d-1}}\n\\
    &\times(\fr{d}{2}+2\ii\wt{\nu}_A+2n)\mathcal{C}_{2\ii\wt{\nu}_A+d/2+2n,d}^{p_1p_2}\mb{F}_{2\ii\wt{\nu}_A+d/2+2n,d}^{p_1}(r_1)\mb{F}_{2\ii\wt{\nu}_A+d/2+2n,d}^{p_2}(r_2)(r_1r_2)^{2n}\n \\
    &\times\frac{f_{AA}(\ii d/2-2\wt{\nu}_A+2\ii n)}{\big(\frac d2+\ii\wt{\nu}_A-1\big)^2\big(\frac d2-\ii\wt{\nu}_A-1\big)^2}\n \\
    &+\text{c.c.},
\end{align}
\begin{align}
    \mathcal{J}_{AA,\text{LS}}^{p_1p_2}=&\frac{(r_1/r_2)^{d/2+2\ii\wt{\nu}_A}\sin[\pi(\fr{d}{2}+2\ii \wt{\nu}_A)]}{512\pi^{d/2}\Gamma\big(\tfrac d2\big)\sin^2(\pi\ii\wt{\nu}_A)}\sum_{n=0}^\infty\FR{(1+n)_{\frac d2-1} \big[(1+\ii\wt\nu_A+n)_{\frac d2-1}\big]^2(1+2\ii\wt\nu_A+n)_{\frac d2-1}}{(1+2\ii\wt\nu_A+2n)_{d-1}}\n\\
    &\times(\fr{d}{2}+2\ii\wt{\nu}_A+2n)\mathcal{C}_{2\ii\wt{\nu}_A+d/2+2n,d}^{p_1p_2}\mb{F}_{2\ii\wt{\nu}_A+d/2+2n,d}^{p_1}(r_1)\mb{F}_{-2\ii\wt{\nu}_A-d/2-2n,d}^{p_2}(r_2)\bigg(\frac{r_1}{r_2}\bigg)^{2n}\n \\
    &\times\frac{f_{AA}(\ii d/2-2\wt{\nu}_A+2\ii n)}{\big(\frac d2+\ii\wt{\nu}_A-1\big)^2\big(\frac d2-\ii\wt{\nu}_A-1\big)^2}\n \\
    &+\text{c.c.}.
\end{align}
The expression of background is
\begin{equation}
    \mathcal{J}_{AA,\text{BG}}^{p_1p_2}=\mathcal{J}_{AA,(\text{B})}^{p_1p_2}+\mathcal{J}_{AA,(\text{3A})}^{p_1p_2}+\mathcal{J}_{AA,(\text{3B})}^{p_1p_2}+\mathcal{J}_{AA,(\text{3C})}^{p_1p_2}+\text{local term},
\end{equation}
with
\begin{align}
    \mathcal{J}_{AA,(\text{B})}^{p_1p_2}=&\frac{\sin(\fr{\pi d}{2})}{256\pi^{d/2}\Gamma\big(\tfrac d2\big)\sin^2(\pi\ii\wt{\nu}_A)}\sum_{n=0}^\infty\FR{\big[(1+n)_{\frac d2-1}\big]^2 (1+\ii\wt\nu_A+n)_{\frac d2-1}(1-\ii\wt\nu_A+n)_{\frac d2-1}}{(1+2n)_{d-1}}\n\\
    &\times(\fr{d}{2}+2n)\mathcal{C}_{d/2+2n,d}^{p_1p_2}\mb{F}_{d/2+2n,d}^{p_1}(r_1)r_1^{2n+d/2}\big(\mb{F}_{d/2+2n,d}^{p_2}(r_2)r_2^{2n+d/2}-\mb{F}_{-d/2-2n,d}^{p_2}(r_2)r_2^{-2n-d/2}\big)\n \\
    &\times\frac{f_{AA}(\ii d/2+2\ii n)}{\big(\frac d2+\ii\wt{\nu}_A-1\big)^2\big(\frac d2-\ii\wt{\nu}_A-1\big)^2},
\end{align}
\begin{align}
    \mathcal{J}_{AA,(\text{3A})}^{p_1p_2}=&-\sum_{n=0}^\infty\sum_{\ell,j=0}^\infty\frac{(-1)^\ell\big(\frac d2-2\ii\wt{\nu}_A+2n\big)\mathcal{C}_{\frac d2-2\ii\wt{\nu}_A+2n,d}^{p_1p_2}\sin^2\big[\pi\big(\frac d2-2\ii\wt{\nu}_A\big)\big](\ell+1)_{2j+d+p_{12}+1}}{2^{2j+d+7}\pi^{\frac d2}\sin^2(\pi\ii\wt{\nu}_A)\Gamma\big(\frac d2\big)n!\big(\frac{\ell+2\ii\wt{\nu}_A+p_2+1}2-n\big)_{j+1}\big(\frac{\ell+d-2\ii\wt{\nu}_A+p_2+1}2+n\big)_{j+1}}\n \\
    &\times\Gamma\Bigg[\begin{matrix}
        \frac d2+n, \frac d2-\ii\wt{\nu}_A+n, \frac d2-2\ii\wt{\nu}_A+n, -\ii\wt{\nu}_A+\frac12+n \\
        \frac{d+1}2-\ii\wt{\nu}_A+n, -2\ii\wt{\nu}_A+1+n, -\ii\wt{\nu}_A+1+n
    \end{matrix}\Bigg]\n \\
    &\times r_1^{2j+d+p_{12}+2}\bigg(\frac{r_1}{r_2}\bigg)^\ell\frac{f_{AA}(\ii d/2+2\wt{\nu}_A+2\ii n)}{\big(\frac d2+\ii\wt{\nu}_A-1\big)^2\big(\frac d2-\ii\wt{\nu}_A-1\big)^2},
\end{align}
\begin{align}
    \mathcal{J}_{AA,(\text{3B})}^{p_1p_2}=&\sum_{n=0}^\infty\sum_{\ell,j=0}^\infty\frac{(-1)^\ell\big(\frac d2+2n\big)\mathcal{C}_{\frac d2+2n,d}^{p_1p_2}\sin^2\big(\frac{\pi d}2\big)(\ell+1)_{2j+d+p_{12}+1}}{2^{2j+d+6}\pi^{\frac d2}\sin^2(\pi\ii\wt{\nu}_A)\Gamma\big(\frac d2\big)n!\big(\frac{\ell+p_2+1}2-n\big)_{j+1}\big(\frac{\ell+d+p_2+1}2+n\big)_{j+1}}\n \\
    &\times\Gamma\Bigg[\begin{matrix}
        \frac d2+\ii\wt{\nu}_A+n, \frac d2+n, \frac d2-\ii\wt{\nu}_A+n, \frac12+n \\
        \frac{d+1}2+n, -\ii\wt{\nu}_A+1+n, \ii\wt{\nu}_A+1+n
    \end{matrix}\Bigg]\n \\
    &\times r_1^{2m+d+p_{12}+2}\bigg(\frac{r_1}{r_2}\bigg)^\ell\frac{f_{AA}(\ii d/2+2\ii n)}{\big(\frac d2+\ii\wt{\nu}_A-1\big)^2\big(\frac d2-\ii\wt{\nu}_A-1\big)^2},
\end{align}
\begin{align}
    \mathcal{J}_{AA,(\text{3C})}^{p_1p_2}=&-\sum_{n=0}^\infty\sum_{\ell,j=0}^\infty\frac{(-1)^\ell\big(\frac d2+2\ii\wt{\nu}_A+2n\big)\mathcal{C}_{\frac d2+2\ii\wt{\nu}_A+2n,d}^{p_1p_2}\sin^2\big[\pi\big(\frac d2+2\ii\wt{\nu}_A\big)\big](\ell+1)_{2j+d+p_{12}+1}}{2^{2j+d+7}\pi^{\frac d2}\sin^2(\pi\ii\wt{\nu}_A)\Gamma\big(\frac d2\big)n!\big(\frac{\ell-2\ii\wt{\nu}_A+p_2+1}2-n\big)_{j+1}\big(\frac{\ell+d+2\ii\wt{\nu}_A+p_2+1}2+n\big)_{j+1}}\n \\
    &\times\Gamma\Bigg[\begin{matrix}
        \frac d2+2\ii\wt{\nu}_A+n, \frac d2+\ii\wt{\nu}_A+n, \frac d2+n, \ii\wt{\nu}_A+\frac12+n \\
        \frac{d+1}2+\ii\wt{\nu}_A+n, \ii\wt{\nu}_A+1+n, 2\ii\wt{\nu}_A+1+n
    \end{matrix}\Bigg]\n \\
    &\times r_1^{2m+d+p_{12}+2}\bigg(\frac{r_1}{r_2}\bigg)^\ell\frac{f_{AA}(\ii d/2-2\wt{\nu}_A+2\ii n)}{\big(\frac d2+\ii\wt{\nu}_A-1\big)^2\big(\frac d2-\ii\wt{\nu}_A-1\big)^2}.
\end{align}
These expressions of the last three terms diverge when $d\rightarrow3$, due to series of $n$ when $j=0,1,2$. The asymptotic expansions of these series for large $(n+1)$ are:
\begin{align}\label{133}
    &\sum_{\ell=0}^\infty\frac{(-1)^\ell\big(\frac d2-2\ii\wt{\nu}_A+2n\big)\mathcal{C}_{\frac d2-2\ii\wt{\nu}_A,d}^{p_1p_2}\sin^2\big[\pi\big(\frac d2-2\ii\wt{\nu}_A\big)\big](\ell+1)_{d+p_{12}+1}}{2^{d+7}\pi^{\frac d2}\sin^2(\pi\ii\wt{\nu}_A)\Gamma\big(\frac d2\big)n!\big(\frac{\ell+2\ii\wt{\nu}_A+p_2+1}2-n\big)\big(\frac{\ell+d-2\ii\wt{\nu}_A+p_2+1}2+n\big)}\n \\
    &\times\Gamma\Bigg[\begin{matrix}
        \frac d2+n, \frac d2-\ii\wt{\nu}_A+n, \frac d2-2\ii\wt{\nu}_A+n, -\ii\wt{\nu}_A+\frac12+n \\
        \frac{d+1}2-\ii\wt{\nu}_A+n, -2\ii\wt{\nu}_A+1+n, -\ii\wt{\nu}_A+1+n
    \end{matrix}\Bigg]\n \\
    &\times r_1^{d+p_{12}+2}\bigg(\frac{r_1}{r_2}\bigg)^\ell f_{AA}(\ii d/2+2\wt{\nu}_A+2\ii n)\n \\
    \sim&\sum_{\ell=0}^\infty\frac{(-1)^\ell\mathcal{C}_{\frac d2-2\ii\wt{\nu}_A,d}^{p_1p_2}\sin^2\big[\pi\big(\frac d2-2\ii\wt{\nu}_A\big)\big](\ell+1)_{d+p_{12}+1}}{2^{d+2}\pi^{\frac d2}\sin^2(\pi\ii\wt{\nu}_A)\Gamma\big(\frac d2\big)}r_1^{d+p_{12}+2}\bigg(\frac{r_1}{r_2}\bigg)^\ell\n \\
    &\times\bigg(-(n+1)^{d-3}\sum_{\iota=-1}^3\Upsilon_{AA,d,\text{A},\ell,0,\iota}^{p_2}(n+1)^\iota\bigg),
\end{align}
\begin{align}\label{134}
    &\sum_{\ell=0}^\infty\frac{(-1)^\ell\big(\frac d2-2\ii\wt{\nu}_A+2n\big)\mathcal{C}_{\frac d2-2\ii\wt{\nu}_A,d}^{p_1p_2}\sin^2\big[\pi\big(\frac d2-2\ii\wt{\nu}_A\big)\big](\ell+1)_{d+p_{12}+3}}{2^{d+9}\pi^{\frac d2}\sin^2(\pi\ii\wt{\nu}_A)\Gamma\big(\frac d2\big)n!\big(\frac{\ell+2\ii\wt{\nu}_A+p_2+1}2-n\big)_2\big(\frac{\ell+d-2\ii\wt{\nu}_A+p_2+1}2+n\big)_2}\n \\
    &\times\Gamma\Bigg[\begin{matrix}
        \frac d2+n, \frac d2-\ii\wt{\nu}_A+n, \frac d2-2\ii\wt{\nu}_A+n, -\ii\wt{\nu}_A+\frac12+n \\
        \frac{d+1}2-\ii\wt{\nu}_A+n, -2\ii\wt{\nu}_A+1+n, -\ii\wt{\nu}_A+1+n
    \end{matrix}\Bigg]\n \\
    &\times r_1^{d+p_{12}+4}\bigg(\frac{r_1}{r_2}\bigg)^\ell f_{AA}(\ii d/2+2\wt{\nu}_A+2\ii n)\n \\
    \sim&\sum_{\ell=0}^\infty\frac{(-1)^\ell\mathcal{C}_{\frac d2-2\ii\wt{\nu}_A,d}^{p_1p_2}\sin^2\big[\pi\big(\frac d2-2\ii\wt{\nu}_A\big)\big](\ell+1)_{d+p_{12}+3}}{2^{d+4}\pi^{\frac d2}\sin^2(\pi\ii\wt{\nu}_A)\Gamma\big(\frac d2\big)}r_1^{d+p_{12}+4}\bigg(\frac{r_1}{r_2}\bigg)^\ell\n \\
    &\times\bigg((n+1)^{d-3}\Big[\Upsilon_{AA,d,\text{A},\ell,1,1}^{p_2}(n+1)+\Upsilon_{AA,d,\text{A},\ell,1,0}^{p_2}+\Upsilon_{AA,d,\text{A},\ell,1,-1}^{p_2}/(n+1)\Big]\bigg),
\end{align}
\begin{align}\label{135}
    &\sum_{\ell=0}^\infty\frac{(-1)^\ell\big(\frac d2-2\ii\wt{\nu}_A+2n\big)\mathcal{C}_{\frac d2-2\ii\wt{\nu}_A,d}^{p_1p_2}\sin^2\big[\pi\big(\frac d2-2\ii\wt{\nu}_A\big)\big](\ell+1)_{d+p_{12}+5}}{2^{d+11}\pi^{\frac d2}\sin^2(\pi\ii\wt{\nu}_A)\Gamma\big(\frac d2\big)n!\big(\frac{\ell+2\ii\wt{\nu}_A+p_2+1}2-n\big)_3\big(\frac{\ell+d-2\ii\wt{\nu}_A+p_2+1}2+n\big)_3}\n \\
    &\times\Gamma\Bigg[\begin{matrix}
        \frac d2+n, \frac d2-\ii\wt{\nu}_A+n, \frac d2-2\ii\wt{\nu}_A+n, -\ii\wt{\nu}_A+\frac12+n \\
        \frac{d+1}2-\ii\wt{\nu}_A+n, -2\ii\wt{\nu}_A+1+n, -\ii\wt{\nu}_A+1+n
    \end{matrix}\Bigg]\n \\
    &\times r_1^{d+p_{12}+6}\bigg(\frac{r_1}{r_2}\bigg)^\ell f_{AA}(\ii d/2+2\wt{\nu}_A+2\ii n)\n \\
    \sim&\sum_{\ell=0}^\infty\frac{(-1)^\ell\mathcal{C}_{\frac d2-2\ii\wt{\nu}_A,d}^{p_1p_2}\sin^2\big[\pi\big(\frac d2-2\ii\wt{\nu}_A\big)\big](\ell+1)_{d+p_{12}+3}}{2^{d+6}\pi^{\frac d2}\sin^2(\pi\ii\wt{\nu}_A)\Gamma\big(\frac d2\big)}r_1^{d+p_{12}+4}\bigg(\frac{r_1}{r_2}\bigg)^\ell\bigg(-16(n+1)^{d-4}\bigg),
\end{align}
\begin{align}\label{136}
    &\sum_{\ell=0}^\infty\frac{(-1)^\ell\big(\frac d2+2n\big)\mathcal{C}_{\frac d2,d}^{p_1p_2}\sin^2\big(\frac{\pi d}2\big)(\ell+1)_{d+p_{12}+1}}{2^{d+6}\pi^{\frac d2}\sin^2(\pi\ii\wt{\nu}_A)\Gamma\big(\frac d2\big)n!\big(\frac{\ell+p_2+1}2-n\big)\big(\frac{\ell+d+p_2+1}2+n\big)}\n \\
    &\times\Gamma\Bigg[\begin{matrix}
        \frac d2+\ii\wt{\nu}_A+n, \frac d2+n, \frac d2-\ii\wt{\nu}_A+n, \frac12+n \\
        \frac{d+1}2+n, -\ii\wt{\nu}_A+1+n, \ii\wt{\nu}_A+1+n
    \end{matrix}\Bigg]\n \\
    &\times r_1^{d+p_{12}+2}\bigg(\frac{r_1}{r_2}\bigg)^\ell f_{AA}(\ii d/2+2\ii n)\n \\
    \sim&\sum_{\ell=0}^\infty\frac{(-1)^\ell\mathcal{C}_{\frac d2,d}^{p_1p_2}\sin^2\big(\frac{\pi d}2\big)(\ell+1)_{d+p_{12}+1}}{2^{d+1}\pi^{\frac d2}\sin^2(\pi\ii\wt{\nu}_A)\Gamma\big(\frac d2\big)}r_1^{d+p_{12}+2}\bigg(\frac{r_1}{r_2}\bigg)^\ell\n \\
    &\times\bigg(-(n+1)^{d-3}\sum_{\iota=-1}^3\Upsilon_{AA,d,\text{B},\ell,0,\iota}^{p_2}(n+1)^\iota\bigg),
\end{align}
\begin{align}\label{137}
    &\sum_{\ell=0}^\infty\frac{(-1)^\ell\big(\frac d2+2n\big)\mathcal{C}_{\frac d2,d}^{p_1p_2}\sin^2\big(\frac{\pi d}2\big)(\ell+1)_{d+p_{12}+3}}{2^{d+8}\pi^{\frac d2}\sin^2(\pi\ii\wt{\nu}_A)\Gamma\big(\frac d2\big)n!\big(\frac{\ell+p_2+1}2-n\big)_2\big(\frac{\ell+d+p_2+1}2+n\big)_2}\n \\
    &\times\Gamma\Bigg[\begin{matrix}
        \frac d2+\ii\wt{\nu}_A+n, \frac d2+n, \frac d2-\ii\wt{\nu}_A+n, \frac12+n \\
        \frac{d+1}2+n, -\ii\wt{\nu}_A+1+n, \ii\wt{\nu}_A+1+n
    \end{matrix}\Bigg]\n \\
    &\times r_1^{d+p_{12}+4}\bigg(\frac{r_1}{r_2}\bigg)^\ell f_{AA}(\ii d/2+2\ii n)\n \\
    \sim&\sum_{\ell=0}^\infty\frac{(-1)^\ell\mathcal{C}_{\frac d2,d}^{p_1p_2}\sin^2\big(\frac{\pi d}2\big)(\ell+1)_{d+p_{12}+3}}{2^{d+3}\pi^{\frac d2}\sin^2(\pi\ii\wt{\nu}_A)\Gamma\big(\frac d2\big)}r_1^{d+p_{12}+4}\bigg(\frac{r_1}{r_2}\bigg)^\ell\n \\
    &\times\bigg((n+1)^{d-3}\Big[\Upsilon_{AA,d,\text{B},\ell,1,1}^{p_2}(n+1)+\Upsilon_{AA,d,\text{B},\ell,1,0}^{p_2}+\Upsilon_{AA,d,\text{B},\ell,1,-1}^{p_2}/(n+1)\Big]\bigg),
\end{align}
\begin{align}\label{138}
    &\sum_{\ell=0}^\infty\frac{(-1)^\ell\big(\frac d2+2n\big)\mathcal{C}_{\frac d2,d}^{p_1p_2}\sin^2\big(\frac{\pi d}2\big)(\ell+1)_{d+p_{12}+5}}{2^{d+10}\pi^{\frac d2}\sin^2(\pi\ii\wt{\nu}_A)\Gamma\big(\frac d2\big)n!\big(\frac{\ell+p_2+1}2-n\big)_3\big(\frac{\ell+d+p_2+1}2+n\big)_3}\n \\
    &\times\Gamma\Bigg[\begin{matrix}
        \frac d2+\ii\wt{\nu}_A+n, \frac d2+n, \frac d2-\ii\wt{\nu}_A+n, \frac12+n \\
        \frac{d+1}2+n, -\ii\wt{\nu}_A+1+n, \ii\wt{\nu}_A+1+n
    \end{matrix}\Bigg]\n \\
    &\times r_1^{d+p_{12}+6}\bigg(\frac{r_1}{r_2}\bigg)^\ell f_{AA}(\ii d/2+2\ii n)\n \\
    \sim&\sum_{\ell=0}^\infty\frac{(-1)^\ell\mathcal{C}_{\frac d2,d}^{p_1p_2}\sin^2\big(\frac{\pi d}2\big)(\ell+1)_{d+p_{12}+5}}{2^{d+5}\pi^{\frac d2}\sin^2(\pi\ii\wt{\nu}_A)\Gamma\big(\frac d2\big)}r_1^{d+p_{12}+6}\bigg(\frac{r_1}{r_2}\bigg)^\ell\bigg(-16(n+1)^{d-4}\bigg).
\end{align}
If taking the transformation $\wt{\nu}\rightarrow(-\wt{\nu})$, the summations in $\mathcal{J}_{AA,(\text{3A})}$ are changed into those in $\mathcal{J}_{AA,(\text{3C})}$. The dimensional regularization gives the expression of $\mathcal{J}_{AA,(\text{3A})}$:
\begin{align}
    \mathcal{J}_{AA,(\text{3A})}^{p_1p_2}=&-\sum_{n=0}^\infty\sum_{\ell,j=0}^\infty\frac{(-1)^\ell\big(\frac d2-2\ii\wt{\nu}_A+2n\big)\mathcal{C}_{\frac d2-2\ii\wt{\nu}_A+2n,d}^{p_1p_2}\sin^2\big[\pi\big(\frac d2-2\ii\wt{\nu}_A\big)\big](\ell+1)_{2j+d+p_{12}+1}}{2^{2j+d+7}\pi^{\frac d2}\sin^2(\pi\ii\wt{\nu}_A)\Gamma\big(\frac d2\big)n!\big(\frac{\ell+2\ii\wt{\nu}_A+p_2+1}2-n\big)_{j+1}\big(\frac{\ell+d-2\ii\wt{\nu}_A+p_2+1}2+n\big)_{j+1}}\n \\
    &\times\Gamma\Bigg[\begin{matrix}
        \frac d2+n, \frac d2-\ii\wt{\nu}_A+n, \frac d2-2\ii\wt{\nu}_A+n, -\ii\wt{\nu}_A+\frac12+n \\
        \frac{d+1}2-\ii\wt{\nu}_A+n, -2\ii\wt{\nu}_A+1+n, -\ii\wt{\nu}_A+1+n
    \end{matrix}\Bigg]\n \\
    &\times r_1^{2j+d+p_{12}+2}\bigg(\frac{r_1}{r_2}\bigg)^\ell\frac{f_{AA}(\ii d/2+2\wt{\nu}_A+2\ii n)}{\big(\frac d2+\ii\wt{\nu}_A-1\big)^2\big(\frac d2-\ii\wt{\nu}_A-1\big)^2}\n \\
    =&\sum_{\ell=0}^\infty\Gamma\Bigg[\begin{matrix}
        \frac d2+n, \frac d2-\ii\wt{\nu}_A+n, \frac d2-2\ii\wt{\nu}_A+n, -\ii\wt{\nu}_A+\frac12+n \\
        \frac{d+1}2-\ii\wt{\nu}_A+n, -2\ii\wt{\nu}_A+1+n, -\ii\wt{\nu}_A+1+n
    \end{matrix}\Bigg]\n \\
    &\times\frac{(-1)^\ell\mathcal{C}_{\frac d2-2\ii\wt{\nu}_A,d}^{p_1p_2}\sin^2\big[\pi\big(\frac d2-2\ii\wt{\nu}_A\big)\big]}{2^{d+2}\pi^{\frac d2}\sin^2(\pi\ii\wt{\nu}_A)\Gamma\big(\frac d2\big)\big(\frac d2+\ii\wt{\nu}_A-1\big)^2\big(\frac d2-\ii\wt{\nu}_A-1\big)^2}\n \\
    &\times\Bigg\{-\Bigg[\sum_{n=0}^\infty\sum_{j=3}^\infty\frac{\big(\frac d4-\ii\wt{\nu}_A+n\big)(\ell+1)_{2j+d+p_{12}+1}}{2^{2j}n!\big(\frac{\ell+2\ii\wt{\nu}_A+p_2+1}2-n\big)_{j+1}\big(\frac{\ell+d-2\ii\wt{\nu}_A+p_2+1}2+n\big)_{j+1}}\n \\
    &\times r_1^{2j+d+p_{12}+2}\bigg(\frac{r_1}{r_2}\bigg)^\ell\frac1{16}f_{AA}(\ii d/2+2\wt{\nu}_A+2\ii n)\Bigg]\n \\
    &-r_1^{d+p_{12}+2}\bigg(\frac{r_1}{r_2}\bigg)^\ell(\ell+1)_{d+p_{12}+1}\n \\
    &\times\Bigg[\sum_{n=0}^\infty\Bigg(\frac{\frac d4-\ii\wt{\nu}_A+n}{n!\big(\frac{\ell+2\ii\wt{\nu}_A+p_2+1}2-n\big)\big(\frac{\ell+d-2\ii\wt{\nu}_A+p_2+1}2+n\big)}\frac1{16}f_{AA}(\ii d/2+2\wt{\nu}_A+2\ii n)\n \\
    &+(n+1)^{d-3}\sum_{\iota=-1}^3\Upsilon_{AA,d,\text{A},\ell,0,\iota}^{p_2}(n+1)^\iota\Bigg)-\sum_{\iota=-1}^3\Upsilon_{AA,d,\text{A},\ell,0,\iota}^{p_2}\zeta(-d+3-\iota)\Bigg]\n \\
    &-\frac14r_1^{d+p_{12}+4}\bigg(\frac{r_1}{r_2}\bigg)^\ell(\ell+1)_{d+p_{12}+3}\n \\
    &\times\Bigg[\sum_{n=0}^\infty\Bigg(\frac{\frac d4-\ii\wt{\nu}_A+n}{n!\big(\frac{\ell+2\ii\wt{\nu}_A+p_2+1}2-n\big)_2\big(\frac{\ell+d-2\ii\wt{\nu}_A+p_2+1}2+n\big)_2}\frac1{16}f_{AA}(\ii d/2+2\wt{\nu}_A+2\ii n)\n \\
    &-(n+1)^{d-3}\sum_{\iota=-1}^1\Upsilon_{AA,d,\text{A},\ell,1,\iota}^{p_2}(n+1)^\iota\Bigg)+\sum_{\iota=-1}^1\Upsilon_{AA,d,\text{A},\ell,1,\iota}^{p_2}\zeta(-d+3-\iota)\Bigg]\n \\
    &-\frac1{16}r_1^{d+p_{12}+6}\bigg(\frac{r_1}{r_2}\bigg)^\ell(\ell+1)_{d+p_{12}+5}\Bigg[\sum_{n=0}^\infty\Bigg(\frac{\frac d4-\ii\wt{\nu}_A+n}{n!\big(\frac{\ell+2\ii\wt{\nu}_A+p_2+1}2-n\big)_3\big(\frac{\ell+d-2\ii\wt{\nu}_A+p_2+1}2+n\big)_3}\n \\
    &\frac1{16}f_{AA}(\ii d/2+2\wt{\nu}_A+2\ii n)+16(n+1)^{d-4}\Bigg)-16\zeta(-d+4)\Bigg]\Bigg\}.
\end{align}
The divergence of $\mathcal{J}_{AA,(\text{3A})}$ is:
\begin{align}
    \text{Div}\mathcal{J}_{AA,(\text{3A})}^{p_1p_2}=&\sum_{\ell=0}^\infty\frac{(-1)^\ell\mathcal{C}_{\frac d2-2\ii\wt{\nu}_A,3}^{p_1p_2}\sin^2\big[\pi\big(\frac d2-2\ii\wt{\nu}_A\big)\big](\ell+1)_{d+p_{12}+1}}{\pi^2\sin^2(\pi\ii\wt{\nu}_A)\big(\frac d2+\ii\wt{\nu}_A-1\big)^2\big(\frac d2-\ii\wt{\nu}_A-1\big)^2}r_1^{d+p_{12}+2}\bigg(\frac{r_1}{r_2}\bigg)^\ell\n \\
    &\times\Bigg[\bigg(8\bigg(\Big(\tfrac d2+p_2+\ell+1\Big)^4-\tfrac{d^2}2\Big(\tfrac d2+p_2+\ell+1\Big)^2+\tfrac{d^4}{16}\bigg)\n \\
    &+(-20+48\wt{\nu}_A^2)\bigg(\Big(\tfrac d2+p_2+\ell+1\Big)^2-\tfrac{d^2}4\bigg)+33+168\wt{\nu}_A^2+144\wt{\nu}_A^4\bigg)/(128(3-d))\Bigg]\n \\
    &+\sum_{\ell=0}^\infty\frac{(-1)^\ell\mathcal{C}_{\frac d2-2\ii\wt{\nu}_A,3}^{p_1p_2}\sin^2\big[\pi\big(\frac d2-2\ii\wt{\nu}_A\big)\big](\ell+1)_{d+p_{12}+3}}{4\pi^2\sin^2(\pi\ii\wt{\nu}_A)\big(\frac d2+\ii\wt{\nu}_A-1\big)^2\big(\frac d2-\ii\wt{\nu}_A-1\big)^2}r_1^{d+p_{12}+4}\bigg(\frac{r_1}{r_2}\bigg)^\ell\n \\
    &\times\Bigg[-\bigg(4\Big[\Big(\tfrac d2+p_2+\ell+2\Big)^2+1-\tfrac{d^2}4\Big]-5+12\wt{\nu}_A^2\bigg)/(8(3-d))\Bigg]\n \\
    &+\sum_{\ell=0}^\infty\frac{(-1)^\ell\mathcal{C}_{\frac d2-2\ii\wt{\nu}_A,3}^{p_1p_2}\sin^2\big[\pi\big(\frac d2-2\ii\wt{\nu}_A\big)\big](\ell+1)_{d+p_{12}+5}}{16\pi^2\sin^2(\pi\ii\wt{\nu}_A)\big(\frac d2+\ii\wt{\nu}_A-1\big)^2\big(\frac d2-\ii\wt{\nu}_A-1\big)^2}r_1^{d+p_{12}+6}\bigg(\frac{r_1}{r_2}\bigg)^\ell\frac1{3-d}.
\end{align}
The divergences of $\mathcal{J}_{AA,(\text{3B})}^{p_1p_2}$ and $\mathcal{J}_{AA,(\text{3C})}^{p_1p_2}$ can be computed in the same way. Sum up the divergences of $\mathcal{J}_{AA,(\text{3A})}^{p_1p_2}$, $\mathcal{J}_{AA,(\text{3B})}^{p_1p_2}$ and $\mathcal{J}_{AA,(\text{3C})}^{p_1p_2}$, and then the divergence of the seed integral is
\begin{align}\label{140}
    \text{Div}\mathcal{J}_{AA}^{p_1p_2}=&-\sum_{\ell=0}^\infty\frac{(-1)^\ell\sin\big[\tfrac\pi2(d+p_{12})\big](\ell+1)_{d+p_{12}+1}}{2\pi^2\big(\frac d2+\ii\wt{\nu}_A-1\big)^2\big(\frac d2-\ii\wt{\nu}_A-1\big)^2}r_1^{d+p_{12}+2}\bigg(\frac{r_1}{r_2}\bigg)^\ell\n \\
    &\times\Bigg[\bigg(8\bigg(\Big(\tfrac d2+p_2+\ell+1\Big)^4-\tfrac{d^2}2\Big(\tfrac d2+p_2+\ell+1\Big)^2+\tfrac{d^4}{16}\bigg)\n \\
    &+(-20+48\wt{\nu}_A^2)\bigg(\Big(\tfrac d2+p_2+\ell+1\Big)^2-\tfrac{d^2}4\bigg)+33+168\wt{\nu}_A^2+144\wt{\nu}_A^4\bigg)/(128(3-d))\Bigg]\n \\
    &-\sum_{\ell=0}^\infty\frac{(-1)^\ell\sin\big[\tfrac\pi2(d+p_{12})\big](\ell+1)_{d+p_{12}+3}}{8\pi^2\big(\frac d2+\ii\wt{\nu}_A-1\big)^2\big(\frac d2-\ii\wt{\nu}_A-1\big)^2}r_1^{d+p_{12}+4}\bigg(\frac{r_1}{r_2}\bigg)^\ell\n \\
    &\times\Bigg[-\bigg(4\Big[\Big(\tfrac d2+p_2+\ell+2\Big)^2+1-\tfrac{d^2}4\Big]-5+12\wt{\nu}_A^2\bigg)/(8(3-d))\Bigg]\n \\
    &-\sum_{\ell=0}^\infty\frac{(-1)^\ell(\ell+1)_{d+p_{12}+5}}{32\pi^2\big(\frac d2+\ii\wt{\nu}_A-1\big)^2\big(\frac d2-\ii\wt{\nu}_A-1\big)^2}r_1^{d+p_{12}+6}\bigg(\frac{r_1}{r_2}\bigg)^\ell\frac1{3-d}.
\end{align}

The counter term of $AA$ model is
\begin{align}\label{141}
    \Delta\mathscr{L}=&-\frac1{24}\delta_4a^{d+1}\nabla^2(a^{-2-p_1}\phi_c^2)\nabla^2(a^{-2-p_2}\phi_c^2)-\frac1{24}\delta_2a^{d-1-p_1}\phi_c^2\nabla^2(a^{-2-p_2}\phi_c^2)-\frac1{24}\delta_0a^{d-3-p_{12}}\phi_c^4,
\end{align}
which is the same as (\ref{91}). From (\ref{140}), it is found that the coefficients $\delta_4,\delta_2,\delta_0$ are
\begin{equation}
    \frac13\delta_4=\frac1{64\pi^2m_A^4}\frac1{3-d},\quad\frac13\delta_2=\frac{-5+12\wt{\nu}_A^2}{128\pi^2m_A^4}\frac1{3-d},\quad\frac13\delta_0=\frac{33+168\wt{\nu}_A^2+144\wt{\nu}_A^4}{512\pi^2m_A^4}\frac1{3-d},
\end{equation}
to make sure that the expression of the background converges when $d\rightarrow3$. In the flat space limit the coefficients are:
\begin{equation}
    \frac13\delta_4=\frac1{64\pi^2m_A^4}\frac1{3-d},\quad\frac13\delta_2=\frac3{32\pi^2m_A^2}\frac1{3-d},\quad\frac13\delta_0=\frac9{32\pi^2}\frac1{3-d}.
\end{equation}
The result is different from (\ref{206}) as expected. Next, the local term in (\ref{125}) will be calculated, and the coefficient $\delta_0$ will be corrected.

From (\ref{119}), it can be known that the difference between
\begin{equation}
    g^{\mu\nu}g^{\mu'\nu'}D_{\wt{\nu}_A,\mu\mu',\pm\pm}(y,y')D_{\wt{\nu}_A,\nu\nu',\pm\pm}(y,y')
\end{equation}
and
\begin{align}
    &g^{\mu\nu}g^{\mu'\nu'}D_{\wt{\nu}_A,\mu\mu',\mp\pm}(y,y')D_{\wt{\nu}_A,\nu\nu',\mp\pm}(y,y')\theta(\tau-\tau')\n \\
    +&g^{\mu\nu}g^{\mu'\nu'}D_{\wt{\nu}_A,\mu\mu',\pm\mp}(y,y')D_{\wt{\nu}_A,\nu\nu',\pm\mp}(y,y')\theta(\tau'-\tau)
\end{align}
is
\begin{align}\label{146}
    &\pm\frac{2\ii}{m_A^2}(-\tau)^{d+3}[D_{\wt{\nu}_A,00',\mp\pm}(y,y')\delta(y-y')\theta(\tau-\tau')+D_{\wt{\nu}_A,\mu\mu',\pm\mp}(y,y')\delta(y-y')\theta(\tau'-\tau)]\n \\
    &-\frac1{m_A^4}(-\tau)^{2d+2}[\delta(y-y')]^2.
\end{align}
Eq. (\ref{146}) gives a correction to the seed integral:
\begin{align}\label{147}
    &-\frac12k_s^{d+2+p_{12}}\sum_{\mathsf{a}=\pm}\int_{-\infty}^0\ud\tau(-\tau)^{d+3+p_{12}}e^{\ii k_{1234}\tau}\bigg[\frac1{m_A^2}2\mathsf{a}\ii D_{\wt{\nu}_A,00',\bar{\mathsf{a}}\mathsf{a}}(y,y'=y)\theta(\tau-\tau')\n \\
    &+\frac1{m_A^2}2\mathsf{a}\ii D_{\wt{\nu}_A,00',\mathsf{a}\bar{\mathsf{a}}}(y,y'=y)\theta(\tau'-\tau)-\frac1{m_A^4}(-\tau)^{d-1}\delta(y-y)\bigg].
\end{align}
Delta function $\delta(y-y)$ vanishes in dimensional regularization. The expression of $D_{\wt{\nu}_A,\mu\nu',-+}(y,y')$ is \cite{10.1063/1.4879496}
\begin{equation}\label{148}
    D_{\wt{\nu}_A,-+,\mu\nu'}(y,y')=\frac1{(4\pi)^{(d+1)/2}m_A^2}K_{\mu\nu'}(y,y'),
\end{equation}
where
\begin{align}
    K_{\mu\nu'}(y,y')=&(dI'_{\wt{\nu}_A}(Z)+ZI''_{\wt{\nu}_A}(Z))(\partial_\mu Z)(\partial_{\nu'}Z)\n \\
    &+(-dZI'_{\wt{\nu}_A}(Z)+(1-Z^2)I''_{\wt{\nu}_A}(Z))\partial_\mu\partial_{\nu'}Z,
\end{align}
with
\begin{equation}
    I_{\wt{\nu}_A}(Z)={}_2\mathcal{F}_1\left[\begin{matrix}
        \frac d2+\ii\wt{\nu}, \frac d2-\ii\wt{\nu} \\
        \frac{d+1}2
    \end{matrix}\middle|\frac{1+Z}2\right].
\end{equation}
The embedding distance $Z$ is defined in (\ref{101}). The expression (\ref{148}) is simplified if $Z=1$:
\begin{equation}
     D_{\wt{\nu}_A,-+,\mu\nu'}(y,y'=y)=-\frac d{2(4\pi)^{\frac{d+1}2}m_A^2}g_{\mu\nu'}{}_2\mathcal{F}_1\left[\begin{matrix}
        \frac d2+\ii\wt{\nu}_A+1, \frac d2-\ii\wt{\nu}_A+1 \\
        \frac{d+3}2
    \end{matrix}\middle|1\right].
\end{equation}
Then, dimensional regularization gives
\begin{align}
    &-\frac d{2(4\pi)^{\frac{d+1}2}m_A^2}{}_2\mathcal{F}_1\left[\begin{matrix}
        \frac d2+\ii\wt{\nu}_A+1, \frac d2-\ii\wt{\nu}_A+1 \\
        \frac{d+3}2
    \end{matrix}\middle|1\right]\n \\
    =&-\frac d{2(4\pi)^{\frac{d+1}2}m_A^2}\Gamma\Bigg[\begin{matrix}
        \frac d2+\ii\wt{\nu}_A+1, \frac d2-\ii\wt{\nu}_A+1, -\frac{d+1}2 \\
        \ii\wt{\nu}_A+\frac12, -\ii\wt{\nu}_A+\frac12
    \end{matrix}\Bigg]\n \\
    \sim&-\frac3{32\pi^2m_A^2}\frac1{16(3-d)}(9+40\wt{\nu}_A^2+16\wt{\nu}_A^4)+\frac1{32\pi^2}\frac1{64}(9+40\wt{\nu}_A^2+16\wt{\nu}_A^4)\n \\
    &\times\big[-5+6\gamma_E-6\log(4\pi)+6\psi\big(\ii\wt{\nu}_A+\tfrac52\big)+6\psi\big(-\ii\wt{\nu}_A+\tfrac52\big)\big].
\end{align}
With the above expression, function (\ref{147}) is calculated as
\begin{align}
    &-2\sin\big[\tfrac\pi2(d+p_{12})\big]\Gamma(d+p_{12}+2)\bigg(\frac{r_1r_2}{r_1+r_2}\bigg)^{d+p_{12}+2}\n \\
    &\times\bigg(-\frac3{32\pi^2m_A^4}\frac1{16(3-d)}(9+40\wt{\nu}_A^2+16\wt{\nu}_A^4)+\frac1{32\pi^2m_A^4}\frac1{64}(9+40\wt{\nu}_A^2+16\wt{\nu}_A^4)\n \\
    &\times\big[-5+6\gamma_E-6\log(4\pi)+6\psi\big(\ii\wt{\nu}_A+\tfrac52\big)+6\psi\big(-\ii\wt{\nu}_A+\tfrac52\big)\big]\bigg)+\mathcal{O}(3-d).
\end{align}
After considering the correction of the diagram from the local term, the coefficients of the three divergent contact terms are:
\begin{equation}\label{154}
    \frac13\delta_4=\frac1{64\pi^2m_A^4}\frac1{3-d},\quad\frac13\delta_2=\frac{-5+12\wt{\nu}_A^2}{128\pi^2m_A^4}\frac1{3-d},\quad\frac13\delta_0=\frac{3+24\wt{\nu}_A^2+48\wt{\nu}_A^4}{256\pi^2m_A^4}\frac1{3-d}.
\end{equation}
In the flat space limit, they are:
\begin{equation}
    \frac13\delta_4=\frac1{64\pi^2m_A^4}\frac1{3-d},\quad\frac13\delta_2=\frac3{32\pi^2m_A^2}\frac1{3-d},\quad\frac13\delta_0=\frac3{16\pi^2}\frac1{3-d}.
\end{equation}
The result is the same as (\ref{206}).

MS subtraction can be taken to the background, and the expression of background $\wh{\mathcal{J}}_{AA,\text{BG}}^{p_1p_2}$ is listed in App. \ref{AppF}.
\section{Conclusions and Outlooks}\label{Sec4}
In this work, the dimensional regularization for the bubble diagrams in de Sitter spacetime is finished. It can be finished for correlators with spin-0 and spin-1 internal lines of arbitrary mass. 
 The key to the dimensional regularization in dS is to extract the sum that results in divergence when $d\rightarrow3$, and then take the sum into the Riemann zeta function. With this procedure, the expressions of various kinds of correlators, including 4-point and 2-point, are derived with the minimal subtraction as the renormalization condition.

Eqs. (\ref{106}) and (\ref{154}) show that the coefficients of the counter terms of $\nabla\sigma\nabla\sigma$ model and $AA$ model contain the Hubble parameter $H$. Therefore, there exist interactions between $\phi_c$ or other scalar fields and the curvature of spacetime in these two models. The Hubble parameter $H$ comes from the curvature tensor $R^\mu_{\ \,\nu\rho\sigma}$, the Ricci tensor $R_{\mu\nu}$, and the Ricci scalar $R$. However, it makes no difference in de Sitter spacetime.

The dimensional regularization method used in this work may be used in other expressions of loop diagrams such as the expression of background of the correlator with one bubble loop in \cite{Qin:2024gtr}. It should be expected that the dimensional regularization can also be taken to correlators containing loops with more complicated topologies. However, the complete expressions for such correlators remain to be derived. The expressions of correlators with these loops should be worked out first before regularization can be applied.

This work mainly focuses on correlators with purely bosonic propagators. In some models of correlators in de Sitter spacetime, some internal lines are fermionic propagators. With the expressions of the propagators of fermions \cite{Candelas:1975du,Allen:1986qj}, the expressions of the correlators of these models of arbitrary spatial dimension may be computed. Then dimensional regularization can be applied to these correlators.

\paragraph{Acknowledgments} I am grateful to Qi Chen and Zhehan Qin for useful discussions. I also thank Qi Chen, Zhehan Qin and Haoyuan Liu for comments on the draft. I thank Zhong-Zhi Xianyu specially for the valuable suggestion on the draft. This work is supported by NSFC under Grant No. 12275146, the National Key R\&D Program of China (2021YFC2203100), and the Dushi Program of Tsinghua University.

\newpage

\begin{appendix}

\section*{Appendix}
\section{Useful Functions}\label{AppA}
In this appendix, functions used frequently in this work are listed. First, here are the shorthand notations for the products of Euler $\Gamma$ functions:
\begin{align}
    \Gamma\left[ z_1,\cdots,z_m \right]
    \equiv&~ \Gamma(z_1)\cdots \Gamma(z_m) ,\\
    \Gamma\left[\bgm z_1,\cdots,z_m \\w_1,\cdots, w_n\edm\right]
    \equiv&~\FR{\Gamma(z_1)\cdots \Gamma(z_m)}{\Gamma(w_1)\cdots \Gamma(w_n)}.
\end{align}
Asymptotic expansions of the ratio of Gamma functions for $z\rightarrow\infty$ are:
\begin{equation}
    \frac{\Gamma(z+a)}{\Gamma(z+b)}\sim z^{a-b}\sum_{k=0}^\infty\bigg(\begin{matrix}
        a-b \\
        k
    \end{matrix}\bigg)\mathrm{B}^{(a-b+1)}_k(a)z^{-k},
\end{equation}
here $\bigg(\begin{matrix}
    n \\
    m
\end{matrix}\bigg)$ is the binomial coefficient and $\mathrm{B}^{(\ell)}_n(x)$ is the generalized Bernoulli polynomial which is defined by
\begin{equation}
    \bigg(\frac t{e^t-1}\bigg)^\ell e^{xt}=\sum_{n=0}^{\infty}\mathrm{B}^{(\ell)}_n(x)\frac{t^n}{n!}.
\end{equation}

The definition of the pochhammer symbol $(z)_n$ is
\begin{align}
      (z)_n\equiv\Gamma\left[\bgm z+n \\ z\edm\right].
\end{align}

The generalized hypergeometric function is defined by:
\begin{align}
    {}_p\mathrm{F}_q\left[\bgm a_1,\cdots,a_p \\ b_1,\cdots,b_q \edm  \middle| z \right]=\sum_{n=0}^\infty\FR{(a_1)_n\cdots (a_p)_n}{(b_1)_n\cdots (b_q)_n}\FR{z^n}{n!}.
\end{align}
Sometimes the regularized hypergeometric function ${}_p\wt{\mathrm{F}}_q$ is used:
\begin{equation}
    {}_p\wt{\mathrm{F}}_q\left[\begin{matrix}
        a_1, \cdots, a_p \\
        b_1, \cdots, b_q
    \end{matrix}\middle|z\right]=\frac1{\Gamma\left[b_1, \cdots, b_q\right]}{}_p\mathrm{F}_q\left[\begin{matrix}
        a_1, \cdots, a_p \\
        b_1, \cdots, b_q
    \end{matrix}\middle|z\right].
\end{equation}
The ``dressed" hypergeometric function is also used in this work, which is defined below:
\begin{equation}\label{162}
    {}_p\mathcal{F}_q\left[\begin{matrix}
        a_1, \cdots, a_p \\
        b_1, \cdots, b_q
    \end{matrix}\middle|z\right]=\Gamma\left[\begin{matrix}
        a_1, \cdots, a_p \\
        b_1, \cdots, b_q
    \end{matrix}\right]{}_p\mathrm{F}_q\left[\begin{matrix}
        a_1, \cdots, a_p \\
        b_1, \cdots, b_q
    \end{matrix}\middle|z\right].
\end{equation}
There are considerable transformations of hypergeometric functions. Here are some transformations which are used in this work \cite{nist:dlmf,b172722b-1796-3f88-b31b-57af959043e8}:
\begin{align}\label{163}
    {}_2\mathrm{F}_1\left[\begin{matrix}
        a, b \\
        c
    \end{matrix}\middle|z\right]=&\Gamma\bigg[\begin{matrix}
        c, c-a-b \\
        c-a, c-b
    \end{matrix}\bigg]z^{-a}{}_2\mathrm{F}_1\left[\begin{matrix}
        a, a-c+1 \\
        a+b-c+1
    \end{matrix}\middle|1-\frac1z\right]\n \\
    &+\Gamma\bigg[\begin{matrix}
        c, a+b-c \\
        a, b
    \end{matrix}\bigg]z^{a-c}(1-z)^{c-a-b}{}_2\mathrm{F}_1\left[\begin{matrix}
        c-a, 1-a \\
        c-a-b+1
    \end{matrix}\middle|1-\frac1z\right],
\end{align}
\begin{equation}\label{164}
    {}_2\mathrm{F}_1\left[\begin{matrix}
        a, b \\
        2b
    \end{matrix}\middle|z\right]=\bigg(1-\frac12z\bigg)^{-a}{}_2\mathrm{F}_1\left[\begin{matrix}
        \frac12a, \frac12a+\frac12 \\
        b+\frac12
    \end{matrix}\middle|\bigg(\frac z{2-z}\bigg)^2\right],
\end{equation}
\begin{align}\label{165}
    {}_{p+1}\mathcal{F}_p\left[\begin{matrix}
        a_1, a_2, \cdots, a_{p+1} \\
        b_1, b_2, \cdots, b_p
    \end{matrix}\middle|z\right]=&\sum_{k=0}^\infty A^{(p)}_k{}_2\mathcal{F}_1\left[\begin{matrix}
        a_1, a_2 \\
        s+a_1+a_2+k
    \end{matrix}\middle|z\right]\n \\
    =&\sum_{k=0}^\infty A^{(p)}_k\Bigg(\Gamma\bigg[\begin{matrix}
        a_1, a_2, s+k \\
        a_1+s+k, a_2+s+k
    \end{matrix}\bigg]z^{-a_1}{}_2\mathrm{F}_1\left[\begin{matrix}
        a_1, -a_2-s-k+1 \\
        -s-k+1
    \end{matrix}\middle|1-\frac1z\right]\n \\
    &+\Gamma(-s-k)z^{-a_2-s-k}(1-z)^{s+k}{}_2\mathrm{F}_1\left[\begin{matrix}
        a_2+s+k, -a_1+1 \\
        s+k+1
    \end{matrix}\middle|1-\frac1z\right]\Bigg),
\end{align}
with
\begin{align}
    A^{(p)}_k=&\sum_{k_2=0}^k\frac{(b_p+b_{p-1}+\cdots+b_2-a_{p+1}-a_p-\cdots-a_3+k_2)_{k-k_2}(b_1-a_3)_{k-k_2}}{(k-k_2)!}\n \\
    &\times\sum_{k_3=0}^{k_2}\frac{(b_p+b_{p-1}+\cdots+b_3-a_{p+1}-a_p-\cdots-a_4+k_3)_{k_2-k_3}(b_2-a_4)_{k_2-k_3}}{(k_2-k_3)!}\n \\
    &\times\cdots\n \\
    &\times\sum_{k_{p-1}=0}^{k_{p-2}}\frac{(b_p+b_{p-1}-a_{p+1}-a_p+k_{p-1})_{k_{p-2}-k_{p-1}}(b_{p-2}-a_p)_{k_{p-2}-k_{p-1}}}{(k_{p-2}-k_{p-1})!}\n \\
    &\times\frac{(b_p-a_{p+1})_{k_{p-1}}(b_{p-1}-a_{p+1})_{k_{p-1}}}{k_{p-1}!},
\end{align}
\begin{equation}
    A^{(2)}_k=\frac{(b_2-a_3)_k(b_1-a_3)_k}{k!},
\end{equation}
and
\begin{equation}
    s\equiv\sum_{i=1}^pb_i-\sum_{i=1}^{p+1}a_i.
\end{equation}

The hypergeometric functions $_2\mathrm{F}_1$ with argument unity can be simplified as
\begin{equation}\label{168}
    {}_2\mathrm{F}_1\left[\begin{matrix}
        a, b \\
        c
    \end{matrix}\middle|1\right]=\Gamma\bigg[\begin{matrix}
        c, c-a-b \\
        c-a, c-b
    \end{matrix}\bigg]
\end{equation}
if $\text{Re}(c-a-b)>0$. 

Riemann Zeta function is used in this work to take the dimensional regularization. It is defined as:
\begin{equation}
    \zeta(s)=\sum_{n=1}^\infty\frac1{n^s},
\end{equation}
for $\text{Re}\;s>1$. Then $\zeta(s)$ is defined by analytic continuation. The only simple pole in $\mathbb{C}$ is $s=1$ with residue 1 \cite{nist:dlmf}.
\section{Pole Structure of the Tree Seed Integral}\label{AppB}
In this section, the pole structure of the tree seed integral will be studied. It can be found that although the pole structure of nonlocal signal, local signal and background are slightly difficult, the pole structure of the full seed integral is simple.

First, consider the nonlocal signal. For the term shown in (\ref{15}) directly, poles in the lower-half $\wt{\nu}'$-plane are of interest, while for the hidden term, poles in the upper-half $\wt{\nu}'$-plane are considered. Poles of (\ref{16}) and (\ref{17}) are studied in the same way. The considered poles of the nonlocal signal are:
\begin{align}
  &\text{Poles}   
  &&\text{terms shown directly}
  &&\text{hidden terms} \n \\
  &\text{Set 1A:} &&\wt{\nu}'=-\ii n,  &&\wt{\nu}'=+\ii n,
\end{align}
In this appendix $n$ goes over all nonnegative integers.

The pole structure of the local signal is slightly difficult. The considered poles of (\ref{16}) are:
\begin{align}
  &\text{Poles}   
  &&\text{terms shown directly}
  &&\text{hidden terms} \n\\
  &\text{Set 2A:} &&\wt{\nu}'=-\ii n,  &&\wt{\nu}'=+\ii n,\\
  &\text{Set 2B:} &&\wt\nu'=-\ii(d/2+p_2+1+n), &&\wt\nu'=\ii(d/2+p_2+1+n).
\end{align}
The poles of Set 1A and Set 2A are second-order, from the factor $\csc^2(\pi\ii\wt{\nu}')$. However, the poles of the sum of the nonlocal and local signals at $\wt{\nu}'=\mp\ii n$ are simple poles. Notice that
\begin{align}
    r^{-n}\mb{F}_{-n,d}^{p}(r)=&(2r)^{p+d/2+1}r^{-n}{}_2\mathcal{F}_1\left[ \begin{matrix}
        \tfrac d4+\tfrac12+\tfrac{p}2-\tfrac n2, \tfrac d4+1+\tfrac{p}2-\tfrac n2 \\
        1-n
    \end{matrix}\middle|r^2\right]\n \\
    =&(2r)^{p+d/2+1}\sum_{\ell=0}^\infty\frac1{\ell!}\Gamma\Bigg[\begin{matrix}
        \tfrac d4+\tfrac12+\tfrac{p}2-\tfrac n2+\ell, \tfrac d4+1+\tfrac{p}2-\tfrac n2+\ell \\
        1-n+\ell
    \end{matrix}\Bigg]r^{2\ell-n}.
\end{align}
The summand vanishes if $\ell<n$ due to the $\Gamma(1-n+\ell)$ in the denominator. Therefore,
\begin{align}
    r^{-n}\mb{F}_{-n,d}^{p}(r)=&(2r)^{p+d/2+1}\sum_{\ell=n}^\infty\frac1{\ell!}\Gamma\Bigg[\begin{matrix}
        \tfrac d4+\tfrac12+\tfrac{p}2-\tfrac n2+\ell, \tfrac d4+1+\tfrac{p}2-\tfrac n2+\ell \\
        1-n+\ell
    \end{matrix}\Bigg]r^{2\ell-n}\n \\
    \xlongequal{\ell\rightarrow\ell+n}&(2r)^{p+d/2+1}\sum_{\ell=0}^\infty\frac1{(\ell+n)!}\Gamma\Bigg[\begin{matrix}
        \tfrac d4+\tfrac12+\tfrac{p}2+\tfrac n2+\ell, \tfrac d4+1+\tfrac{p}2+\tfrac n2+\ell \\
        1+\ell
    \end{matrix}\Bigg]r^{2\ell+n}\n \\
    =&r^n\mb{F}_{n,d}^{p}(r).
\end{align}
This cancellation makes that the poles of the sum of the nonlocal and local signals at $\wt{\nu}'=\mp\ii n$ simpler. If $\sinh(\pi\wt{\nu}')$ from the spectral function of K\"all\'en-Lehmann representation is considered, there is no pole at $\wt{\nu}'=\mp\ii n$ in the integrand of (\ref{13}).

Here are the considered poles of (\ref{17}):
\begin{align}
  &\text{Poles}   
  &&\text{terms shown directly}
  &&\text{hidden terms} \n \\
  &\text{Set 3A:} &&\wt\nu'=-\ii(d/2+p_2+1+n), &&\wt\nu'=\ii(d/2+p_2+1+n).
\end{align}
To find out the residues of (\ref{17}), it is better to rewrite the expression in the following way:
\begin{align}
    \mathcal{I}_{\text{BG},\wt\nu}^{p_1p_2}(r_1,r_2)=&-\sum_{\ell,j=0}^\infty\frac{(-1)^{\ell+1}\mathcal{C}_{\ii\wt{\nu},d}^{p_1p_2}\sin(\pi\ii\wt{\nu})(\ell+1)_{2j+d+p_{12}+1}}{2^{2j-1}\big(\tfrac{\ell-\ii \wt{\nu}+p_2+1}2+\tfrac d4\big)_{j+1}\big(\tfrac{\ell+\ii \wt{\nu}+p_2+1}2+\tfrac d4\big)_{j+1}}r_1^{2j+d+p_{12}+2}\Big(\frac{r_1}{r_2}\Big)^{\ell}+(\wt{\nu}\rightarrow(-\wt{\nu}))\n \\
    =&\Bigg(\sum_{\ell,j=0}^\infty\frac{\mathcal{C}_{\ii\wt{\nu},d}^{p_1p_2}\sin(\pi\ii\wt{\nu})}{2^{2j-1}(2\ell)!}r_1^{2j+d+p_{12}+2}\bigg(\frac{r_1}{r_2}\bigg)^{2\ell}\n \\
    &\times\Gamma\Bigg[\begin{matrix}
        2\ell+2j+d+p_{12}+2, \frac{-\ii\wt{\nu}+p_2+1}2+\frac d4+\ell, \frac{\ii\wt{\nu}+p_2+1}2+\frac d4+\ell \\
        \frac{-\ii\wt{\nu}+p_2+3}2+\frac d4+\ell+j, \frac{\ii\wt{\nu}+p_2+3}2+\frac d4+\ell+j
    \end{matrix}\Bigg]\n \\
    &-\sum_{\ell,j=0}^\infty\frac{\mathcal{C}_{\ii\wt{\nu},d}^{p_1p_2}\sin(\pi\ii\wt{\nu})}{2^{2j-1}(2\ell+1)!}r_1^{2j+d+p_{12}+2}\bigg(\frac{r_1}{r_2}\bigg)^{2\ell+1}\n \\
    &\times\Gamma\Bigg[\begin{matrix}
        2\ell+2j+d+p_{12}+3, \frac{-\ii\wt{\nu}+p_2}2+1+\frac d4+\ell, \frac{\ii\wt{\nu}+p_2}2+1+\frac d4+\ell \\
        \frac{-\ii\wt{\nu}+p_2}2+2+\frac d4+\ell+j, \frac{\ii\wt{\nu}+p_2}2+2+\frac d4+\ell+j
    \end{matrix}\Bigg]\Bigg)\n \\
    &+(\wt{\nu}\rightarrow(-\wt{\nu}))\n \\
    \xlongequal{\ell+j\rightarrow j}&\Bigg(\sum_{j=0}^\infty\sum_{\ell=0}^j\frac{\mathcal{C}_{\ii\wt{\nu},d}^{p_1p_2}\sin(\pi\ii\wt{\nu})}{2^{2j-2\ell-1}(2\ell)!}r_1^{2j+d+p_{12}+2}r_2^{-2\ell}\n \\
    &\times\Gamma\Bigg[\begin{matrix}
        2j+d+p_{12}+2, \frac{-\ii\wt{\nu}+p_2+1}2+\frac d4+\ell, \frac{\ii\wt{\nu}+p_2+1}2+\frac d4+\ell \\
        \frac{-\ii\wt{\nu}+p_2+3}2+\frac d4+j, \frac{\ii\wt{\nu}+p_2+3}2+\frac d4+j
    \end{matrix}\Bigg]\n \\
    &-\sum_{j=0}^\infty\sum_{\ell=0}^j\frac{\mathcal{C}_{\ii\wt{\nu},d}^{p_1p_2}\sin(\pi\ii\wt{\nu})}{2^{2j-2\ell-1}(2\ell+1)!}r_1^{2j+d+p_{12}+3}r_2^{-2\ell-1}\n \\
    &\times\Gamma\Bigg[\begin{matrix}
        2j+d+p_{12}+3, \frac{-\ii\wt{\nu}+p_2}2+1+\frac d4+\ell, \frac{\ii\wt{\nu}+p_2}2+1+\frac d4+\ell \\
        \frac{-\ii\wt{\nu}+p_2}2+2+\frac d4+j, \frac{\ii\wt{\nu}+p_2}2+2+\frac d4+j
    \end{matrix}\Bigg]\Bigg)\n \\
    &+(\wt{\nu}\rightarrow(-\wt{\nu})).
\end{align}
If $n$ is even, the residues of poles of Set 3A are
\begin{align}
    &\text{Res}\Big[\ii\mathcal{I}'^{p_1p_2}_{\text{BG},\wt\nu'}(r_1,r_2)\Big]_{\wt{\nu}'=-\ii(d/2+p_2+1+n)}\n \\
    =&\sum_{j=\max\{0,\frac n2\}}^\infty\sum_{\ell=0}^{\min\{j,\frac n2\}}\frac{(-1)^{\frac n2-\ell}\mathcal{C}_{d/2+p_2+1+n,d}^{p_1p_2}\sin\big[\pi\big(\frac d2+p_2\big)\big]}{2^{2j-2\ell-2}\big(\frac n2-\ell\big)!(2\ell)!}r_1^{2j+d+p_{12}+2}r_2^{-2\ell}\n \\
    &\times\Gamma\Bigg[\begin{matrix}
        2j+d+p_{12}+2, p_2+\frac d2+\frac n2+\ell+1 \\
        j-\frac n2+1, p_2+\frac d2+\frac n2+j+2
    \end{matrix}\Bigg].
\end{align}
The parameter $j$ is not smaller than $n/2$, and $n$ is not smaller than $0$, therefore,
\begin{align}
    &\text{Res}\Big[\ii\mathcal{I}'^{p_1p_2}_{\text{BG},\wt\nu'}(r_1,r_2)\Big]_{\wt{\nu}'=-\ii(d/2+p_2+1+n)}\n \\
    =&\sum_{j=\frac n2}^\infty\sum_{\ell=0}^{\frac n2}\frac{(-1)^{\frac n2-\ell}\mathcal{C}_{d/2+p_2+1+n,d}^{p_1p_2}\sin\big[\pi\big(\frac d2+p_2\big)\big]}{2^{2j-2\ell-2}\big(\frac n2-\ell\big)!(2\ell)!}r_1^{2j+d+p_{12}+2}r_2^{-2\ell}\n \\
    &\times\Gamma\Bigg[\begin{matrix}
        2j+d+p_{12}+2, p_2+\frac d2+\frac n2+\ell+1 \\
        j-\frac n2+1, p_2+\frac d2+\frac n2+j+2
    \end{matrix}\Bigg]\n \\
    \xlongequal{j\rightarrow j+n/2}&\sum_{j=0}^\infty\sum_{\ell=0}^{\frac n2}\frac{(-1)^{\frac n2-\ell}\mathcal{C}_{d/2+p_2+1+n,d}^{p_1p_2}\sin\big[\pi\big(\frac d2+p_2\big)\big]}{2^{2j-2\ell+n-2}\big(\frac n2-\ell\big)!(2\ell)!}r_1^{2j+n+d+p_{12}+2}r_2^{-2\ell}\n \\
    &\times\Gamma\Bigg[\begin{matrix}
        2j+n+d+p_{12}+2, p_2+\frac d2+\frac n2+\ell+1 \\
        j+1, p_2+\frac d2+n+j+2
    \end{matrix}\Bigg]\n \\
    \xlongequal{\ell\rightarrow n/2-\ell}&\sum_{j=0}^\infty\sum_{\ell=0}^{\frac n2}\frac{(-1)^\ell\mathcal{C}_{d/2+p_2+1+n,d}^{p_1p_2}\sin\big[\pi\big(\frac d2+p_2\big)\big]}{2^{2j+2\ell-2}\ell!(n-2\ell)!}r_1^{2j+n+d+p_{12}+2}r_2^{-n+2\ell}\n \\
    &\times\Gamma\Bigg[\begin{matrix}
        2j+n+d+p_{12}+2, p_2+\frac d2+n-\ell+1 \\
        j+1, p_2+\frac d2+n+j+2
    \end{matrix}\Bigg]\n \\
    =&-\sum_{j=0}^\infty\sum_{\ell=0}^{\frac n2}\frac{2^{d+p_{12}+3+n}\pi^{\frac12}\mathcal{C}_{d/2+p_2+1+n,d}^{p_1p_2}\big(-\frac n2\big)_\ell\big(\frac{-n+1}2\big)_\ell}{\ell!n!}r_1^{2j+n+d+p_{12}+2}r_2^{-n+2\ell}\n \\
    &\times\Gamma\Bigg[\begin{matrix}
        \frac{n+d+p_{12}}2+j+1, \frac{n+d+p_{12}+3}2+j \\
        j+1, p_2+\frac d2+n+j+2, -p_2-\frac d2-n+\ell
    \end{matrix}\Bigg]\n \\
    =&-2^{d+p_{12}+3+n}\pi^{\frac12}\mathcal{C}_{d/2+p_2+1+n,d}^{p_1p_2}r_1^{n+d+p_{12}+2}r_2^{-n}{}_2\mathcal{F}_1\left[\begin{matrix}
        \frac{n+d+p_{12}}2+1, \frac{n+d+p_{12}+3}2 \\
        p_2+\frac d2+n+2
    \end{matrix}\middle|r_1^2\right]\n \\
    &\times\frac1{n!}{}_2\wt{\mathrm{F}}_1\left[\begin{matrix}
        -\frac{n}2, \frac{-n+1}2 \\
        -p_2-\frac d2-n
    \end{matrix}\middle|r_2^2\right]\n \\
    =&-2^{\frac d2+p_2+2+n}\pi^{\frac12}\mathcal{C}_{d/2+p_2+1+n,d}^{p_1p_2}r_1^{\frac d2+p_2+1+n}\mb{F}_{d/2+p_2+1+n,d}^{p_1}(r_1)\frac1{n!}{}_2\wt{\mathrm{F}}_1\left[\begin{matrix}
        -\frac{n}2, \frac{-n+1}2 \\
        -p_2-\frac d2-n
    \end{matrix}\middle|r_2^2\right].
\end{align}
If $n$ is odd, the residues are
\begin{align}
    &\text{Res}\Big[\ii\mathcal{I}'^{p_1p_2}_{\text{BG},\wt\nu'}(r_1,r_2)\Big]_{\wt{\nu}'=-\ii(d/2+p_2+1+n)}\n \\
    =&\sum_{j=\max\{0,\frac{n-1}2\}}^\infty\sum_{\ell=0}^{\min\{j,\frac{n-1}2\}}\frac{(-1)^{\frac n2-\frac12-\ell}\mathcal{C}_{d/2+p_2+1+n,d}^{p_1p_2}\sin\big[\pi\big(\frac d2+p_2\big)\big]}{2^{2j-2\ell-2}\big(\frac{n-1}2-\ell\big)!(2\ell+1)!}r_1^{2j+d+p_{12}+3}r_2^{-2\ell-1}\n \\
    &\times\Gamma\Bigg[\begin{matrix}
        2j+d+p_{12}+3, p_2+\frac d2+\frac n2+\ell+\frac32 \\
        j-\frac n2+\frac32, p_2+\frac d2+\frac n2+j+\frac52
    \end{matrix}\Bigg].
\end{align}
It is easily to find that $0\leqslant(n-1)/2\leqslant j$, therefore,
\begin{align}
    &\text{Res}\Big[\ii\mathcal{I}'^{p_1p_2}_{\text{BG},\wt\nu'}(r_1,r_2)\Big]_{\wt{\nu}'=-\ii(d/2+p_2+1+n)}\n \\
    =&\sum_{j=\frac{n-1}2}^\infty\sum_{\ell=0}^{\frac{n-1}2}\frac{(-1)^{\frac n2-\frac12-\ell}\mathcal{C}_{d/2+p_2+1+n,d}^{p_1p_2}\sin\big[\pi\big(\frac d2+p_2\big)\big]}{2^{2j-2\ell-2}\big(\frac{n-1}2-\ell\big)!(2\ell+1)!}r_1^{2j+d+p_{12}+3}r_2^{-2\ell-1}\n \\
    &\times\Gamma\Bigg[\begin{matrix}
        2j+d+p_{12}+3, p_2+\frac d2+\frac n2+\ell+\frac32 \\
        j-\frac n2+\frac32, p_2+\frac d2+\frac n2+j+\frac52
    \end{matrix}\Bigg]\n \\
    \xlongequal{j\rightarrow j+(n-1)/2}&\sum_{j=0}^\infty\sum_{\ell=0}^{\frac{n-1}2}\frac{(-1)^{\frac n2-\frac12-\ell}\mathcal{C}_{d/2+p_2+1+n,d}^{p_1p_2}\sin\big[\pi\big(\frac d2+p_2\big)\big]}{2^{2j-2\ell+n-3}\big(\frac{n-1}2-\ell\big)!(2\ell+1)!}r_1^{2j+n+d+p_{12}+2}r_2^{-2\ell-1}\n \\
    &\times\Gamma\Bigg[\begin{matrix}
        2j+n+d+p_{12}+2, p_2+\frac d2+\frac n2+\ell+\frac32 \\
        j+1, p_2+\frac d2+n+j+2
    \end{matrix}\Bigg]\n \\
    \xlongequal{\ell\rightarrow(n-1)/2-\ell}&\sum_{j=0}^\infty\sum_{\ell=0}^{\frac{n-1}2}\frac{(-1)^{\ell}\mathcal{C}_{d/2+p_2+1+n,d}^{p_1p_2}\sin\big[\pi\big(\frac d2+p_2\big)\big]}{2^{2j-2\ell+n-3}\ell!(n-2\ell)!}r_1^{2j+n+d+p_{12}+2}r_2^{-n+2\ell}\n \\
    &\times\Gamma\Bigg[\begin{matrix}
        2j+n+d+p_{12}+2, p_2+\frac d2+n-\ell+1 \\
        j+1, p_2+\frac d2+n+j+2
    \end{matrix}\Bigg]\n \\
    =&\sum_{j=0}^\infty\sum_{\ell=0}^{\frac n2}\frac{2^{d+p_{12}+3+n}\pi^{\frac12}\mathcal{C}_{d/2+p_2+1+n,d}^{p_1p_2}\big(-\frac n2\big)_\ell\big(\frac{-n+1}2\big)_\ell}{\ell!n!}r_1^{2j+n+d+p_{12}+2}r_2^{-n+2\ell}\n \\
    &\times\Gamma\Bigg[\begin{matrix}
        \frac{n+d+p_{12}}2+j+1, \frac{n+d+p_{12}+3}2+j \\
        j+1, p_2+\frac d2+n+j+2, -p_2-\frac d2-n+\ell
    \end{matrix}\Bigg]\n \\
    =&2^{d+p_{12}+3+n}\pi^{\frac12}\mathcal{C}_{d/2+p_2+1+n,d}^{p_1p_2}r_1^{n+d+p_{12}+2}r_2^{-n}{}_2\mathcal{F}_1\left[\begin{matrix}
        \frac{n+d+p_{12}}2+1, \frac{n+d+p_{12}+3}2 \\
        p_2+\frac d2+n+2
    \end{matrix}\middle|r_1^2\right]\n \\
    &\times\frac1{n!}{}_2\wt{\mathrm{F}}_1\left[\begin{matrix}
        -\frac{n}2, \frac{-n+1}2 \\
        -p_2-\frac d2-n
    \end{matrix}\middle|r_2^2\right]\n \\
    =&2^{\frac d2+p_2+2+n}\pi^{\frac12}\mathcal{C}_{d/2+p_2+1+n,d}^{p_1p_2}r_1^{\frac d2+p_2+1+n}\mb{F}_{d/2+p_2+1+n,d}^{p_1}(r_1)\frac1{n!}{}_2\wt{\mathrm{F}}_1\left[\begin{matrix}
        -\frac{n}2, \frac{-n+1}2 \\
        -p_2-\frac d2-n
    \end{matrix}\middle|r_2^2\right].
\end{align}
Therefore, the residues of poles which are considered are
\begin{align}
    &\text{Res}\Big[\ii\mathcal{I}'^{p_1p_2}_{\text{BG},\wt\nu'}(r_1,r_2)\Big]_{\wt{\nu}'=-\ii(d/2+p_2+1+n)}\n \\
    =&-(-1)^n2^{\frac d2+p_2+2+n}\pi^{\frac12}\mathcal{C}_{d/2+p_2+1+n,d}^{p_1p_2}r_1^{\frac d2+p_2+1+n}\mb{F}_{d/2+p_2+1+n,d}^{p_1}(r_1)\frac1{n!}{}_2\wt{\mathrm{F}}_1\left[\begin{matrix}
        -\frac{n}2, \frac{-n+1}2 \\
        -p_2-\frac d2-n
    \end{matrix}\middle|r_2^2\right].
\end{align}

The residues of poles of Set 2B are
\begin{align}
    &\text{Res}\Big[\ii\mathcal{I}'^{p_1p_2}_{\text{L},\wt\nu'}(r_1,r_2)\Big]_{\wt{\nu}'=-\ii(d/2+p_2+1+n)}\n \\
    =&(-1)^n2^{\frac d2+p_2+2+n}\pi^{\frac12}\mathcal{C}_{d/2+p_2+1+n,d}^{p_1p_2}r_1^{\frac d2+p_2+1+n}\mb{F}_{d/2+p_2+1+n,d}^{p_1}(r_1)\frac1{n!}{}_2\wt{\mathrm{F}}_1\left[\begin{matrix}
        -\frac{n}2, \frac{-n+1}2 \\
        -p_2-\frac d2-n
    \end{matrix}\middle|r_2^2\right].
\end{align}

The residues of poles of Set 2B and Set 3A can be canceled. Therefore, the poles shown in this section do not appear in the calculation of the integration (\ref{13}).
\section{2-point Correlator at Tree Level}\label{AppC}
In this appendix, the folded limit $r\rightarrow1$ is taken to the tree seed integral $\mathcal{I}_{\wt\nu}^{p_1p_2}(r,1)$. The expression of the signal of the integral is:
\begin{align}\label{183}
    &\mathcal{I}_{\text{S},\wt\nu}^{p_1p_2}(r(u),1)\n \\
    =&\Bigg(-2^{1-p_2-d/2}\sin(\pi\ii\wt{\nu})\Gamma\Bigg[\begin{matrix}
        p_2+\frac d2+1+\ii\wt{\nu}, p_2+\frac d2+1-\ii\wt{\nu} \\
        p_2+\frac d2+\frac32
    \end{matrix}\Bigg]\mathcal{C}_{\ii\wt\nu,d}^{p_1p_2}r^{\ii\wt{\nu}}\mb{F}_{\ii\wt\nu,d}^{p_1}(r)\Bigg)+(\wt{\nu}\rightarrow(-\wt{\nu}))\n \\
    =&\Bigg(-2^{1-p_{12}-d}\mathcal{C}_{\ii\wt\nu,d}^{p_1p_2}\Gamma\Bigg[\begin{matrix}
        p_1+\frac d2+1+\ii\wt{\nu}, p_1+\frac d2+1-\ii\wt{\nu}, p_2+\frac d2+1+\ii\wt{\nu}, p_2+\frac d2+1-\ii\wt{\nu} \\
        p_1+\frac d2+\frac32, p_2+\frac d2+\frac32
    \end{matrix}\Bigg]\n \\
    &\times\frac{\sin(\pi\ii\wt{\nu})\sin\big[\pi\big(p_1+\frac d2-\ii\wt{\nu}\big)\big]}{\sin\big[\pi\big(p_1+\frac d2+\frac12\big)\big]}u^{p_1+d/2+1/2}{}_2\mathrm{F}_1\left[\begin{matrix}
        \frac12+\ii\wt{\nu}, \frac12-\ii\wt{\nu} \\
        p_1+\frac d2+\frac32
    \end{matrix}\middle|1-\frac1u\right]\n \\
    &-2^{1-p_{12}-d}\mathcal{C}_{\ii\wt\nu,d}^{p_1p_2}\sin(\pi\ii\wt{\nu})\Gamma\Bigg[\begin{matrix}
        p_2+\frac d2+1+\ii\wt{\nu}, p_2+\frac d2+1-\ii\wt{\nu} \\
        p_2+\frac d2+\frac32
    \end{matrix}\Bigg]u^{p_1+d/2+1/2}(1-u)^{-p_1-d/2-1/2}\n \\
    &\times\Gamma\big(p_1+\tfrac d2+\tfrac12\big){}_2\mathrm{F}_1\left[\begin{matrix}
        \frac12+\ii\wt{\nu}, \frac12-\ii\wt{\nu} \\
        -p_1-\frac d2+\frac12
    \end{matrix}\middle|1-\frac1u\right]\Bigg)+(\wt{\nu}\rightarrow(-\wt{\nu})).
\end{align}
Function (\ref{163}) is used in the above formula. For the background, by using (\ref{165}), the expression is changed as
\begin{align}
    \mathcal{I}_{\text{BG},\wt\nu}^{p_1p_2}(r(u),1)=&2^{-p_{12}-d+1}\mathcal{C}_{\ii\wt{\nu},d}^{p_1p_2}\sin(\pi\ii\wt{\nu})\Gamma\big[p_{12}+d+2, \ii\wt{\nu}+p_2+1+\tfrac d2, -\ii\wt{\nu}+p_2+1+\tfrac d2\big]\n \\
    &\times u^{p_{12}+d+2}{}_3\wt{\mathrm{F}}_2\left[\begin{matrix}
        p_{12}+d+2, p_2+\tfrac{d+3}2, 1 \\
        \ii\wt{\nu}+p_2+2+\tfrac d2, -\ii\wt{\nu}+p_2+2+\tfrac d2
    \end{matrix}\middle|u\right]+(\wt{\nu}\rightarrow(-\wt{\nu}))\n \\
    =&\Bigg[2^{-p_{12}-d+1}u^{p_{12}+d+2}\mathcal{C}_{\ii\wt{\nu},d}^{p_1p_2}\sin(\pi\ii\wt{\nu})\Gamma\Bigg[\begin{matrix}
        1+p_2+\frac d2+\ii\wt{\nu}, 1+p_2+\frac d2-\ii\wt{\nu} \\
        p_2+\tfrac d2+\tfrac32
    \end{matrix}\Bigg]\n \\
    &\times\sum_{j=0}^\infty\Bigg(\frac{\big(p_2+\frac d2+1+\ii\wt{\nu}\big)_j\big(p_2+\frac d2+1-\ii\wt{\nu}\big)_j}{j!}\Gamma\Bigg[\begin{matrix}
        p_2+\frac d2+\frac32, p_{12}+d+2, -p_1-\frac d2-\frac12+j \\
        -p_1+p_2+1+j, p_2+\frac d2+\frac32+j
    \end{matrix}\Bigg]\n \\
    &\times u^{-p_2-\frac d2-\frac32}{}_2\mathrm{F}_1\left[\begin{matrix}
        p_2+\frac d2+\frac32, -p_2-\frac d2-\frac12-j \\
        p_1+\frac d2+\frac32-j
    \end{matrix}\middle|1-\frac1u\right]\n \\
    &+\frac{\big(p_2+\frac d2+1+\ii\wt{\nu}\big)_j\big(p_2+\frac d2+1-\ii\wt{\nu}\big)_j}{j!}\Gamma\big(p_1+\tfrac d2+\tfrac12-j\big)u^{-p_2-\frac d2-\frac32}(1-u)^{-p_1-d/2-1/2}\n \\
    &\times\bigg(\frac1u-1\bigg)^j{}_2\mathrm{F}_1\left[\begin{matrix}
        p_2+\frac d2+\frac32+j, -p_2-\frac d2-\frac12 \\
        -p_1-\frac d2+\frac12+j
    \end{matrix}\middle|1-\frac1u\right]\Bigg)\Bigg]+(\wt{\nu}\rightarrow(-\wt{\nu}))\n \\
    =&\Bigg[2^{-p_{12}-d+1}u^{p_{12}+d+2}\mathcal{C}_{\ii\wt{\nu},d}^{p_1p_2}\sin(\pi\ii\wt{\nu})\Gamma\Bigg[\begin{matrix}
        1+p_2+\frac d2+\ii\wt{\nu}, 1+p_2+\frac d2-\ii\wt{\nu} \\
        p_2+\tfrac d2+\tfrac32
    \end{matrix}\Bigg]\n \\
    &\times\sum_{j=0}^\infty\Bigg(\frac{\big(p_2+\frac d2+1+\ii\wt{\nu}\big)_j\big(p_2+\frac d2+1-\ii\wt{\nu}\big)_j}{j!}\Gamma\Bigg[\begin{matrix}
        p_2+\frac d2+\frac32, p_{12}+d+2, -p_1-\frac d2-\frac12+j \\
        -p_1+p_2+1+j, p_2+\frac d2+\frac32+j
    \end{matrix}\Bigg]\n \\
    &\times u^{-p_2-\frac d2-\frac32}{}_2\mathrm{F}_1\left[\begin{matrix}
        p_2+\frac d2+\frac32, -p_2-\frac d2-\frac12-j \\
        p_1+\frac d2+\frac32-j
    \end{matrix}\middle|1-\frac1u\right]\n \\
    &+\sum_{n=0}^\infty\frac{\big(p_2+\frac d2+1+\ii\wt{\nu}\big)_j\big(p_2+\frac d2+1-\ii\wt{\nu}\big)_j}{j!}\Gamma\big(p_1+\tfrac d2+\tfrac12-j\big)u^{-p_2-\frac d2-\frac32}\n \\
    &\times(1-u)^{-p_1-d/2-1/2}\bigg(\frac1u-1\bigg)^j\frac{(p_2+\frac d2+\frac32+j)_n(-p_2-\frac d2-\frac12)_n}{n!(-p_1-\frac d2+\frac12+j)_n}\bigg(1-\frac1u\bigg)^n\Bigg)\Bigg]\n \\
    &+(\wt{\nu}\rightarrow(-\wt{\nu})).
\end{align}
For the second term, change the summation variable $n'=n+j$ and then $n'\rightarrow n$:
\begin{align}\label{185}
    \mathcal{I}_{\text{BG},\wt\nu}^{p_1p_2}(r(u),1)=&\Bigg[2^{-p_{12}-d+1}u^{p_{12}+d+2}\mathcal{C}_{\ii\wt{\nu},d}^{p_1p_2}\sin(\pi\ii\wt{\nu})\Gamma\Bigg[\begin{matrix}
        1+p_2+\frac d2+\ii\wt{\nu}, 1+p_2+\frac d2-\ii\wt{\nu} \\
        p_2+\tfrac d2+\tfrac32
    \end{matrix}\Bigg]\n \\
    &\times\Bigg(\sum_{j=0}^\infty\sum_{\ell=0}^\infty\frac{\big(p_2+\frac d2+1+\ii\wt{\nu}\big)_j\big(p_2+\frac d2+1-\ii\wt{\nu}\big)_j}{j!}\n \\
    &\times\Gamma\Bigg[\begin{matrix}
        p_2+\frac d2+\frac32+\ell, p_{12}+d+2, -p_1-\frac d2-\frac12+j-\ell \\
        -p_1+p_2+1+j, p_2+\frac d2+\frac32+j-\ell
    \end{matrix}\Bigg]u^{-p_2-\frac d2-\frac32}\frac1{\ell!}\bigg(1-\frac1u\bigg)^\ell\n \\
    &+\sum_{n=0}^\infty\sum_{j=0}^n\frac{\big(p_2+\frac d2+1+\ii\wt{\nu}\big)_j\big(p_2+\frac d2+1-\ii\wt{\nu}\big)_j}{j!}\Gamma\big(p_1+\tfrac d2+\tfrac12-j\big)u^{-p_2-\frac d2-\frac32}\n \\
    &\times(1-u)^{-p_1-d/2-1/2}\frac{\big(p_2+\frac d2+\frac32+j\big)_{n-j}\big(-p_2-\frac d2-\frac12\big)_{n-j}}{(n-j)!\big(-p_1-\frac d2+\frac12+k\big)_{n-j}}(-1)^j\bigg(1-\frac1u\bigg)^n\Bigg)\Bigg]\n \\
    &+(\wt{\nu}\rightarrow(-\wt{\nu}))\n \\
    =&\Bigg[2^{-p_{12}-d+1}u^{p_{12}+d+2}\mathcal{C}_{\ii\wt{\nu},d}^{p_1p_2}\sin(\pi\ii\wt{\nu})\Gamma\Bigg[\begin{matrix}
        1+p_2+\frac d2+\ii\wt{\nu}, 1+p_2+\frac d2-\ii\wt{\nu} \\
        p_2+\tfrac d2+\tfrac32
    \end{matrix}\Bigg]\n \\
    &\times\Bigg(\sum_{\ell=0}^\infty\Gamma\Bigg[\begin{matrix}
        p_2+\frac d2+\frac32+\ell, p_{12}+d+2, -p_1-\frac d2-\frac12-\ell \\
        -p_1+p_2+1, p_2+\frac d2+\frac32-\ell
    \end{matrix}\Bigg]u^{-p_2-\frac d2-\frac32}\frac1{\ell!}\bigg(1-\frac1u\bigg)^\ell\n \\
    &\times{}_3\mathrm{F}_2\left[\begin{matrix}
        p_2+\frac d2+1+\ii\wt{\nu}, p_2+\frac d2+1-\ii\wt{\nu}, -p_1-\frac d2-\frac12-\ell \\
        -p_1+p_2+1, p_2+\frac d2+\frac32-\ell
    \end{matrix}\middle|1\right]\n \\
    &+\sum_{n=0}^\infty\frac1{n!}\Gamma\Bigg[\begin{matrix}
        p_2+\frac d2+\frac32+n \\
        -p_1-\frac d2+\frac12+n, p_2+\frac d2+\frac32, -p_2-\frac d2-\frac12, p_2+\frac d2+\frac32-n
    \end{matrix}\Bigg]\n \\
    &\times\frac{\pi^2}{\sin\big[\pi\big(p_2+\frac d2+\frac32-n\big)\big]\sin\big[\pi\big(-p_1-\frac d2+\frac12\big)\big]}\n \\
    &\times{}_3\mathrm{F}_2\left[\begin{matrix}
        p_2+\frac d2+1+\ii\wt{\nu}, p_2+\frac d2+1-\ii\wt{\nu}, -n \\
        p_2+\frac d2+\frac32, p_2+\frac d2+\frac32-n
    \end{matrix}\middle|1\right]\n \\
    &\times u^{-p_2-\frac d2-\frac32}(1-u)^{-p_1-d/2-1/2}\bigg(1-\frac1u\bigg)^n\Bigg)\Bigg]+(\wt{\nu}\rightarrow(-\wt{\nu})).
\end{align}
Using the following formula \cite{nist:dlmf,Slater:1966},
\begin{equation}
    {}_3\mathrm{F}_2\left[\begin{matrix}
        a, b, -n \\
        c, 1+a+b-c-n
    \end{matrix}\middle|1\right]=\frac{(c-a)_n(c-b)_n}{(c)_n(c-a-b)_n},
\end{equation}
the expression (\ref{185}) becomes
\begin{align}\label{187}
    \mathcal{I}_{\text{BG},\wt\nu}^{p_1p_2}(r(u),1)=&\Bigg[2^{-p_{12}-d+1}u^{p_{12}+d+2}\mathcal{C}_{\ii\wt{\nu},d}^{p_1p_2}\sin(\pi\ii\wt{\nu})\Gamma\Bigg[\begin{matrix}
        1+p_2+\frac d2+\ii\wt{\nu}, 1+p_2+\frac d2-\ii\wt{\nu} \\
        p_2+\tfrac d2+\tfrac32
    \end{matrix}\Bigg]\n \\
    &\times\Bigg(\sum_{\ell=0}^\infty\Gamma\Bigg[\begin{matrix}
        p_2+\frac d2+\frac32+\ell, p_{12}+d+2, -p_1-\frac d2-\frac12-\ell \\
        -p_1+p_2+1, p_2+\frac d2+\frac32-\ell
    \end{matrix}\Bigg]u^{-p_2-\frac d2-\frac32}\frac1{\ell!}\bigg(1-\frac1u\bigg)^\ell\n \\
    &\times{}_3\mathrm{F}_2\left[\begin{matrix}
        p_2+\frac d2+1+\ii\wt{\nu}, p_2+\frac d2+1-\ii\wt{\nu}, -p_1-\frac d2-\frac12-\ell \\
        -p_1+p_2+1, p_2+\frac d2+\frac32-\ell
    \end{matrix}\middle|1\right]\n \\
    &+\sum_{n=0}^\infty\frac1{n!}\Gamma\Bigg[\begin{matrix}
        p_2+\frac d2+\frac32+n \\
        -p_1-\frac d2+\frac12+n, p_2+\frac d2+\frac32, -p_2-\frac d2-\frac12, p_2+\frac d2+\frac32-n
    \end{matrix}\Bigg]\n \\
    &\times\frac{\pi^2}{\sin\big[\pi\big(p_2+\frac d2+\frac32-n\big)\big]\sin\big[\pi\big(-p_1-\frac d2+\frac12\big)\big]}\frac{\big(\frac12+\ii\wt{\nu}\big)_n\big(\frac12-\ii\wt{\nu}\big)_n}{\big(p_2+\frac d2+\frac32\big)_n\big(-p_2-\frac d2-\frac12\big)_n}\n \\
    &\times u^{-p_2-\frac d2-\frac32}(1-u)^{-p_1-d/2-1/2}\bigg(1-\frac1u\bigg)^n\Bigg)\Bigg]+(\wt{\nu}\rightarrow(-\wt{\nu}))\n \\
    =&\Bigg[2^{-p_{12}-d+1}u^{p_{12}+d+2}\mathcal{C}_{\ii\wt{\nu},d}^{p_1p_2}\sin(\pi\ii\wt{\nu})\Gamma\Bigg[\begin{matrix}
        1+p_2+\frac d2+\ii\wt{\nu}, 1+p_2+\frac d2-\ii\wt{\nu} \\
        p_2+\tfrac d2+\tfrac32
    \end{matrix}\Bigg]\n \\
    &\times\Bigg(\sum_{\ell=0}^\infty\Gamma\Bigg[\begin{matrix}
        p_2+\frac d2+\frac32+\ell, p_{12}+d+2, -p_1-\frac d2-\frac12-\ell \\
        -p_1+p_2+1, p_2+\frac d2+\frac32-\ell
    \end{matrix}\Bigg]u^{-p_2-\frac d2-\frac32}\frac1{\ell!}\bigg(1-\frac1u\bigg)^\ell\n \\
    &\times{}_3\mathrm{F}_2\left[\begin{matrix}
        p_2+\frac d2+1+\ii\wt{\nu}, p_2+\frac d2+1-\ii\wt{\nu}, -p_1-\frac d2-\frac12-\ell \\
        -p_1+p_2+1, p_2+\frac d2+\frac32-\ell
    \end{matrix}\middle|1\right]\n \\
    &+\Gamma\big(p_1+\tfrac d2+\tfrac12\big)u^{-p_2-\frac d2-\frac32}(1-u)^{-p_1-d/2-1/2}{}_2\mathrm{F}_1\left[\begin{matrix}
        \frac12+\ii\wt{\nu}, \frac12-\ii\wt{\nu} \\
        -p_1-\frac d2+\frac12
    \end{matrix}\middle|1-\frac1u\right]\Bigg)\Bigg]+(\wt{\nu}\rightarrow(-\wt{\nu})).
\end{align}
After combining (\ref{183}) and (\ref{187}), the expression of the seed integral is
\begin{align}
    \mathcal{I}_{\wt\nu}^{p_1p_2}(r(u),1)=&\mathcal{I}_{\text{S},\wt\nu}^{p_1p_2}(r(u),1)+\mathcal{I}_{\text{BG},\wt\nu}^{p_1p_2}(r(u),1)\n \\
    =&\Bigg(-2^{1-p_{12}-d}\mathcal{C}_{\ii\wt\nu,d}^{p_1p_2}\Gamma\Bigg[\begin{matrix}
        p_1+\frac d2+1+\ii\wt{\nu}, p_1+\frac d2+1-\ii\wt{\nu}, p_2+\frac d2+1+\ii\wt{\nu}, p_2+\frac d2+1-\ii\wt{\nu} \\
        p_1+\frac d2+\frac32, p_2+\frac d2+\frac32
    \end{matrix}\Bigg]\n \\
    &\times\frac{\sin(\pi\ii\wt{\nu})\sin\big[\pi\big(p_1+\frac d2-\ii\wt{\nu}\big)\big]}{\sin\big[\pi\big(p_1+\frac d2+\frac12\big)\big]}u^{p_1+d/2+1/2}{}_2\mathrm{F}_1\left[\begin{matrix}
        \frac12+\ii\wt{\nu}, \frac12-\ii\wt{\nu} \\
        p_1+\frac d2+\frac32
    \end{matrix}\middle|1-\frac1u\right]\n \\
    &+2^{-p_{12}-d+1}u^{p_1+\frac d2+\frac12}\mathcal{C}_{\ii\wt{\nu},d}^{p_1p_2}\sin(\pi\ii\wt{\nu})\Gamma\Bigg[\begin{matrix}
        1+p_2+\frac d2+\ii\wt{\nu}, 1+p_2+\frac d2-\ii\wt{\nu} \\
        p_2+\tfrac d2+\tfrac32
    \end{matrix}\Bigg]\n \\
    &\times\sum_{\ell=0}^\infty\Gamma\Bigg[\begin{matrix}
        p_2+\frac d2+\frac32+\ell, p_{12}+d+2, -p_1-\frac d2-\frac12-\ell \\
        -p_1+p_2+1, p_2+\frac d2+\frac32-\ell
    \end{matrix}\Bigg]\frac1{\ell!}\bigg(1-\frac1u\bigg)^\ell\n \\
    &\times{}_3\mathrm{F}_2\left[\begin{matrix}
        p_2+\frac d2+1+\ii\wt{\nu}, p_2+\frac d2+1-\ii\wt{\nu}, -p_1-\frac d2-\frac12-\ell \\
        -p_1+p_2+1, p_2+\frac d2+\frac32-\ell
    \end{matrix}\middle|1\right]\Bigg)+(\wt{\nu}\rightarrow(-\wt{\nu})).
\end{align}
Take the folded limit, then the seed integral becomes
\begin{align}\label{189}
    \mathcal{I}_{\wt\nu}^{p_1p_2}(1,1)=&\Bigg(-2^{1-p_{12}-d}\mathcal{C}_{\ii\wt\nu,d}^{p_1p_2}\frac{\sin(\pi\ii\wt{\nu})\sin\big[\pi\big(p_1+\frac d2-\ii\wt{\nu}\big)\big]}{\sin\big[\pi\big(p_1+\frac d2+\frac12\big)\big]}\n \\
    &\times\Gamma\Bigg[\begin{matrix}
        p_1+\frac d2+1+\ii\wt{\nu}, p_1+\frac d2+1-\ii\wt{\nu}, p_2+\frac d2+1+\ii\wt{\nu}, p_2+\frac d2+1-\ii\wt{\nu} \\
        p_1+\frac d2+\frac32, p_2+\frac d2+\frac32
    \end{matrix}\Bigg]\n \\
    &+2^{-p_{12}-d+1}\mathcal{C}_{\ii\wt{\nu},d}^{p_1p_2}\sin(\pi\ii\wt{\nu})\Gamma\big[1+p_2+\tfrac d2+\ii\wt{\nu}, 1+p_2+\tfrac d2-\ii\wt{\nu}\big]\n \\
    &\times\Gamma\big[p_{12}+d+2, -p_1-\tfrac d2-\tfrac12\big]\n \\
    &\times{}_3\mathrm{\wt{F}}_2\left[\begin{matrix}
        p_2+\frac d2+1+\ii\wt{\nu}, p_2+\frac d2+1-\ii\wt{\nu}, -p_1-\frac d2-\frac12 \\
        -p_1+p_2+1, p_2+\frac d2+\frac32
    \end{matrix}\middle|1\right]\Bigg)+(\wt{\nu}\rightarrow(-\wt{\nu})).
\end{align}
Using the following identity \cite{Slater:1966}
\begin{align}
    &{}_3\mathrm{F}_2\left[\begin{matrix}
        a, b, e+f-a-b-1 \\
        e, f
    \end{matrix}\middle|1\right]=\Gamma\bigg[\begin{matrix}
        e, f, e-a-b, f-a-b \\
        e-a, e-b, f-a, f-b
    \end{matrix}\bigg]\n \\
    &+\frac1{a+b-e}\Gamma\Bigg[\begin{matrix}
        e, f \\
        a, b, e+f-a-b
    \end{matrix}\Bigg]{}_3\mathrm{F}_2\left[\begin{matrix}
        e-a, e-b, 1 \\
        1+e-a-b, e+f-a-b
    \end{matrix}\middle|1\right]
\end{align}
with
\begin{align}
    &a=p_2+\frac d2+1-\ii\wt{\nu},&&b=-p_1-\frac d2-\frac12\n \\
    &e=p_2+\frac d2+\frac32,&&f=-p_1+p_2+1,
\end{align}
Equation (\ref{189}) becomes
\begin{align}
    \mathcal{I}_{\wt\nu}^{p_1p_2}(1,1)=&\Bigg(-2^{-p_{12}-d+1}\mathcal{C}_{\ii\wt{\nu},d}^{p_1p_2}\sin(\pi\ii\wt{\nu})\Gamma\big[p_{12}+d+2, \ii\wt{\nu}+p_1+1+\tfrac d2, \ii\wt{\nu}+p_2+1+\tfrac d2\big]\n \\
    &\times{}_3\mathrm{\wt{F}}_2\left[\begin{matrix}
        p_{12}+d+2, \frac12+\ii\wt{\nu}, 1 \\
        p_1+\frac d2+2+\ii\wt{\nu}, p_2+\frac d2+2+\ii\wt{\nu}
    \end{matrix}\middle|1\right]\Bigg)+(\wt{\nu}\rightarrow(-\wt{\nu})).
\end{align}

The expression holds if $\wt{\nu}$ is real, or $\wt{\nu}$ is complex and $\ii\wt{\nu}\leqslant\frac12$. The hypergeometric functions in the expression diverge if $\frac12<\ii\wt{\nu}<\frac d2$. Consider the following identity:
\begin{equation}
    {}_3\mathrm{F}_2\left[\begin{matrix}
        a, b, c \\
        e, f
    \end{matrix}\middle|1\right]=\Gamma\bigg[\begin{matrix}
        e, f, s \\
        a, s+b, s+c
    \end{matrix}\bigg]{}_3\mathrm{F}_2\left[\begin{matrix}
        e-a, f-a, s \\
        s+b, s+c
    \end{matrix}\middle|1\right]
\end{equation}
where $s=e+f-a-b-c$. Set
\begin{align}
    &a=1,&&b=p_{12}+d+2,&&c=\frac12+\ii\wt{\nu},\n \\
    &e=p_1+\frac d2+2+\ii\wt{\nu},&&f=-p_2+\frac d2+2+\ii\wt{\nu},
\end{align}
and the expression becomes
\begin{align}
    \mathcal{I}_{\wt\nu}^{p_1p_2}(1,1)=&\Bigg(-2^{-p_{12}-d+1}\mathcal{C}_{\ii\wt{\nu},d}^{p_1p_2}\sin(\pi\ii\wt{\nu})\Gamma(p_{12}+d+2)\n \\
    &\times{}_3\mathcal{F}_2\left[\begin{matrix}
        \frac12+\ii\wt{\nu}, p_1+\frac d2+1+\ii\wt{\nu}, p_2+\frac d2+1+\ii\wt{\nu} \\
        p_{12}+d+\frac52+\ii\wt{\nu}, 1+2\ii\wt{\nu}
    \end{matrix}\middle|1\right]\Bigg)+(\wt{\nu}\rightarrow(-\wt{\nu})).
\end{align}
This expression converges for arbitrary $\wt{\nu}$. However, the above expression is not suitable to be one part of the integrand of the K\"all\'en-Lehmann integral. It is hard to find the divergence of $d\rightarrow3$.
\section{One-Loop Bubble Diagrams in Flat Spacetime}\label{AppD}
In this appendix, the diagrams related to $\nabla\sigma\nabla\sigma$ model and $AA$ model are calculated in flat spacetime with dimension $D=d+1$ of spacetime. These diagrams are computed to compare the coefficients of divergences with those in de Sitter spacetime. It is expected that the coefficients in dS should be the same as those in flat spacetime after taking flat space limit $H\rightarrow0$.
\subsection{$\nabla\sigma\nabla\sigma$}
The diagram related to $\nabla\sigma\nabla\sigma$ model is
\begin{align}
    &-\frac12\int\frac{\ud^Dp}{(2\pi)^D}\frac{-[p\cdot(p+k)]^2}{(p^2+m^2-\ii\epsilon)[(k+p)^2+m^2-\ii\epsilon]}\n \\
    =&\frac12\int_0^1\ud x\int\frac{\ud^Dp}{(2\pi)^D}\frac{[p\cdot(p+k)]^2}{[(p+xk)^2+m^2+x(1-x)k^2-\ii\epsilon]^2}\n \\
    \xlongequal{p\rightarrow p-xk}&\frac12\int_0^1\ud x\int\frac{\ud^Dp}{(2\pi)^D}\frac{\{(p-xk)\cdot[p+(1-x)k]\}^2}{[p^2+m^2+x(1-x)k^2-\ii\epsilon]^2}\n \\
    =&\frac12\int_0^1\ud x\int\frac{\ud^Dp}{(2\pi)^D}\frac{(p^2)^2+(1-2x)^2(p\cdot k)^2-2x(1-x)p^2k^2+x^2(1-x)^2(k^2)^2}{[p^2+m^2+x(1-x)k^2-\ii\epsilon]^2}.
\end{align}
Then take the Wick rotation:
\begin{align}
    &\frac{\ii}2\int_0^1\ud x\int\frac{(\ud^Dp)_E}{(2\pi)^D}\frac{(p^2)^2+(1-2x)^2(p\cdot k)^2-2x(1-x)p^2k^2+x^2(1-x)^2(k^2)^2}{[p^2+m^2+x(1-x)k^2]^2}\n \\
    =&\frac{\ii}2\int_0^1\ud x\int\frac{(\ud^Dp)_E}{(2\pi)^D}\frac{(p^2)^2+\big[\frac1D-\big(\frac4D+2\big)x(1-x)\big]p^2k^2+x^2(1-x)^2(k^2)^2}{[p^2+m^2+x(1-x)k^2]^2}\n \\
    =&\frac{\ii}2\int_0^1\ud x\int_0^\infty\frac{\ud\kappa\Omega_D\kappa^{D-1}}{(2\pi)^D}\frac{\kappa^4+\big[\frac1D-\big(\frac4D+2\big)x(1-x)\big]\kappa^2k^2+x^2(1-x)^2(k^2)^2}{[\kappa^2+m^2+x(1-x)k^2]^2},
\end{align}
with
\begin{equation}
    \Omega_D=\frac{2\pi^{\frac D2}}{\Gamma\big(\frac D2\big)}.
\end{equation}
Using the following formula:
\begin{equation}
    \int_0^\infty\ud\kappa\frac{\kappa^{\ell-1}}{(\kappa^2+\nu^2)^m}=\nu^{\ell-2m}\frac{\Gamma\big(\frac\ell2\big)\Gamma\big(m-\frac\ell2\big)}{2\Gamma(m)},
\end{equation}
the result is
\begin{align}
    &-\frac12\int\frac{\ud^Dp}{(2\pi)^D}\frac{-[p\cdot(k+p)]^2}{(p^2+m^2-\ii\epsilon)[(k+p)^2+m^2-\ii\epsilon]}\n \\
    =&\frac{\ii}2\int_0^1\ud x\frac{\Omega_D}{(2\pi)^D}\bigg\{[m^2+x(1-x)k^2]^{\frac D2}\frac{\Gamma\big(\frac D2+2\big)\Gamma\big(-\frac D2\big)}2+\bigg[\frac1D-\big(\frac4D+2\big)x(1-x)\bigg]\n \\
    &\times k^2[m^2+x(1-x)k^2]^{\frac D2-1}\frac{\Gamma\big(\frac D2+1\big)\Gamma\big(1-\frac D2\big)}2\n \\
    &+x^2(1-x)^2(k^2)^2[m^2+x(1-x)k^2]^{\frac D2-2}\frac{\Gamma\big(\frac D2\big)\Gamma\big(2-\frac D2\big)}2\bigg\}.
\end{align}
For the limit in $D\rightarrow4$
\begin{align}
    &-\frac12\int\frac{\ud^Dp}{(2\pi)^D}\frac{-[p\cdot(k+p)]^2}{(p^2+m^2-\ii\epsilon)[(k+p)^2+m^2-\ii\epsilon]}\n \\
    =&\frac{\ii}{16\pi^2}\frac1{4-D}\bigg(\frac{(k^2)^2}4+\frac{3k^2m^2}2+3m^4\bigg)+\mathcal{O}((4-D)^0).
\end{align}
Therefore, the coefficients in (\ref{91}) should be
\begin{equation}\label{202}
    \frac13\delta_4=\frac1{64\pi^2}\frac1{4-D},\quad\frac13\delta_2=\frac{3m^2}{32\pi^2}\frac1{4-D},\quad\frac13\delta_0=\frac{3m^4}{16\pi^2}\frac1{4-D}.
\end{equation}

\subsection{$AA$}
The expression of the diagram in $AA$ model is
\begin{align}
    &-\frac12\int\frac{\ud^Dp}{(2\pi)^D}\frac{-\big(g_{\mu\nu}+\frac{p_\mu p_\nu}{m_A^2}\big)\big(g^{\mu\nu}+\frac{(p^\mu+k^\mu)(p^\nu+k^\nu)}{m_A^2}\big)}{(p^2+m_A^2-\ii\epsilon)[(k+p)^2+m_A^2-\ii\epsilon]}\n \\
    =&\frac1{2m_A^4}\int\frac{\ud^Dp}{(2\pi)^D}\frac{Dm_A^4+m_A^2(2p^2+2p\cdot k+k^2)+(p^2+p\cdot k)^2}{(p^2+m_A^2-\ii\epsilon)[(k+p)^2+m_A^2-\ii\epsilon]}\n \\
    =&\frac1{2m_A^4}\int_0^1\ud x\int\frac{\ud^Dp}{(2\pi)^D}\frac{Dm_A^4+m_A^2(2p^2+2p\cdot k+k^2)+(p^2+p\cdot k)^2}{[(p+xk)^2+m_A^2+x(1-x)k^2-\ii\epsilon]^2}\n \\
    \xlongequal{p\rightarrow p-xk}&\frac1{2m_A^4}\int_0^1\ud x\int\frac{\ud^Dp}{(2\pi)^D}\frac1{[p^2+m^2+x(1-x)k^2-\ii\epsilon]^2}\big(Dm_A^4\n \\
    &+m_A^2\{2(p-xk)\cdot[p+(1-x)k]+k^2\}+\{(p-xk)\cdot[p+(1-x)k]\}^2\big)\n \\
    =&\frac1{2m_A^4}\int_0^1\ud x\int\frac{\ud^Dp}{(2\pi)^D}\frac1{[p^2+m_A^2+x(1-x)k^2-\ii\epsilon]^2}\big(Dm_A^4\n \\
    &+m_A^2\{2p^2+[1-2x(1-x)]k^2\}+(p^2)^2+(1-2x)^2(p\cdot k)^2\n \\
    &-2x(1-x)p^2k^2+x^2(1-x)^2(k^2)^2\big)
\end{align}
After taking the Wick rotation, the expression becomes
\begin{align}
    &\frac{\ii}{2m_A^4}\int_0^1\ud x\int\frac{(\ud^Dp)_E}{(2\pi)^D}\frac1{[p^2+m_A^2+x(1-x)k^2]^2}\big(Dm_A^4\n \\
    &+m_A^2\{2p^2+[1-2x(1-x)]k^2\}+(p^2)^2+(1-2x)^2(p\cdot k)^2\n \\
    &-2x(1-x)p^2k^2+x^2(1-x)^2(k^2)^2\big)\n \\
    =&\frac{\ii}{2m_A^4}\int_0^1\ud x\int\frac{(\ud^Dp)_E}{(2\pi)^D}\frac1{[p^2+m_A^2+x(1-x)k^2]^2}\Big(Dm_A^4\n \\
    &+m_A^2\{2p^2+[1-2x(1-x)]k^2\}+(p^2)^2+\big[\tfrac1D-\big(\tfrac4D+2\big)x(1-x)\big]p^2k^2\n \\
    &+x^2(1-x)^2(k^2)^2\Big)\n \\
    =&\frac{\ii}{2m_A^4}\int_0^1\ud x\int_0^\infty\frac{\ud\kappa\Omega_D\kappa^{D-1}}{(2\pi)^D}\frac1{[\kappa^2+m_A^2+x(1-x)k^2]^2}\Big(Dm_A^4\n \\
    &+m_A^2\{2\kappa^2+[1-2x(1-x)]k^2\}+(\kappa^2)^2+\big[\tfrac1D-\big(\tfrac4D+2\big)x(1-x)\big]\kappa^2k^2\n \\
    &+x^2(1-x)^2(k^2)^2\Big)
\end{align}
In the limit $D\rightarrow4$
\begin{align}
    &-\frac12\int\frac{\ud^Dp}{(2\pi)^D}\frac{-\big(g_{\mu\nu}+\frac{p_\mu p_\nu}{m_A^2}\big)\big(g^{\mu\nu}+\frac{(p^\mu+k^\mu)(p^\nu+k^\nu)}{m_A^2}\big)}{(p^2+m_A^2-\ii\epsilon)[(k+p)^2+m_A^2-\ii\epsilon]}\n \\
    =&\frac{\ii}{16\pi^2}\frac1{4-D}\bigg(\frac{(k^2)^2}{4m_A^4}+\frac{3k^2}{2m_A^2}+3\bigg)+\mathcal{O}((4-D)^0)
\end{align}
The coefficients in (\ref{141}) are
\begin{equation}\label{206}
    \frac13\delta_4=\frac1{64\pi^2m_A^4}\frac1{4-D},\quad\frac13\delta_2=\frac3{32\pi^2m_A^2}\frac1{4-D},\quad\frac13\delta_0=\frac3{16\pi^2}\frac1{4-D}.
\end{equation}
\section{Expressions of Spectral Function}\label{AppE}
In this appendix, the spectral function $\rho_{\nabla\sigma\nabla\sigma}^{\mathcal{P},0}(\wt{\nu}')$ and $\rho_{AA}^{\mathcal{P},0}(\wt{\nu}')$ are computed. The computation is finished in embedding space. Therefore, the embedding formalism of dS is introduced at first. Then the inverse formulae of K\"all\'en-Lehmann representation are introduced, which are essential to get the expression of the spectral functions. Finally, the explicit expressions of the spectral functions are computed.
\subsection{Embedding Formalism}
The introduction in this subsection closely follows \cite{Loparco:2023rug, Costa:2014kfa}.

A $(d+1)$-dimensional de Sitter spacetime can be considered as a hypersurface in $(d+2)$-dimensional Minkowski spacetime
\begin{equation}
    -Y_0^2+Y_1^2+\cdots+Y_{d+1}^2=1.
\end{equation}
The parametrization of the Poincar\'e patch of dS$_{d+1}$ is given by:
\begin{equation}
    Y^0=-\frac{1-\tau^2+\mathbf{y}^2}{2\tau},\quad Y^i=-\frac{y^i}\tau,\quad Y^{d+1}=-\frac{1+\tau^2-\mathbf{y}^2}{2\tau}.
\end{equation}

Consider a symmetric traceless tensor $T_{A_1\cdots A_J}(Y)$ in the embedding space. The tensor satisfies the tangential condition:
\begin{equation}
    Y^{A_1}T_{A_1\cdots A_J}(Y)=0.
\end{equation}
The components of a tensor in dS$_{d+1}$ are obtained if the projection is taken
\begin{equation}
    T_{\mu_1\cdots\mu_J}(y)=\pdd{Y^{A_1}}{y^{\mu_1}}\cdots\pdd{Y^{A_J}}{y^{\mu_J}}T_{A_1\cdots A_J}(Y).
\end{equation}
The indices of the tensor can be contracted with a null vector $W^A$
\begin{equation}
    T(Y,W)=W^{A_1}\cdots W^{A_J}T_{A_1\cdots A_J}(Y),
\end{equation}
with $Y\cdot W=0$.

To recover the tensor, a differential operator is introduced:
\begin{align}
    K_A=&\bigg(\frac{d-1}2\bigg)\bigg[\pdd{}{W^A}-Y_A\bigg(Y\cdot\pdd{}{W}\bigg)\bigg]+\bigg(W\cdot\pdd{}{W}\bigg)\pdd{}{W^A}-Y_A\bigg(Y\cdot\pdd{}{W}\bigg)\bigg(W\cdot\pdd{}{W}\bigg)\n \\
    &-\frac12W_A\bigg[\pdd{^2}{W\cdot\partial W}-\bigg(Y\cdot\pdd{}{W}\bigg)\bigg(Y\cdot\pdd{}{W}\bigg)\bigg].
\end{align}
The differential operator acts on a monomial of $W^A$ as
\begin{equation}
    K_{A_1}\cdots K_{A_J}W^{B_1}\cdots W^{B_J}=J!\bigg(\frac{d-1}2\bigg)_J\bigg[\frac1{J!}\sum_{\pi}G_{A_{\pi_1}}^{\ \ B_1}\cdots G_{A_{\pi_J}}^{\ \ B_J}-\text{traces}\bigg]
\end{equation}
with
\begin{equation}
    G_{AB}=\eta_{AB}-Y_AY_B,
\end{equation}
the sum $\pi$ is over all permutations of the indices $A_j$, the traces are subtracted using $G_{AB}$.

The embedding differential operator is
\begin{equation}
    \nabla_A=\pdd{}{Y^A}-Y_A\bigg(Y\cdot\pdd{}{Y}\bigg)-W_A\bigg(Y\cdot\pdd{}{W}\bigg).
\end{equation}

\subsection{Inversion Formulae}
K\"all\'en-Lehmann representation of operator $\mathcal{O}^{(J)}$ is \cite{Loparco:2023rug}
\begin{align}
    G_{\mathcal{O}^{(J)}}(Y_1,Y_2;W_1,W_2)=&\sum_{\ell=0}^J\int_{-\infty}^\infty\ud\wt{\nu}'\rho_{\mathcal{O}^{(J)}}^{\mathcal{P},\ell}(\wt{\nu}')[(W_1\cdot\nabla_1)(W_2\cdot\nabla_2)]^{J-\ell}G_{\wt{\nu}',\ell}(Y_1,Y_2;W_1,W_2)\n \\
    &+\cdots.
\end{align}

Before writing the inversion formulae, it is convenient to introduce the Wick rotation to Euclidean Anti de Sitter space. The coordinates transform as
\begin{equation}
    Y_1(\tau_1,\mathbf{y}_1)\xrightarrow{\tau_1\rightarrow\tau_1e^{\ii\pi/2}}X_1(\tau_1,\mathbf{y}_1),\quad Y_2(\tau_2,\mathbf{y}_2)\xrightarrow{\tau_2\rightarrow\tau_2e^{-\ii\pi/2}}X_2(\tau_2,\mathbf{y}_2).
\end{equation}

Wightman functions in dS are changed to harmonic functions in EAdS:
\begin{equation}
    G_{\wt{\nu},\ell}(Y_1,Y_2;W_1,W_2)\rightarrow\Gamma[\ii\wt{\nu},-\ii\wt{\nu}]\Omega_{\wt{\nu},\ell}(X_1,X_2;W_1,W_2).
\end{equation}

The inversion formulae of K\"all\'en-Lehmann representation is
\begin{equation}
    \rho_{\mathcal{O}^{(J)}}^{\mathcal{P},\ell}(\wt{\nu}')=\frac1{\mathcal{N}_{J,\ell}}\int_{X_1}\Omega_{\wt{\nu}',\ell}(X_2,X_1;K_2,K_1)[(K_1\cdot\nabla_1)(K_2\cdot\nabla_2)]^{J-\ell}G_{\mathcal{O}^{(J)}}(X_1,X_2;W_1,W_2).
\end{equation}
The formula is integrating $X_1$ over EAdS. The coefficient $\mathcal{N}_{J,\ell}$ is
\begin{equation}
    \mathcal{N}_{J,\ell}=\frac{(2\ell+d-2)(\ell+d-3)!\big[\ell!(J-\ell)!\big(\frac{d-1}2\big)_\ell(d+2\ell-1)_{J-\ell}\Gamma\big(\frac d2+J+\ii\wt{\nu}'\big)\Gamma\big(\frac d2+J-\ii\wt{\nu}'\big)\big]^2}{4^{J-\ell}(4\pi)^{\frac{d+1}2}(d-2)!\ell!\Gamma\big(\frac{d+1}2\big)\Gamma\big(\frac d2+\ell+\ii\wt{\nu}'\big)\Gamma\big(\frac d2+\ell-\ii\wt{\nu}'\big)\prod_{t=0}^{\ell-1}\big[\big(\frac d2+\ii\wt{\nu}'+t-1\big)\big(\frac d2-\ii\wt{\nu}'+t-1\big)\big]}
\end{equation}
for $d\geqslant3$.

\subsection{Explicit Expressions}
Split representation of propagators in EAdS is \cite{Costa:2014kfa}
\begin{equation}
    \Omega_{\wt{\nu},\ell}(X_1,X_2;W_1,W_2)=\frac{\wt{\nu}^2}{\pi\ell!\big(\frac d2-1\big)_\ell}\int_P\Pi_{\frac d2+\ii\wt{\nu},\ell}(X_1,P;W_1,D_Z)\Pi_{\frac d2-\ii\wt{\nu},\ell}(X_2,P;W_2,Z),
\end{equation}
where $\Pi_{\frac d2+\ii\wt{\nu},\ell}(X_1,P;W_1,D_Z)$ is a bulk-to-boundary propagator in EAdS, in which $P$ is a null vector of $(d+2)$-dimensional Minkowski spacetime. $Z$ is an auxiliary vector of $P$, while $D_Z$ is defined as
\begin{equation}
    D_Z^A=\bigg(\frac d2-1+Z\cdot\pdd{}{Z}\bigg)\pdd{}{Z_A}-\frac12Z^A\pdd{^2}{Z\cdot\partial Z}.
\end{equation}
Notation ``$\int_P$" means
\begin{equation}
    \int_P(\cdots)\equiv\frac2{\text{Vol GL}(1,\mathbb{R})^+}\int_{P^0>0}\ud^{d+2}P\delta(P^2)(\cdots).
\end{equation}
The expression of bulk-to-boundary propagators are given by
\begin{equation}
    \Pi_{\Delta,\ell}(X,P;W,Z)=\mathcal{C}_{\Delta,\ell}\frac{((-2P\cdot X)(W\cdot Z)+2(W\cdot P)(Z\cdot X))^\ell}{(-2P\cdot X)^{\Delta+J}},
\end{equation}
with
\begin{equation}
    \mathcal{C}_{\Delta,\ell}=\frac{(\ell+\Delta-1)\Gamma(\Delta)}{2\pi^{\frac d2}(\Delta-1)\Gamma\big(\Delta+1-\frac d2\big)}.
\end{equation}

\subsubsection{$\nabla\sigma\nabla\sigma$}
Spectral density $\rho_{\nabla\sigma\nabla\sigma}^{\mathcal{P},0}(\wt{\nu}')$ is
\begin{align}
    \rho_{\nabla\sigma\nabla\sigma}^{\mathcal{P},0}(\wt{\nu}')=&\frac1{\mathcal{N}_{0,0}\big(\frac{d-1}2\big)^2}\int_{X_1}\Omega_{\wt{\nu}',0}(X_2,X_1)\nabla_{1A_1}\nabla_{2A_2}G_{\wt{\nu},0}(X_1,X_2)K_1^{\ A_1}K_2^{\ A_2}(W_1\cdot\nabla_1)(W_2\cdot\nabla_2)G_{\wt{\nu},0}(X_1,X_2)\n \\
    =&\frac{(\Gamma[\ii\wt{\nu},-\ii\wt{\nu}])^2}{\mathcal{N}_{0,0}\big(\frac{d-1}2\big)^2}\int_{X_1}\Omega_{\wt{\nu}',0}(X_2,X_1)\nabla_{1A_1}\nabla_{2A_2}\Omega_{\wt{\nu},0}(X_1,X_2)\n \\
    &\times K_1^{\ A_1}K_2^{\ A_2}(W_1\cdot\nabla_1)(W_2\cdot\nabla_2)\Omega_{\wt{\nu},0}(X_1,X_2)\n \\
    =&\frac{\wt{\nu}'^2\wt{\nu}^4(\Gamma[\ii\wt{\nu},-\ii\wt{\nu}])^2}{\pi^3\mathcal{N}_{0,0}\big(\frac{d-1}2\big)^2}\int_{X_1}\int_{P_1}\int_{P_2}\int_{P_3}\Pi_{\Delta',0}(X_1,P_3)\Pi_{\bar{\Delta}',0}(X_2,P_3)\nabla_{1A_1}\Pi_{\Delta,0}(X_1,P_1)\n \\
    &\times \nabla_{2A_2}\Pi_{\bar{\Delta},0}(X_2,P_1)K_1^{\ A_1}(W_1\cdot\nabla_1)\Pi_{\Delta,0}(X_1,P_2)K_2^{\ A_2}(W_2\cdot\nabla_2)\Pi_{\bar{\Delta},0}(X_2,P_2).
\end{align}
Due to the following integration \cite{Costa:2014kfa}:
\begin{align}
    &\frac1{J!\big(\frac{d-1}2\big)_J}\int_X\Pi_{\Delta_2,0}(X,P_2)\Pi_{\Delta_3,J}(X,P_3;K,Z)(W\cdot\nabla)^J\Pi_{\Delta_1,0}(X,P_1)\n \\
    =&b(\Delta_1,\Delta_2,\Delta_3,J)\frac{[(Z\cdot P_1)P_{23}-(Z\cdot P_2)P_{13}]^J}{(P_{12})^{\Delta_{123,0}+J/2}(P_{13})^{\Delta_{132,0}+J/2}(P_{23})^{\Delta_{231,0}+J/2}},
\end{align}
the integration of $X_1$ can be computed directly:
\begin{align}
    &\frac1{\big(\frac{d-1}2\big)}\int_{X_1}\Pi_{\Delta',0}(X_1,P_3)\nabla_{1A_1}\Pi_{\Delta,0}(X_1,P_1)K_1^{\ A_1}(W_1\cdot\nabla_1)\Pi_{\Delta,0}(X_1,P_2)\n \\
    =&\mathcal{C}_{\Delta',0}\mathcal{C}_{\Delta,0}\mathcal{C}_{\Delta,0}\int_{X_1}\frac1{(-2P_3\cdot X_1)^{\Delta'}}\Bigg[\frac{4\Delta^2P_1\cdot P_2}{(-2P_1\cdot X_1)^{\Delta+1}(-2P_2\cdot X_1)^{\Delta+1}}+\frac{\Delta^2}{(-2P_1\cdot X_1)^{\Delta}(-2P_2\cdot X_1)^{\Delta}}\Bigg]\n \\
    =&\Delta^2\frac{-2\frac{\mathcal{C}_{\bar{\Delta},0}\mathcal{C}_{\bar{\Delta},0}}{\mathcal{C}_{\bar{\Delta}+1,0}\mathcal{C}_{\bar{\Delta}+1,0}}b(\Delta+1,\Delta+1,\Delta',0)+b(\Delta,\Delta,\Delta',0)}{P_{31}^{\frac{\Delta'}2}P_{32}^{\frac{\Delta'}2}P_{12}^{\frac{2\Delta-\Delta'}2}},
\end{align}
with the notation $P_{ij}=-2P_i\cdot P_j$ and
\begin{equation}
    \Delta_{ijl,k}\equiv\frac{\Delta_i+k_i+\Delta_j+k_j-\Delta_l-k_l}2.
\end{equation}
The constant $b(\Delta_1,\Delta_2,\Delta_3,J)$ is
\begin{equation}
    b(\Delta_1,\Delta_2,\Delta_3,J)=\frac{\Gamma\big(\frac{\Delta_1+\Delta_2+\Delta_3-d+J}2\big)\Gamma\big(\frac{\Delta_1+\Delta_2-\Delta_3+J}2\big)\Gamma\big(\frac{\Delta_1-\Delta_2+\Delta_3+J}2\big)\Gamma\big(\frac{-\Delta_1+\Delta_2+\Delta_3+J}2\big)}{2^{1-J}\pi^{-\frac d2}\Gamma(\Delta_1)\Gamma(\Delta_2)\Gamma(\Delta_3+J)}\mathcal{C}_{\Delta_1,0}\mathcal{C}_{\Delta_2,0}\mathcal{C}_{\Delta_3,J}.
\end{equation}
The integration of $P_i$ is finished by \cite{Loparco:2023rug} with the help of the following formula:
\begin{equation}
    \int_{P_1}\int_{P_2}\int_{P_3}\frac{\prod_{j=1}^3\Pi_{\bar{\Delta}_j-k_j,0}(X_2,P_j)}{(P_{12})^{\Delta_{123,k}}(P_{13})^{\Delta_{132,k}}(P_{23})^{\Delta_{231,k}}}\equiv\mathcal{I}^{\text{QFT}}_{k_1,k_2,k_3}(\wt{\nu}_1,\wt{\nu}_2),
\end{equation}
with
\begin{equation}
    \mathcal{I}^{\text{QFT}}_{k_1,k_2,k_3}=\frac{\Gamma\big(\frac d2-\Delta_{123,k}\big)\Gamma\big(\frac d2-\Delta_{132,k}\big)\Gamma\big(\frac d2-\Delta_{231,k}\big)\Gamma\big(d-\sum_{j=1}^3\frac{\Delta_j+k_j}2\big)}{8\Gamma(d)\Gamma(1-\ii\wt{\nu}_1-k_1)\Gamma(1-\ii\wt{\nu}_2-k_2)\Gamma(1-\ii\wt{\nu}_3-k_3)}.
\end{equation}
In this case, the integration of $P_i$ is
\begin{align}
    &\frac1{\big(\frac{d-1}2\big)}\int_{P_1}\int_{P_2}\int_{P_3}\frac{\Pi_{\bar{\Delta}',0}(X_2,P_3)\nabla_{2A_2}\Pi_{\bar{\Delta},0}(X_2,P_1)K_2^{\ A_2}(W_2\cdot\nabla_2)\Pi_{\bar{\Delta},0}(X_2,P_2)}{P_{31}^{\frac{\Delta'}2}P_{32}^{\frac{\Delta'}2}P_{12}^{\frac{2\Delta-\Delta'}2}}\n \\
    =&-2\bar{\Delta}^2\frac{\mathcal{C}_{\bar{\Delta},0}\mathcal{C}_{\bar{\Delta},0}}{\mathcal{C}_{\bar{\Delta}+1,0}\mathcal{C}_{\bar{\Delta}+1,0}}\mathcal{I}^{\text{QFT}}_{-1,-1,0}(\wt{\nu},\wt{\nu})+\bar{\Delta}^2\mathcal{I}^{\text{QFT}}_{0,0,0}(\wt{\nu},\wt{\nu}).
\end{align}
Therefore,
\begin{align}
    \rho_{\nabla\sigma\nabla\sigma}^{\mathcal{P},0}(\wt{\nu}')=&\frac{\wt{\nu}'^2\wt{\nu}^4(\Gamma[\ii\wt{\nu},-\ii\wt{\nu}])^2}{\pi^3\mathcal{N}_{0,0}}\bigg(-2\bar{\Delta}^2\frac{\mathcal{C}_{\bar{\Delta},0}\mathcal{C}_{\bar{\Delta},0}}{\mathcal{C}_{\bar{\Delta}+1,0}\mathcal{C}_{\bar{\Delta}+1,0}}\mathcal{I}^{\text{QFT}}_{-1,-1,0}(\wt{\nu},\wt{\nu})+\bar{\Delta}^2\mathcal{I}^{\text{QFT}}_{0,0,0}(\wt{\nu},\wt{\nu})\bigg)\n \\
    &\times\Delta^2\bigg(-2\frac{\mathcal{C}_{\bar{\Delta},0}\mathcal{C}_{\bar{\Delta},0}}{\mathcal{C}_{\bar{\Delta}+1,0}\mathcal{C}_{\bar{\Delta}+1,0}}b(\Delta+1,\Delta+1,\Delta',0)+b(\Delta,\Delta,\Delta',0)\bigg)\n \\
    =&\frac{\wt{\nu}'^2\wt{\nu}^4(\Gamma[\ii\wt{\nu},-\ii\wt{\nu}])^2}{2^7\pi^{3+d}\Gamma(d)\mathcal{N}_{0,0}}\Bigg[\frac{\big(\frac d2+2\ii\wt{\nu}+\ii\wt{\nu}'\big)\big(\frac d2+2\ii\wt{\nu}-\ii\wt{\nu}'\big)}{2\big(\frac d2+\ii\wt{\nu}\big)^2}-1\Bigg]\Bigg[\frac{\big(\frac d2-2\ii\wt{\nu}+\ii\wt{\nu}'\big)\big(\frac d2-2\ii\wt{\nu}-\ii\wt{\nu}'\big)}{2\big(\frac d2-\ii\wt{\nu}\big)^2}-1\Bigg]\n \\
    &\times\bigg(\frac d2+\ii\wt{\nu}\bigg)^2\bigg(\frac d2-\ii\wt{\nu}\bigg)^2\frac{\prod_{\pm,\pm,\pm}\Gamma\Big(\frac{\frac d2\pm\ii\wt{\nu}\pm\ii\wt{\nu}\pm\ii\wt{\nu}'}2\Big)}{\big(\Gamma(1+\ii\wt{\nu})\Gamma(1-\ii\wt{\nu})\big)^2\Gamma(1+\ii\wt{\nu}')\Gamma(1-\ii\wt{\nu}')}\n \\
    =&\frac{\Gamma\big(\frac{d+1}2\big)\wt{\nu}'\sinh(\pi\wt{\nu}')}{2^{8-d}\pi^{\frac{d+7}2}\Gamma(d)\Gamma\big(\frac d2+\ii\wt{\nu}'\big)\Gamma\big(\frac d2-\ii\wt{\nu}'\big)}\bigg(\frac{d^2}4+2\wt{\nu}-\wt{\nu}'^2\bigg)^2\prod_{\pm,\pm,\pm}\Gamma\Big(\frac{\frac d2\pm\ii\wt{\nu}\pm\ii\wt{\nu}\pm\ii\wt{\nu}'}2\Big).
\end{align}

\subsubsection{$AA$}
By definition, the expression of $\rho^{\mathcal{P},0}_{AA}(\wt{\nu'})$ is
\begin{align}
    \rho^{\mathcal{P},0}_{AA}(\wt{\nu'})=&\frac1{\mathcal{N}_{0,0}\big(\frac{d-1}2\big)^2}\int_{X_1}\Omega_{\wt{\nu}',0}(X_2,X_1)G_{\wt{\nu}_A,1}(X_1,X_2;K_1,K_2)G_{\wt{\nu}_A,1}(X_1,X_2;W_1,W_2)\n \\
    =&\frac{(\Gamma[\ii\wt{\nu}_A,-\ii\wt{\nu}_A])^2}{\mathcal{N}_{0,0}\big(\frac{d-1}2\big)^2}\int_{X_1}\Omega_{\wt{\nu}',0}(X_2,X_1)\Omega_{\wt{\nu}_A,1}(X_1,X_2;K_1,K_2)\Omega_{\wt{\nu}_A,1}(X_1,X_2;W_1,W_2)\n \\
    =&\frac{\wt{\nu}'^2\wt{\nu}_A^4(\Gamma[\ii\wt{\nu}_A,-\ii\wt{\nu}_A])^2}{\pi^3\mathcal{N}_{0,0}\big(\frac{d-1}2\big)^2\big(\frac d2-1\big)^2}\int_{X_1}\int_{P_1}\int_{P_2}\int_{P_3}\Pi_{\Delta',0}(X_1,P_3)\Pi_{\bar{\Delta}',0}(X_2,P_3)\Pi_{\Delta_A,1}(X_1,P_1;K_1,D_{Z_1})\n \\
    &\times\Pi_{\bar{\Delta}_A,1}(X_2,P_1;K_2,Z_1)\Pi_{\Delta_A,1}(X_1,P_2;W_1,D_{Z_2})\Pi_{\bar{\Delta}_A,1}(X_2,P_2;W_2,Z_2).
\end{align}

The integration of $X_1$ is
\begin{align}
    &\frac1{\big(\frac{d-1}2\big)}\int_{X_1}\Pi_{\Delta',0}(X_1,P_3)\Pi_{\Delta_A,1}(X_1,P_1;K_1,D_{Z_1})\Pi_{\Delta_A,1}(X_1,P_2;W_1,D_{Z_2})\n \\
    =&\mathcal{C}_{\Delta',0}\mathcal{C}_{\Delta_A,1}\mathcal{C}_{\Delta_A,1}\int_{X_1}\frac1{(-2P_3\cdot X_1)^{\Delta'}(-2P_1\cdot X_1)^{\Delta_A+1}(-2P_2\cdot X_1)^{\Delta_A+1}}\n \\
    &\times[(-2P_1\cdot X_1)(-2P_2\cdot X_1)(D_{Z_1}\cdot D_{Z_2})+2(-2P_1\cdot X_1)(D_{Z_2}\cdot X_1)(D_{Z_1}\cdot P_2)\n \\
    &+2(-2P_2\cdot X_1)(D_{Z_1}\cdot X_1)(D_{Z_2}\cdot P_1)+4(D_{Z_1}\cdot X_1)(D_{Z_2}\cdot X_1)(P_1\cdot P_2)]\n \\
    =&\frac1{\Delta_A^2}\frac{\mathcal{C}_{\Delta_A,1}\mathcal{C}_{\Delta_A,1}}{\mathcal{C}_{\Delta_A,0}\mathcal{C}_{\Delta_A,0}}\big[\Delta_A^2(D_{Z_1}\cdot D_{Z_2})+\Delta_A(D_{Z_1}\cdot P_2)(D_{Z_2}\cdot \partial_{P_2})+\Delta_A(D_{Z_2}\cdot P_1)(D_{Z_1}\cdot\partial_{P_1})\n \\
    &+(P_1\cdot P_2)(D_{Z_1}\cdot\partial_{P_1})(D_{Z_2}\cdot\partial_{P_2})\big]\frac{b(\Delta_A,\Delta_A,\Delta',0)}{P_{31}^{\frac{\Delta'}2}P_{32}^{\frac{\Delta'}2}P_{12}^{\frac{2\Delta_A-\Delta'}2}}.
\end{align}
Then $P_i$ is integrated, the result is
\begin{align}
    &\frac1{\big(\frac{d-1}2\big)}\int_{P_1}\int_{P_2}\int_{P_3}\Bigg[\frac1{\Delta_A^2}\frac{\mathcal{C}_{\Delta_A,1}\mathcal{C}_{\Delta_A,1}}{\mathcal{C}_{\Delta_A,0}\mathcal{C}_{\Delta_A,0}}\big[\Delta_A^2(D_{Z_1}\cdot D_{Z_2})+\Delta_A(D_{Z_1}\cdot P_2)(D_{Z_2}\cdot \partial_{P_2})\n \\
    &+\Delta_A(D_{Z_2}\cdot P_1)(D_{Z_1}\cdot\partial_{P_1})+(P_1\cdot P_2)(D_{Z_1}\cdot\partial_{P_1})(D_{Z_2}\cdot\partial_{P_2})\big]\frac{b(\Delta_A,\Delta_A,\Delta',0)}{P_{31}^{\frac{\Delta'}2}P_{32}^{\frac{\Delta'}2}P_{12}^{\frac{2\Delta_A-\Delta'}2}}\Bigg]\n \\
    &\times\Pi_{\bar{\Delta}',0}(X_2,P_3)\Pi_{\bar{\Delta}_A,1}(X_2,P_1;K_2,Z_1)\Pi_{\bar{\Delta}_A,1}(X_2,P_2;W_2,Z_2)\n \\
    =&\frac1{\big(\frac{d-1}2\big)}\int_{P_1}\int_{P_2}\int_{P_3}\frac{b(\Delta_A,\Delta_A,\Delta',0)}{\Delta_A^2}\frac{\mathcal{C}_{\Delta_A,1}\mathcal{C}_{\Delta_A,1}}{\mathcal{C}_{\Delta_A,0}\mathcal{C}_{\Delta_A,0}}\Bigg[\frac{\big(\Delta_A^2-\Delta_A+\frac12\Delta'\big)(D_{Z_1}\cdot D_{Z_2})}{P_{31}^{\frac{\Delta'}2}P_{32}^{\frac{\Delta'}2}P_{12}^{\frac{2\Delta_A-\Delta'}2}}\n \\
    &+\frac{\frac12\Delta'^2(D_{Z_1}\cdot P_2)(D_{Z_2}\cdot P_3)}{P_{31}^{\frac{\Delta'}2}P_{32}^{\frac{\Delta'}2+1}P_{12}^{\frac{2\Delta_A-\Delta'}2}}+\frac{(2\Delta_A-\Delta')(\Delta_A+\frac12\Delta'-1)(D_{Z_1}\cdot P_2)(D_{Z_2}\cdot P_1)}{P_{31}^{\frac{\Delta'}2}P_{32}^{\frac{\Delta'}2}P_{12}^{\frac{2\Delta_A-\Delta'}2+1}}\n \\
    &+\frac{\frac12\Delta'^2(D_{Z_2}\cdot P_1)(D_{Z_1}\cdot P_3)}{P_{31}^{\frac{\Delta'}2+1}P_{32}^{\frac{\Delta'}2}P_{12}^{\frac{2\Delta_A-\Delta'}2}}-\frac{\frac12\Delta'^2(D_{Z_2}\cdot P_3)(D_{Z_1}\cdot P_3)}{P_{31}^{\frac{\Delta'}2+1}P_{32}^{\frac{\Delta'}2+1}P_{12}^{\frac{2\Delta_A-\Delta'}2-1}}\Bigg]\n \\
    &\times\Pi_{\bar{\Delta}',0}(X_2,P_3)\Pi_{\bar{\Delta}_A,1}(X_2,P_1;K_2,Z_1)\Pi_{\bar{\Delta}_A,1}(X_2,P_2;W_2,Z_2)\n \\
    =&\int_{P_1}\int_{P_2}\int_{P_3}\frac{b(\Delta_A,\Delta_A,\Delta',0)}{\Delta_A^2}\frac{\mathcal{C}_{\Delta_A,1}\mathcal{C}_{\Delta_A,1}}{\mathcal{C}_{\Delta_A,0}\mathcal{C}_{\Delta_A,0}}\Bigg[\frac{\big(\Delta_A^2-\Delta_A+\frac12\Delta'\big)(D_{Z_1}\cdot D_{Z_2})}{P_{31}^{\frac{\Delta'}2}P_{32}^{\frac{\Delta'}2}P_{12}^{\frac{2\Delta_A-\Delta'}2}}\n \\
    &+\frac{\frac12\Delta'^2(D_{Z_1}\cdot P_2)(D_{Z_2}\cdot P_3)}{P_{31}^{\frac{\Delta'}2}P_{32}^{\frac{\Delta'}2+1}P_{12}^{\frac{2\Delta_A-\Delta'}2}}+\frac{(2\Delta_A-\Delta')(\Delta_A+\frac12\Delta'-1)(D_{Z_1}\cdot P_2)(D_{Z_2}\cdot P_1)}{P_{31}^{\frac{\Delta'}2}P_{32}^{\frac{\Delta'}2}P_{12}^{\frac{2\Delta_A-\Delta'}2+1}}\n \\
    &+\frac{\frac12\Delta'^2(D_{Z_2}\cdot P_1)(D_{Z_1}\cdot P_3)}{P_{31}^{\frac{\Delta'}2+1}P_{32}^{\frac{\Delta'}2}P_{12}^{\frac{2\Delta_A-\Delta'}2}}-\frac{\frac12\Delta'^2(D_{Z_2}\cdot P_3)(D_{Z_1}\cdot P_3)}{P_{31}^{\frac{\Delta'}2+1}P_{32}^{\frac{\Delta'}2+1}P_{12}^{\frac{2\Delta_A-\Delta'}2-1}}\Bigg]\n \\
    &\times\mathcal{C}_{\bar{\Delta}',0}\mathcal{C}_{\bar{\Delta}_A,1}\mathcal{C}_{\bar{\Delta}_A,1}\frac1{(-2P_3\cdot X_2)^{\bar{\Delta}'}(-2P_1\cdot X_2)^{\bar{\Delta}_A+1}(-2P_2\cdot X_2)^{\bar{\Delta}_A+1}}\n \\
    &\times\big[(-2P_1\cdot X_2)(-2P_2\cdot X_2)(Z_1\cdot Z_2)+2(-2P_1\cdot X_2)(Z_2\cdot X_2)(Z_1\cdot P_2)\n \\
    &+2(-2P_2\cdot X_2)(Z_1\cdot X_2)(Z_2\cdot P_1)+4(Z_1\cdot X_2)(Z_2\cdot X_2)(P_1\cdot P_2)\big]\n \\
    =&\bigg(\frac d2-1\bigg)^2\frac{b(\Delta_A,\Delta_A,\Delta',0)}{\Delta_A^2}\frac{\mathcal{C}_{\Delta_A,1}\mathcal{C}_{\Delta_A,1}}{\mathcal{C}_{\Delta_A,0}\mathcal{C}_{\Delta_A,0}}\bigg[\bigg(\Delta_A^2-\Delta_A+\frac12\Delta'\bigg)\n \\
    &\times\Bigg\{d\Bigg(\frac{\mathcal{C}_{\bar{\Delta}_A,1}\mathcal{C}_{\bar{\Delta}_A,1}}{\mathcal{C}_{\bar{\Delta}_A,0}\mathcal{C}_{\bar{\Delta}_A,0}}\Bigg)\mathcal{I}^{\text{QFT}}_{0,0,0}(\wt{\nu}_A,\wt{\nu}_A)+2\Bigg(\frac{\mathcal{C}_{\bar{\Delta}_A,1}\mathcal{C}_{\bar{\Delta}_A,1}}{\mathcal{C}_{\bar{\Delta}_A+1,0}\mathcal{C}_{\bar{\Delta}_A+1,0}}\Bigg)\mathcal{I}^{\text{QFT}}_{-1,-1,0}(\wt{\nu}_A,\wt{\nu}_A)\bigg]\n \\
    &+\frac14\Delta'^2\Bigg[-2\Bigg(\frac{\mathcal{C}_{\bar{\Delta}_A,1}\mathcal{C}_{\bar{\Delta}_A,1}}{\mathcal{C}_{\bar{\Delta}_A,0}\mathcal{C}_{\bar{\Delta}_A,0}}\Bigg)\mathcal{I}^{\text{QFT}}_{0,0,0}+\Bigg(\frac{\mathcal{C}_{\bar{\Delta}_A,1}\mathcal{C}_{\bar{\Delta}_A,1}}{\mathcal{C}_{\bar{\Delta}_A+1,0}\mathcal{C}_{\bar{\Delta}_A-1,0}}\Bigg)\mathcal{I}^{\text{QFT}}_{-1,1,0}(\wt{\nu}_A,\wt{\nu}_A)\n \\
    &+\Bigg(\frac{\mathcal{C}_{\bar{\Delta}_A,1}\mathcal{C}_{\bar{\Delta}_A,1}}{\mathcal{C}_{\bar{\Delta}_A-1,0}\mathcal{C}_{\bar{\Delta}_A+1,0}}\Bigg)\mathcal{I}^{\text{QFT}}_{1,-1,0}(\wt{\nu}_A,\wt{\nu}_A)-2\Bigg(\frac{\mathcal{C}_{\bar{\Delta}_A,1}\mathcal{C}_{\bar{\Delta}_A,1}\mathcal{C}_{\bar{\Delta}',0}}{\mathcal{C}_{\bar{\Delta}_A+1,0}\mathcal{C}_{\bar{\Delta}_A,0}\mathcal{C}_{\bar{\Delta}'-1,0}}\Bigg)\mathcal{I}^{\text{QFT}}_{-1,0,1}(\wt{\nu}_A,\wt{\nu}_A)\n \\
    &-2\Bigg(\frac{\mathcal{C}_{\bar{\Delta}_A,1}\mathcal{C}_{\bar{\Delta}_A,1}\mathcal{C}_{\bar{\Delta}',0}}{\mathcal{C}_{\bar{\Delta}_A,0}\mathcal{C}_{\bar{\Delta}_A+1,0}\mathcal{C}_{\bar{\Delta}'-1,0}}\Bigg)\mathcal{I}^{\text{QFT}}_{0,-1,1}(\wt{\nu}_A,\wt{\nu}_A)+\Bigg(\frac{\mathcal{C}_{\bar{\Delta}_A,1}\mathcal{C}_{\bar{\Delta}_A,1}\mathcal{C}_{\bar{\Delta}',0}}{\mathcal{C}_{\bar{\Delta}_A+1,0}\mathcal{C}_{\bar{\Delta}_A+1,0}\mathcal{C}_{\bar{\Delta}'-2,0}}\Bigg)\mathcal{I}^{\text{QFT}}_{-1,-1,2}(\wt{\nu}_A,\wt{\nu}_A)\bigg]\n \\
    &-(2\Delta_A-\Delta')\big(\Delta_A+\tfrac12\Delta'-1\big)\Bigg(\frac{\mathcal{C}_{\bar{\Delta}_A,1}\mathcal{C}_{\bar{\Delta}_A,1}}{\mathcal{C}_{\bar{\Delta}_A,0}\mathcal{C}_{\bar{\Delta}_A,0}}\Bigg)\mathcal{I}^{\text{QFT}}_{0,0,0}(\wt{\nu}_A,\wt{\nu}_A)\Bigg\}\n \\
    =&\bigg(\frac d2-1\bigg)^2\frac{b(\Delta_A,\Delta_A,\Delta',0)}{64\big(\frac d2+\ii\wt{\nu}_A-1\big)^2\big(\frac d2-\ii\wt{\nu}_A-1\big)^2}(8d^3-11d^4+4d^5-32d\wt{\nu}'^2+24d^2\wt{\nu}'^2+16\wt{\nu}'^4\n \\
    &+64d\wt{\nu}_A^2-80d^2\wt{\nu}_A^2+32d^3\wt{\nu}_A^2-64\wt{\nu}'^2\wt{\nu}_A^2+64d\wt{\nu}_A^4)\mathcal{I}^{\text{QFT}}_{0,0,0}.
\end{align}
The result is
\begin{align}
    \rho^{\mathcal{P},0}_{AA}(\wt{\nu}')=&\frac{\wt{\nu}'\sinh(\pi\wt{\nu}')\Gamma\big(\frac{d+1}2\big)}{2^{12-d}\pi^{\frac{d+7}2}\Gamma(d)\big(\frac d2+\ii\wt{\nu}_A-1\big)^2\big(\frac d2-\ii\wt{\nu}_A-1\big)^2\Gamma\big(\frac d2+\ii\wt{\nu}'\big)\Gamma\big(\frac d2-\ii\wt{\nu}'\big)}\n \\
    &\times\prod_{\pm,\pm,\pm}\Gamma\Big(\frac{\frac d2\pm\ii\wt{\nu}\pm\ii\wt{\nu}\pm\ii\wt{\nu}'}2\Big)(8d^3-11d^4+4d^5-32d\wt{\nu}'^2+24d^2\wt{\nu}'^2\n \\
    &+16\wt{\nu}'^4+64d\wt{\nu}_A^2-80d^2\wt{\nu}_A^2+32d^3\wt{\nu}_A^2-64\wt{\nu}'^2\wt{\nu}_A^2+64d\wt{\nu}_A^4).
\end{align}
\section{Explict Expressions}\label{AppF}
In this appendix, some expressions which are too difficult for the main text are listed.
\subsection{Useful Asymptotic Expansions for Combinations of Gamma Functions}
In this subsection, some useful asymptotic expansions for large $(n+1)$ are collected, which are useful in subsection \ref{Sec3.2} and \ref{Sec3.3}.
\begin{align}
    &\frac{\big(\frac d4-\ii\wt{\nu}+n\big)}{n!\big(\frac{\ell+2\ii\wt{\nu}+p_2+1}2-n\big)\big(\frac{\ell+d-2\ii\wt{\nu}+p_2+1}2+n\big)}\bigg(\frac d4-\ii\wt{\nu}+n\bigg)^4\n \\
    &\times\Gamma\Bigg[\begin{matrix}
        \frac d2+n, \frac d2-\ii\wt{\nu}+n, \frac d2-2\ii\wt{\nu}+n, -\ii\wt{\nu}+\frac12+n \\
        \frac{d+1}2-\ii\wt{\nu}+n, -2\ii\wt{\nu}+1+n, -\ii\wt{\nu}+1+n
    \end{matrix}\Bigg]\n \\
    \sim&-(n+1)^{d-3}\Big[(n+1)^3+\tfrac{d(d-4-4\ii\wt{\nu})}4(n+1)^2+\big(24\big(\tfrac d2+p_2+\ell+1\big)^2+\n \\
    &+3d^4-28d^3+81d^2-62d-48(d^2-2d+2)\wt{\nu}^2-24d(d-4)(d-1)\ii\wt{\nu}\big)(n+1)/96\n \\
    &+(d-2)(d-4-4\ii\wt{\nu})\big(24\big(\tfrac d2+p_2+\ell+1\big)^2+d^4-10d^3+33d^2-30d\n \\
    &-16(d^2-4d+6)\wt{\nu}^2-8d(d-4)(d-1)\ii\wt{\nu}\big)/384\n \\
    &+\Big(5760\big[\big(\tfrac d2+p_2+\ell+1\big)^4-\tfrac{d^2}2\big(\tfrac d2+p_2+\ell+1\big)^2+\tfrac{d^4}{16}\big]\n \\
    &+240(3d^4-40d^3+207d^2-398d+288-48(d-4)(d-2)\wt{\nu}^2)\big[\big(\tfrac d2+p_2+\ell+1\big)^2-\tfrac{d^2}4\big]\n \\
    &+15d^8-360d^7+3890d^6-23508d^5+83075d^4-163380d^3+163220d^2-61872d\n \\
    &-480(d-4)(d-2)(3d^4-28d^3+99d^2-134d+72)\wt{\nu}^2+3840(d-4)(d-2)(d^2-6d+12)\wt{\nu}^4\n \\
    &-240d(d-4)(d-3)(d-2)\big[24\big(\tfrac d2+p_2+\ell+1\big)^2+d^4-10d^3+33d^2-30d\big]\ii\wt{\nu}\n \\
    &+3840(d-4)(d-3)(d-2)(d^2-4d+6)\ii\wt{\nu}^3\Big)/(92160(n+1))\Big],
\end{align}
\begin{align}
    &\frac{\big(\frac d4-\ii\wt{\nu}+n\big)}{n!\big(\frac{\ell+2\ii\wt{\nu}+p_2+1}2-n\big)\big(\frac{\ell+d-2\ii\wt{\nu}+p_2+1}2+n\big)}\bigg(\frac d4-\ii\wt{\nu}+n\bigg)^2\n \\
    &\times\Gamma\Bigg[\begin{matrix}
        \frac d2+n, \frac d2-\ii\wt{\nu}+n, \frac d2-2\ii\wt{\nu}+n, -\ii\wt{\nu}+\frac12+n \\
        \frac{d+1}2-\ii\wt{\nu}+n, -2\ii\wt{\nu}+1+n, -\ii\wt{\nu}+1+n
    \end{matrix}\Bigg]\n \\
    \sim&-(n+1)^{d-3}\Big[(n+1)+\tfrac{(d-2)(d-4-4\ii\wt{\nu})}4+\Big(24\big[\big(\tfrac d2+p_2+\ell+1\big)^2-\tfrac{d^2}4\big]\n \\
    &+3d^4-40d^3+201d^2-398d+288-48(d-4)(d-2)\wt{\nu}^2\n \\
    &-24(d-4)(d-3)(d-2)\ii\wt{\nu}\Big)/(96(n+1))\Big],
\end{align}
\begin{align}
    &\frac{\big(\frac d4-\ii\wt{\nu}+n\big)}{n!\big(\frac{\ell+2\ii\wt{\nu}+p_2+1}2-n\big)\big(\frac{\ell+d-2\ii\wt{\nu}+p_2+1}2+n\big)}\n \\
    &\times\Gamma\Bigg[\begin{matrix}
        \frac d2+n, \frac d2-\ii\wt{\nu}+n, \frac d2-2\ii\wt{\nu}+n, -\ii\wt{\nu}+\frac12+n \\
        \frac{d+1}2-\ii\wt{\nu}+n, -2\ii\wt{\nu}+1+n, -\ii\wt{\nu}+1+n
    \end{matrix}\Bigg]\n \\
    \sim&-(n+1)^{d-4},
\end{align}
\begin{align}
    &\frac{\big(\frac d4-\ii\wt{\nu}+n\big)}{n!\big(\frac{\ell+2\ii\wt{\nu}+p_2+1}2-n\big)_2\big(\frac{\ell+d-2\ii\wt{\nu}+p_2+1}2+n\big)_2}\bigg(\frac d4-\ii\wt{\nu}+n\bigg)^4\n \\
    &\times\Gamma\Bigg[\begin{matrix}
        \frac d2+n, \frac d2-\ii\wt{\nu}+n, \frac d2-2\ii\wt{\nu}+n, -\ii\wt{\nu}+\frac12+n \\
        \frac{d+1}2-\ii\wt{\nu}+n, -2\ii\wt{\nu}+1+n, -\ii\wt{\nu}+1+n
    \end{matrix}\Bigg]\n \\
    \sim&(n+1)^{d-3}\Big[(n+1)+\tfrac{(d-2)(d-4-4\ii\wt{\nu})}4+\big(48\big[\big(\tfrac d2+p_2+\ell+2\big)^2+1-\tfrac{d^2}4\big]+\n \\
    &+3d^4-40d^3+207d^2-398d+288-48(d-4)(d-2)\wt{\nu}^2\n \\
    &-24(d-4)(d-3)(d-2)\ii\wt{\nu}\big)/(96(n+1))\Big],
\end{align}
\begin{align}
    &\frac{\big(\frac d4-\ii\wt{\nu}+n\big)}{n!\big(\frac{\ell+2\ii\wt{\nu}+p_2+1}2-n\big)_2\big(\frac{\ell+d-2\ii\wt{\nu}+p_2+1}2+n\big)_2}\bigg(\frac d4-\ii\wt{\nu}+n\bigg)^2\n \\
    &\times\Gamma\Bigg[\begin{matrix}
        \frac d2+n, \frac d2-\ii\wt{\nu}+n, \frac d2-2\ii\wt{\nu}+n, -\ii\wt{\nu}+\frac12+n \\
        \frac{d+1}2-\ii\wt{\nu}+n, -2\ii\wt{\nu}+1+n, -\ii\wt{\nu}+1+n
    \end{matrix}\Bigg]\n \\
    \sim&(n+1)^{d-4},
\end{align}
\begin{align}
    &\frac{\big(\frac d4-\ii\wt{\nu}+n\big)}{n!\big(\frac{\ell+2\ii\wt{\nu}+p_2+1}2-n\big)_3\big(\frac{\ell+d-2\ii\wt{\nu}+p_2+1}2+n\big)_3}\bigg(\frac d4-\ii\wt{\nu}+n\bigg)^4\n \\
    &\times\Gamma\Bigg[\begin{matrix}
        \frac d2+n, \frac d2-\ii\wt{\nu}+n, \frac d2-2\ii\wt{\nu}+n, -\ii\wt{\nu}+\frac12+n \\
        \frac{d+1}2-\ii\wt{\nu}+n, -2\ii\wt{\nu}+1+n, -\ii\wt{\nu}+1+n
    \end{matrix}\Bigg]\n \\
    \sim&-(n+1)^{d-4},
\end{align}
\begin{align}
    &\frac{\big(\frac d4+n\big)}{n!\big(\frac{\ell+p_2+1}2-n\big)\big(\frac{\ell+d+p_2+1}2+n\big)}\bigg(\frac d4+n\bigg)^4\n \\
    &\times\Gamma\Bigg[\begin{matrix}
        \frac d2+\ii\wt{\nu}+n, \frac d2+n, \frac d2-\ii\wt{\nu}+n, \frac12+n \\
        \frac{d+1}2+n, -\ii\wt{\nu}+1+n, \ii\wt{\nu}+1+n
    \end{matrix}\Bigg]\n \\
    \sim&-(n+1)^{d-3}\Big[(n+1)^3+\tfrac{d(d-4)}4+\big[24\big(\tfrac d2+p_2+\ell+1\big)^2\n \\
    &+3d^4-28d^3+81d^2-62d+48(d-2)\wt{\nu}^2\big](n+1)/96\n \\
    &+(d-4)(d-2)\big(24\big(\tfrac d2+p_2+\ell+1\big)^2+d^4-10d^3+33d^2-30d+48(d-2)\wt{\nu}^2\big)/384\n \\
    &+\Big(5760\big[\big(\tfrac d2+p_2+\ell+1\big)^4-\tfrac{d^2}2\big(\tfrac d2+p_2+\ell+1\big)^2+\tfrac{d^4}{16}\big]\n \\
    &+240(3d^4-40d^3+207d^2-398d+288+48(d-2)\wt{\nu}^2)\big[\big(\tfrac d2+p_2+\ell+1\big)^2-\tfrac{d^2}4\big]\n \\
    &+15d^8-360d^7+3890d^6-23508d^5+83075d^4-163380d^3+163220d^2-61872d\n \\
    &+480(d-2)^2(3d^3-34d^2+139d-144)\wt{\nu}^2+11520(d-4)(d-2)\wt{\nu}^4\Big)/(92160(n+1))\Big],
\end{align}
\begin{align}
    &\frac{\big(\frac d4+n\big)}{n!\big(\frac{\ell+p_2+1}2-n\big)\big(\frac{\ell+d+p_2+1}2+n\big)}\bigg(\frac d4+n\bigg)^2\n \\
    &\times\Gamma\Bigg[\begin{matrix}
        \frac d2+\ii\wt{\nu}+n, \frac d2+n, \frac d2-\ii\wt{\nu}+n, \frac12+n \\
        \frac{d+1}2+n, -\ii\wt{\nu}+1+n, \ii\wt{\nu}+1+n
    \end{matrix}\Bigg]\n \\
    \sim&-(n+1)^{d-3}\Big[(n+1)+\tfrac{(d-4)(d-2)}4+\Big(24\big[\big(\tfrac d2+p_2+\ell+1\big)^2-\tfrac{d^2}4\big]\n \\
    &+3d^4-40d^3+201d^2-398d+288+48(d-2)\wt{\nu}^2\Big)/(96(n+1))\Big],
\end{align}
\begin{align}
    &\frac{\big(\frac d4+n\big)}{n!\big(\frac{\ell+p_2+1}2-n\big)\big(\frac{\ell+d+p_2+1}2+n\big)}\n \\
    &\times\Gamma\Bigg[\begin{matrix}
        \frac d2+\ii\wt{\nu}+n, \frac d2+n, \frac d2-\ii\wt{\nu}+n, \frac12+n \\
        \frac{d+1}2+n, -\ii\wt{\nu}+1+n, \ii\wt{\nu}+1+n
    \end{matrix}\Bigg]\n \\
    \sim&-(n+1)^{d-4},
\end{align}
\begin{align}
    &\frac{\big(\frac d4+n\big)}{n!\big(\frac{\ell+p_2+1}2-n\big)_2\big(\frac{\ell+d+p_2+1}2+n\big)_2}\bigg(\frac d4+n\bigg)^4\n \\
    &\times\Gamma\Bigg[\begin{matrix}
        \frac d2+\ii\wt{\nu}+n, \frac d2+n, \frac d2-\ii\wt{\nu}+n, \frac12+n \\
        \frac{d+1}2+n, -\ii\wt{\nu}+1+n, \ii\wt{\nu}+1+n
    \end{matrix}\Bigg]\n \\
    \sim&(n+1)^{d-3}\Big[(n+1)+\frac{(d-4)(d-2)}4+\Big(48\big[\big(\tfrac d2+p_2+\ell+2\big)^2+1-\tfrac{d^2}4\big]+\n \\
    &+3d^4-40d^3+207d^2-398d+288+48(d-2)\wt{\nu}^2\Big)/(96(n+1))\Big],
\end{align}
\begin{align}
    &\frac{\big(\frac d4+n\big)}{n!\big(\frac{\ell+p_2+1}2-n\big)_2\big(\frac{\ell+d+p_2+1}2+n\big)_2}\bigg(\frac d4+n\bigg)^2\n \\
    &\times\Gamma\Bigg[\begin{matrix}
        \frac d2+\ii\wt{\nu}+n, \frac d2+n, \frac d2-\ii\wt{\nu}+n, \frac12+n \\
        \frac{d+1}2+n, -\ii\wt{\nu}+1+n, \ii\wt{\nu}+1+n
    \end{matrix}\Bigg]\n \\
    \sim&(n+1)^{d-4},
\end{align}
\begin{align}
    &\frac{\big(\frac d4+n\big)}{n!\big(\frac{\ell+p_2+1}2-n\big)_3\big(\frac{\ell+d+p_2+1}2+n\big)_3}\bigg(\frac d4+n\bigg)^4\n \\
    &\times\Gamma\Bigg[\begin{matrix}
        \frac d2+\ii\wt{\nu}+n, \frac d2+n, \frac d2-\ii\wt{\nu}+n, \frac12+n \\
        \frac{d+1}2+n, -\ii\wt{\nu}+1+n, \ii\wt{\nu}+1+n
    \end{matrix}\Bigg]\n \\
    \sim&(n+1)^{d-4}.
\end{align}

Therefore, the coefficients in (\ref{82}), (\ref{83}), (\ref{85}), (\ref{86}), (\ref{133}), (\ref{134}), (\ref{136}), (\ref{137}) are
\begin{align}
    \Upsilon_{\nabla\sigma\nabla\sigma,d,\text{A},\ell,0,3}^{p_2}=&1, \\
    \Upsilon_{\nabla\sigma\nabla\sigma,d,\text{A},\ell,0,2}^{p_2}=&\tfrac{d(d-4-4\ii\wt{\nu})}4, \\
    \Upsilon_{\nabla\sigma\nabla\sigma,d,\text{A},\ell,0,1}^{p_2}=&\big[24\big(\tfrac d2+p_2+\ell+1\big)^2+3d^4-28d^3+93d^2\n \\
    &-62d-48d(d-2)\wt{\nu}^2-24d(d-4)(d-1)\ii\wt{\nu}\big]/96, \\
    \Upsilon_{\nabla\sigma\nabla\sigma,d,\text{A},\ell,0,0}^{p_2}=&(d-2)(d-4-4\ii\wt{\nu})\big[24\big(\tfrac d2+p_2+\ell+1\big)^2+d^4-10d^3+45d^2-30d\n \\
    &-16d(d-4)\wt{\nu}^2-8d(d-4)(d-1)\ii\wt{\nu}\big]/384, \\
    \Upsilon_{\nabla\sigma\nabla\sigma,d,\text{A},\ell,0,-1}^{p_2}=&\Big(5760\big[\big(\tfrac d2+p_2+\ell+1\big)^4-\tfrac{d^2}2\big(\tfrac d2+p_2+\ell+1\big)^2+\tfrac{d^4}{16}\big]\n \\
    &+240[3d^4-40d^3+219d^2-398d+288-48(d^2-6d+6)\wt{\nu}^2]\big[\big(\tfrac d2+p_2+\ell+1\big)^2-\tfrac{d^2}4\big]\n \\
    &+15d^8-360d^7+4250d^6-28308d^5+107555d^4-211140d^3+197780d^2-61872d\n \\
    &-480(3d^6-46d^5+297d^4-944d^3+1350d^2-708d)\wt{\nu}^2\n \\
    &+3840(d^4-12d^3+44d^2-48d+6)\wt{\nu}^4\n \\
    &-240(d-4)(d-3)(d-2)\big[24\big(\tfrac d2+p_2+\ell+1\big)^2+d^4-10d^3+45d^2-30d\big]\ii\wt{\nu}\n \\
    &+3840d(d-4)^2(d-3)(d-2)\ii\wt{\nu}^3\Big)/92160,
\end{align}
\begin{align}
    \Upsilon_{\nabla\sigma\nabla\sigma,d,\text{A},\ell,1,1}^{p_2}=&1, \\
    \Upsilon_{\nabla\sigma\nabla\sigma,d,\text{A},\ell,1,0}^{p_2}=&\tfrac{(d-2)(d-4-4\ii\wt{\nu})}4, \\
    \Upsilon_{\nabla\sigma\nabla\sigma,d,\text{A},\ell,1,-1}^{p_2}=&\Big(48\big[\big(\tfrac d2+p_2+\ell+2\big)^2+1-\tfrac{d^2}4\big]\n \\
    &+3d^4-40d^3+219d^2-398d+288-48(d^2-6d+6)\wt{\nu}^2\n \\
    &-24(d-4)(d-3)(d-2)\ii\wt{\nu}\Big)/96,
\end{align}
\begin{align}
    \Upsilon_{\nabla\sigma\nabla\sigma,d,\text{B},\ell,0,3}^{p_2}=&1, \\
    \Upsilon_{\nabla\sigma\nabla\sigma,d,\text{B},\ell,0,2}^{p_2}=&\tfrac{d(d-4)}4, \\
    \Upsilon_{\nabla\sigma\nabla\sigma,d,\text{B},\ell,0,1}^{p_2}=&\big[24\big(\tfrac d2+p_2+\ell+1\big)^2+3d^4-28d^3+93d^2-62d+48d\wt{\nu}^2\big]/96, \\
    \Upsilon_{\nabla\sigma\nabla\sigma,d,\text{B},\ell,0,0}^{p_2}=&(d-4)(d-2)\big[24\big(\tfrac d2+p_2+\ell+1\big)^2+d^4-10d^3+45d^2-30d+48d\wt{\nu}^2\big]/384, \\
    \Upsilon_{\nabla\sigma\nabla\sigma,d,\text{B},\ell,0,-1}^{p_2}=&\Big(5760\big[\big(\tfrac d2+p_2+\ell+1\big)^4-\tfrac{d^2}2\big(\tfrac d2+p_2+\ell+1\big)^2+\tfrac{d^4}{16}\big]\n \\
    &+240[3d^4-40d^3+219d^2-398d+288+48d\wt{\nu}^2]\big[\big(\tfrac d2+p_2+\ell+1\big)^2-\tfrac{d^2}4\big]\n \\
    &+15d^8-360d^7+4250d^6-28308d^5+107555d^4-211140d^3+197780d^2-61872d\n \\
    &+480d(3d^4-40d^3+219d^2-446d+336)\wt{\nu}^2+11520(d^2-2d+2)\wt{\nu}^4\Big)/92160,
\end{align}
\begin{align}
    \Upsilon_{\nabla\sigma\nabla\sigma,d,\text{B},\ell,1,1}^{p_2}=&1, \\
    \Upsilon_{\nabla\sigma\nabla\sigma,d,\text{B},\ell,1,0}^{p_2}=&\tfrac{(d-4)(d-2)}4, \\
    \Upsilon_{\nabla\sigma\nabla\sigma,d,\text{B},\ell,1,-1}^{p_2}=&\Big(48\big[\big(\tfrac d2+p_2+\ell+2\big)^2+1-\tfrac{d^2}4\big]\n \\
    &+3d^4-40d^3+219d^2-398d+288+48d\wt{\nu}^2\Big)/96,
\end{align}
\begin{align}
    \Upsilon_{AA,d,\text{A},\ell,0,3}^{p_2}=&16, \\
    \Upsilon_{AA,d,\text{A},\ell,0,2}^{p_2}=&4d(d-4-4\ii\wt{\nu}_A), \\
    \Upsilon_{AA,d,\text{A},\ell,0,1}^{p_2}=&\big[24\big(\tfrac d2+p_2+\ell+1\big)^2+3d^4-28d^3+45d^2-14d\n \\
    &-48d(d-2)\wt{\nu}_A^2-24d(d-4)(d-1)\ii\wt{\nu}_A\big]/6, \\
    \Upsilon_{AA,d,\text{A},\ell,0,0}^{p_2}=&(d-2)(d-4-4\ii\wt{\nu}_A)\big[24\big(\tfrac d2+p_2+\ell+1\big)^2+d^4-10d^3-3d^2+18d\n \\
    &-16d(d-4)\wt{\nu}_A^2-8d(d-4)(d-1)\ii\wt{\nu}_A\big]/24, \\
    \Upsilon_{AA,d,\text{A},\ell,0,-1}^{p_2}=&\Big(5760\big[\big(\tfrac d2+p_2+\ell+1\big)^4-\tfrac{d^2}2\big(\tfrac d2+p_2+\ell+1\big)^2+\tfrac{d^4}{16}\big]\n \\
    &+240[3d^4-40d^3+171d^2-350d+288-48(d^2-6d+6)\wt{\nu}_A^2\n \\
    &-24(d-4)(d-3)(d-2)\ii\wt{\nu}_A]\big[\big(\tfrac d2+p_2+\ell+1\big)^2-\tfrac{d^2}4\big]\n \\
    &+15d^8-360d^7+2810d^6-6228d^5-12445d^4+79260d^3-131500d^2+76368d\n \\
    &-480d(3d^5-46d^4+249d^3-632d^2+750d-372)\wt{\nu}_A^2\n \\
    &+3840d(d^3-12d^2+44d-42)\wt{\nu}_A^4\n \\
    &-240d(d-4)(d-3)(d-2)(d^3-10d^2+3d+18)\ii\wt{\nu}_A\n \\
    &+3840d(d-4)^2(d-3)(d-2)\ii\wt{\nu}_A^3\Big)/5760,
\end{align}
\begin{align}
    \Upsilon_{AA,d,\text{A},\ell,1,1}^{p_2}=&16, \\
    \Upsilon_{AA,d,\text{A},\ell,1,0}^{p_2}=&4(d-2)(d-4-4\ii\wt{\nu}_A), \\
    \Upsilon_{AA,d,\text{A},\ell,1,-1}^{p_2}=&\Big(48\big[\big(\tfrac d2+p_2+\ell+2\big)^2+1-\tfrac{d^2}4\big]+\n \\
    &+(d-2)(3d^3-34d^2+103d-144)-48(d^2-6d+6)\wt{\nu}^2\n \\
    &-24(d-4)(d-3)(d-2)\ii\wt{\nu}\Big)/6,
\end{align}
\begin{align}
    \Upsilon_{AA,d,\text{B},\ell,0,3}^{p_2}=&16, \\
    \Upsilon_{AA,d,\text{B},\ell,0,2}^{p_2}=&4d(d-4)(n+1)^2, \\
    \Upsilon_{AA,d,\text{B},\ell,0,1}^{p_2}=&\big[24\big(\tfrac d2+p_2+\ell+1\big)^2+3d^4-28d^3+45d^2-14d+48d\wt{\nu}^2\big]/6, \\
    \Upsilon_{AA,d,\text{B},\ell,0,0}^{p_2}=&(d-4)(d-2)\big[24\big(\tfrac d2+p_2+\ell+1\big)^2+d^4-10d^3-3d^2+18d+48d\wt{\nu}^2\big]/24, \\
    \Upsilon_{AA,d,\text{B},\ell,0,-1}^{p_2}=&\Big(5760\big[\big(\tfrac d2+p_2+\ell+1\big)^4-\tfrac{d^2}2\big(\tfrac d2+p_2+\ell+1\big)^2+\tfrac{d^4}{16}\big]\n \\
    &+240[3d^4-40d^3+171d^2-350d+288+48d\wt{\nu}^2]\big[\big(\tfrac d2+p_2+\ell+1\big)^2-\tfrac{d^2}4\big]\n \\
    &+15d^8-360d^7+2810d^6-6228d^5-12445d^4+79260d^3-131500d^2+76368d\n \\
    &+480d(3d^4-40d^3+195d^2-374d+288)\wt{\nu}^2+11520d^2\wt{\nu}^4\Big)/5760,
\end{align}
\begin{align}
    \Upsilon_{AA,d,\text{B},\ell,1,1}^{p_2}=&16, \\
    \Upsilon_{AA,d,\text{B},\ell,1,0}^{p_2}=&4(d-4)(d-2), \\
    \Upsilon_{AA,d,\text{B},\ell,1,-1}^{p_2}=&\Big(48\big[\big(\tfrac d2+p_2+\ell+2\big)^2+1-\tfrac{d^2}4\big]+\n \\
    &+3d^4-40d^3+171d^2-350d+288+48d\wt{\nu}^2\Big)/6.
\end{align}

\subsection{Expressions of Background in $d=3$}
In this subsection, the expressions of background of $\nabla\sigma\nabla\sigma$ model and $AA$ are listed.
\begin{equation}
    \wh{\mathcal{J}}_{\nabla\sigma\nabla\sigma,\text{BG}}^{p_1p_2}=\mathcal{J}_{\nabla\sigma\nabla\sigma,(\text{B})}^{p_1p_2}+\wh{\mathcal{J}}_{\nabla\sigma\nabla\sigma,(\text{3A})}^{p_1p_2}+\wh{\mathcal{J}}_{\nabla\sigma\nabla\sigma,(\text{3B})}^{p_1p_2}+\wh{\mathcal{J}}_{AA,(\text{3C})}^{p_1p_2},
\end{equation}
\begin{align}
    \wh{\mathcal{J}}_{\nabla\sigma\nabla\sigma,(\text{3A})}^{p_1p_2}=&-\sum_{n,\ell=0}^\infty\sum_{j=3}^\infty\frac{(-1)^\ell\big(\frac32-2\ii\wt{\nu}+2n\big)\mathcal{C}_{\frac32-2\ii\wt{\nu}+2n,3}^{p_1p_2}\cos^2\big(2\pi\ii\wt{\nu}\big)(\ell+1)_{2j+p_{12}+4}}{2^{2j+5}\pi^2\sin^2(\pi\ii\wt{\nu})n!\big(\frac{\ell+2\ii\wt{\nu}+p_2+1}2-n\big)_{j+1}\big(\frac{\ell-2\ii\wt{\nu}+p_2+4}2+n\big)_{j+1}}\n \\
    &\times\Gamma\Bigg[\begin{matrix}
        \frac32+n, -\ii\wt{\nu}+\frac32+n, -2\ii\wt{\nu}+\frac32+n, -\ii\wt{\nu}+\frac12+n \\
        -\ii\wt{\nu}+2+n, -2\ii\wt{\nu}+1+n, -\ii\wt{\nu}+1+n
    \end{matrix}\Bigg]\n \\
    &\times r_1^{2j+p_{12}+5}\bigg(\frac{r_1}{r_2}\bigg)^\ell\big[\tfrac94+2\wt{\nu}^2+\big(\tfrac32-2\ii\wt{\nu}+2n\big)^2\big]^2\n \\
    &-\sum_{\ell=0}^\infty\frac{(-1)^\ell\mathcal{C}_{\frac32-2\ii\wt{\nu},3}^{p_1p_2}\cos^2\big(2\pi\ii\wt{\nu}\big)(\ell+1)_{p_{12}+4}}{\pi^2\sin^2(\pi\ii\wt{\nu})}r_1^{p_{12}+5}\bigg(\frac{r_1}{r_2}\bigg)^\ell\n \\
    &\times\Bigg[\sum_{n=0}^\infty\Bigg(\frac{\frac34-\ii\wt{\nu}+n}{n!\big(\frac{\ell+2\ii\wt{\nu}+p_2+1}2-n\big)\big(\frac{\ell-2\ii\wt{\nu}+p_2+4}2+n\big)}\n \\
    &\times\Gamma\Bigg[\begin{matrix}
        \frac32+n, -\ii\wt{\nu}+\frac32+n, -2\ii\wt{\nu}+\frac32+n, -\ii\wt{\nu}+\frac12+n \\
        -\ii\wt{\nu}+2+n, -2\ii\wt{\nu}+1+n, -\ii\wt{\nu}+1+n
    \end{matrix}\Bigg]\big[\tfrac9{16}+\tfrac{\wt{\nu}^2}{2}+\big(\tfrac34-\ii\wt{\nu}+n\big)^2\big]^2\n \\
    &+\Big[(n+1)^3+\tfrac{3(-1-4\ii\wt{\nu})}4(n+1)^2+\big[4\big(p_2+\ell+\tfrac52\big)^2+23-24\wt{\nu}^2+24\ii\wt{\nu}\big](n+1)/16\n \\
    &+(-1-4\ii\wt{\nu})\big[4\big(p_2+\ell+\tfrac52\big)^2+21+8\wt{\nu}^2+8\ii\wt{\nu}\big]/64\n \\
    &+\Big(8\big[\big(p_2+\ell+\tfrac52\big)^4-\tfrac92\big(p_2+\ell+\tfrac52\big)^2+\tfrac{81}{16}\big]\n \\
    &+(76+48\wt{\nu}^2)\big[\big(p_2+\ell+\tfrac52\big)^2-\tfrac94\big]+189+264\wt{\nu}^2+80\wt{\nu}^4\Big)/(128(n+1))\Big]\Bigg)\n \\
    &+\big[20\big(p_2+\ell+\tfrac52\big)^2+107-120\wt{\nu}^2+120\ii\wt{\nu}\big]/960\n \\
    &+(-1-4\ii\wt{\nu})\big[4\big(p_2+\ell+\tfrac52\big)^2+21+8\wt{\nu}^2+8\ii\wt{\nu}\big]/128\n \\
    &-\Big(8\big[\big(p_2+\ell+\tfrac52\big)^4-\tfrac92\big(p_2+\ell+\tfrac52\big)^2+\tfrac{81}{16}\big]\n \\
    &+(76+48\wt{\nu}^2)\big[\big(p_2+\ell+\tfrac52\big)^2-\tfrac94\big]+189+264\wt{\nu}^2+80\wt{\nu}^4\Big)\n \\
    &\times\Bigg(\gamma_E+\log\mu_R+\tfrac12\log4\pi+\psi\big(\tfrac32\big)-\frac{\pi\sin\big[\pi\big(-2\ii\wt{\nu}+\tfrac{p_{12}}2\big)\big]}{16\cos^2(2\pi\ii\wt{\nu})\mathcal{C}_{\frac32-2\ii\wt{\nu},3}^{p_1p_2}}\Bigg)/128\n \\
    &+\frac{9+40\wt{\nu}^2+16\wt{\nu}^4}{512}\big[-3+2\gamma_E-4\log\mu_R-2\log(4\pi)+2\psi\big(\ii\wt{\nu}+\tfrac52\big)+2\psi\big(-\ii\wt{\nu}+\tfrac52\big)\big]\n \\
    &+\Big((-140+60\ii\wt{\nu})\big[\big(p_2+\ell+\tfrac52\big)^2-\tfrac94\big]-917-420\wt{\nu}^2+450\ii\wt{\nu}+120\ii\wt{\nu}^3\Big)/960\Bigg]\n \\
    &-\sum_{\ell=0}^\infty\frac{(-1)^\ell\mathcal{C}_{\frac32-2\ii\wt{\nu},3}^{p_1p_2}\cos^2\big(2\pi\ii\wt{\nu}\big)(\ell+1)_{p_{12}+6}}{4\pi^2\sin^2(\pi\ii\wt{\nu})}r_1^{p_{12}+7}\bigg(\frac{r_1}{r_2}\bigg)^\ell\n \\
    &\times\Bigg[\sum_{n=0}^\infty\Bigg(\frac{\frac34-\ii\wt{\nu}+n}{n!\big(\frac{\ell+2\ii\wt{\nu}+p_2+1}2-n\big)_2\big(\frac{\ell-2\ii\wt{\nu}+p_2+4}2+n\big)_2}\n \\
    &\times\Gamma\Bigg[\begin{matrix}
        \frac32+n, -\ii\wt{\nu}+\frac32+n, -2\ii\wt{\nu}+\frac32+n, -\ii\wt{\nu}+\frac12+n \\
        -\ii\wt{\nu}+2+n, -2\ii\wt{\nu}+1+n, -\ii\wt{\nu}+1+n
    \end{matrix}\Bigg]\big[\tfrac9{16}+\tfrac{\wt{\nu}^2}{2}+\big(\tfrac34-\ii\wt{\nu}+n\big)^2\big]^2\n \\
    &-\Big[(n+1)+\tfrac{(-1-4\ii\wt{\nu})}4+\Big(4\big[\big(p_2+\ell+\tfrac72\big)^2-\tfrac54\big]+19+12\wt{\nu}^2\Big)/(8(n+1))\Big]\Bigg)\n \\
    &+\tfrac{1+12\ii\wt{\nu}}{24}+\Big(4\big[\big(p_2+\ell+\tfrac72\big)^2-\tfrac54\big]+19+12\wt{\nu}^2\Big)\n \\
    &\times\Bigg(\gamma_E+\log\mu_R+\tfrac12\log4\pi+\psi\big(\tfrac32\big)-\frac{\pi\sin\big[\pi\big(-2\ii\wt{\nu}+\tfrac{p_{12}}2\big)\big]}{16\cos^2(2\pi\ii\wt{\nu})\mathcal{C}_{\frac32-2\ii\wt{\nu},3}^{p_1p_2}}\Bigg)/8+(7-3\ii\wt{\nu})/{12}\Bigg]\n \\
    &-\sum_{\ell=0}^\infty\frac{(-1)^\ell\mathcal{C}_{\frac32-2\ii\wt{\nu},3}^{p_1p_2}\cos^2\big(2\pi\ii\wt{\nu}\big)(\ell+1)_{p_{12}+8}}{16\pi^2\sin^2(\pi\ii\wt{\nu})}r_1^{p_{12}+9}\bigg(\frac{r_1}{r_2}\bigg)^\ell\n \\
    &\times\Bigg[\sum_{n=0}^\infty\Bigg(\frac{\frac34-\ii\wt{\nu}+n}{n!\big(\frac{\ell+2\ii\wt{\nu}+p_2+1}2-n\big)_3\big(\frac{\ell-2\ii\wt{\nu}+p_2+4}2+n\big)_3}\n \\
    &\times\Gamma\Bigg[\begin{matrix}
        \frac32+n, -\ii\wt{\nu}+\frac32+n, -2\ii\wt{\nu}+\frac32+n, -\ii\wt{\nu}+\frac12+n \\
        -\ii\wt{\nu}+2+n, -2\ii\wt{\nu}+1+n, -\ii\wt{\nu}+1+n
    \end{matrix}\Bigg]\big[\tfrac9{16}+\tfrac{\wt{\nu}^2}{2}+\big(\tfrac34-\ii\wt{\nu}+n\big)^2\big]^2\n \\
    &+\frac1{n+1}\Bigg)-\Bigg(\gamma_E+\log\mu_R+\tfrac12\log4\pi+\psi\big(\tfrac32\big)-\frac{\pi\sin\big[\pi\big(-2\ii\wt{\nu}+\tfrac{p_{12}}2\big)\big]}{16\cos^2(2\pi\ii\wt{\nu})\mathcal{C}_{\frac32-2\ii\wt{\nu},3}^{p_1p_2}}\Bigg)\Bigg],
\end{align}
\begin{align}
    \wh{\mathcal{J}}_{\nabla\sigma\nabla\sigma,(\text{3B})}^{p_1p_2}=&\sum_{n,\ell=0}^\infty\sum_{j=3}^\infty\frac{(-1)^\ell\big(\frac32+2n\big)\mathcal{C}_{\frac32+2n,3}^{p_1p_2}(\ell+1)_{2j+p_{12}+4}}{2^{2j+4}\pi^2\sin^2(\pi\ii\wt{\nu})n!\big(\frac{\ell+p_2+1}2-n\big)_{j+1}\big(\frac{\ell+p_2+4}2+n\big)_{j+1}}\n \\
    &\times\Gamma\Bigg[\begin{matrix}
        \ii\wt{\nu}+\frac32+n, \frac32+n, -\ii\wt{\nu}+\frac32+n, \frac12+n \\
        2+n, -\ii\wt{\nu}+1+n, \ii\wt{\nu}+1+n
    \end{matrix}\Bigg]\n \\
    &\times r_1^{2j+p_{12}+5}\bigg(\frac{r_1}{r_2}\bigg)^\ell\big[\tfrac94+2\wt{\nu}^2+\big(\tfrac32+2n\big)^2\big]^2\n \\
    &+\sum_{\ell=0}^\infty\frac{(-1)^\ell\mathcal{C}_{\frac32,3}^{p_1p_2}(\ell+1)_{p_{12}+4}}{2^{-1}\pi^2\sin^2(\pi\ii\wt{\nu})}r_1^{p_{12}+5}\bigg(\frac{r_1}{r_2}\bigg)^\ell\Bigg[\sum_{n=0}^\infty\Bigg(\frac{\frac34+n}{n!\big(\frac{\ell+p_2+1}2-n\big)\big(\frac{\ell+p_2+4}2+n\big)}\n \\
    &\times\Gamma\Bigg[\begin{matrix}
        \ii\wt{\nu}+\frac32+n, \frac32+n, -\ii\wt{\nu}+\frac32+n, \frac12+n \\
        2+n, -\ii\wt{\nu}+1+n, \ii\wt{\nu}+1+n
    \end{matrix}\Bigg]\big[\tfrac9{16}+\tfrac{\wt{\nu}^2}{2}+\big(\tfrac34+n\big)^2\big]^2\n \\
    &+\Big[(n+1)^3-\tfrac34(n+1)^2+\big[4\big(p_2+\ell+\tfrac52\big)^2+23+24\nu^2](n+1)/16\n \\
    &-\big[4\big(p_2+\ell+\tfrac52\big)^2+21+24\wt{\nu}^2\big]/64\n \\
    &+\Big(8\big[\big(p_2+\ell+\tfrac52\big)^4-\tfrac92\big(p_2+\ell+\tfrac52\big)^2+\tfrac{81}{16}\big]\n \\
    &+(76+48\wt{\nu}^2)\big[\big(p_2+\ell+\tfrac52\big)^2-\tfrac94\big]+189+264\wt{\nu}^2+80\wt{\nu}^4\Big)/(128(n+1))\Big]\Bigg)\n \\
    &-\big[20\big(p_2+\ell+\tfrac52\big)^2+101+120\wt{\nu}^2\big]/1920\n \\
    &-\bigg(8\big[\big(p_2+\ell+\tfrac52\big)^4-\tfrac92\big(p_2+\ell+\tfrac52\big)^2+\tfrac{81}{16}\big]\n \\
    &+(76+48\wt{\nu}^2)\big[\big(p_2+\ell+\tfrac52\big)^2-\tfrac94\big]+189+264\wt{\nu}^2+80\wt{\nu}^4\bigg)\n \\
    &\times\Bigg(\gamma_E+\log\mu_R+\tfrac12\log4\pi+\psi\big(\tfrac32\big)-\frac{\pi\sin\big(\tfrac{\pi p_{12}}2\big)}{16\mathcal{C}_{\frac32,3}^{p_1p_2}}\Bigg)/128\n \\
    &+\frac{9+40\wt{\nu}^2+16\wt{\nu}^4}{512}\big[-3+2\gamma_E-4\log\mu_R-2\log(4\pi)+2\psi\big(\ii\wt{\nu}+\tfrac52\big)+2\psi\big(-\ii\wt{\nu}+\tfrac52\big)\big]\n \\
    &+\Big((-140+120\nu^2)\big[\big(p_2+\ell+\tfrac52\big)^2-\tfrac94\big]-917+540\nu^2+480\wt{\nu}^4\Big)/960\Bigg]\n \\
    &+\sum_{\ell=0}^\infty\frac{(-1)^\ell\mathcal{C}_{\frac32,3}^{p_1p_2}(\ell+1)_{p_{12}+6}}{2^{-1}\pi^2\sin^2(\pi\ii\wt{\nu})}r_1^{p_{12}+7}\bigg(\frac{r_1}{r_2}\bigg)^\ell\Bigg[\sum_{n=0}^\infty\Bigg(\frac{\frac34+n}{n!\big(\frac{\ell+p_2+1}2-n\big)_2\big(\frac{\ell+p_2+4}2+n\big)_2}\n \\
    &\times\Gamma\Bigg[\begin{matrix}
        \ii\wt{\nu}+\frac32+n, \frac32+n, -\ii\wt{\nu}+\frac32+n, \frac12+n \\
        2+n, -\ii\wt{\nu}+1+n, \ii\wt{\nu}+1+n
    \end{matrix}\Bigg]\big[\tfrac9{16}+\tfrac{\wt{\nu}^2}{2}+\big(\tfrac34+n\big)^2\big]^2\n \\
    &-\Big[(n+1)-\tfrac14+\Big(4\big[\big(p_2+\ell+\tfrac72\big)^2+1-\tfrac{d^2}4\big]+19+12\wt{\nu}^2\Big)/(8(n+1))\Big]\Bigg)\n \\
    &+\tfrac1{24}+\Big(4\big[\big(p_2+\ell+\tfrac72\big)^2+1-\tfrac{d^2}4\big]+19+12\wt{\nu}^2\Big)\n \\
    &\times\Bigg(\gamma_E+\log\mu_R+\tfrac12\log4\pi+\psi\big(\tfrac32\big)-\frac{\pi\sin\big(\tfrac{\pi p_{12}}2\big)}{16\mathcal{C}_{\frac32,3}^{p_1p_2}}\Bigg)/8+(7-6\wt{\nu}^2)/{12}\Bigg]\n \\
    &+\sum_{\ell=0}^\infty\frac{(-1)^\ell\mathcal{C}_{\frac32,3}^{p_1p_2}(\ell+1)_{p_{12}+8}}{2^{-1}\pi^2\sin^2(\pi\ii\wt{\nu})}r_1^{p_{12}+9}\bigg(\frac{r_1}{r_2}\bigg)^\ell\Bigg[\sum_{n=0}^\infty\Bigg(\frac{\frac34+n}{n!\big(\frac{\ell+p_2+1}2-n\big)_3\big(\frac{\ell+p_2+4}2+n\big)_3}\n \\
    &\times\Gamma\Bigg[\begin{matrix}
        \ii\wt{\nu}+\frac32+n, \frac32+n, -\ii\wt{\nu}+\frac32+n, \frac12+n \\
        2+n, -\ii\wt{\nu}+1+n, \ii\wt{\nu}+1+n
    \end{matrix}\Bigg]\big[\tfrac9{16}+\tfrac{\wt{\nu}^2}{2}+\big(\tfrac34+n\big)^2\big]^2\n \\
    &+\frac1{n+1}\Bigg)-\Bigg(\gamma_E+\log\mu_R+\tfrac12\log4\pi+\psi\big(\tfrac32\big)-\frac{\pi\sin\big(\tfrac{\pi p_{12}}2\big)}{16\mathcal{C}_{\frac32,3}^{p_1p_2}}\Bigg)\Bigg],
\end{align}
\begin{align}
    \wh{\mathcal{J}}_{\nabla\sigma\nabla\sigma,(\text{3C})}^{p_1p_2}=&-\sum_{n,\ell=0}^\infty\sum_{j=3}^\infty\frac{(-1)^\ell\big(\frac32+2\ii\wt{\nu}+2n\big)\mathcal{C}_{\frac32+2\ii\wt{\nu}+2n,3}^{p_1p_2}\cos^2\big(2\pi\ii\wt{\nu}\big)(\ell+1)_{2j+p_{12}+4}}{2^{2j+5}\pi^2\sin^2(\pi\ii\wt{\nu})n!\big(\frac{\ell-2\ii\wt{\nu}+p_2+1}2-n\big)_{j+1}\big(\frac{\ell+2\ii\wt{\nu}+p_2+4}2+n\big)_{j+1}}\n \\
    &\times\Gamma\Bigg[\begin{matrix}
        2\ii\wt{\nu}+\frac32+n, \ii\wt{\nu}+\frac32+n, \frac32+n, \ii\wt{\nu}+\frac12+n \\
        \ii\wt{\nu}+2+n, \ii\wt{\nu}+1+n, 2\ii\wt{\nu}+1+n
    \end{matrix}\Bigg]\n \\
    &\times r_1^{2j+p_{12}+5}\bigg(\frac{r_1}{r_2}\bigg)^\ell\big[\tfrac94+2\wt{\nu}^2+\big(\tfrac32+2\ii\wt{\nu}+2n\big)^2\big]^2\n \\
    &-\sum_{\ell=0}^\infty\frac{(-1)^\ell\mathcal{C}_{\frac32+2\ii\wt{\nu},3}^{p_1p_2}\cos^2\big(2\pi\ii\wt{\nu}\big)(\ell+1)_{p_{12}+4}}{\pi^2\sin^2(\pi\ii\wt{\nu})}r_1^{p_{12}+5}\bigg(\frac{r_1}{r_2}\bigg)^\ell\n \\
    &\times\Bigg[\sum_{n=0}^\infty\Bigg(\frac{\frac34+\ii\wt{\nu}+n}{n!\big(\frac{\ell-2\ii\wt{\nu}+p_2+1}2-n\big)\big(\frac{\ell+2\ii\wt{\nu}+p_2+4}2+n\big)}\n \\
    &\times\Gamma\Bigg[\begin{matrix}
        2\ii\wt{\nu}+\frac32+n, \ii\wt{\nu}+\frac32+n, \frac32+n, \ii\wt{\nu}+\frac12+n \\
        \ii\wt{\nu}+2+n, \ii\wt{\nu}+1+n, 2\ii\wt{\nu}+1+n
    \end{matrix}\Bigg]\big[\tfrac9{16}+\tfrac{\wt{\nu}^2}{2}+\big(\tfrac34+\ii\wt{\nu}+n\big)^2\big]^2\n \\
    &+\Big[(n+1)^3+\tfrac{3(-1+4\ii\wt{\nu})}4(n+1)^2+\big[4\big(p_2+\ell+\tfrac52\big)^2+23-24\wt{\nu}^2-24\ii\wt{\nu}\big](n+1)/16\n \\
    &+(-1+4\ii\wt{\nu})\big[4\big(p_2+\ell+\tfrac52\big)^2+21+8\wt{\nu}^2-8\ii\wt{\nu}\big]/64\n \\
    &+\Big(8\big[\big(p_2+\ell+\tfrac52\big)^4-\tfrac92\big(p_2+\ell+\tfrac52\big)^2+\tfrac{81}{16}\big]\n \\
    &+(76+48\wt{\nu}^2)\big[\big(p_2+\ell+\tfrac52\big)^2-\tfrac94\big]+189+264\wt{\nu}^2+80\wt{\nu}^4\Big)/(128(n+1))\Big]\Bigg)\n \\
    &+\big[20\big(p_2+\ell+\tfrac52\big)^2+107-120\wt{\nu}^2-120\ii\wt{\nu}\big]/960\n \\
    &+(-1+4\ii\wt{\nu})\big[4\big(p_2+\ell+\tfrac52\big)^2+21+8\wt{\nu}^2-8\ii\wt{\nu}\big]/128\n \\
    &-\Big(8\big[\big(p_2+\ell+\tfrac52\big)^4-\tfrac92\big(p_2+\ell+\tfrac52\big)^2+\tfrac{81}{16}\big]\n \\
    &+(76+48\wt{\nu}^2)\big[\big(p_2+\ell+\tfrac52\big)^2-\tfrac94\big]+189+264\wt{\nu}^2+80\wt{\nu}^4\Big)\n \\
    &\times\Bigg(\gamma_E+\log\mu_R+\tfrac12\log4\pi+\psi\big(\tfrac32\big)-\frac{\pi\sin\big[\pi\big(2\ii\wt{\nu}+\tfrac{p_{12}}2\big)\big]}{16\cos^2(2\pi\ii\wt{\nu})\mathcal{C}_{\frac32+2\ii\wt{\nu},3}^{p_1p_2}}\Bigg)/128\n \\
    &+\frac{9+40\wt{\nu}^2+16\wt{\nu}^4}{512}\big[-3+2\gamma_E-4\log\mu_R-2\log(4\pi)+2\psi\big(\ii\wt{\nu}+\tfrac52\big)+2\psi\big(-\ii\wt{\nu}+\tfrac52\big)\big]\n \\
    &+\Big((-140-60\ii\wt{\nu})\big[\big(p_2+\ell+\tfrac52\big)^2-\tfrac94\big]-917-420\wt{\nu}^2-450\ii\wt{\nu}-120\ii\wt{\nu}^3\Big)/960\Bigg]\n \\
    &-\sum_{\ell=0}^\infty\frac{(-1)^\ell\mathcal{C}_{\frac32+2\ii\wt{\nu},3}^{p_1p_2}\cos^2\big(2\pi\ii\wt{\nu}\big)(\ell+1)_{p_{12}+6}}{4\pi^2\sin^2(\pi\ii\wt{\nu})}r_1^{p_{12}+7}\bigg(\frac{r_1}{r_2}\bigg)^\ell\n \\
    &\times\Bigg[\sum_{n=0}^\infty\Bigg(\frac{\frac34+\ii\wt{\nu}+n}{n!\big(\frac{\ell-2\ii\wt{\nu}+p_2+1}2-n\big)_2\big(\frac{\ell+2\ii\wt{\nu}+p_2+4}2+n\big)_2}\n \\
    &\times\Gamma\Bigg[\begin{matrix}
        2\ii\wt{\nu}+\frac32+n, \ii\wt{\nu}+\frac32+n, \frac32+n, \ii\wt{\nu}+\frac12+n \\
        \ii\wt{\nu}+2+n, \ii\wt{\nu}+1+n, 2\ii\wt{\nu}+1+n
    \end{matrix}\Bigg]\big[\tfrac9{16}+\tfrac{\wt{\nu}^2}{2}+\big(\tfrac34+\ii\wt{\nu}+n\big)^2\big]^2\n \\
    &-\Big[(n+1)+\tfrac{(-1+4\ii\wt{\nu})}4+\Big(4\big[\big(p_2+\ell+\tfrac72\big)^2+1-\tfrac{d^2}4\big]+19+12\wt{\nu}^2\Big)/(8(n+1))\Big]\Bigg)\n \\
    &+\tfrac{1-12\ii\wt{\nu}}{24}+\Big(4\big[\big(p_2+\ell+\tfrac72\big)^2+1-\tfrac{d^2}4\big]+19+12\wt{\nu}^2\Big)\n \\
    &\times\Bigg(\gamma_E+\log\mu_R+\tfrac12\log4\pi+\psi\big(\tfrac32\big)-\frac{\pi\sin\big[\pi\big(2\ii\wt{\nu}+\tfrac{p_{12}}2\big)\big]}{16\cos^2(2\pi\ii\wt{\nu})\mathcal{C}_{\frac32+2\ii\wt{\nu},3}^{p_1p_2}}\Bigg)/8+(7+3\ii\wt{\nu})/{12}\Bigg]\n \\
    &-\sum_{\ell=0}^\infty\frac{(-1)^\ell\mathcal{C}_{\frac32+2\ii\wt{\nu},3}^{p_1p_2}\cos^2\big(2\pi\ii\wt{\nu}\big)(\ell+1)_{p_{12}+8}}{16\pi^2\sin^2(\pi\ii\wt{\nu})}r_1^{p_{12}+9}\bigg(\frac{r_1}{r_2}\bigg)^\ell\n \\
    &\times\Bigg[\sum_{n=0}^\infty\Bigg(\frac{\frac34+\ii\wt{\nu}+n}{n!\big(\frac{\ell-2\ii\wt{\nu}+p_2+1}2-n\big)_3\big(\frac{\ell+2\ii\wt{\nu}+p_2+4}2+n\big)_3}\n \\
    &\times\Gamma\Bigg[\begin{matrix}
        2\ii\wt{\nu}+\frac32+n, \ii\wt{\nu}+\frac32+n, \frac32+n, \ii\wt{\nu}+\frac12+n \\
        \ii\wt{\nu}+2+n, \ii\wt{\nu}+1+n, 2\ii\wt{\nu}+1+n
    \end{matrix}\Bigg]\big[\tfrac9{16}+\tfrac{\wt{\nu}^2}{2}+\big(\tfrac34+\ii\wt{\nu}+n\big)^2\big]^2\n \\
    &+\frac1{n+1}\Bigg)-\Bigg(\gamma_E+\log\mu_R+\tfrac12\log4\pi+\psi\big(\tfrac32\big)-\frac{\pi\sin\big[\pi\big(2\ii\wt{\nu}+\tfrac{p_{12}}2\big)\big]}{16\cos^2(2\pi\ii\wt{\nu})\mathcal{C}_{\frac32+2\ii\wt{\nu},3}^{p_1p_2}}\Bigg)\Bigg].
\end{align}

\begin{equation}
    \wh{\mathcal{J}}_{AA,\text{BG}}^{p_1p_2}=\mathcal{J}_{AA,(\text{B})}^{p_1p_2}+\wh{\mathcal{J}}_{AA,(\text{3A})}^{p_1p_2}+\wh{\mathcal{J}}_{AA,(\text{3B})}^{p_1p_2}+\wh{\mathcal{J}}_{AA,(\text{3C})}^{p_1p_2},
\end{equation}
\begin{align}
    \wh{\mathcal{J}}_{AA,(\text{3A})}^{p_1p_2}=&-\sum_{n,\ell=0}^\infty\sum_{j=3}^\infty\frac{(-1)^\ell\big(\frac32-2\ii\wt{\nu}_A+2n\big)\mathcal{C}_{\frac32-2\ii\wt{\nu}_A+2n,3}^{p_1p_2}\cos^2(2\ii\wt{\nu}_A)(\ell+1)_{2j+p_{12}+4}}{2^{2j+9}\pi^2\sin^2(\pi\ii\wt{\nu}_A)n!\big(\frac{\ell+2\ii\wt{\nu}_A+p_2+1}2-n\big)_{j+1}\big(\frac{\ell-2\ii\wt{\nu}_A+p_2+4}2+n\big)_{j+1}}\n \\
    &\times\Gamma\Bigg[\begin{matrix}
        \frac32+n, -\ii\wt{\nu}_A+\frac32+n, -2\ii\wt{\nu}_A+\frac32+n, -\ii\wt{\nu}_A+\frac12+n \\
        -\ii\wt{\nu}_A+2+n, -2\ii\wt{\nu}_A+1+n, -\ii\wt{\nu}_A+1+n
    \end{matrix}\Bigg]\n \\
    &\times r_1^{2j+p_{12}+5}\bigg(\frac{r_1}{r_2}\bigg)^\ell\frac{f_{AA}(3\ii/2+2\wt{\nu}_A+2\ii n)\big|_{d=3}}{\big(\ii\wt{\nu}_A+\frac12\big)^2\big(-\ii\wt{\nu}_A+\frac12\big)^2}\n \\
    &-\sum_{\ell=0}^\infty\frac{(-1)^\ell\mathcal{C}_{\frac32-2\ii\wt{\nu}_A,3}^{p_1p_2}\cos^2(2\pi\ii\wt{\nu}_A)(\ell+1)_{p_{12}+4}}{16\pi^2\sin^2(\pi\ii\wt{\nu}_A)\big(\ii\wt{\nu}_A+\frac12\big)^2\big(-\ii\wt{\nu}_A+\frac12\big)^2}r_1^{p_{12}+5}\bigg(\frac{r_1}{r_2}\bigg)^\ell\n \\
    &\times\Bigg[\sum_{n=0}^\infty\Bigg(\frac{\frac34-\ii\wt{\nu}_A+n}{n!\big(\frac{\ell+2\ii\wt{\nu}_A+p_2+1}2-n\big)\big(\frac{\ell-2\ii\wt{\nu}_A+p_2+4}2+n\big)}\n \\
    &\times\Gamma\Bigg[\begin{matrix}
        \frac32+n, -\ii\wt{\nu}_A+\frac32+n, -2\ii\wt{\nu}_A+\frac32+n, -\ii\wt{\nu}_A+\frac12+n \\
        -\ii\wt{\nu}_A+2+n, -2\ii\wt{\nu}_A+1+n, -\ii\wt{\nu}_A+1+n
    \end{matrix}\Bigg]\frac1{16}f_{AA}(3\ii/2+2\wt{\nu}_A+2\ii n)\big|_{d=3}\n \\
    &+\Big[16(n+1)^3+12(-1-4\ii\wt{\nu}_A)(n+1)^2+\big[4\big(p_2+\ell+\tfrac52\big)^2-25-24\wt{\nu}_A^2+24\ii\wt{\nu}_A\big](n+1)\n \\
    &+(-1-4\ii\wt{\nu}_A)\big[24\big(p_2+\ell+\tfrac52\big)^2-27+8\wt{\nu}_A^2+8\ii\wt{\nu}_A\big]/4\n \\
    &+\Big(8\big[\big(p_2+\ell+\tfrac52\big)^4-\tfrac92\big(p_2+\ell+\tfrac52\big)^2+\tfrac{81}{16}\big]\n \\
    &+(-20+48\wt{\nu}_A^2)\big[\big(p_2+\ell+\tfrac52\big)^2-\tfrac94\big]+33+168\wt{\nu}_A^2+144\wt{\nu}_A^4\Big)/(8(n+1))\Big]\Bigg)\n \\
    &+\big[20\big(p_2+\ell+\tfrac52\big)^2-133-120\wt{\nu}_A^2+120\ii\wt{\nu}_A\big]/60\n \\
    &+(-1-4\ii\wt{\nu}_A)\big[24\big(p_2+\ell+\tfrac52\big)^2-27+8\wt{\nu}_A^2+8\ii\wt{\nu}_A\big]/8\n \\
    &-\Big(8\big[\big(p_2+\ell+\tfrac52\big)^4-\tfrac92\big(p_2+\ell+\tfrac52\big)^2+\tfrac{81}{16}\big]\n \\
    &+(-20+48\wt{\nu}_A^2)\big[\big(p_2+\ell+\tfrac52\big)^2-\tfrac94\big]+33+168\wt{\nu}_A^2+144\wt{\nu}_A^4\Big)\n \\
    &\times\Bigg(\gamma_E+\log\mu_R+\tfrac12\log4\pi+\psi\big(\tfrac32\big)-\frac{\pi\sin\big[\pi\big(-2\ii\wt{\nu}_A+\tfrac{p_{12}}2\big)\big]}{16\cos^2(2\pi\ii\wt{\nu}_A)\mathcal{C}_{\frac32-2\ii\wt{\nu}_A,3}^{p_1p_2}}\Bigg)/8\n \\
    &-\frac{9+40\wt{\nu}_A^2+16\wt{\nu}_A^4}{32}\big[-5+6\gamma_E-12\log\mu_R-6\log(4\pi)+6\psi\big(\ii\wt{\nu}_A+\tfrac52\big)+6\psi\big(-\ii\wt{\nu}_A+\tfrac52\big)\big]\n \\
    &+\Big((-190+30\ii\wt{\nu}_A)\big[\big(p_2+\ell+\tfrac52\big)^2-\tfrac94\big]+119+210\wt{\nu}_A^2+120\wt{\nu}_A^4\Big)/30\Bigg]\n \\
    &-\sum_{\ell=0}^\infty\frac{(-1)^\ell\mathcal{C}_{\frac32-2\ii\wt{\nu}_A,3}^{p_1p_2}\cos^2(2\pi\ii\wt{\nu}_A)(\ell+1)_{p_{12}+6}}{64\pi^2\sin^2(\pi\ii\wt{\nu}_A)\big(\ii\wt{\nu}_A+\frac12\big)^2\big(-\ii\wt{\nu}_A+\frac12\big)^2}r_1^{p_{12}+7}\bigg(\frac{r_1}{r_2}\bigg)^\ell\n \\
    &\times\Bigg[\sum_{n=0}^\infty\Bigg(\frac{\frac34-\ii\wt{\nu}_A+n}{n!\big(\frac{\ell+2\ii\wt{\nu}_A+p_2+1}2-n\big)_2\big(\frac{\ell-2\ii\wt{\nu}_A+p_2+4}2+n\big)_2}\n \\
    &\times\Gamma\Bigg[\begin{matrix}
        \frac32+n, -\ii\wt{\nu}_A+\frac32+n, -2\ii\wt{\nu}_A+\frac32+n, -\ii\wt{\nu}_A+\frac12+n \\
        -\ii\wt{\nu}_A+2+n, -2\ii\wt{\nu}_A+1+n, -\ii\wt{\nu}_A+1+n
    \end{matrix}\Bigg]\frac1{16}f_{AA}(3\ii/2+2\wt{\nu}_A+2\ii n)\big|_{d=3}\n \\
    &-\Big[16(n+1)+4(-1-4\ii\wt{\nu}_A)+\Big(8\big[\big(p_2+\ell+\tfrac72\big)^2-\tfrac54\big]-10+24\wt{\nu}_A^2\Big)/((n+1))\Big]\Bigg)\n \\
    &+\tfrac{2+24\ii\wt{\nu}_A}3+\Big(8\big[\big(p_2+\ell+\tfrac72\big)^2-\tfrac54\big]-10+24\wt{\nu}_A^2\Big)\n \\
    &\times\Bigg(\gamma_E+\log\mu_R+\tfrac12\log4\pi+\psi\big(\tfrac32\big)-\frac{\pi\sin\big[\pi\big(-2\ii\wt{\nu}_A+\tfrac{p_{12}}2\big)\big]}{16\cos^2(2\pi\ii\wt{\nu}_A)\mathcal{C}_{\frac32-2\ii\wt{\nu}_A,3}^{p_1p_2}}\Bigg)+(76-12\ii\wt{\nu}_A)/3\Bigg]\n \\
    &-\sum_{\ell=0}^\infty\frac{(-1)^\ell\mathcal{C}_{\frac32-2\ii\wt{\nu}_A,d}^{p_1p_2}\cos^2(2\pi\ii\wt{\nu}_A)(\ell+1)_{p_{12}+8}}{256\pi^2\sin^2(\pi\ii\wt{\nu}_A)\big(\ii\wt{\nu}_A+\frac12\big)^2\big(-\ii\wt{\nu}_A+\frac12\big)^2}r_1^{d+p_{12}+6}\bigg(\frac{r_1}{r_2}\bigg)^\ell\n \\
    &\times\Bigg[\sum_{n=0}^\infty\Bigg(\frac{\frac34-\ii\wt{\nu}_A+n}{n!\big(\frac{\ell+2\ii\wt{\nu}_A+p_2+1}2-n\big)_3\big(\frac{\ell-2\ii\wt{\nu}_A+p_2+4}2+n\big)_3}\n \\
    &\times\Gamma\Bigg[\begin{matrix}
        \frac32+n, -\ii\wt{\nu}_A+\frac32+n, -2\ii\wt{\nu}_A+\frac32+n, -\ii\wt{\nu}_A+\frac12+n \\
        -\ii\wt{\nu}_A+2+n, -2\ii\wt{\nu}_A+1+n, -\ii\wt{\nu}_A+1+n
    \end{matrix}\Bigg]\frac1{16}f_{AA}(3\ii/2+2\wt{\nu}_A+2\ii n)\big|_{d=3}\n \\
    &+\frac{16}{n+1}\Bigg)-16\Bigg(\gamma_E+\log\mu_R+\tfrac12\log4\pi+\psi\big(\tfrac32\big)-\frac{\pi\sin\big[\pi\big(-2\ii\wt{\nu}_A+\tfrac{p_{12}}2\big)\big]}{16\cos^2(2\pi\ii\wt{\nu}_A)\mathcal{C}_{\frac32-2\ii\wt{\nu}_A,3}^{p_1p_2}}\Bigg)\Bigg],
\end{align}
\begin{align}
    \wh{\mathcal{J}}_{AA,(\text{3B})}^{p_1p_2}=&\sum_{n,\ell=0}^\infty\sum_{j=3}^\infty\frac{(-1)^\ell\big(\frac32+2n\big)\mathcal{C}_{\frac32+2n,3}^{p_1p_2}(\ell+1)_{2j+p_{12}+4}}{2^{2j+8}\pi^2\sin^2(\pi\ii\wt{\nu}_A)n!\big(\frac{\ell+p_2+1}2-n\big)_{j+1}\big(\frac{\ell+p_2+4}2+n\big)_{j+1}}\n \\
    &\times\Gamma\Bigg[\begin{matrix}
        \ii\wt{\nu}_A+\frac32+n, \frac32+n, -\ii\wt{\nu}_A+\frac32+n, \frac12+n \\
        2+n, -\ii\wt{\nu}_A+1+n, \ii\wt{\nu}_A+1+n
    \end{matrix}\Bigg]\n \\
    &\times r_1^{2j+p_{12}+5}\bigg(\frac{r_1}{r_2}\bigg)^\ell\frac{f_{AA}(3\ii/2+2\wt{\nu}_A+2\ii n)\big|_{d=3}}{\big(\ii\wt{\nu}_A+\frac12\big)^2\big(-\ii\wt{\nu}_A+\frac12\big)^2}\n \\
    &+\sum_{\ell=0}^\infty\frac{(-1)^\ell\mathcal{C}_{\frac32,3}^{p_1p_2}(\ell+1)_{p_{12}+4}}{8\pi^2\sin^2(\pi\ii\wt{\nu}_A)}r_1^{p_{12}+5}\bigg(\frac{r_1}{r_2}\bigg)^\ell\n \\
    &\times\Bigg[\sum_{n=0}^\infty\Bigg(\frac{\frac34+n}{n!\big(\frac{\ell+p_2+1}2-n\big)\big(\frac{\ell+p_2+4}2+n\big)}\n \\
    &\times\Gamma\Bigg[\begin{matrix}
        \ii\wt{\nu}_A+\frac32+n, \frac32+n, -\ii\wt{\nu}_A+\frac32+n, \frac12+n \\
        2+n, -\ii\wt{\nu}_A+1+n, \ii\wt{\nu}_A+1+n
    \end{matrix}\Bigg]\frac1{16}f_{AA}(3\ii/2+2\ii n)\big|_{d=3}\n \\
    &+\Big[16(n+1)^3-12(n+1)^2+\big[4\big(p_2+\ell+\tfrac52\big)^2-25-24\wt{\nu}_A^2\big](n+1)\n \\
    &-\big[4\big(p_2+\ell+\tfrac52\big)^2-27+24\wt{\nu}_A^2\big]/4\n \\
    &+\Big(8\big[\big(p_2+\ell+\tfrac52\big)^4-\tfrac92\big(p_2+\ell+\tfrac52\big)^2+\tfrac{81}{16}\big]\n \\
    &+(-20+48\wt{\nu}_A^2)\big[\big(p_2+\ell+\tfrac52\big)^2-\tfrac94\big]+33+168\wt{\nu}_A^2+144\wt{\nu}_A^4\Big)/(8(n+1))\Big]\Bigg)\n \\
    &-\tfrac2{15}+\big[4\big(p_2+\ell+\tfrac52\big)^2-25-24\wt{\nu}_A^2\big]/12\n \\
    &-\big[320\big(p_2+\ell+\tfrac52\big)^2-139+600\wt{\nu}_A^2\big]/120\n \\
    &-\Big(8\big[\big(p_2+\ell+\tfrac52\big)^4-\tfrac92\big(p_2+\ell+\tfrac52\big)^2+\tfrac{81}{16}\big]\n \\
    &+(-20+48\wt{\nu}_A^2)\big[\big(p_2+\ell+\tfrac52\big)^2-\tfrac94\big]+33+168\wt{\nu}_A^2+144\wt{\nu}_A^4\Big)\n \\
    &\times\Bigg(\gamma_E+\log\mu_R+\tfrac12\log4\pi+\psi\big(\tfrac32\big)-\frac{\pi\sin\big(\tfrac{\pi p_{12}}2\big)}{16\mathcal{C}_{\frac32,3}^{p_1p_2}}\Bigg)/8\n \\
    &-\frac{9+40\wt{\nu}_A^2+16\wt{\nu}_A^4}{32}\big[-5+6\gamma_E-12\log\mu_R-6\log(4\pi)+6\psi\big(\ii\wt{\nu}_A+\tfrac52\big)+6\psi\big(-\ii\wt{\nu}_A+\tfrac52\big)\big]\n \\
    &+\Big((-190+60\wt{\nu}_A^2)\big[\big(p_2+\ell+\tfrac52\big)^2-\tfrac94\big]+119-30\wt{\nu}_A^2+360\wt{\nu}_A^4\Big)/30\Bigg]\n \\
    &+\sum_{\ell=0}^\infty\frac{(-1)^\ell\mathcal{C}_{\frac32,3}^{p_1p_2}(\ell+1)_{p_{12}+6}}{32\pi^2\sin^2(\pi\ii\wt{\nu}_A)}r_1^{p_{12}+7}\bigg(\frac{r_1}{r_2}\bigg)^\ell\n \\
    &\times\Bigg[\sum_{n=0}^\infty\Bigg(\frac{\frac34+n}{n!\big(\frac{\ell+p_2+1}2-n\big)_2\big(\frac{\ell+p_2+4}2+n\big)_2}\n \\
    &\times\Gamma\Bigg[\begin{matrix}
        \ii\wt{\nu}_A+\frac32+n, \frac32+n, -\ii\wt{\nu}_A+\frac32+n, \frac12+n \\
        2+n, -\ii\wt{\nu}_A+1+n, \ii\wt{\nu}_A+1+n
    \end{matrix}\Bigg]\frac1{16}f_{AA}(3\ii/2+2\ii n)\big|_{d=3}\n \\
    &-\Big[16(n+1)-4+\Big(8\big[\big(p_2+\ell+\tfrac72\big)^2-\tfrac54\big]-10+24\wt{\nu}_A^2\Big)/((n+1))\Big]\Bigg)\n \\
    &+\tfrac23+\Big(8\big[\big(p_2+\ell+\tfrac72\big)^2-\tfrac54\big]-10+24\wt{\nu}_A^2\Big)\n \\
    &\times\Bigg(\gamma_E+\log\mu_R+\tfrac12\log4\pi+\psi\big(\tfrac32\big)-\frac{\pi\sin\big(\tfrac{\pi p_{12}}2\big)}{16\mathcal{C}_{\frac32,3}^{p_1p_2}}\Bigg)+(76-24\wt{\nu}_A^2)/3\Bigg]\n \\
    &+\sum_{\ell=0}^\infty\frac{(-1)^\ell\mathcal{C}_{\frac32,3}^{p_1p_2}(\ell+1)_{p_{12}+8}}{128\pi^2\sin^2(\pi\ii\wt{\nu}_A)}r_1^{p_{12}+9}\bigg(\frac{r_1}{r_2}\bigg)^\ell\n \\
    &\times\Bigg[\sum_{n=0}^\infty\Bigg(\frac{\frac34+n}{n!\big(\frac{\ell+p_2+1}2-n\big)_3\big(\frac{\ell+p_2+4}2+n\big)_3}\n \\
    &\times\Gamma\Bigg[\begin{matrix}
        \ii\wt{\nu}_A+\frac32+n, \frac32+n, -\ii\wt{\nu}_A+\frac32+n, \frac12+n \\
        2+n, -\ii\wt{\nu}_A+1+n, \ii\wt{\nu}_A+1+n
    \end{matrix}\Bigg]\frac1{16}f_{AA}(3\ii/2+2\ii n)\big|_{d=3}\n \\
    &+\frac{16}{n+1}\Bigg)-16\Bigg(\gamma_E+\log\mu_R+\tfrac12\log4\pi+\psi\big(\tfrac32\big)-\frac{\pi\sin\big(\tfrac{\pi p_{12}}2\big)}{16\mathcal{C}_{\frac32,3}^{p_1p_2}}\Bigg)\Bigg],
\end{align}
\begin{align}
    \wh{\mathcal{J}}_{AA,(\text{3C})}^{p_1p_2}=&-\sum_{n,\ell=0}^\infty\sum_{j=3}^\infty\frac{(-1)^\ell\big(\frac32+2\ii\wt{\nu}_A+2n\big)\mathcal{C}_{\frac32+2\ii\wt{\nu}_A+2n,3}^{p_1p_2}\cos^2(2\ii\wt{\nu}_A)(\ell+1)_{2j+p_{12}+4}}{2^{2j+9}\pi^2\sin^2(\pi\ii\wt{\nu}_A)n!\big(\frac{\ell-2\ii\wt{\nu}_A+p_2+1}2-n\big)_{j+1}\big(\frac{\ell+2\ii\wt{\nu}_A+p_2+4}2+n\big)_{j+1}}\n \\
    &\times\Gamma\Bigg[\begin{matrix}
        2\ii\wt{\nu}_A+\frac32+n, \ii\wt{\nu}_A+\frac32+n, \frac32+n, \ii\wt{\nu}_A+\frac12+n \\
        \ii\wt{\nu}_A+2+n, \ii\wt{\nu}_A+1+n, 2\ii\wt{\nu}_A+1+n
    \end{matrix}\Bigg]\n \\
    &\times r_1^{2j+p_{12}+5}\bigg(\frac{r_1}{r_2}\bigg)^\ell\frac{f_{AA}(3\ii/2-2\wt{\nu}_A+2\ii n)\big|_{d=3}}{\big(\ii\wt{\nu}_A+\frac12\big)^2\big(-\ii\wt{\nu}_A+\frac12\big)^2}\n \\
    &-\sum_{\ell=0}^\infty\frac{(-1)^\ell\mathcal{C}_{\frac32+2\ii\wt{\nu}_A,3}^{p_1p_2}\cos^2(2\pi\ii\wt{\nu}_A)(\ell+1)_{p_{12}+4}}{16\pi^2\sin^2(\pi\ii\wt{\nu}_A)\big(\ii\wt{\nu}_A+\frac12\big)^2\big(-\ii\wt{\nu}_A+\frac12\big)^2}r_1^{p_{12}+5}\bigg(\frac{r_1}{r_2}\bigg)^\ell\n \\
    &\times\Bigg[\sum_{n=0}^\infty\Bigg(\frac{\frac34+\ii\wt{\nu}_A+n}{n!\big(\frac{\ell-2\ii\wt{\nu}_A+p_2+1}2-n\big)\big(\frac{\ell+2\ii\wt{\nu}_A+p_2+4}2+n\big)}\n \\
    &\times\Gamma\Bigg[\begin{matrix}
        2\ii\wt{\nu}_A+\frac32+n, \ii\wt{\nu}_A+\frac32+n, \frac32+n, \ii\wt{\nu}_A+\frac12+n \\
        \ii\wt{\nu}_A+2+n, \ii\wt{\nu}_A+1+n, 2\ii\wt{\nu}_A+1+n
    \end{matrix}\Bigg]\frac1{16}f_{AA}(3\ii/2-2\wt{\nu}_A+2\ii n)\big|_{d=3}\n \\
    &+\Big[16(n+1)^3+12(-1+4\ii\wt{\nu}_A)(n+1)^2+\big[4\big(p_2+\ell+\tfrac52\big)^2-25-24\wt{\nu}_A^2-24\ii\wt{\nu}_A\big](n+1)\n \\
    &+(-1+4\ii\wt{\nu}_A)\big[24\big(p_2+\ell+\tfrac52\big)^2-27+8\wt{\nu}_A^2-8\ii\wt{\nu}_A\big]/4\n \\
    &+\Big(8\big[\big(p_2+\ell+\tfrac52\big)^4-\tfrac92\big(p_2+\ell+\tfrac52\big)^2+\tfrac{81}{16}\big]\n \\
    &+(-20+48\wt{\nu}_A^2)\big[\big(p_2+\ell+\tfrac52\big)^2-\tfrac94\big]+33+168\wt{\nu}_A^2+144\wt{\nu}_A^4\Big)/(8(n+1))\Big]\Bigg)\n \\
    &+\big[20\big(p_2+\ell+\tfrac52\big)^2-133-120\wt{\nu}_A^2-120\ii\wt{\nu}_A\big]/60\n \\
    &+(-1+4\ii\wt{\nu}_A)\big[24\big(p_2+\ell+\tfrac52\big)^2-27+8\wt{\nu}_A^2-8\ii\wt{\nu}_A\big]/8\n \\
    &-\Big(8\big[\big(p_2+\ell+\tfrac52\big)^4-\tfrac92\big(p_2+\ell+\tfrac52\big)^2+\tfrac{81}{16}\big]\n \\
    &+(-20+48\wt{\nu}_A^2)\big[\big(p_2+\ell+\tfrac52\big)^2-\tfrac94\big]+33+168\wt{\nu}_A^2+144\wt{\nu}_A^4\Big)\n \\
    &\times\Bigg(\gamma_E+\log\mu_R+\tfrac12\log4\pi+\psi\big(\tfrac32\big)-\frac{\pi\sin\big[\pi\big(2\ii\wt{\nu}_A+\tfrac{p_{12}}2\big)\big]}{16\cos^2(2\pi\ii\wt{\nu}_A)\mathcal{C}_{\frac32+2\ii\wt{\nu}_A,3}^{p_1p_2}}\Bigg)/8\n \\
    &-\frac{9+40\wt{\nu}_A^2+16\wt{\nu}_A^4}{32}\big[-5+6\gamma_E-12\log\mu_R-6\log(4\pi)+6\psi\big(\ii\wt{\nu}_A+\tfrac52\big)+6\psi\big(-\ii\wt{\nu}_A+\tfrac52\big)\big]\n \\
    &+\Big((-190-30\ii\wt{\nu}_A)\big[\big(p_2+\ell+\tfrac52\big)^2-\tfrac94\big]+119+210\wt{\nu}_A^2+120\wt{\nu}_A^4\Big)/30\Bigg]\n \\
    &-\sum_{\ell=0}^\infty\frac{(-1)^\ell\mathcal{C}_{\frac32+2\ii\wt{\nu}_A,3}^{p_1p_2}\cos^2(2\pi\ii\wt{\nu}_A)(\ell+1)_{p_{12}+6}}{64\pi^2\sin^2(\pi\ii\wt{\nu}_A)\big(\ii\wt{\nu}_A+\frac12\big)^2\big(-\ii\wt{\nu}_A+\frac12\big)^2}r_1^{p_{12}+7}\bigg(\frac{r_1}{r_2}\bigg)^\ell\n \\
    &\times\Bigg[\sum_{n=0}^\infty\Bigg(\frac{\frac34+\ii\wt{\nu}_A+n}{n!\big(\frac{\ell-2\ii\wt{\nu}_A+p_2+1}2-n\big)_2\big(\frac{\ell+2\ii\wt{\nu}_A+p_2+4}2+n\big)_2}\n \\
    &\times\Gamma\Bigg[\begin{matrix}
        2\ii\wt{\nu}_A+\frac32+n, \ii\wt{\nu}_A+\frac32+n, \frac32+n, \ii\wt{\nu}_A+\frac12+n \\
        \ii\wt{\nu}_A+2+n, \ii\wt{\nu}_A+1+n, 2\ii\wt{\nu}_A+1+n
    \end{matrix}\Bigg]\frac1{16}f_{AA}(3\ii/2-2\wt{\nu}_A+2\ii n)\big|_{d=3}\n \\
    &-\Big[16(n+1)+4(-1+4\ii\wt{\nu}_A)+\Big(8\big[\big(p_2+\ell+\tfrac72\big)^2-\tfrac54\big]-10+24\wt{\nu}_A^2\Big)/((n+1))\Big]\Bigg)\n \\
    &+\tfrac{2-24\ii\wt{\nu}_A}3+\Big(8\big[\big(p_2+\ell+\tfrac72\big)^2-\tfrac54\big]-10+24\wt{\nu}_A^2\Big)\n \\
    &\times\Bigg(\gamma_E+\log\mu_R+\tfrac12\log4\pi+\psi\big(\tfrac32\big)-\frac{\pi\sin\big[\pi\big(2\ii\wt{\nu}_A+\tfrac{p_{12}}2\big)\big]}{16\cos^2(2\pi\ii\wt{\nu}_A)\mathcal{C}_{\frac32+2\ii\wt{\nu}_A,3}^{p_1p_2}}\Bigg)+(76+12\ii\wt{\nu}_A)/3\Bigg]\n \\
    &-\sum_{\ell=0}^\infty\frac{(-1)^\ell\mathcal{C}_{\frac32+2\ii\wt{\nu}_A,d}^{p_1p_2}\cos^2(2\pi\ii\wt{\nu}_A)(\ell+1)_{p_{12}+8}}{256\pi^2\sin^2(\pi\ii\wt{\nu}_A)\big(\ii\wt{\nu}_A+\frac12\big)^2\big(-\ii\wt{\nu}_A+\frac12\big)^2}r_1^{d+p_{12}+6}\bigg(\frac{r_1}{r_2}\bigg)^\ell\n \\
    &\times\Bigg[\sum_{n=0}^\infty\Bigg(\frac{\frac34+\ii\wt{\nu}_A+n}{n!\big(\frac{\ell-2\ii\wt{\nu}_A+p_2+1}2-n\big)_3\big(\frac{\ell+2\ii\wt{\nu}_A+p_2+4}2+n\big)_3}\n \\
    &\times\Gamma\Bigg[\begin{matrix}
        2\ii\wt{\nu}_A+\frac32+n, \ii\wt{\nu}_A+\frac32+n, \frac32+n, \ii\wt{\nu}_A+\frac12+n \\
        \ii\wt{\nu}_A+2+n, \ii\wt{\nu}_A+1+n, 2\ii\wt{\nu}_A+1+n
    \end{matrix}\Bigg]\frac1{16}f_{AA}(3\ii/2-2\wt{\nu}_A+2\ii n)\big|_{d=3}\n \\
    &+\frac{16}{n+1}\Bigg)-16\Bigg(\gamma_E+\log\mu_R+\tfrac12\log4\pi+\psi\big(\tfrac32\big)-\frac{\pi\sin\big[\pi\big(2\ii\wt{\nu}_A+\tfrac{p_{12}}2\big)\big]}{16\cos^2(2\pi\ii\wt{\nu}_A)\mathcal{C}_{\frac32+2\ii\wt{\nu}_A,3}^{p_1p_2}}\Bigg)\Bigg].
\end{align}
\section{More on Spectral Functions}\label{AppG}
In this appendix, the relation between K\"all\'en-Lehmann spectral density \cite{Loparco:2023rug} and spectral function in \cite{Marolf:2010zp, Xianyu:2022jwk} is studied.

The spectral function in \cite{Marolf:2010zp} is derived from $\rho_{\sigma_1\sigma_2}(L)$ in \cite{Marolf:2010zp} by setting $\sigma_1=-d/2+\ii\wt{\nu}_1, \sigma_2=-d/2+\ii\wt{\nu}_2, L=-d/2+\ii\wt{\nu}'$. The expression is
\begin{align}
    \rho_{\wt{\nu}_1,\wt{\nu}_2}^\text{dS}(\wt{\nu}')=&~\Bigg\{\frac1{2(4\pi)^{(d+1)/2}}\frac{\cos[\pi(\frac d2-\ii\wt{\nu}_1)]}{\sin(-\pi \ii \wt{\nu}_1)}\Gamma\left[\begin{matrix}
        \frac{3-d}{2}, \frac{d}{2}-\ii\wt\nu_1 \\
         \frac{2-d}{2}-\ii\wt\nu_1
    \end{matrix}\right]\n \\
    &~\times {}_7\mathcal{F}_6\left[\begin{matrix}
        \fr{2-d}2+\ii \wt{\nu}'-\ii \wt{\nu}_1, \tfrac{3-d/2+\ii \wt{\nu}'-\ii \wt{\nu}_1}2,  \fr{2-d}2, \fr{2-d}2-\ii \wt{\nu}_1, \fr{2-d}2+\ii \wt{\nu}', \tfrac{\ii \wt{\nu}'-\ii \wt{\nu}_1-\ii \wt{\nu}_2+d/2}2, \tfrac{\ii \wt{\nu}'-\ii \wt{\nu}_1+\ii \wt{\nu}_2+d/2}2 \\
        \tfrac{1-d/2+\ii \wt{\nu}'-\ii \wt{\nu}_1}2, 1+\ii \wt{\nu}'-\ii \wt{\nu}_1, 1+\ii \wt{\nu}', 1-\ii \wt{\nu}_1, \tfrac{4+\ii \wt{\nu}'-\ii \wt{\nu}_1+\ii \wt{\nu}_2-3d/2}2, \tfrac{4+\ii \wt{\nu}'-\ii \wt{\nu}_1-\ii \wt{\nu}_2-3d/2}2
    \end{matrix}\middle|1\right]\n\\
    &~+(\wt\nu_1\to-\wt\nu_1)\Bigg\}+(\wt{\nu}_1\leftrightarrow\wt{\nu}_2),
\end{align}
The following formula is used:
\begin{align}
    {}_7\mathrm{F}_6\left[\begin{matrix}
        \mathbf{A} \\
        \mathbf{B}
    \end{matrix}\middle| 1\right]=\Gamma\bigg[\begin{matrix}
        \mathbf{C} \\
        \mathbf{D}
    \end{matrix}\bigg]-\Gamma\bigg[\begin{matrix}
        \mathbf{E} \\
        \mathbf{F}
    \end{matrix}\bigg]{}_7\mathrm{F}_6\left[\begin{matrix}
        \mathbf{G} \\
        \mathbf{H}
    \end{matrix}\middle| 1\right],
\end{align}
where {\allowdisplaybreaks
\begin{align}
    \mathbf{A}=&~\{a, 1+\tfrac a2, b, c, e, f, g\},\n \\
    \mathbf{B}=&~\{\tfrac a2, 1+a-b, 1+a-c, 1+a-e, 1+a-f, 1+a-g\},\n \\
    \mathbf{C}=&~\{1+a-c, 1+a-e, 1+a-f, 1+a-g, b+c-a, b+e-a, b+f-a, b+g-a\},\n \\
    \mathbf{D}=&~\{1+a, b-a, 1+a-e-f, 1+a-c-f, 1+a-c-e, \n\\
    &~~\,1+a-c-g, 1+a-e-g, 1+a-f-g\},\n \\
    \mathbf{E}=&~\{1+2b-a, b+c-a, b+e-a, b+f-a, b+g-a, a-b, \n\\
    &~~\,1+a-c, 1+a-e, 1+a-f, 1+a-g\},\n \\
    \mathbf{F}=&~\{1+b-c, 1+b-e, 1+b-f, 1+b-g, b-a, 1+a, c, e, f, g\},\n \\
    \mathbf{G}=&~\{2b-a, 1+b-\tfrac a2, b, b+c-a, b+e-a, b+f-a, b+g-a\},\n \\
    \mathbf{H}=&~\{b-\tfrac a2, 1+b-a, 1+b-c, 1+b-e, 1+b-f ,1+b-g\}.
\end{align} }
The parameters in the above identity are set as:
\begin{align}
    &a=1-\frac d2+\ii\wt{\nu}'-\ii\wt{\nu}_1, &&b=\frac{d/2+\ii\wt{\nu}'-\ii\wt{\nu}_1-\ii\wt{\nu}_2}2, &&c=1-\frac d2-\ii\wt{\nu}_1, \\
    &e=\frac{d/2+\ii\wt{\nu}'-\ii\wt{\nu}_1+\ii\wt{\nu}_2}2, &&f=1-\frac d2, &&g=1-\frac d2+\ii\wt{\nu}'.
\end{align}
Then $\rho_{\wt{\nu}_1,\wt{\nu}_2}^\text{dS}(\wt{\nu}')$ can be written as
\begin{align}
    &\rho_{\wt{\nu}_1,\wt{\nu}_2}^\text{dS}(\wt{\nu}')\n \\
    =&\Bigg\{\Bigg[\frac1{16\pi^{d/2+3}}\frac{\cos\big[\pi\big(\frac d2-\ii\wt{\nu}_1\big)\big]\sin\big[\frac\pi2\big(-\frac{3d}2+\ii\wt{\nu}'-\ii\wt{\nu}_1+\ii\wt{\nu}_2\big)\big]\sin\big[\frac\pi2\big(\frac d2-\ii\wt{\nu}'+\ii\wt{\nu}_1+\ii\wt{\nu}_2)\big]}{\sin(-\pi\ii\wt{\nu}_1)\sin\big[\pi\big(\frac d2-\ii\wt{\nu}'\big)\big]}\n \\
    &\times\sin\big[\tfrac\pi2\big(\tfrac d2-\ii\wt{\nu}'-\ii\wt{\nu}_1+\ii\wt{\nu}_2)\big]\sin\big[\tfrac\pi2\big(\tfrac d2+\ii\wt{\nu}'+\ii\wt{\nu}_1+\ii\wt{\nu}_2)\big]\n \\
    &\times\Gamma\Bigg[\begin{matrix}
        2-d, -1+d \\
        \frac d2, \frac d2+\ii\wt{\nu}', \frac d2-\ii\wt{\nu}'
    \end{matrix}\Bigg]\prod_{\pm,\pm,\pm}\Gamma\bigg(\frac{\frac d2\pm\ii\wt{\nu}\pm\ii\wt{\nu}\pm\ii\wt{\nu}'}2\bigg)\n \\
    &+\frac1{2(4\pi)^{(d+1)/2}}\frac{\cos[\pi(\frac d2-\ii\wt{\nu}_1)]}{\sin(-\pi \ii \wt{\nu}_1)}\Gamma\left[\begin{matrix}
        \frac{3-d}{2}, \frac{d}{2}-\ii\wt\nu_1 \\
         \frac{2-d}{2}-\ii\wt\nu_1
    \end{matrix}\right]\n \\
    &\times{}_7\mathcal{F}_6\left[\begin{matrix}
        -1+d-\ii\wt{\nu}_2, \frac{1+d-\ii\wt{\nu}_2}2, \frac{d/2+\ii\wt{\nu}'-\ii\wt{\nu}_1-\ii\wt{\nu}_2}2, \frac{d/2-\ii\wt{\nu}'-\ii\wt{\nu}_1-\ii\wt{\nu}_2}2, -1+d, \frac{d/2-\ii\wt{\nu}'+\ii\wt{\nu}_1-\ii\wt{\nu}_2}2, \frac{d/2+\ii\wt{\nu}'+\ii\wt{\nu}_1-\ii\wt{\nu}_2}2 \\
        \frac{-1+d-\ii\wt{\nu}_2}2, \frac{3d/2-\ii\wt{\nu}'+\ii\wt{\nu}_1-\ii\wt{\nu}_2}2, \frac{3d/2+\ii\wt{\nu}'+\ii\wt{\nu}_1-\ii\wt{\nu}_2}2, 1-\ii\wt{\nu}_2, \frac{3d/2+\ii\wt{\nu}'-\ii\wt{\nu}_1-\ii\wt{\nu}_2}2, \frac{3d/2-\ii\wt{\nu}'-\ii\wt{\nu}_1-\ii\wt{\nu}_2}2
    \end{matrix}\middle|1\right]\Bigg]\n \\
    &+(\wt\nu_1\to-\wt\nu_1)\Bigg\}+(\wt{\nu}_1\leftrightarrow\wt{\nu}_2).
\end{align}
The difference between $\rho_{\wt{\nu}_1,\wt{\nu}_2}^\text{dS}(\wt{\nu}')$ and $\rho_{\wt{\nu}_1,\wt{\nu}_2}^\text{dS}(-\wt{\nu}')$ is
\begin{align}
    &\rho_{\wt{\nu}_1,\wt{\nu}_2}^\text{dS}(\wt{\nu}')-\rho_{\wt{\nu}_1,\wt{\nu}_2}^\text{dS}(-\wt{\nu}')\n \\
    =&\Bigg\{\frac1{16\pi^{d/2+3}}\Bigg[\frac{\cos\big[\pi\big(\frac d2-\ii\wt{\nu}_1\big)\big]\sin\big[\frac\pi2\big(-\frac{3d}2+\ii\wt{\nu}'-\ii\wt{\nu}_1+\ii\wt{\nu}_2\big)\big]\sin\big[\frac\pi2\big(\frac d2-\ii\wt{\nu}'+\ii\wt{\nu}_1+\ii\wt{\nu}_2)\big]}{\sin(-\pi\ii\wt{\nu}_1)\sin\big[\pi\big(\frac d2-\ii\wt{\nu}'\big)\big]}\n \\
    &\times\sin\big[\tfrac\pi2\big(\tfrac d2-\ii\wt{\nu}'-\ii\wt{\nu}_1+\ii\wt{\nu}_2)\big]\sin\big[\tfrac\pi2\big(\tfrac d2+\ii\wt{\nu}'+\ii\wt{\nu}_1+\ii\wt{\nu}_2)\big]-(\wt{\nu}'\rightarrow-\wt{\nu'})\Bigg]\n \\
    &\times\Gamma\Bigg[\begin{matrix}
        2-d, -1+d \\
        \frac d2, \frac d2+\ii\wt{\nu}', \frac d2-\ii\wt{\nu}'
    \end{matrix}\Bigg]\prod_{\pm,\pm,\pm}\Gamma\bigg(\frac{\frac d2\pm\ii\wt{\nu}_1\pm\ii\wt{\nu}_2\pm\ii\wt{\nu}'}2\bigg)+(\wt\nu_1\to-\wt\nu_1)\Bigg\}+(\wt{\nu}_1\leftrightarrow\wt{\nu}_2)\n \\
    =&\Bigg\{\frac1{16\pi^{d/2+3}}\Bigg[\Bigg(\frac{\cos\big[\pi\big(\frac d2-\ii\wt{\nu}_1\big)\big]\sin\big[\frac\pi2\big(-\frac{3d}2+\ii\wt{\nu}'-\ii\wt{\nu}_1+\ii\wt{\nu}_2\big)\big]\sin\big[\frac\pi2\big(\frac d2-\ii\wt{\nu}'+\ii\wt{\nu}_1+\ii\wt{\nu}_2)\big]}{\sin(-\pi\ii\wt{\nu}_1)\sin\big[\pi\big(\frac d2-\ii\wt{\nu}'\big)\big]}\n \\
    &\times\sin\big[\tfrac\pi2\big(\tfrac d2-\ii\wt{\nu}'-\ii\wt{\nu}_1+\ii\wt{\nu}_2)\big]\sin\big[\tfrac\pi2\big(\tfrac d2+\ii\wt{\nu}'+\ii\wt{\nu}_1+\ii\wt{\nu}_2)\big]\n \\
    &+\frac{\cos\big[\pi\big(\frac d2+\ii\wt{\nu}_1\big)\big]\sin\big[\frac\pi2\big(-\frac{3d}2+\ii\wt{\nu}'+\ii\wt{\nu}_1+\ii\wt{\nu}_2\big)\big]\sin\big[\frac\pi2\big(\frac d2-\ii\wt{\nu}'-\ii\wt{\nu}_1+\ii\wt{\nu}_2)\big]}{\sin(\pi\ii\wt{\nu}_1)\sin\big[\pi\big(\frac d2-\ii\wt{\nu}'\big)\big]}\n \\
    &\times\sin\big[\tfrac\pi2\big(\tfrac d2-\ii\wt{\nu}'+\ii\wt{\nu}_1+\ii\wt{\nu}_2)\big]\sin\big[\tfrac\pi2\big(\tfrac d2+\ii\wt{\nu}'-\ii\wt{\nu}_1+\ii\wt{\nu}_2)\big]\Bigg)-(\wt{\nu}'\rightarrow-\wt{\nu'})\Bigg]\n \\
    &\times\Gamma\Bigg[\begin{matrix}
        2-d, -1+d \\
        \frac d2, \frac d2+\ii\wt{\nu}', \frac d2-\ii\wt{\nu}'
    \end{matrix}\Bigg]\prod_{\pm,\pm,\pm}\Gamma\bigg(\frac{\frac d2\pm\ii\wt{\nu}_1\pm\ii\wt{\nu}_2\pm\ii\wt{\nu}'}2\bigg)\Bigg\}+(\wt{\nu}_1\leftrightarrow\wt{\nu}_2)\n \\
    =&\Bigg\{\frac1{16\pi^{d/2+3}}\Bigg[\frac{\sin\big[\frac\pi2\big(\frac d2-\ii\wt{\nu}'+\ii\wt{\nu}_1+\ii\wt{\nu}_2)\big]\sin\big[\tfrac\pi2\big(\tfrac d2-\ii\wt{\nu}'-\ii\wt{\nu}_1+\ii\wt{\nu}_2)\big]}{\sin(-\pi\ii\wt{\nu}_1)\sin\big[\pi\big(\frac d2-\ii\wt{\nu}'\big)\big]}\n \\
    &\times\Big(\cos\big[\pi\big(\tfrac d2-\ii\wt{\nu}_1\big)\big]\sin\big[\tfrac\pi2\big(-\tfrac{3d}2+\ii\wt{\nu}'-\ii\wt{\nu}_1+\ii\wt{\nu}_2\big)\big]\sin\big[\tfrac\pi2\big(\tfrac d2+\ii\wt{\nu}'+\ii\wt{\nu}_1+\ii\wt{\nu}_2)\big]\n \\
    &-\cos\big[\pi\big(\tfrac d2+\ii\wt{\nu}_1\big)\big]\sin\big[\tfrac\pi2\big(-\tfrac{3d}2+\ii\wt{\nu}'+\ii\wt{\nu}_1+\ii\wt{\nu}_2\big)\big]\sin\big[\tfrac\pi2\big(\tfrac d2+\ii\wt{\nu}'-\ii\wt{\nu}_1+\ii\wt{\nu}_2)\big]\Big)\Bigg]\n \\
    &\times\Gamma\Bigg[\begin{matrix}
        2-d, -1+d \\
        \frac d2, \frac d2+\ii\wt{\nu}', \frac d2-\ii\wt{\nu}'
    \end{matrix}\Bigg]\prod_{\pm,\pm,\pm}\Gamma\bigg(\frac{\frac d2\pm\ii\wt{\nu}_1\pm\ii\wt{\nu}_2\pm\ii\wt{\nu}'}2\bigg)-(\wt\nu'\to-\wt\nu')\Bigg\}+(\wt{\nu}_1\leftrightarrow\wt{\nu}_2)\n \\
    =&\Bigg\{\frac1{32\pi^{d/2+3}}\Bigg[\frac{\sin\big[\frac\pi2\big(\frac d2-\ii\wt{\nu}'+\ii\wt{\nu}_1+\ii\wt{\nu}_2)\big]\sin\big[\tfrac\pi2\big(\tfrac d2-\ii\wt{\nu}'-\ii\wt{\nu}_1+\ii\wt{\nu}_2)\big]}{\sin(-\pi\ii\wt{\nu}_1)\sin\big[\pi\big(\frac d2-\ii\wt{\nu}'\big)\big]}\n \\
    &\times\Big[\cos\big[\pi\big(\tfrac d2-\ii\wt{\nu}_1\big)\big]\Big(\cos\big[\pi\big(d+\ii\wt{\nu}_1\big)\big]-\cos\big[\pi\big(-\tfrac d2+\ii\wt{\nu}'+\ii\wt{\nu}_2\big)\big]\Big)\n \\
    &-\cos\big[\pi\big(\tfrac d2+\ii\wt{\nu}_1\big)\big]\Big(\cos\big[\pi\big(d-\ii\wt{\nu}_1\big)\big]-\cos\big[\pi\big(-\tfrac d2+\ii\wt{\nu}'+\ii\wt{\nu}_2\big)\big]\Big)\Big]\Bigg]\n \\
    &\times\Gamma\Bigg[\begin{matrix}
        2-d, -1+d \\
        \frac d2, \frac d2+\ii\wt{\nu}', \frac d2-\ii\wt{\nu}'
    \end{matrix}\Bigg]\prod_{\pm,\pm,\pm}\Gamma\bigg(\frac{\frac d2\pm\ii\wt{\nu}_1\pm\ii\wt{\nu}_2\pm\ii\wt{\nu}'}2\bigg)-(\wt\nu'\to-\wt\nu')\Bigg\}+(\wt{\nu}_1\leftrightarrow\wt{\nu}_2)\n \\
    =&\Bigg\{\frac1{32\pi^{d/2+3}}\Bigg[\frac{\sin\big[\frac\pi2\big(\frac d2-\ii\wt{\nu}'+\ii\wt{\nu}_1+\ii\wt{\nu}_2)\big]\sin\big[\tfrac\pi2\big(\tfrac d2-\ii\wt{\nu}'-\ii\wt{\nu}_1+\ii\wt{\nu}_2)\big]}{\sin(-\pi\ii\wt{\nu}_1)\sin\big[\pi\big(\frac d2-\ii\wt{\nu}'\big)\big]}\n \\
    &\times\Big[\cos\big[\pi\big(\tfrac d2-\ii\wt{\nu}_1\big)\big]\Big(\cos\big(\tfrac{\pi d}2\big)\cos\big[\pi\big(\tfrac d2+\ii\wt{\nu}_1\big)\big]-\sin\big(\tfrac{\pi d}2\big)\sin\big[\pi\big(\tfrac d2+\ii\wt{\nu}_1\big)\big]-\cos\big[\pi\big(-\tfrac d2+\ii\wt{\nu}'+\ii\wt{\nu}_2\big)\big]\Big)\n \\
    &-\cos\big[\pi\big(\tfrac d2+\ii\wt{\nu}_1\big)\big]\Big(\cos\big(\tfrac{\pi d}2\big)\cos\big[\pi\big(\tfrac d2-\ii\wt{\nu}_1\big)\big]-\sin\big(\tfrac{\pi d}2\big)\sin\big[\pi\big(\tfrac d2-\ii\wt{\nu}_1\big)\big]-\cos\big[\pi\big(-\tfrac d2+\ii\wt{\nu}'+\ii\wt{\nu}_2\big)\big]\Big)\Big]\Bigg]\n \\
    &\times\Gamma\Bigg[\begin{matrix}
        2-d, -1+d \\
        \frac d2, \frac d2+\ii\wt{\nu}', \frac d2-\ii\wt{\nu}'
    \end{matrix}\Bigg]\prod_{\pm,\pm,\pm}\Gamma\bigg(\frac{\frac d2\pm\ii\wt{\nu}_1\pm\ii\wt{\nu}_2\pm\ii\wt{\nu}'}2\bigg)-(\wt\nu'\to-\wt\nu')\Bigg\}+(\wt{\nu}_1\leftrightarrow\wt{\nu}_2)\n \\
    =&\Bigg\{\frac1{16\pi^{d/2+3}}\Bigg[\frac{\sin\big[\frac\pi2\big(\frac d2-\ii\wt{\nu}'+\ii\wt{\nu}_1+\ii\wt{\nu}_2)\big]\sin\big[\tfrac\pi2\big(\tfrac d2-\ii\wt{\nu}'-\ii\wt{\nu}_1+\ii\wt{\nu}_2)\big]}{\sin\big[\pi\big(\frac d2-\ii\wt{\nu}'\big)\big]}\n \\
    &\times\sin\big(\tfrac{\pi d}2\big)\Big(\cos(\pi\ii\wt{\nu}_1)+\cos\big[\pi\big(-\tfrac d2+\ii\wt{\nu}'+\ii\wt{\nu}_2\big)\big]\Big)\n \\
    &\times\Gamma\Bigg[\begin{matrix}
        2-d, -1+d \\
        \frac d2, \frac d2+\ii\wt{\nu}', \frac d2-\ii\wt{\nu}'
    \end{matrix}\Bigg]\prod_{\pm,\pm,\pm}\Gamma\bigg(\frac{\frac d2\pm\ii\wt{\nu}\pm\ii\wt{\nu}\pm\ii\wt{\nu}'}2\bigg)-(\wt\nu'\to-\wt\nu')\Bigg\}+(\wt{\nu}_1\leftrightarrow\wt{\nu}_2)\n \\
    =&\Bigg\{\frac1{32\pi^{d/2+3}}\frac{\sin\big(\tfrac{\pi d}2\big)\Big(\cos(\pi\ii\wt{\nu}_1)-\cos\big[\pi\big(\tfrac d2-\ii\wt{\nu}'+\ii\wt{\nu}_2\big)\big]\Big)\Big(\cos(\pi\ii\wt{\nu}_1)+\cos\big[\pi\big(-\tfrac d2+\ii\wt{\nu}'+\ii\wt{\nu}_2\big)\big]\Big)}{\sin\big[\pi\big(\frac d2-\ii\wt{\nu}'\big)\big]}\n \\
    &\times\Gamma\Bigg[\begin{matrix}
        2-d, -1+d \\
        \frac d2, \frac d2+\ii\wt{\nu}', \frac d2-\ii\wt{\nu}'
    \end{matrix}\Bigg]\prod_{\pm,\pm,\pm}\Gamma\bigg(\frac{\frac d2\pm\ii\wt{\nu}_1\pm\ii\wt{\nu}_2\pm\ii\wt{\nu}'}2\bigg)-(\wt\nu'\to-\wt\nu')\Bigg\}+(\wt{\nu}_1\leftrightarrow\wt{\nu}_2)\n \\
    =&\Bigg\{\frac1{32\pi^{d/2+3}}\frac{\sin\big(\tfrac{\pi d}2\big)\Big(\cos^2(\pi\ii\wt{\nu}_1)+2\cos(\pi\ii\wt{\nu}_1)\sin\big[\pi\big(\frac d2-\ii\wt{\nu}'\big)\big]\sin(\pi\ii\wt{\nu}_2)-\cos^2(\pi\ii\wt{\nu}_2)+\sin^2\big[\pi\big(\frac d2-\ii\wt{\nu}'\big)\big]\Big)}{\sin\big[\pi\big(\frac d2-\ii\wt{\nu}'\big)\big]}\n \\
    &\times\Gamma\Bigg[\begin{matrix}
        2-d, -1+d \\
        \frac d2, \frac d2+\ii\wt{\nu}', \frac d2-\ii\wt{\nu}'
    \end{matrix}\Bigg]\prod_{\pm,\pm,\pm}\Gamma\bigg(\frac{\frac d2\pm\ii\wt{\nu}_1\pm\ii\wt{\nu}_2\pm\ii\wt{\nu}'}2\bigg)-(\wt\nu'\to-\wt\nu')\Bigg\}+(\wt{\nu}_1\leftrightarrow\wt{\nu}_2)\n \\
    =&\frac1{16\pi^{d/2+3}}\sin\big(\tfrac{\pi d}2\big)\Big(\sin\big[\pi\big(\tfrac d2-\ii\wt{\nu}'\big)\big]-\sin\big[\pi\big(\tfrac d2+\ii\wt{\nu}'\big)\big]\Big)\Gamma\Bigg[\begin{matrix}
        2-d, -1+d \\
        \frac d2, \frac d2+\ii\wt{\nu}', \frac d2-\ii\wt{\nu}'
    \end{matrix}\Bigg]\prod_{\pm,\pm,\pm}\Gamma\bigg(\frac{\frac d2\pm\ii\wt{\nu}_1\pm\ii\wt{\nu}_2\pm\ii\wt{\nu}'}2\bigg)\n \\
    =&-\frac1{8\pi^{d/2+3}}\sin\big(\tfrac{\pi d}2\big)\cos\big(\tfrac{\pi d}2\big)\sin(\pi\ii\wt{\nu}')\Gamma\Bigg[\begin{matrix}
        2-d, -1+d \\
        \frac d2, \frac d2+\ii\wt{\nu}', \frac d2-\ii\wt{\nu}'
    \end{matrix}\Bigg]\prod_{\pm,\pm,\pm}\Gamma\bigg(\frac{\frac d2\pm\ii\wt{\nu}_1\pm\ii\wt{\nu}_2\pm\ii\wt{\nu}'}2\bigg)\n \\
    =&\frac1{16\pi^{d/2+2}}\sin(\pi\ii\wt{\nu}')\frac1{\Gamma\big[\frac d2, \frac d2+\ii\wt{\nu}', \frac d2-\ii\wt{\nu}'\big]}\prod_{\pm,\pm,\pm}\Gamma\bigg(\frac{\frac d2\pm\ii\wt{\nu}_1\pm\ii\wt{\nu}_2\pm\ii\wt{\nu}'}2\bigg).
\end{align}
Compare the above formula with $\rho_{\sigma_1\sigma_2}^{\mathcal{P},0}(\wt{\nu}')$ \cite{Loparco:2023rug}
\begin{equation}
    \rho_{\sigma_1\sigma_2}^{\mathcal{P},0}(\wt{\nu}')=\frac{\wt{\nu}'\sinh(\pi\wt{\nu}')\Gamma\big(\frac{d+1}2\big)}{2^{6-d}\pi^{(d+7)/2}\Gamma\big[d, \frac d2+\ii\wt{\nu}', \frac d2-\ii\wt{\nu}'\big]}\prod_{\pm,\pm,\pm}\Gamma\bigg(\frac{\frac d2\pm\ii\wt{\nu}_1\pm\ii\wt{\nu}_2\pm\ii\wt{\nu}'}2\bigg),
\end{equation}
it can be derived that
\begin{equation}
    \rho_{\sigma_1\sigma_2}^{\mathcal{P},0}(\wt{\nu}')=\frac{\wt{\nu}'}{2\pi\ii}\big(\rho_{\wt{\nu}_1,\wt{\nu}_2}^\text{dS}(\wt{\nu}')-\rho_{\wt{\nu}_1,\wt{\nu}_2}^\text{dS}(-\wt{\nu}')\big).
\end{equation}

In the flat space limit $H\rightarrow0$, the spectral density $\rho_{\sigma\sigma}^{\mathcal{P},0}(\wt{\nu}')$ appearing in the K\"all\'en-Lehmann decomposition for $d=3$ becomes \cite{Cespedes:2025dnq}
\begin{equation}
    \rho_\sigma(m'^2)^{\text{mink}}=\lim_{H\rightarrow0}\frac1{\wt{\nu}'}\rho_{\sigma\sigma}^{\mathcal{P},0}(\wt{\nu}')=\frac1{16\pi^2}\bigg(1-\frac{4m^2}{m'^2}\bigg)^{\frac12}\theta(m'^2-4m^2),
\end{equation}
where $m$ and $m'$ are masses corresponding to $\wt{\nu}$ and $\wt{\nu}'$.
\section{Convergence of Series of 1-loop 2-point Correlators}\label{AppH}
In this appendix, the convergence of series of 1-loop 2-point correlators is proved. First, consider the double series in the expression of $\wh{\mathcal{J}}_{\sigma\sigma,(\text{A})}^{p_1p_2}(1,1)$ shown in (\ref{61}):
\begin{align}\label{301}
    &\sum_{n=0}^\infty\sum_{j=1}^\infty\frac{\big(\frac32-2\ii\wt{\nu}+2n\big)\cos^2(2\pi\ii\wt{\nu})\Gamma(p_{12}+5+j)\big(2-2\ii\wt{\nu}+2n\big)_j}{n!(-2\ii\wt{\nu}+p_1+4+2n)_{j+1}(-2\ii\wt{\nu}+p_2+4+2n)_{j+1}}\n \\
    &\times\frac{\mathcal{C}_{\frac32-2\ii\wt{\nu}+2n,d}^{p_1p_2}}{2^{p_{12}+6}\pi^2\sin^2(\pi\ii\wt{\nu})}\Gamma\Bigg[\begin{matrix}
        \frac32+n, -\ii\wt{\nu}+\frac32+n, -2\ii\wt{\nu}+\frac32+n, -\ii\wt{\nu}+\frac12+n \\
        -\ii\wt{\nu}+2+n, -2\ii\wt{\nu}+1+n, -\ii\wt{\nu}+1+n
    \end{matrix}\Bigg]\n \\
    \equiv&\sum_{n=0}^\infty\sum_{j=1}^\infty\frac{\cos^2(2\pi\ii\wt{\nu})\mathcal{C}_{\frac32-2\ii\wt{\nu}+2n,d}^{p_1p_2}}{2^{p_{12}+6}\pi^2\sin^2(\pi\ii\wt{\nu})}\mathcal{A}_{\text{A},n,j}.
\end{align}
Each row-series and each column-series corresponding to $\sum_{n=0}^\infty\sum_{j=1}^\infty|\mathcal{A}_{\text{A},n,j}|$ are convergent. Then consider the ratios $\mathcal{A}_{\text{A},n,j+1}/\mathcal{A}_{\text{A},n+1,j}$ and $\mathcal{A}_{\text{A},n+1,j}/\mathcal{A}_{\text{A},n,j}$:
\begin{align}
    \frac{\mathcal{A}_{\text{A},n,j+1}}{\mathcal{A}_{\text{A},n,j}}=&\frac{(p_{12}+5+j)(2-2\ii\wt{\nu}+2n+j)}{(-2\ii\wt{\nu}+p_1+5+2n+j)(-2\ii\wt{\nu}+p_2+5+2n+j)}\n \\
    =&1-\frac{(-2\ii\wt{\nu}+3+2n)j+\mathcal{O}(j^0)}{(-2\ii\wt{\nu}+p_1+5+2n+j)(-2\ii\wt{\nu}+p_2+5+2n+j)},
\end{align}
\begin{align}
    \frac{\mathcal{A}_{\text{A},n+1,j}}{\mathcal{A}_{\text{A},n,j}}=&\frac{\big(\frac72-2\ii\wt{\nu}+2n\big)(2-2\ii\wt{\nu}+2n+j)_2n(-2\ii\wt{\nu}+p_1+4+2n)_2}{\big(\frac32-2\ii\wt{\nu}+2n\big)(2-2\ii\wt{\nu}+2n)_2(n+1)(-2\ii\wt{\nu}+p_1+5+2n+j)_2}\n \\
    &\times\frac{(-2\ii\wt{\nu}+p_2+4+2n)_2\big(\frac52+n\big)\big(-\ii\wt{\nu}+\frac52+n\big)\big(-2\ii\wt{\nu}+\frac52+n\big)}{(-2\ii\wt{\nu}+p_2+5+2n+j)_2\big(\frac32+n\big)\big(-\ii\wt{\nu}+\frac12+n\big)\big(-2\ii\wt{\nu}+\frac32+n\big)}\n \\
    &\times\frac{\big(-\ii\wt{\nu}+1+n\big)\big(-2\ii\wt{\nu}+1+n\big)}{\big(-\ii\wt{\nu}+3+n\big)\big(-2\ii\wt{\nu}+2+n\big)}\n \\
    =&1-128(j+1)n^{12}/\Big(\big(\tfrac32-2\ii\wt{\nu}+2n\big)(2-2\ii\wt{\nu}+2n)_2(n+1)(-2\ii\wt{\nu}+p_1+5+2n+j)_2\n \\
    &\times(-2\ii\wt{\nu}+p_2+5+2n+j)_2\big(\tfrac32+n\big)\big(-\ii\wt{\nu}+\tfrac12+n\big)\big(-2\ii\wt{\nu}+\tfrac32+n\big)\n \\
    &\times\big(-\ii\wt{\nu}+3+n\big)\big(-2\ii\wt{\nu}+2+n\big)\Big).
\end{align}
Therefore, for large $j$ and $n$, the ratios are
\begin{equation}
    \frac{\mathcal{A}_{\text{A},n,j+1}}{\mathcal{A}_{\text{A},n+1,j}}\sim1-\frac{-2\ii\wt{\nu}+3+2n}{j},
\end{equation}
\begin{equation}
    \frac{\mathcal{A}_{\text{A},n+1,j}}{\mathcal{A}_{\text{A},n,j}}\sim1-\frac{j+1}n.
\end{equation}
According to Raabe's test for double series \cite{book}, the double series (\ref{301}) is convergent if $\text{Re}(-2\ii\wt{\nu}+3)>1$. The condition satisfies if $\wt{\nu}$ is real, or $\wt{\nu}$ is complex and $\ii\wt{\nu}<1$. If $\ii\wt{\nu}=1$, the summation of $n$ starts at $n=1$ and it is convergent. The divergence which appears for $1<\ii\wt{\nu}<\frac32$ is from the divergence of the 2-point correlator of tree level (See App. \ref{AppC} for details). However, the divergence of the correlator is merely a flaw in the expression. The expression of $\wh{\mathcal{J}}_{\sigma\sigma,(\text{A})}^{p_1p_2}(1,1)$ can be replaced by the following formula:
\begin{align}
    \wh{\mathcal{J}}_{\sigma\sigma,(\text{A})}^{p_1p_2}(1,1)=&\sum_{n=1}^\infty\sum_{j=1}^\infty\frac{\big(\frac32-2\ii\wt{\nu}+2n\big)\cos^2(2\pi\ii\wt{\nu})\Gamma(p_{12}+5+j)\big(2-2\ii\wt{\nu}+2n\big)_j}{n!(-2\ii\wt{\nu}+p_1+4+2n)_{j+1}(-2\ii\wt{\nu}+p_2+4+2n)_{j+1}}\n \\
    &\times\frac{\mathcal{C}_{\frac32-2\ii\wt{\nu}+2n,d}^{p_1p_2}}{2^{p_{12}+6}\pi^2\sin^2(\pi\ii\wt{\nu})}\Gamma\Bigg[\begin{matrix}
        \frac32+n, -\ii\wt{\nu}+\frac32+n, -2\ii\wt{\nu}+\frac32+n, -\ii\wt{\nu}+\frac12+n \\
        -\ii\wt{\nu}+2+n, -2\ii\wt{\nu}+1+n, -\ii\wt{\nu}+1+n
    \end{matrix}\Bigg]\n \\
    &+\frac{\mathcal{C}_{\frac32-2\ii\wt{\nu},d}^{p_1p_2}\cos^2(2\pi\ii\wt{\nu})\big(\frac32-2\ii\wt{\nu}\big)}{2^{p_{12}+6}\pi^2\sin^2(\pi\ii\wt{\nu})}\Gamma\Bigg[\begin{matrix}
        \frac32, -\ii\wt{\nu}+\frac32, -2\ii\wt{\nu}+\frac32, -\ii\wt{\nu}+\frac12 \\
        -\ii\wt{\nu}+2, -2\ii\wt{\nu}+1, -\ii\wt{\nu}+1
    \end{matrix}\Bigg]\n \\
    &\times\Bigg[\sum_{j=1}^\infty\Bigg(\frac{\Gamma(p_{12}+5+j)\big(2-2\ii\wt{\nu}\big)_j}{(-2\ii\wt{\nu}+p_1+4)_{j+1}(-2\ii\wt{\nu}+p_2+4)_{j+1}}\n \\
    &-\Gamma\bigg[\begin{matrix}
        -2\ii\wt{\nu}+p_1+4, -2\ii\wt{\nu}+p_2+4 \\
        2-2\ii\wt{\nu}
    \end{matrix}\bigg]j^{2\ii\wt{\nu}-3}\Bigg)\n \\
    &+\Gamma\bigg[\begin{matrix}
        -2\ii\wt{\nu}+p_1+4, -2\ii\wt{\nu}+p_2+4 \\
        2-2\ii\wt{\nu}
    \end{matrix}\bigg]\zeta(-2\ii\wt{\nu}+3)\Bigg]\n \\
    &+\frac{\mathcal{C}_{\frac32-2\ii\wt{\nu},d}^{p_1p_2}\cos^2(2\pi\ii\wt{\nu})\Gamma(p_{12}+5)}{2^{p_{12}+6}\pi^2\sin^2(\pi\ii\wt{\nu})}\Bigg[\sum_{n=0}^\infty\Bigg(\frac{\big(\frac32-2\ii\wt{\nu}+2n\big)}{(-2\ii\wt{\nu}+p_1+4+2n)(-2\ii\wt{\nu}+p_2+4+2n)}\n \\
    &\times\Gamma\Bigg[\begin{matrix}
        \frac32+n, -\ii\wt{\nu}+\frac32+n, -2\ii\wt{\nu}+\frac32+n, -\ii\wt{\nu}+\frac12+n \\
        1+n, -\ii\wt{\nu}+2+n, -2\ii\wt{\nu}+1+n, -\ii\wt{\nu}+1+n
    \end{matrix}\Bigg]-\frac1{2(n+1)}\Bigg)\n \\
    &+\frac12\bigg(\gamma_E+\log\mu_R+\tfrac12\log4\pi+\psi\big(\tfrac32\big)-\frac{\pi\sin\big[\pi\big(-2\ii\wt{\nu}+\tfrac{p_{12}}2\big)\big]}{16\cos^2(2\pi\ii\wt{\nu})\mathcal{C}_{\frac32-2\ii\wt{\nu},3}^{p_1p_2}}\bigg)\Bigg].
\end{align}

Next, consider the following series:
\begin{align}\label{307}
    &\sum_{n=0}^\infty\Bigg(\frac{\big(\frac32-2\ii\wt{\nu}+2n\big)}{(-2\ii\wt{\nu}+p_1+4+2n)(-2\ii\wt{\nu}+p_2+4+2n)}\n \\
    &\times\Gamma\Bigg[\begin{matrix}
        \frac32+n, -\ii\wt{\nu}+\frac32+n, -2\ii\wt{\nu}+\frac32+n, -\ii\wt{\nu}+\frac12+n \\
        1+n, -\ii\wt{\nu}+2+n, -2\ii\wt{\nu}+1+n, -\ii\wt{\nu}+1+n
    \end{matrix}\Bigg]-\frac1{2(n+1)}\Bigg)\n \\
    \equiv&\sum_{n=0}^\infty\mathcal{B}_{\text{A},n}.
\end{align}
For large $n$, the ratio $\mathcal{B}_{\text{A},n+1}/\mathcal{B}_{\text{A},n}$ becomes
\begin{align}
    \frac{\mathcal{B}_{\text{A},n+1}}{\mathcal{B}_{\text{A},n}}\sim&\bigg(\frac{-9-2p_{12}+4\ii\wt{\nu}}{8n^2}+\frac{101+p_{12}(29-8\ii\wt{\nu})+2p_{12}^2-2p_1p_2-68\ii\wt{\nu}-4\wt{\nu}^2}{16n^3}+\mathcal{O}(n^{-4})\bigg)\n \\
    &\bigg/\bigg(\frac{-9-2p_{12}+4\ii\wt{\nu}}{8n^2}+\frac{65+p_{12}(21-8\ii\wt{\nu})+2p_{12}^2-2p_1p_2-52\ii\wt{\nu}-4\wt{\nu}^2}{16n^3}+\mathcal{O}(n^{-4})\bigg)\n \\
    \sim&1-\frac2n+\mathcal{O}(n^{-2}).
\end{align}
According to Raabe's test, the series (\ref{307}) is convergent. In summary, the expression of $\wh{\mathcal{J}}_{\sigma\sigma,(\text{A})}^{p_1p_2}(1,1)$ is convergent, or it can be written in a convergent way. The convergence of $\wh{\mathcal{J}}_{\sigma\sigma,(\text{B})}^{p_1p_2}(1,1)$ and $\wh{\mathcal{J}}_{\sigma\sigma,(\text{C})}^{p_1p_2}(1,1)$ can be proved in the same way.

\end{appendix}

\newpage
\bibliography{ref} 
\bibliographystyle{utphys}

\end{document}